\documentclass[12pt]{article}
\pdfoutput=1
\usepackage{amssymb,amsmath,dsfont,mathrsfs}
\usepackage{mathtools}
\usepackage{array}
\usepackage{longtable}
\usepackage{physics}
\usepackage{slashed}
\usepackage[normalem]{ulem}
\usepackage{epsfig}
\usepackage{epstopdf}
\usepackage{latexsym}
\usepackage{tikz}
\usepackage{tikz-cd, graphicx}
\usetikzlibrary{calc, shapes.geometric, positioning, arrows.meta, decorations.pathreplacing}
\usepackage{graphicx}
\usepackage{booktabs}
\usepackage{bbm}
\usepackage{xspace}
\usepackage{cancel}
\usepackage[utf8]{inputenc}
\usepackage{csquotes}
\usepackage{multirow}
\usepackage{enumitem}
\usepackage{braket}
\usepackage[numbers,compress]{natbib}
\usepackage[T1]{fontenc}
\usepackage[margin=20pt,small]{caption}
\usepackage[toc]{appendix}
\usepackage{color}
\usepackage{float}
\usepackage{yfonts}
\usepackage{datetime}
\usepackage[normalem]{ulem}
\usepackage[
colorlinks=false,
linkcolor=darkblue,  
urlcolor=blue,    
filecolor=blue,     
citecolor=red,
linktocpage=true,
pdfstartview=FitV,
bookmarksopen=true,
pdfencoding=auto,
pagebackref]{hyperref}
\numberwithin{equation}{section}
\usepackage{anyfontsize}
\usepackage{subcaption}
\usepackage{diagbox}

\usepackage[scr=dutchcal]{mathalpha}

\definecolor{cardinal}{rgb}{0.6,0,0}
\definecolor{darkgreen}{rgb}{0,0.5,0}
\definecolor{golden}{rgb}{0.92, 0.7, 0}
\definecolor{midnight}{rgb}{0, 0, 0.5}
\definecolor{darkblue}{rgb}{0.2, 0, 0.8}

\usepackage[most]{tcolorbox}

\begin{document}  

	\begin{titlepage}
\vskip .3in \noindent
		\bigskip
		\begin{center} 
              {\fontsize{22pt}{0pt} \bf Non-invertible SPT phases 
               from fusion rings \\  \vspace{0.5cm} via Tannaka duality}

			\vskip30pt

			{\large Antonio Santaniello, Roberto Valandro, Massimo Zorzenon}
            
			\vskip 20pt
                Dipartimento di Fisica, Universit\`{a} di Trieste, Strada Costiera 11, 34151 Trieste, Italy\\
               
                \vskip 2mm

            	 and INFN, Sezione di Trieste, Via Valerio 2, 34127 Trieste, Italy\\
                    \vskip 7mm

			\texttt{ \scriptsize antonio dot santaniello at phd dot units dot it, \, roberto dot valandro at ts dot infn dot it \\
           massimo dot zorzenon at phd dot units dot it}\\
		\end{center}
		
		\bigskip
		\bigskip

\begin{abstract} 

In a (1+1)-dimensional QFT with non-invertible symmetry, the fusion rules of topological lines are often the most accessible data, yet they do not fully determine whether the symmetry is anomaly-free, i.e. whether it admits trivially gapped phases. Such phases are characterized by fiber functors of the corresponding fusion category. Building on Tannaka duality, we extract fiber functors from the fusion ring alone by solving simultaneously for monoidal structure maps and tensor automorphisms. This allows us to find categorifications admitting a fiber functor without solving the pentagon equations directly. We test this procedure on the known Tambara--Yamagami ring of \(\mathbb{Z}_2\times\mathbb{Z}_2\). Our main result concerns the rank-$11$ fusion ring shared by
the representation categories of the two extraspecial groups of order $27$, which has a multiplicity $3$ channel and is relevant
to self-triality. In both cases we find more than one fiber functor, i.e. several non-invertible SPT phases. We use the explicit data of these fiber functors to derive the \(F\)-symbols of the two representation categories in a compact form depending on a nondegenerate alternating bicharacter and an invertible object \(\epsilon\). Using the reconstructed data, we compute the anomalous algebra on the interface between two different SPT phases.

\end{abstract}

	\end{titlepage}

	\setcounter{tocdepth}{2}	

\tableofcontents

\section{Introduction}\label{sec:introduction}

Discrete internal bosonic symmetries of two-dimensional quantum field
theories (QFTs) are implemented by topological line defects and are
mathematically described by fusion categories
\cite{Frohlich:2006ch,Gaiotto:2014kfa,Bhardwaj:2017xup}. The fusion of simple
lines is encoded by the fusion ring $\mathcal K(\mathcal C)$ of the symmetry
category $\mathcal C$. A symmetry is non-invertible when at least one simple
line has no fusion inverse; otherwise it is an ordinary, possibly anomalous,
group-like symmetry.
The fusion ring does not specify the symmetry completely. A fusion category
also includes $F$-symbols, which encode the associativity of line fusion.
Inequivalent choices of $F$-symbols, modulo changes of basis and automorphisms of the fusion ring, therefore define distinct categorifications of the same fusion ring.

A symmetry category $\mathcal{C}$ is called anomaly-free if it admits a monoidal fiber functor \cite{Thorngren:2019iar,Zhang:2023wlu,Choi:2023xjw}
$$
    \mathcal{F}:\mathcal{C} \rightarrow \text{Vec} \ .
$$
The existence of this functor depends both on the fusion ring $\mathcal{K}(\mathcal{C})$ and on a compatibility condition between the $F$-symbols of $\mathcal{C}$ and the monoidal structure of the functor $\mathcal{F}$. Every fusion category admitting a fiber functor is equivalent to
$\operatorname{Rep}(H)$ for a finite-dimensional semisimple Hopf algebra
$H$~\cite{etingof2015tensor}.  When $H$ is noncommutative,
$\operatorname{Rep}(H)$ contains non-invertible simple objects.

In $1+1$ dimensions, a monoidal fiber functor of $\mathcal C$ corresponds to a
$\mathcal C$-symmetric gapped phase with a unique bulk vacuum
\cite{Thorngren:2019iar,Inamura:2021wuo}. Conversely, the absence of a fiber functor is an
obstruction to the existence of such a phase. A given symmetry may admit
several inequivalent symmetric gapped phases that cannot be connected by a
symmetry-preserving deformation without crossing a phase transition. These
are symmetry-protected topological (SPT) phases \cite{PhysRevLett.50.1153,PhysRevLett.59.799,Pollmann_2012,Chen:2010gda,Chen:2011pg,Schuch:2011niz}. 
For a unitary, bosonic symmetry group
$G$, two-dimensional $G$-SPT phases are classified by the second group cohomology
$H^2(G,U(1))$~\cite{Chen:2011pg}. For a non-invertible symmetry, this
classification is generalized by fiber functors: the monoidally inequivalent
fiber functors on $\mathcal C$ label the distinct $\mathcal C$-symmetric SPT
phases.  
Explicit realizations of
non-invertible SPT phases have been constructed in lattice models based on
cluster states, anyon chains, and tensor networks \cite{Bhardwaj:2024kvy,Seifnashri:2024dsd,Jia:2024bng,Inamura:2024jke,Meng:2024nxx,Li:2025wes,Warman:2024lir}. In many of these constructions, the monoidal
structure maps of the fiber functor are crucial for explicitly constructing the
corresponding $\mathcal C$-symmetric state. 

The relation between fusion rules, associators, and fiber functors leads to
a concrete reconstruction problem: starting from the fusion rules alone, can
one reconstruct the categorifications that admit a fiber functor, including
their associators, and determine their inequivalent fiber functors? In this
work we develop and apply a bootstrap that answers this question and thereby
determines the associated non-invertible
SPT phases for the categories we reconstruct.
The only categorical input is the fusion ring: neither a categorification
$\mathcal C$ nor a representative of its $F$-symbols is assumed. This makes
the bootstrap physically useful: fusion rules can often be obtained directly
from the composition of topological lines, whereas $F$-symbols require access
to their junction data and are typically computed in more controlled
settings, such as rational conformal field theories and free theories
\cite{Chang:2018iay,Thorngren:2021yso,Lu:2022ver,Lu:2024lzf,Damia:2024xju,Kim:2025zdy}.
The procedure is a constructive application of Tannaka duality
\cite{Tannaka1938}.\footnote{For other applications of Tannaka duality in
physics, see also
\cite{Dimofte:2024bwe,AliAhmad:2025bnd,McNamara:2026isz}.}
Standard Tannaka reconstruction starts from a specified fusion category
$\mathcal C$, including its monoidal structure, together with the forgetful fiber
functor, and reconstructs the corresponding finite-dimensional
semisimple Hopf algebra. When
$\mathcal C=\operatorname{Rep}(G)$, the finite group $G$ is recovered as the
group of tensor automorphisms of the forgetful functor. In this setting, the monoidal
structure maps of the forgetful functor are the Clebsch--Gordan maps, or
$3j$-symbols, while the $F$-symbols are the corresponding $6j$-symbols and
can be reconstructed from the monoidal structure maps.

We reverse this logic. Starting from the fusion ring alone, we treat the
monoidal structure maps and their tensor automorphisms as unknowns and solve
the coupled consistency squares. When the square equations leave part of the 
monoidal structure undetermined, we additionally require that the maps admit 
a consistent set of $F$-symbols, whose existence is otherwise guaranteed.
Solving this full system modulo the allowed
changes of basis determines all compatible fiber functors. The resulting monoidal structure maps determine the $F$-symbols,
and Tannaka reconstruction yields the corresponding finite group
or Hopf algebra.

When the tensor automorphisms form a group of the maximal order allowed by the
fusion ring, the solutions reconstruct a category
$\operatorname{Rep}(G)$. For the group representation categories considered below, the
resulting numbers of monoidally inequivalent fiber functors agree with the
classification of~\cite{ostrik2006modulecategoriesdrinfelddouble}, and
we determine explicitly how the different solutions are related by monoidal
equivalences.

We apply the procedure to three fusion rings.
The fusion ring of $\operatorname{Rep}(S_3)$ provides a first check: the
bootstrap reproduces the expected unique solution and reconstructs $S_3$ as
its group of tensor automorphisms. We then consider the Tambara--Yamagami
fusion ring associated with $\mathbb Z_2\times\mathbb Z_2$, which appears
physically in self-duality symmetries
\cite{Tambara:1998vmj,Thorngren:2021yso,Antinucci:2022vyk,Niro:2022ctq,
Antinucci:2023ezl}. The solutions satisfying the maximality condition give
four labeled parametric forms: three reconstruct
$\operatorname{Rep}(D_8)$ and one reconstructs
$\operatorname{Rep}(Q_8)$. At fixed associator, they yield three and one
inequivalent fiber functors, respectively. Relaxing maximality also recovers
$\operatorname{Rep}(H_8)$ and its unique fiber functor, with $H_8$ the
Kac--Paljutkin Hopf algebra. This shows that the method recovers representation categories of both finite groups and semisimple Hopf algebras that are not group algebras.
The first two cases allow us to check the reconstruction against
known results~\cite{Barter_2022,Diatlyk:2023fwf,vercleyen2024lowrankmultiplicityfreefusioncategories,Perez-Lona:2023djo,Perez-Lona:2024sds}.

Our main application is the fusion ring shared by the representation
categories of the two extraspecial groups of order $27$, with exponents $3$
and $9$, respectively. It is a Jordan--Larson fusion ring with eleven simple
objects and multiplicity-three fusion channels, relevant to self-triality
symmetries \cite{Jordan_2009,Lu:2024ytl,Maeda:2025rxc,Kaidi:2025hyr}.
For this ring, we restrict the reconstruction to the maximal
tensor-automorphism case; categorifications arising in the non-maximal
case are left to future work. The consistency squares produce eighteen
labeled parametric forms. From their reconstructed $F$-symbols we extract
two invariants, $\epsilon$ and $\iota$, unchanged by fusion-space basis
transformations. With the simple-object labels fixed, these invariants
distinguish all eighteen forms. Under relabeling, the forms fall into two
orbits, corresponding to the two extraspecial representation categories.
In group-theoretic terms, $\epsilon$ identifies the determinant of a chosen
three-dimensional simple representation, while $\iota$ records its
nontrivial central character.

For a fixed associator, the compatible fiber functors are labeled by
characters $\vartheta\in\widehat{\Gamma}$, where
$\Gamma\simeq\mathbb Z_3^2$ is the group of invertible objects.
The associator selects a distinguished element $\epsilon\in\Gamma$
and imposes the compatibility condition $\vartheta(\epsilon)=1$.
For $\operatorname{Rep}(\mathrm{Esp}_3)$ one has $\epsilon=1$, so all nine
characters are allowed, whereas for
$\operatorname{Rep}(\mathrm{Esp}_9)$ the element $\epsilon$ is nontrivial
and only three characters satisfy the selection rule. We obtain uniform expressions for the complete $F$-symbols in terms of
$(\epsilon,\iota)$, including all multiplicity-three blocks.
\footnote{Another example of explicit associator data with fusion
multiplicities is the rank-four Haagerup category $\mathcal H_1$,
whose $F$-symbols were obtained through Morita equivalence and the
module tube category, starting from a known fusion category and
module category~\cite{Barter_2022}.}
We check the pentagon identities exactly for all eighteen labeled forms.
For $\epsilon=1$, our associator is gauge equivalent to the triality
solution of Ref.~\cite{Kaidi:2025hyr}, while for $\epsilon\neq1$ we obtain
new explicit associator data for the representation category of the
extraspecial group of exponent $9$.

We use the reconstructed monoidal structure maps to compute
the algebras acting on interfaces between the corresponding SPT phases. For interfaces
between a chosen reference phase and any distinct phase, we obtain
$M_3(\mathbb C)^{\oplus3}$ in both cases. The three irreducible sectors
each carry a symmetry-protected triplet. The explicit algebra
representations determine the symmetry action on these interface states,
including their invertible charges.

For the two extraspecial groups of order $p^3$, with $p$ an arbitrary
odd prime, the explicit $p=3$ reconstruction suggests that the
maximal branch contains $p^2(p-1)$ labeled parametric forms,
classified by $(\epsilon,\iota)\in\Gamma_p\times\mathbb F_p^*$.
Independently, the group-theoretical classification establishes
$p^2$ and $p$ inequivalent fiber functors for the representation
categories of the groups of exponents $p$ and $p^2$, respectively.
The classification of parametric forms and the proposed
fixed-associator parametrization by characters satisfying
$\vartheta(\epsilon)=1$ remain conjectural for $p>3$.

The paper is organized as follows. In Section~\ref{sec:FusCat} we review the
required notions of fusion categories, fiber functors, and standard Tannaka
reconstruction. In Section~\ref{sec:Bootstrap} we formulate the bootstrap
procedure and illustrate it using the fusion ring of
$\operatorname{Rep}(S_3)$. Section~\ref{sec:TYZ2} treats the
Tambara--Yamagami fusion ring shared by $\operatorname{Rep}(D_8)$ and
$\operatorname{Rep}(Q_8)$ and discusses the additional
$\operatorname{Rep}(H_8)$ solution. In Section~\ref{sec:extraspecial} we
reconstruct the fiber functors and $F$-symbols associated with the two
extraspecial groups of order $27$. 
Section~\ref{sec:physics} reviews the physical interpretation of the
reconstructed data and computes the interface algebras and protected
multiplets for the extraspecial SPT phases.
Section~\ref{sec:conclusions} presents our conclusions and outlook.
Appendix~\ref{app:Tambara-comparison} relates our
Tambara--Yamagami solutions to Tambara's classification.
Appendix~\ref{app:twists_and_fiber_functors} describes
the realization of fiber functors by twists.
Explicit results for the extraspecial categories are collected in
Appendix~\ref{app:p3-mixed-matrices}, with the complete data
provided in the companion Mathematica notebooks.

\section{Fusion categories, fiber functors and Tannaka duality}\label{sec:FusCat}

In this section we provide a concise overview of the fundamental concepts of category theory that are used in the rest of the paper. 
For a more rigorous exposition of these topics see \cite{etingof2015tensor}.

\subsection{Fusion Categories}

Let \(\mathcal C\) be a fusion category with a finite set of simple objects
\(\mathscr{I}\). Any object of
\(\mathcal C\) can be decomposed as a direct sum
\(
    \bigoplus_{a\in\mathscr{I}} n^a a,
\)
where the \(n^a\) are non-negative integers. The rank of
\(\mathcal C\) is the number~\(|\mathscr{I}|\) of simple objects.

A fusion category is equipped with a tensor product
\begin{equation}\label{eq:product}
    \otimes:\mathcal C\times\mathcal C\longrightarrow\mathcal C.
\end{equation}
The tensor product of two simple objects decomposes as
\begin{equation}\label{eq:fusionring}
    a\otimes b
    =
    \bigoplus_{c\in\mathscr{I}}c^{\oplus N_{ab}^{c}},
\end{equation}
where the non-negative integers \(N_{ab}^{c}\) are the fusion coefficients.

Among the simple objects there is a distinguished tensor unit, denoted by
\(1\), satisfying
\begin{equation}
    1\otimes a=a\otimes1=a.
\end{equation}
Moreover, every simple object \(a\) has a unique dual
\(\overline a \), characterized by
\(
    N_{a\overline a}^{1}
    =
    N_{\overline a a}^{1}
    =
    1.
\)

The fusion coefficients determine the fusion ring
\(\mathcal K(\mathcal C)\).
An automorphism of this ring is a permutation \(\sigma\) of the simple objects satisfying
\begin{equation}
N_{\sigma(a)\sigma(b)}^{\sigma(c)}=N_{ab}^{c}
\qquad\text{for all }a,b,c\in\mathscr{I}\:.
\end{equation}

For each allowed fusion channel
\(a\otimes b\to c\), the corresponding fusion space is a complex vector space
\begin{equation}
    V_{ab}^{c}
    :=
    \operatorname{Hom}(a\otimes b,c)\,,
    \qquad
    \dim_\mathbb{C} V_{ab}^{c}=N_{ab}^{c}\:.
\end{equation}
After choosing a basis, its elements are represented graphically by
\begin{center}
    \tikzset{every picture/.style={line width=0.75pt}} 

\begin{tikzpicture}[x=0.75pt,y=0.75pt,yscale=-1,xscale=1]

\draw    (172.33,46) -- (172.33,80) ;
\draw  [draw opacity=0][fill={rgb, 255:red, 0; green, 0; blue, 0 }  ,fill opacity=1 ] (169.07,81.36) .. controls (169.07,79.43) and (170.54,77.86) .. (172.37,77.86) .. controls (174.19,77.86) and (175.67,79.43) .. (175.67,81.36) .. controls (175.67,83.3) and (174.19,84.86) .. (172.37,84.86) .. controls (170.54,84.86) and (169.07,83.3) .. (169.07,81.36) -- cycle ;
\draw    (171.37,80.36) -- (195.33,111) ;
\draw    (172.37,81.36) -- (148.33,110) ;

\draw (138,113) node [anchor=north west][inner sep=0.75pt]   [align=left] {$\displaystyle a$};
\draw (197.33,113) node [anchor=north west][inner sep=0.75pt]   [align=left] {$\displaystyle b$};
\draw (179.33,29) node [anchor=north west][inner sep=0.75pt]   [align=left] {$\displaystyle c$};
\draw (151.33,64) node [anchor=north west][inner sep=0.75pt]   [align=left] {$\displaystyle \alpha $};
\draw (227,70) node [anchor=north west][inner sep=0.75pt]   [align=left] {$\displaystyle \in \ \text{Hom}( a\otimes b,c) \qquad \mbox{with }\qquad\alpha=1,...,N_{ab}^c \: .$};

\end{tikzpicture}
\end{center}
A different choice of basis in \(V_{ab}^{c}\) is implemented by a matrix
\begin{equation}
    M_{ab}^{c}\in GL(N_{ab}^{c},\mathbb C)
\end{equation}
and acts graphically as
\begin{center}
    \tikzset{every picture/.style={line width=0.75pt}} 

\begin{tikzpicture}[x=0.75pt,y=0.75pt,yscale=-1,xscale=1]

\draw    (115.33,47) -- (115.33,81) ;
\draw  [draw opacity=0][fill={rgb, 255:red, 0; green, 0; blue, 0 }  ,fill opacity=1 ] (112.07,82.36) .. controls (112.07,80.43) and (113.54,78.86) .. (115.37,78.86) .. controls (117.19,78.86) and (118.67,80.43) .. (118.67,82.36) .. controls (118.67,84.3) and (117.19,85.86) .. (115.37,85.86) .. controls (113.54,85.86) and (112.07,84.3) .. (112.07,82.36) -- cycle ;
\draw    (114.37,81.36) -- (138.33,112) ;
\draw    (115.37,82.36) -- (91.33,111) ;
\draw    (331.33,47) -- (331.33,81) ;
\draw  [draw opacity=0][fill={rgb, 255:red, 0; green, 0; blue, 0 }  ,fill opacity=1 ] (328.07,82.36) .. controls (328.07,80.43) and (329.54,78.86) .. (331.37,78.86) .. controls (333.19,78.86) and (334.67,80.43) .. (334.67,82.36) .. controls (334.67,84.3) and (333.19,85.86) .. (331.37,85.86) .. controls (329.54,85.86) and (328.07,84.3) .. (328.07,82.36) -- cycle ;
\draw    (330.37,81.36) -- (354.33,112) ;
\draw    (331.37,82.36) -- (307.33,111) ;

\draw (81,114) node [anchor=north west][inner sep=0.75pt]   [align=left] {$\displaystyle a$};
\draw (140.33,115) node [anchor=north west][inner sep=0.75pt]   [align=left] {$\displaystyle b$};
\draw (122.33,30) node [anchor=north west][inner sep=0.75pt]   [align=left] {$\displaystyle c$};
\draw (92.33,65) node [anchor=north west][inner sep=0.75pt]   [align=left] {$\displaystyle \alpha '$};
\draw (297,114) node [anchor=north west][inner sep=0.75pt]   [align=left] {$\displaystyle a$};
\draw (356.33,115) node [anchor=north west][inner sep=0.75pt]   [align=left] {$\displaystyle b$};
\draw (338.33,30) node [anchor=north west][inner sep=0.75pt]   [align=left] {$\displaystyle c$};
\draw (310.33,65) node [anchor=north west][inner sep=0.75pt]   [align=left] {$\displaystyle \alpha $};
\draw (168,48) node [anchor=north west][inner sep=0.75pt]   [align=left] {$\displaystyle =\sum _{\alpha =1}^{N_{ab}^{c}}\left[ M_{ab}^{c}\right]_{\alpha '\alpha } \ \ \ \ \ \ \ \ \ \ \ \ \ \ \ \ \ \ \ \ \ \ \ \ \ .$};

\end{tikzpicture}
\end{center}
Such basis transformations do not change the fusion category, but only its
presentation in terms of matrix data.

The tensor product must be associative. At the level of the fusion
coefficients, this requires
\begin{equation}
    \sum_{e\in\mathscr{I}}N_{ab}^{e}N_{ec}^{d}
    =
    \sum_{f\in\mathscr{I}}N_{af}^{d}N_{bc}^{f}\:.
\end{equation}
This relation ensures that the two parenthesizations of a triple tensor
product have the same decomposition into simple objects. At the categorical
level, they are related by a natural\footnote{In category theory, the adjective natural means compatible with morphisms.} isomorphism, called the associator:
\begin{equation}
    \alpha_{a,b,c}:
    (a\otimes b)\otimes c
    \longrightarrow
    a\otimes(b\otimes c)\:.
\end{equation}
After choosing bases in the fusion spaces, the components of the associator
are represented by matrices
\begin{equation}
    [F_{abc}^{d}]_{(e,\alpha,\beta)(f,\gamma,\delta)}\:,
\end{equation}
called \(F\)-symbols and represented graphically by
\begin{center}
    \tikzset{every picture/.style={line width=0.75pt}} 

\begin{tikzpicture}[x=0.75pt,y=0.75pt,yscale=-1,xscale=1]

\draw  [draw opacity=0][fill={rgb, 255:red, 0; green, 0; blue, 0 }  ,fill opacity=1 ] (111.07,125.36) .. controls (111.07,123.43) and (112.54,121.86) .. (114.37,121.86) .. controls (116.19,121.86) and (117.67,123.43) .. (117.67,125.36) .. controls (117.67,127.3) and (116.19,128.86) .. (114.37,128.86) .. controls (112.54,128.86) and (111.07,127.3) .. (111.07,125.36) -- cycle ;
\draw    (114.37,125.36) -- (139.33,156) ;
\draw    (156.33,74.67) -- (89.33,155) ;
\draw  [draw opacity=0][fill={rgb, 255:red, 0; green, 0; blue, 0 }  ,fill opacity=1 ] (133.07,99.36) .. controls (133.07,97.43) and (134.54,95.86) .. (136.37,95.86) .. controls (138.19,95.86) and (139.67,97.43) .. (139.67,99.36) .. controls (139.67,101.3) and (138.19,102.86) .. (136.37,102.86) .. controls (134.54,102.86) and (133.07,101.3) .. (133.07,99.36) -- cycle ;
\draw    (136.37,99.36) -- (184.33,156) ;
\draw  [draw opacity=0][fill={rgb, 255:red, 0; green, 0; blue, 0 }  ,fill opacity=1 ] (460.07,125.36) .. controls (460.07,123.43) and (461.54,121.86) .. (463.37,121.86) .. controls (465.19,121.86) and (466.67,123.43) .. (466.67,125.36) .. controls (466.67,127.3) and (465.19,128.86) .. (463.37,128.86) .. controls (461.54,128.86) and (460.07,127.3) .. (460.07,125.36) -- cycle ;
\draw    (463.37,125.36) -- (441.33,156) ;
\draw    (461.33,74.67) -- (394.33,155) ;
\draw  [draw opacity=0][fill={rgb, 255:red, 0; green, 0; blue, 0 }  ,fill opacity=1 ] (438.07,99.36) .. controls (438.07,97.43) and (439.54,95.86) .. (441.37,95.86) .. controls (443.19,95.86) and (444.67,97.43) .. (444.67,99.36) .. controls (444.67,101.3) and (443.19,102.86) .. (441.37,102.86) .. controls (439.54,102.86) and (438.07,101.3) .. (438.07,99.36) -- cycle ;
\draw    (441.37,99.36) -- (489.33,156) ;

\draw (80,158) node [anchor=north west][inner sep=0.75pt]   [align=left] {$\displaystyle a$};
\draw (139.33,158) node [anchor=north west][inner sep=0.75pt]   [align=left] {$\displaystyle b$};
\draw (186.33,158) node [anchor=north west][inner sep=0.75pt]   [align=left] {$\displaystyle c$};
\draw (96.33,109) node [anchor=north west][inner sep=0.75pt]   [align=left] {$\displaystyle \alpha $};
\draw (130.33,111) node [anchor=north west][inner sep=0.75pt]   [align=left] {$\displaystyle e$};
\draw (163.33,56) node [anchor=north west][inner sep=0.75pt]   [align=left] {$\displaystyle d$};
\draw (120.07,82.36) node [anchor=north west][inner sep=0.75pt]   [align=left] {$\displaystyle \beta $};
\draw (192,81) node [anchor=north west][inner sep=0.75pt]   [align=left] {$\displaystyle =\sum _{f\in \mathcal{I}}\sum _{\gamma =1}^{N_{bc}^{f}}\sum _{\delta =1}^{N_{af}^{d}}\left[ F_{abc}^{d}\right]_{( e,\alpha, \beta )( f,\gamma, \delta )} \ \ \ \ \ \ \ \ \ \ \ \ \ \ \ \ \ \ \ \ \ \ \ \ \ \ \ \ .$};
\draw (385,158) node [anchor=north west][inner sep=0.75pt]   [align=left] {$\displaystyle a$};
\draw (440.33,158) node [anchor=north west][inner sep=0.75pt]   [align=left] {$\displaystyle b$};
\draw (491.33,158) node [anchor=north west][inner sep=0.75pt]   [align=left] {$\displaystyle c$};
\draw (472.33,109) node [anchor=north west][inner sep=0.75pt]   [align=left] {$\displaystyle \gamma $};
\draw (435.33,111) node [anchor=north west][inner sep=0.75pt]   [align=left] {$\displaystyle f$};
\draw (468.33,56) node [anchor=north west][inner sep=0.75pt]   [align=left] {$\displaystyle d$};
\draw (425.07,82.36) node [anchor=north west][inner sep=0.75pt]   [align=left] {$\displaystyle \delta $};

\end{tikzpicture}
\end{center}
Here \(e\) and \(f\) label the intermediate simple objects, while
\(\alpha,\beta,\gamma,\delta\) label basis elements in the corresponding
fusion spaces.

The associator must satisfy the pentagon identity,
\begin{center}
    \tikzset{every picture/.style={line width=0.75pt}} 

\begin{tikzpicture}[x=0.75pt,y=0.75pt,yscale=-1,xscale=1]

\draw    (362.33,57.33) -- (410.66,89.23) ;
\draw [shift={(412.33,90.33)}, rotate = 213.42] [color={rgb, 255:red, 0; green, 0; blue, 0 }  ][line width=0.75]    (10.93,-3.29) .. controls (6.95,-1.4) and (3.31,-0.3) .. (0,0) .. controls (3.31,0.3) and (6.95,1.4) .. (10.93,3.29)   ;
\draw    (241.33,58.33) -- (192,91.22) ;
\draw [shift={(190.33,92.33)}, rotate = 326.31] [color={rgb, 255:red, 0; green, 0; blue, 0 }  ][line width=0.75]    (10.93,-3.29) .. controls (6.95,-1.4) and (3.31,-0.3) .. (0,0) .. controls (3.31,0.3) and (6.95,1.4) .. (10.93,3.29)   ;
\draw    (180.33,127.67) -- (202.5,175.85) ;
\draw [shift={(203.33,177.67)}, rotate = 245.3] [color={rgb, 255:red, 0; green, 0; blue, 0 }  ][line width=0.75]    (10.93,-3.29) .. controls (6.95,-1.4) and (3.31,-0.3) .. (0,0) .. controls (3.31,0.3) and (6.95,1.4) .. (10.93,3.29)   ;
\draw    (432.33,127.67) -- (415.04,173.79) ;
\draw [shift={(414.33,175.67)}, rotate = 290.56] [color={rgb, 255:red, 0; green, 0; blue, 0 }  ][line width=0.75]    (10.93,-3.29) .. controls (6.95,-1.4) and (3.31,-0.3) .. (0,0) .. controls (3.31,0.3) and (6.95,1.4) .. (10.93,3.29)   ;
\draw    (270.33,192.67) -- (346.33,192.67) ;
\draw [shift={(348.33,192.67)}, rotate = 180] [color={rgb, 255:red, 0; green, 0; blue, 0 }  ][line width=0.75]    (10.93,-3.29) .. controls (6.95,-1.4) and (3.31,-0.3) .. (0,0) .. controls (3.31,0.3) and (6.95,1.4) .. (10.93,3.29)   ;
\draw    (304.59,141.67) .. controls (276.7,132.01) and (286.55,104.67) .. (308.7,104.67) .. controls (330.4,104.67) and (338.72,127.07) .. (315.1,140.08) ;
\draw [shift={(313.62,140.86)}, rotate = 327.98] [color={rgb, 255:red, 0; green, 0; blue, 0 }  ][line width=0.75]    (10.93,-3.29) .. controls (6.95,-1.4) and (3.31,-0.3) .. (0,0) .. controls (3.31,0.3) and (6.95,1.4) .. (10.93,3.29)   ;

\draw (124,101) node [anchor=north west][inner sep=0.75pt]   [align=left] {$\displaystyle ( a\otimes ( b\otimes c)) \otimes d$};
\draw (362,183) node [anchor=north west][inner sep=0.75pt]   [align=left] {$\displaystyle a\otimes ( b\otimes ( c\otimes d))$};
\draw (147,183) node [anchor=north west][inner sep=0.75pt]   [align=left] {$\displaystyle a\otimes (( b\otimes c) \otimes d)$};
\draw (388,101) node [anchor=north west][inner sep=0.75pt]   [align=left] {$\displaystyle ( a\otimes b) \otimes ( c\otimes d)$};
\draw (249,38) node [anchor=north west][inner sep=0.75pt]   [align=left] {$\displaystyle (( a\otimes b) \otimes c) \otimes d$};

\end{tikzpicture}
\end{center}
which in components gives the \emph{pentagon equations}
\begin{equation}\label{eq:pentagon}
\begin{split}
    \sum_{\eta=1}^{N_{ik}^{e}}
    [F_{abk}^{e}]_{(i,\alpha,\eta)(l,\xi,\rho)}
    &[F_{icd}^{e}]_{(j,\beta,\delta)(k,\gamma,\eta)}
    =
    \\
    &\sum_{m\in\mathscr{I}}
    \sum_{\mu=1}^{N_{bc}^{m}}
    \sum_{\nu=1}^{N_{am}^{j}}
    \sum_{\sigma=1}^{N_{md}^{l}}
    [F_{abc}^{j}]_{(i,\alpha,\beta)(m,\mu,\nu)}
    [F_{amd}^{e}]_{(j,\nu,\delta)(l,\sigma,\rho)}
    [F_{bcd}^{l}]_{(m,\mu,\sigma)(k,\gamma,\xi)}.
\end{split}
\end{equation}
For a generic fusion ring, these equations are difficult to solve and may
not admit any solution. The fusion rules, together with a solution of the
pentagon equations, determine the monoidal structure of the fusion
category.

Two fusion categories with the same fusion ring are monoidally equivalent if and only if their \(F\)-symbols are related by a combination of the following transformations:
\begin{itemize}
   \item an automorphism \(\sigma\) of the fusion ring 
   which acts on the \(F\)-symbols as
    \begin{equation}\label{eq:F_equiv_auto}
        [F_{abc}^{d}]_{(e,\alpha,\beta)(f,\gamma,\delta)}
        \sim
        [F_{\sigma(a)\sigma(b)\sigma(c)}^{\sigma(d)}]
        _{(\sigma(e),\alpha,\beta)(\sigma(f),\gamma,\delta)} \:;
    \end{equation}

   \item a change of basis \(M_{ab}^{c}\in GL(N_{ab}^{c},\mathbb C)\) in the fusion spaces, also referred to as a gauge transformation of the $F$-symbols, which acts as
    \begin{equation}\label{eq:F_equiv_base}
    \begin{split}
        [F_{abc}^{d}]_{(e,\alpha,\beta)(f,\gamma,\delta)}
        \sim
        \sum_{\mu,\nu,\sigma,\tau}
        &[F_{abc}^{d}]_{(e,\mu,\nu)(f,\sigma,\tau)}
        [M_{ab}^{e}]_{\alpha\mu}
        [M_{ec}^{d}]_{\beta\nu}
        \\
        &\times
        [M_{af}^{d}]^{-1}_{\tau\delta}
        [M_{bc}^{f}]^{-1}_{\sigma\gamma} \:.
    \end{split}
    \end{equation}

\end{itemize}
Solutions of the pentagon equations up to these equivalences form a finite
set by Ocneanu rigidity~\cite{Ocneanu1989,ENO2005}.

The fusion coefficients therefore encode only the decategorified
tensor-product structure. In general, inequivalent monoidal categories may
share the same fusion rules. This distinction will play a central role in
the examples below.

\subsection{Fiber Functors}\label{Sec:FiberFunctor}

A fiber functor on a fusion category \(\mathcal C\) is a faithful,
\(\mathbb C\)-linear monoidal functor
\begin{equation}\label{eq:fiber}
    \mathcal F:\mathcal C\longrightarrow\mathrm{Vec}\:,
\end{equation}
where \(\mathrm{Vec}\) is the category of finite-dimensional complex vector
spaces. It assigns to each simple object \(a\) a vector space
\begin{equation}
    V_a:=\mathcal F(a)\:.
\end{equation}

\subsubsection*{The monoidal structure maps \(J\)}

Since \(\mathcal F\) is monoidal, it is equipped with natural isomorphisms
\begin{equation}
    J_{a,b}:
    V_a\otimes V_b
    \xrightarrow{\;\simeq\;}
    \mathcal F(a\otimes b)\:,
\end{equation}
which define its monoidal structure.

Using the decomposition of the tensor product in Eq.~\eqref{eq:fusionring},
and after fixing bases in the corresponding fusion spaces, we identify
$\mathcal F(a\otimes b)
=
\bigoplus_{c\in \mathscr{I}}V_c^{\oplus N_{ab}^{c}}$. 
With respect to this identification, the monoidal structure is represented by
an invertible matrix
\begin{equation}
    J_{a,b}:
    V_a\otimes V_b
    \longrightarrow
    \bigoplus_{c\in \mathscr{I}}V_c^{\oplus N_{ab}^{c}}\:.
    \label{eq:monoidal_structure}
\end{equation}
Here and in the following, we use the same symbol \(J_{a,b}\) for the
abstract monoidal isomorphism and for its matrix representative with respect
to the chosen decompositions.

Projecting Eq.~\eqref{eq:monoidal_structure} onto the individual copies of
\(V_c\), one obtains the multiplication morphisms
\begin{equation}
    [J_{ab}^{c}]_{\alpha}:
    V_a\otimes V_b
    \longrightarrow
    V_c\,,
    \qquad
    \alpha=1,\ldots,N_{ab}^{c}\:.
    \label{eq:multiplication_morphisms}
\end{equation}
Thus, determining all matrices \(J_{a,b}\), modulo the appropriate changes
of basis, amounts to determining the monoidal structure of the fiber
functor.

The matrix representatives depend on the choices of bases in the vector
spaces \(V_a\). A family of invertible maps $\{\phi_a:V_a\longrightarrow V_a \}_{a\in \mathscr{I}}$
transforms the multiplication morphisms as 
\begin{equation}
    [J_{ab}^{c}]_{\alpha}
    \longmapsto
    \phi_c\circ[J_{ab}^{c}]_{\alpha}
    \circ
    \left(\phi_a\otimes\phi_b\right)^{-1}\:.
    \label{eq:fiber_equiv_base}
\end{equation}
Fiber functors related by such a transformation are \emph{monoidally isomorphic}.

The monoidal structure must also be compatible with the associator of
\(\mathcal C\). In terms of the multiplication morphisms, this condition is
\begin{equation}\label{eq:fiber-Fsymb}
\begin{split}
    [J_{ec}^{d}]_{\beta}
    \circ
    \left([J_{ab}^{e}]_{\alpha}\otimes\mathbbm 1_c\right)
    =
    \sum_{f,\gamma,\delta}
    [F_{abc}^{d}]_{(e,\alpha,\beta)(f,\gamma,\delta)}
    [J_{af}^{d}]_{\delta}
    \circ
    \left(\mathbbm 1_a\otimes[J_{bc}^{f}]_{\gamma}\right)\:.
\end{split}
\end{equation}
Equation~\eqref{eq:fiber-Fsymb} expresses the compatibility between the
\(F\)-symbols describing the associator of the fusion category and the
matrices \(J_{a,b}\) describing the monoidal structure of the fiber functor.

To write Eq.~\eqref{eq:fiber-Fsymb}, one must first choose a representative
of the \(F\)-symbols within the equivalence class generated by
Eqs.~\eqref{eq:F_equiv_auto} and~\eqref{eq:F_equiv_base}. For a fixed such
representative, Eq.~\eqref{eq:fiber-Fsymb} is invariant under the changes of
basis in the vector spaces \(V_a\) described in
Eq.~\eqref{eq:fiber_equiv_base}.

By contrast, changing the representative of the \(F\)-symbols through a
basis transformation
\(M_{ab}^{c}\in GL(N_{ab}^{c},\mathbb C)\)
in the fusion spaces also transforms the multiplication morphisms:
\begin{equation}
    [J_{ab}^{c}]_{\alpha}
    \longmapsto
    \sum_{\beta=1}^{N_{ab}^{c}}
    [M_{ab}^{c}]_{\alpha\beta}
    [J_{ab}^{c}]_{\beta}\:.
    \label{eq:fiber_fusion_space_map}
\end{equation}

Similarly, a relabeling \(\sigma\) of the simple objects preserving the fusion
rules acts as 
\begin{equation}
    [J_{ab}^{c}]_{\alpha}
    \longmapsto
    [J_{\sigma(a)\sigma(b)}^{\sigma(c)}]_{\alpha}\:,
    \label{eq:fiber_relabeling}
\end{equation}
up to the corresponding identifications of the fusion spaces.

It is important to distinguish these operations. Changes of basis in the
vector spaces \(V_a\), as in Eq.~\eqref{eq:fiber_equiv_base}, identify
monoidally isomorphic fiber functors. By contrast, changes of basis in the
fusion spaces and relabelings of the simple objects change the chosen
presentation of the fusion category and act simultaneously on its
\(F\)-symbols and on the matrices \(J_{a,b}\).

When such a transformation leaves a chosen representative of the
\(F\)-symbols invariant, it defines a monoidal autoequivalence of the
fusion category and acts on its fiber functors through
Eqs.~\eqref{eq:fiber_fusion_space_map}
and~\eqref{eq:fiber_relabeling}. This action may permute distinct
monoidal-isomorphism classes of fiber functors.

\subsubsection*{Tensor automorphisms $\tau$}

Given a fiber functor \((\mathcal F,J)\), a tensor automorphism is a natural
isomorphism
\(
\tau:\mathcal F\Longrightarrow\mathcal F
\)
whose components
\begin{equation}
\tau_a:V_a\longrightarrow V_a
\end{equation}
are compatible with the monoidal structure, i.e. for every fusion channel
\(a\otimes b\to c\), the following diagram must commute:
\begin{equation}\label{eq:DiagramTannaka}
\begin{tikzcd}
V_a\otimes V_b
    \arrow[rr,"\tau_a\otimes\tau_b"]
    \arrow[d,"{[J_{ab}^{c}]_{\alpha}}"']
&&
V_a\otimes V_b
    \arrow[d,"{[J_{ab}^{c}]_{\alpha}}"]
\\
V_c
    \arrow[rr,"\tau_c"']
&&
V_c 
\end{tikzcd}\:.
\end{equation}
Commutativity of these diagrams gives the tensor-automorphism
compatibility equations, which we will call the \emph{square equations}
\begin{equation}
    [J_{ab}^{c}]_{\alpha}
    \circ
    (\tau_a\otimes\tau_b)
    =
    \tau_c
    \circ
    [J_{ab}^{c}]_{\alpha}\,,
    \qquad
    \alpha=1,\ldots,N_{ab}^{c}\:.
    \label{eq:tannaka_square}
\end{equation}
The tensor automorphisms form a group under componentwise composition,
denoted by $ \text{Aut}^\otimes(\mathcal{F})$. Each vector space \(V_a\) carries a representation of this group.

\subsection{Pointed and representation categories}

We now specialize the preceding discussion to two classes of fusion
categories associated with finite groups. Pointed categories
\(\mathrm{Vec}_G^\varpi\) provide a simple setting in which the associator
and its gauge transformations are described by group cohomology.
Representation categories \(\operatorname{Rep}(G)\), which constitute the
main setting of this work, are equipped with a canonical forgetful fiber
functor whose monoidal structure maps are the Clebsch--Gordan maps.

\subsubsection*{\(\boldsymbol{\mathrm{Vec}_G^\varpi}\)}

Let \(G\) be a finite group. The simple objects of a pointed fusion category
are labeled by the elements \(g\in G\), with fusion rules
\begin{equation}\label{eq:vecG}
    g\otimes h=gh.
\end{equation}
All fusion spaces are one-dimensional. Consequently, the associator is
specified by a function
\begin{equation}
    [F_{g,h,k}^{ghk}]_{gh,hk}
    =
    \varpi(g,h,k)\in\mathbb C^\ast,
\end{equation}
and the pentagon equations are equivalent to the cocycle condition
\(
    \delta\varpi=1.
\)

A change of basis in the fusion space associated with \(g\otimes h\to gh\)
is multiplication by a nonzero scalar
\begin{equation}
    M_{g,h}:=M_{g,h}^{gh}\in\mathbb C^\ast\:.
\end{equation}
The collection \(M=\{M_{g,h}\}\) is therefore a \(2\)-cochain, whose
coboundary determines the corresponding transformation of \(\varpi\).
Thus, pointed fusion categories with these fusion rules are classified, up
to monoidal equivalence, by the orbits of
\(H^3(G,\mathbb C^\ast)\) under \(\operatorname{Aut}(G)\). The category
associated with \(\varpi\) is denoted by \(\mathrm{Vec}_G^\varpi\).

A fiber functor maps every simple object to a one-dimensional vector space,
\(\mathcal F(g)\simeq\mathbb C\). Its monoidal structure maps are therefore
nonzero scalars
\begin{equation}
    J_{gh}=\mu(g,h)\:,
\end{equation}
which form a \(2\)-cochain. Compatibility with the associator relates
\(\delta\mu\) to \(\varpi\), and implies that
\(\mathrm{Vec}_G^\varpi\) admits a fiber functor if and only if
\([\varpi]\) is trivial in \(H^3(G,\mathbb C^\ast)\). Two such fiber
functors are monoidally isomorphic when their \(2\)-cochains differ by a
coboundary. Hence, whenever nonempty, their isomorphism classes form a
torsor over \(H^2(G,\mathbb C^\ast)\).

\subsubsection*{\(\boldsymbol{\operatorname{Rep}(G)}\)}

Let \(G\) be a finite group. The simple objects of
\(\operatorname{Rep}(G)\) are its irreducible representations, and the
fusion coefficients are determined by their tensor-product decompositions.
Thus the fusion ring is determined by the character table of \(G\).
Non-isomorphic groups may have the same character table and hence the same
fusion rules, while their representation categories remain inequivalent as
monoidal categories.

The category \(\operatorname{Rep}(G)\) always admits a fiber functor, namely the forgetful functor
\begin{equation}
    \mathcal F_{\mathrm{forg}}:
    \operatorname{Rep}(G)\longrightarrow\mathrm{Vec}\:,
\end{equation}
which assigns to a representation its underlying vector space. After
choosing bases, its monoidal structure maps
\begin{equation}
    J_{ab}:
    V_a\otimes V_b
    \longrightarrow
    \bigoplus_c V_c^{\oplus N_{ab}^{c}}
\end{equation}
are precisely the Clebsch--Gordan decomposition maps. The forgetful functor is, up to monoidal isomorphism, the unique fiber
functor compatible with the standard symmetric braiding
\cite{etingof2015tensor}.

More general fiber functors on \(\operatorname{Rep}(G)\) are classified by
conjugacy classes of pairs \((A,[\psi])\), where \(A\subseteq G\) and
\([\psi]\in H^2(A,\mathbb C^\ast)\) is non-degenerate. We will return to
this classification in
Section~\ref{ssec:Bootp3-counting-fiber-functors} and to its explicit
realization by twists in Appendix~\ref{app:twists_and_fiber_functors}.

\subsection{Standard Tannaka reconstruction for finite groups}
\label{sec:Tannaka}

Let \(G\) be a finite group and let
\begin{equation}
    \mathcal F_{\mathrm{forg}}:
    \operatorname{Rep}(G)\longrightarrow\mathrm{Vec}
\end{equation}
be the canonical forgetful fiber functor. Tannaka duality
\cite{Tannaka1938} states that \(G\) can be reconstructed as the group of
tensor automorphisms of \(\mathcal F_{\mathrm{forg}}\):
\begin{equation}
    \operatorname{Aut}^\otimes
    (\mathcal F_{\mathrm{forg}})
    \cong G\:.
\end{equation}
Indeed, every \(g\in G\) defines a tensor automorphism whose component on a
representation \(V\) is the representation matrix
\(\tau_V^{(g)}=\rho_V(g)\), and every tensor automorphism of the forgetful
functor arises from a unique element of \(G\).

Restricting to the simple objects
\(a\in\operatorname{Irr}(G)\), a tensor automorphism is described by a
collection of invertible matrices \(\{\tau_a\}\) satisfying the square
equations~\eqref{eq:tannaka_square}. After choosing bases adapted to the
decomposition of tensor products into irreducible representations, the
monoidal structure maps \(J_{ab}\) are the Clebsch--Gordan decomposition
maps. Standard Tannaka reconstruction therefore takes as input both the
fusion coefficients and the monoidal structure maps of a chosen fiber
functor:
\begin{equation}
    \bigl\{N_{ab}^{c},J_{ab}\bigr\}
    \longmapsto
    \operatorname{Aut}^\otimes
    (\mathcal F_{\mathrm{forg}})\:.
\end{equation}

\paragraph{The example of \(\operatorname{Rep}(S_3)\).}

The fusion ring of \(\operatorname{Rep}(S_3)\) has three simple objects:
the trivial representation \(1\), the sign representation \(\psi\), and a
two-dimensional representation \(\pi\). Their fusion rules are
\begin{equation}\label{eq:Fus_S3}
\begin{split}
        \mathsf{1} \otimes \mathsf{1}&= \mathsf{1}\,,\\
        \mathsf{1} \otimes \psi &= \psi \,,\\
        \mathsf{1} \otimes \pi &= \pi \,,
    \end{split} \qquad\qquad\qquad\qquad
    \begin{split}
        \psi \otimes \psi &= \mathsf{1}\,,\\
        \psi \otimes \pi &= \pi \,,\\
        \pi \otimes \pi &= \mathsf{1}\oplus \psi \oplus \pi \ .
    \end{split}
\end{equation}
The forgetful fiber functor assigns the vector spaces
\begin{equation}
    V_\mathsf{1}\simeq\mathbb C\,,
    \qquad
    V_\psi\simeq\mathbb C\,,
    \qquad
    V_\pi\simeq\mathbb C^2\:.
\end{equation}

In standard Tannaka reconstruction, the Clebsch--Gordan maps are part of
the input. In a convenient choice of bases, the only nontrivial independent
maps needed for the reconstruction are
\begin{equation}\label{eq:S3-standard-CG}
    J_{\psi\pi}
    =
    \begin{pmatrix}
        1&0\\
        0&-1
    \end{pmatrix},
    \qquad
    J_{\pi\pi}
    =
    \begin{pmatrix}
        0&1&1&0\\
        0&1&-1&0\\
        0&0&0&1\\
        1&0&0&0
    \end{pmatrix}\:.
\end{equation}
Since the forgetful functor is symmetric monoidal, \(J_{\pi\psi}\) is
determined by \(J_{\psi\pi}\) and does not constitute an independent input.

From these data one can determine the allowed values of the scalars $\tau_\mathsf{1}, \tau_\psi$ as well as those of the $2\times 2$ matrix $\tau_\pi$. The first fusion channels in \eqref{eq:Fus_S3} give the commutative diagrams
\[
\begin{tikzcd}
V_\mathsf{1}\otimes V_\mathsf{1}
    \arrow[rr,"\tau_\mathsf{1}\otimes\tau_\mathsf{1}"]
    \arrow[d,"{J_{\mathsf{1}\mathsf{1}}}"']
&&
V_\mathsf{1}\otimes V_\mathsf{1}
    \arrow[d,"{J_{\mathsf{1}\mathsf{1}}}"]
\\
V_\mathsf{1}
    \arrow[rr,"\tau_\mathsf{1}"']
&&
V_\mathsf{1} 
\end{tikzcd}
\qquad\qquad
\begin{tikzcd}
V_\psi\otimes V_\psi
    \arrow[rr,"\tau_\psi\otimes\tau_\psi"]
    \arrow[d,"{J_{\psi\psi}}"']
&&
V_\psi\otimes V_\psi
    \arrow[d,"{J_{\psi\psi}}"]
\\
V_\mathsf{1}
    \arrow[rr,"\tau_\mathsf{1}"']
&&
V_\mathsf{1}
\end{tikzcd}
\]
which translate into the equations $J_{\mathsf{1}\mathsf{1}}\tau_\mathsf{1}^2=\tau_\mathsf{1}J_{\mathsf{1}\mathsf{1}}$ and $J_{\psi\psi}\tau_\psi^2=\tau_\mathsf{1}J_{\psi\psi}$, respectively.
This immediately gives
\begin{equation}
    \tau_\mathsf{1}=1
    \qquad\mbox{and}\qquad
    \tau_\psi=\pm1\:.
\end{equation}
The remaining nontrivial conditions are $J_{\psi\pi}(\tau_\psi\tau_\pi)=\tau_\pi J_{\psi\pi}$ and $J_{\pi\pi}(\tau_\pi\otimes\tau_\pi)=(\tau_\pi\oplus\tau_\psi\oplus\tau_\mathsf{1}) J_{\pi\pi}$.

Substituting the Clebsch--Gordan maps~\eqref{eq:S3-standard-CG}
into these remaining conditions gives three solutions for \(\tau_\pi\)
for each choice \(\tau_\psi=\pm1\). We therefore obtain six tensor
automorphisms, forming a group isomorphic to \(S_3\) under componentwise
composition. Thus
\begin{equation}
    \operatorname{Aut}^\otimes
    (\mathcal F_{\mathrm{forg}})
    \cong S_3\:.
\end{equation}

This example displays the logic of standard Tannaka reconstruction: the
monoidal structure maps \(J_{ab}\) are supplied as input, while the
components \(\tau_a\) of the tensor automorphisms are the unknowns to be
determined.

\paragraph{The \(D_8/Q_8\) ambiguity.}

The groups \(D_8\) and \(Q_8\) have the same character table and hence the
same fusion ring, namely the Tambara--Yamagami fusion ring
\(\operatorname{TY}(\mathbb Z_2\times\mathbb Z_2)\). Their representation
categories nevertheless have inequivalent monoidal structures, and their
canonical forgetful fiber functors are described by different
Clebsch--Gordan maps. Standard Tannaka reconstruction therefore gives
\[
    \bigl\{N_{ab}^{c},J_{ab}^{D_8}\bigr\}
    \longmapsto D_8,
    \qquad
    \bigl\{N_{ab}^{c},J_{ab}^{Q_8}\bigr\}
    \longmapsto Q_8.
\]
Thus, the fusion rules alone do not determine the reconstructed group: the
result depends essentially on the monoidal structure maps of the chosen
fiber functor.

\section{Bootstrapping fiber functors}\label{sec:Bootstrap}

As reviewed in the previous section, standard Tannaka reconstruction starts
from a fusion category equipped with a fiber functor. In the case of
\(\operatorname{Rep}(G)\), after choosing bases, the monoidal structure maps
\(J_{ab}\) are the Clebsch--Gordan decomposition maps. Once these maps are
known, the group of tensor automorphisms is reconstructed by solving the
square equations~\eqref{eq:tannaka_square}.

Here we reverse this logic. We start from a fusion category \(\mathcal C\), assuming that only its fusion
rules and the existence of at least one fiber functor are known. We take neither a choice of \(F\)-symbols for \(\mathcal C\) nor the
monoidal structure maps of a fiber functor as input. The dimensions of the vector spaces \(V_a\) are determined by requiring
compatibility with the fusion rules, while both the monoidal structure maps \(J_{ab}\) and the
components \(\tau_a\) of the tensor automorphisms are treated as unknown
matrices. We then solve the \emph{square equations} \eqref{eq:tannaka_square}
for every allowed fusion channel, requiring the full maps \(J_{ab}\)
and the matrices \(\tau_a\) to be invertible. 

We first illustrate this procedure for the fusion ring of
\(\operatorname{Rep}(S_3)\).  This simple example displays the main steps of
our \emph{bootstrap approach} (solving simultaneously for $J$ and $\tau$, reconstructing
the associator, and identifying equivalent solutions) while keeping the computations explicit. After this example, we describe the general bootstrap procedure.

\subsection{Example: Rep(\texorpdfstring{$S_3$}{S3})}

For the fusion rules in Eq.~\eqref{eq:Fus_S3}, the target is
the fiber-functor data reconstructed in Section~\ref{sec:Tannaka}.
The uniqueness of the fiber functor on \(\operatorname{Rep}(S_3)\),
up to monoidal isomorphism, provides a check on the bootstrap:
its solutions should form a single class with tensor automorphism
group \(S_3\).

Since a fiber functor preserves direct sums and tensor products, the fusion
rules determine
\begin{equation}
    V_\mathsf{1}\simeq\mathbb C\,,
    \qquad
    V_\psi\simeq\mathbb C\,,
    \qquad
    V_\pi\simeq\mathbb C^2\:.
\end{equation}
They therefore determine the matrix sizes of the monoidal structure
maps \(J_{ab}\).

As recalled in Section~\ref{sec:Tannaka}, once the vector spaces associated
with the simple objects are fixed, a tensor automorphism of a fiber functor
is specified by a collection of invertible maps
$(\tau_\mathsf{1},\tau_\psi,\tau_\pi)$, required to be compatible with the monoidal
structure maps $J_{ab}$ for every fusion channel. This compatibility is encoded by the square equations
\eqref{eq:tannaka_square}, which we solve simultaneously for the
unknown \(J_{ab}\) and \(\tau_a\). For the relevant fusion channels,
these equations read
\begin{align}
    J_{\mathsf{1}\mathsf{1}}\tau_\mathsf{1}^2
    &=\tau_\mathsf{1}J_{\mathsf{1}\mathsf{1}},
    &
    J_{\mathsf{1}\psi}\tau_\mathsf{1}\tau_\psi
    &=\tau_\psi J_{\mathsf{1}\psi},
    \label{eq:BootS3-1}
    \\[2mm]
    J_{\psi \mathsf{1}}\tau_\psi\tau_\mathsf{1}
    &=\tau_\psi J_{\psi \mathsf{1}},
    &
    J_{\psi\psi}\tau_\psi^2
    &=\tau_\mathsf{1}J_{\psi\psi},
    \label{eq:BootS3-2}
    \\[2mm]
    J_{\mathsf{1}\pi}\tau_\pi
    &=\tau_\pi J_{\mathsf{1}\pi},
    &
    J_{\psi\pi}\tau_\pi\tau_\psi
    &=\tau_\pi J_{\psi\pi},
    \label{eq:BootS3-3}
    \\[2mm]
    J_{\pi\psi}\tau_\pi\tau_\psi
    &=\tau_\pi J_{\pi\psi},
    &
    J_{\pi\pi}(\tau_\pi\otimes\tau_\pi)
    &=(\tau_\mathsf{1}\oplus\tau_\psi\oplus\tau_\pi)J_{\pi\pi}.
    \label{eq:BootS3-4}
\end{align}
Since the multiplication maps $J_{\mathsf{1}\mathsf{1}}$, $J_{\mathsf{1}\psi}$, $J_{\psi \mathsf{1}}$, and
$J_{\psi\psi}$ are nonzero scalars, they factor out of the corresponding
equations, which immediately give
\begin{equation}
    \tau_\mathsf{1}=1
    \qquad\mbox{and}\qquad
    \tau_\psi=\pm1\:.
\end{equation}
In particular, the explicit normalization of the multiplication maps does not play any role in the following analysis.

The automorphisms are then of the form $(\tau_\mathsf{1},\tau_\psi,\tau_\pi)=(1,\pm 1,\tau_\pi^\pm)$.
We now need to determine $\tau_\pi^+$ and $\tau_\pi^-$. The equation involving
$J_{\psi\pi}$  becomes
\begin{equation}
    \pm J_{\psi\pi}\tau_\pi^\pm
    =
    \tau_\pi^\pm J_{\psi\pi}.
    \label{eq:BootS3-5}
\end{equation}
Thus, $J_{\psi\pi}$ commutes with $\tau_\pi^+$ and anticommutes with
$\tau_\pi^-$. This implies that $J_{\psi\pi}$ and $\tau_\pi^-$ are both traceless.\footnote{Since both $J_{\psi\pi}$ and $\tau_\pi^-$ are invertible,
the anticommutation relation implies $(\tau_\pi^-)^{-1}J_{\psi\pi}\tau_\pi^-=-J_{\psi\pi}$, and similarly $J_{\psi\pi}^{-1}\tau_\pi^-J_{\psi\pi}=-\tau_\pi^-$. Hence both matrices are similar to their negatives and are therefore
traceless.}  Invertible and traceless $2\times 2$ matrices are diagonalizable. 
We can then choose a basis in which $J_{\psi\pi}$ is diagonal,
\begin{equation}
    J_{\psi\pi}
    =
    \begin{pmatrix}
        \mathrm{x}_\psi & 0\\
        0 & -\mathrm{x}_\psi
    \end{pmatrix},
    \qquad \mathrm{x}_\psi\in\mathbb C^\ast.
    \label{eq:BootS3-Jpsipi}
\end{equation}
In this basis, the commuting branch must be diagonal, whereas the
anticommuting branch must be off-diagonal. Therefore
\begin{equation}
    \tau_{\pi}^+
    =
    \begin{pmatrix}
        t_1 & 0\\
        0 & t_2
    \end{pmatrix}
    \qquad\mbox{and}\qquad
    \tau_{\pi}^-
    =
    \begin{pmatrix}
        0 & \ell_1\\
        \ell_2 & 0
    \end{pmatrix}.
    \label{eq:BootS3-taupi}
\end{equation}
The same argument applied to \eqref{eq:BootS3-3}, and to the analogous
square for $\pi\otimes\mathsf{1}$, shows that $J_{\mathsf{1}\pi}$ and $J_{\pi\mathsf{1}}$ commute with
every $\tau_\pi$ and are therefore multiples of the identity:
\begin{equation}
    J_{\mathsf{1}\pi}=\mathrm{x}_\mathsf{1}\,\mathbbm{1},
    \qquad\qquad
    J_{\pi \mathsf{1}}=\mathrm{x}_\mathsf{1}'\,\mathbbm{1},
    \qquad\qquad \mathrm{x}_\mathsf{1},\mathrm{x}_\mathsf{1}'\in\mathbb C^\ast\:.
\end{equation}
As a final step, we impose the square associated with the fusion
$\pi\otimes\pi$. For the two branches $\tau_\psi=\pm1$, this gives
\begin{align}
    \label{eq:JpisS31}J_{\pi\pi}(\tau_\pi^+ \otimes \tau_\pi^+)
    &=(1\oplus 1 \oplus \tau_\pi^+)J_{\pi\pi}\:,
    \\
    \label{eq:JpisS32}J_{\pi\pi}(\tau_\pi^- \otimes \tau_\pi^-)
    &=(1\oplus(-1)\oplus \tau_\pi^-)J_{\pi\pi}\:.
\end{align}
Solving Eqs.~\eqref{eq:JpisS31} and~\eqref{eq:JpisS32} for an
invertible \(J_{\pi\pi}\) gives the following form, with three
free nonzero complex parameters:
\begin{equation}
    J_{\pi\pi}=
    \begin{pmatrix}
        0&\mathrm{z}_{\mathsf{1}}&\mathrm{z}_{\mathsf{1}}&0\\
        0&\mathrm{z}_{\psi}&-\mathrm{z}_{\psi}&0 \\
        0&0&0&\mathrm{z}_{\pi} \\
        \mathrm{z}_{\pi}&0&0&0
    \end{pmatrix},
    \qquad
    \mathrm{z}_{\mathsf{1}},\mathrm{z}_{\psi},\mathrm{z}_{\pi}\in\mathbb C^\ast .
\end{equation}
Substituting
\(
    \tau_\pi^+
    =
    \begin{pmatrix}t_1&0\\0&t_2\end{pmatrix}
\)
into $\eqref{eq:JpisS31}$ gives
\begin{equation}
    t_1^2=t_2\,,
    \quad
    t_2^2=t_1\,,
    \quad
    t_1t_2=1\,.
\qquad\Longleftrightarrow\qquad    
t_1^3=1\,,
    \qquad
    t_2=t_1^2\:.
\end{equation}
Similarly, substituting
\(
    \tau_\pi^-=
    \begin{pmatrix}0&\ell_1\\ \ell_2&0\end{pmatrix}
\)
into \eqref{eq:JpisS32} gives
\begin{equation}
    \ell_1^2=\ell_2\,,
    \qquad
    \ell_2^2=\ell_1\,,
    \qquad
    \ell_1\ell_2=1\,,
\qquad\Longleftrightarrow\qquad  
    \ell_1^3=1\,,
    \qquad
    \ell_2=\ell_1^2\:.
\end{equation}
Denoting by $\omega=e^{2\pi i/3}$ a primitive third root of unity, we
therefore obtain the following six tensor automorphisms, written as the
triples $(\tau_\mathsf{1},\tau_\psi,\tau_\pi)$:
\begin{equation}
\begin{split}
    &\left(
    1,\,1,\,
    \begin{pmatrix}1&0\\0&1\end{pmatrix}
    \right),
    \qquad
    \left(
    1,\,1,\,
    \begin{pmatrix}\omega&0\\0&\omega^2\end{pmatrix}
    \right),
    \qquad
    \left(
    1,\,1,\,
    \begin{pmatrix}\omega^2&0\\0&\omega\end{pmatrix}
    \right),
    \\[2mm]
    &\left(
    1,\,-1,\,
    \begin{pmatrix}0&1\\1&0\end{pmatrix}
    \right),
    \qquad
    \left(
    1,\,-1,\,
    \begin{pmatrix}0&\omega\\\omega^2&0\end{pmatrix}
    \right),
    \qquad
    \left(
    1,\,-1,\,
    \begin{pmatrix}0&\omega^2\\\omega&0\end{pmatrix}
    \right).
\end{split}
\end{equation}
These six triples are closed under componentwise composition and form a
group isomorphic to $S_3$. In particular, the third component realizes the
two-dimensional irreducible representation $\pi$, while the second
component is the sign representation $\psi$. We thus identify the category as~\(\operatorname{Rep}(S_3)\).

Substituting the six tensor automorphisms above into the remaining mixed
square equation gives
\begin{equation}
 J_{\pi\psi}
 =
 \mathrm{x}_\psi'
 \begin{pmatrix}
  1&0\\
  0&-1
 \end{pmatrix},
 \qquad
 \mathrm{x}_\psi'\in\mathbb C^\ast\:.
\end{equation}
Together with the forms of \(J_{\mathsf{1}\pi}\), \(J_{\pi\mathsf{1}}\),
\(J_{\psi\pi}\), and \(J_{\pi\pi}\) obtained above, this completes the
solution of the square equations for the multiplication maps.\footnote{With a suitable choice of these normalizations and a change of basis in
\(V_\pi\), these maps reproduce the Clebsch--Gordan realization used in
Section~\ref{sec:Tannaka}.}

The nonzero scalar coefficients appearing in these maps, together with
the pointed coefficients \(J_{\mathsf{1}\mathsf{1}}\), \(J_{\mathsf{1}\psi}\), \(J_{\psi\mathsf{1}}\), and
\(J_{\psi\psi}\), are normalization choices. Indeed, all fusion spaces
of \(\operatorname{Rep}(S_3)\) considered here are one-dimensional, so their bases can
be rescaled independently as in Eq.~\eqref{eq:fiber_fusion_space_map}.
The coefficients can therefore be fixed to any convenient nonzero values
and do not label inequivalent fiber functors.

\subsubsection*{A posteriori reconstruction of the \(F\)-symbols}

Once the matrices \(J_{ab}\) have been determined, the associator can be
reconstructed from the monoidal compatibility
equation~\eqref{eq:fiber-Fsymb}.

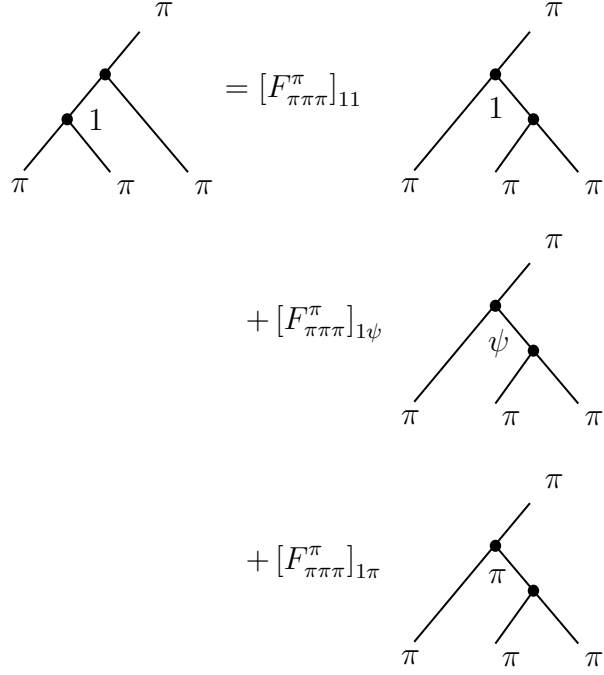
\begin{figure}[ht]
\begin{center}
    \tikzset{every picture/.style={line width=0.75pt}} 

\begin{tikzpicture}[x=0.65pt,y=0.65pt,yscale=-1,xscale=1]

\draw  [draw opacity=0][fill={rgb, 255:red, 0; green, 0; blue, 0 }  ,fill opacity=1 ] (166.07,79.36) .. controls (166.07,77.43) and (167.54,75.86) .. (169.37,75.86) .. controls (171.19,75.86) and (172.67,77.43) .. (172.67,79.36) .. controls (172.67,81.3) and (171.19,82.86) .. (169.37,82.86) .. controls (167.54,82.86) and (166.07,81.3) .. (166.07,79.36) -- cycle ;
\draw    (169.37,79.36) -- (194.33,110) ;
\draw    (211.33,28.67) -- (144.33,109) ;
\draw  [draw opacity=0][fill={rgb, 255:red, 0; green, 0; blue, 0 }  ,fill opacity=1 ] (188.07,53.36) .. controls (188.07,51.43) and (189.54,49.86) .. (191.37,49.86) .. controls (193.19,49.86) and (194.67,51.43) .. (194.67,53.36) .. controls (194.67,55.3) and (193.19,56.86) .. (191.37,56.86) .. controls (189.54,56.86) and (188.07,55.3) .. (188.07,53.36) -- cycle ;
\draw    (191.37,53.36) -- (239.33,110) ;
\draw  [draw opacity=0][fill={rgb, 255:red, 0; green, 0; blue, 0 }  ,fill opacity=1 ] (436.07,79.36) .. controls (436.07,77.43) and (437.54,75.86) .. (439.37,75.86) .. controls (441.19,75.86) and (442.67,77.43) .. (442.67,79.36) .. controls (442.67,81.3) and (441.19,82.86) .. (439.37,82.86) .. controls (437.54,82.86) and (436.07,81.3) .. (436.07,79.36) -- cycle ;
\draw    (439.37,79.36) -- (417.33,110) ;
\draw    (437.33,28.67) -- (370.33,109) ;
\draw  [draw opacity=0][fill={rgb, 255:red, 0; green, 0; blue, 0 }  ,fill opacity=1 ] (414.07,53.36) .. controls (414.07,51.43) and (415.54,49.86) .. (417.37,49.86) .. controls (419.19,49.86) and (420.67,51.43) .. (420.67,53.36) .. controls (420.67,55.3) and (419.19,56.86) .. (417.37,56.86) .. controls (415.54,56.86) and (414.07,55.3) .. (414.07,53.36) -- cycle ;
\draw    (417.37,53.36) -- (465.33,110) ;
\draw  [draw opacity=0][fill={rgb, 255:red, 0; green, 0; blue, 0 }  ,fill opacity=1 ] (436.07,213.36) .. controls (436.07,211.43) and (437.54,209.86) .. (439.37,209.86) .. controls (441.19,209.86) and (442.67,211.43) .. (442.67,213.36) .. controls (442.67,215.3) and (441.19,216.86) .. (439.37,216.86) .. controls (437.54,216.86) and (436.07,215.3) .. (436.07,213.36) -- cycle ;
\draw    (439.37,213.36) -- (417.33,244) ;
\draw    (437.33,162.67) -- (370.33,243) ;
\draw  [draw opacity=0][fill={rgb, 255:red, 0; green, 0; blue, 0 }  ,fill opacity=1 ] (414.07,187.36) .. controls (414.07,185.43) and (415.54,183.86) .. (417.37,183.86) .. controls (419.19,183.86) and (420.67,185.43) .. (420.67,187.36) .. controls (420.67,189.3) and (419.19,190.86) .. (417.37,190.86) .. controls (415.54,190.86) and (414.07,189.3) .. (414.07,187.36) -- cycle ;
\draw    (417.37,187.36) -- (465.33,244) ;
\draw  [draw opacity=0][fill={rgb, 255:red, 0; green, 0; blue, 0 }  ,fill opacity=1 ] (436.07,352.36) .. controls (436.07,350.43) and (437.54,348.86) .. (439.37,348.86) .. controls (441.19,348.86) and (442.67,350.43) .. (442.67,352.36) .. controls (442.67,354.3) and (441.19,355.86) .. (439.37,355.86) .. controls (437.54,355.86) and (436.07,354.3) .. (436.07,352.36) -- cycle ;
\draw    (439.37,352.36) -- (417.33,383) ;
\draw    (437.33,301.67) -- (370.33,382) ;
\draw  [draw opacity=0][fill={rgb, 255:red, 0; green, 0; blue, 0 }  ,fill opacity=1 ] (414.07,326.36) .. controls (414.07,324.43) and (415.54,322.86) .. (417.37,322.86) .. controls (419.19,322.86) and (420.67,324.43) .. (420.67,326.36) .. controls (420.67,328.3) and (419.19,329.86) .. (417.37,329.86) .. controls (415.54,329.86) and (414.07,328.3) .. (414.07,326.36) -- cycle ;
\draw    (417.37,326.36) -- (465.33,383) ;

\draw (135,112) node [anchor=north west][inner sep=0.75pt]   [align=left] {$\displaystyle \pi $};
\draw (196.33,113) node [anchor=north west][inner sep=0.75pt]   [align=left] {$\displaystyle \pi $};
\draw (241.33,113) node [anchor=north west][inner sep=0.75pt]   [align=left] {$\displaystyle \pi $};
\draw (179.83,71.83) node [anchor=north west][inner sep=0.75pt]   [align=left] {$ $$\displaystyle 1$};
\draw (218.33,10) node [anchor=north west][inner sep=0.75pt]   [align=left] {$\displaystyle \pi $};
\draw (259,51) node [anchor=north west][inner sep=0.75pt]   [align=left] {$\displaystyle =\left[ F_{\pi \pi \pi }^{\pi }\right]_{11}$};
\draw (361,112) node [anchor=north west][inner sep=0.75pt]   [align=left] {$\displaystyle \pi $};
\draw (419.33,113) node [anchor=north west][inner sep=0.75pt]   [align=left] {$\displaystyle \pi $};
\draw (467.33,113) node [anchor=north west][inner sep=0.75pt]   [align=left] {$\displaystyle \pi $};
\draw (411.33,65) node [anchor=north west][inner sep=0.75pt]   [align=left] {$\displaystyle 1$};
\draw (444.33,10) node [anchor=north west][inner sep=0.75pt]   [align=left] {$\displaystyle \pi $};
\draw (271,185) node [anchor=north west][inner sep=0.75pt]   [align=left] {$\displaystyle +\left[ F_{\pi \pi \pi }^{\pi }\right]_{1\psi }$};
\draw (361,246) node [anchor=north west][inner sep=0.75pt]   [align=left] {$\displaystyle \pi $};
\draw (419.33,247) node [anchor=north west][inner sep=0.75pt]   [align=left] {$\displaystyle \pi $};
\draw (467.33,247) node [anchor=north west][inner sep=0.75pt]   [align=left] {$\displaystyle \pi $};
\draw (411.33,199) node [anchor=north west][inner sep=0.75pt]   [align=left] {$\displaystyle \psi $};
\draw (444.33,144) node [anchor=north west][inner sep=0.75pt]   [align=left] {$\displaystyle \pi $};
\draw (271,324) node [anchor=north west][inner sep=0.75pt]   [align=left] {$\displaystyle +\left[ F_{\pi \pi \pi }^{\pi }\right]_{1\pi }$};
\draw (361,385) node [anchor=north west][inner sep=0.75pt]   [align=left] {$\displaystyle \pi $};
\draw (419.33,386) node [anchor=north west][inner sep=0.75pt]   [align=left] {$\displaystyle \pi $};
\draw (467.33,386) node [anchor=north west][inner sep=0.75pt]   [align=left] {$\displaystyle \pi $};
\draw (411.33,338) node [anchor=north west][inner sep=0.75pt]   [align=left] {$\displaystyle \pi $};
\draw (444.33,283) node [anchor=north west][inner sep=0.75pt]   [align=left] {$\displaystyle \pi $};

\end{tikzpicture} 
\caption{Fusion channels entering the reconstruction of
\(F_{\pi\pi\pi}^{\pi}\) for \(\operatorname{Rep}(S_3)\). The two sides
represent the two parenthesizations of \(\pi\otimes\pi\otimes\pi\), with
intermediate channels \(\mathsf{1}\), \(\psi\), and \(\pi\).}\label{FigS3Fusion}
\end{center}
\end{figure}

Let us illustrate this explicitly in the present example. For the calculation below, we choose
\begin{equation}
    J_{\mathsf{1}\pi}=J_{\pi\mathsf{1}}=\mathbbm{1}_2,
    \qquad
    \mathrm{x}_\psi'=1,
    \qquad
    \mathrm{z}_{\pi}=\mathrm{z}_{\mathsf{1}}=1,
    \qquad
    \mathrm{z}_{\psi}=-1.
    \label{eq:S3-Fsymb-normalization}
\end{equation}
We then consider the fusion channels shown in Figure~\ref{FigS3Fusion}.
The two different parenthesizations give
\begin{equation}
    \begin{split}
        J_{\mathsf{1}\pi}^\pi \circ (J_{\pi\pi}^\mathsf{1}\otimes \mathbbm{1}_\pi)&=
        \begin{pmatrix}
        0&0&1&0&1&0&0&0\\
        0&0&0&1&0&1&0&0
        \end{pmatrix},\\
        J_{\pi \mathsf{1}}^\pi \circ (\mathbbm{1}_\pi\otimes J_{\pi\pi}^\mathsf{1})&=
        \begin{pmatrix}
        0&1&1&0&0&0&0&0\\
        0&0&0&0&0&1&1&0
        \end{pmatrix},\\
        J_{\pi \psi}^\pi \circ
        (\mathbbm{1}_\pi\otimes J_{\pi\pi}^\psi)&=
        \begin{pmatrix}
        0&-1&1&0&0&0&0&0\\
        0&0&0&0&0&1&-1&0
        \end{pmatrix},\\
        J_{\pi \pi}^\pi \circ
        (\mathbbm{1}_\pi\otimes J_{\pi\pi}^\pi)&=
        \begin{pmatrix}
        0&0&0&0&1&0&0&0\\
        0&0&0&1&0&0&0&0
        \end{pmatrix}.
    \end{split}
\end{equation}
Substitution into Eq.~\eqref{eq:fiber-Fsymb} gives
\begin{align}
    \begin{pmatrix}
        0&0&1&0&1&0&0&0\\
        0&0&0&1&0&1&0&0
    \end{pmatrix}
    &=
    \left(F_{\pi\pi\pi}^{\pi}\right)_{\mathsf{1}\mathsf{1}}
    \begin{pmatrix}
        0&1&1&0&0&0&0&0\\
        0&0&0&0&0&1&1&0
    \end{pmatrix}
    \nonumber\\
    &\quad+
    \left(F_{\pi\pi\pi}^{\pi}\right)_{\mathsf{1}\psi}
    \begin{pmatrix}
        0&-1&1&0&0&0&0&0\\
        0&0&0&0&0&1&-1&0
    \end{pmatrix}
    \\
    &\quad+
    \left(F_{\pi\pi\pi}^{\pi}\right)_{\mathsf{1}\pi}
    \begin{pmatrix}
        0&0&0&0&1&0&0&0\\
        0&0&0&1&0&0&0&0
    \end{pmatrix}.\nonumber
\end{align}
Solving this linear system yields
\begin{equation}
    \left[F_{\pi\pi\pi}^{\pi}\right]_{\mathsf{1}\mathsf{1}}
    =\frac12 \,,
    \qquad
    \left[F_{\pi\pi\pi}^{\pi}\right]_{\mathsf{1}\psi}
    =\frac12 \,,
    \qquad
    \left[F_{\pi\pi\pi}^{\pi}\right]_{\mathsf{1}\pi}
    =1 \:.
\end{equation}
The remaining \(F\)-symbols are obtained in the same way. 

\subsection{General bootstrap procedure}
\label{ssec:bootstrap-code}

Having illustrated the bootstrap for $\operatorname{Rep}(S_3)$, we now
summarize the general procedure that will be applied in the following
sections.

We take as input the simple-object labels and fusion coefficients
\(N_{ab}^{c}\), and ask whether these fusion rules admit a
categorification with a fiber functor. The fusion rules determine the dimensions \(d_a=\dim V_a\) as the
unique positive solution of
\begin{equation}
    d_a d_b=\sum_c N_{ab}^{c}\,d_c \:.
    \label{eq:bootstrap-dimensions}
\end{equation}
A fiber functor can exist only if all \(d_a\) are integers. In that case,
we set \(V_a\simeq\mathbb C^{d_a}\).

The monoidal structure maps $J_{ab}$, the tensor automorphisms
$\tau=\{\tau_a\}_{a \in \mathscr{I}}$, and the square equations
relating them were introduced in Section~\ref{sec:FusCat}. In standard
Tannaka reconstruction, the maps $J_{ab}$ are part of the input and the
square equations are solved for the tensor automorphisms $\tau$. Here we take a different starting point: starting only from the fusion-ring data, we treat both the $J_{ab}$ and the
$\tau_a$ as unknowns and solve the coupled system
Eq.~\eqref{eq:tannaka_square} simultaneously.

We identify solutions related by changes of basis in the vector spaces \(V_a\). The resulting $J$-maps organize into families of matrices depending
on free parameters, subject to invertibility. We call these families \emph{parametric forms}. For each such form, the compatible solutions
\(\tau=\{\tau_a\}_{a\in\mathscr{I}}\) determine a candidate
tensor-automorphism group \(G\). We first consider cases for which
its order equals the sum of the squared dimensions of the \(V_a\):
\begin{equation}
    |G|=\sum_a d_a^2 \:.
    \label{eq:bootstrap-group-order}
\end{equation}
Under condition~\eqref{eq:bootstrap-group-order}, the square
equations ensure that the $V_a$ form a complete set of pairwise
inequivalent irreducible representations of $G$, with the $J_{ab}$
providing their Clebsch--Gordan decompositions.

By the finite-group Tannaka reconstruction reviewed above, this
parametric form therefore reconstructs \(\operatorname{Rep}(G)\).

For forms satisfying Eq.~\eqref{eq:bootstrap-group-order}, that we call the \emph{maximal cases}, the square
equations already provide the fiber-functor data. The corresponding
\(F\)-symbols can be reconstructed from Eq.~\eqref{eq:fiber-Fsymb}
without imposing further conditions on the \(J_{ab}\).
For forms with fewer tensor automorphisms, the preceding argument no
longer identifies the reconstructed category with
\(\operatorname{Rep}(G)\). The square equations alone need not ensure
the existence of a consistent associator, so we additionally require
that Eq.~\eqref{eq:fiber-Fsymb} admit well-defined \(F\)-symbols. 
In any case, the resulting \(F\)-symbols automatically
satisfy the pentagon equations.\footnote{
For fixed simple objects \(a,b,c,d,e\), composing
the invertible \(J\)-maps along each bracketing of
\(a\otimes b\otimes c\otimes d\) gives a linearly
independent family of maps
\(V_a\otimes V_b\otimes V_c\otimes V_d\to V_e\),
indexed by intermediate objects and multiplicity labels.
Equation~\eqref{eq:fiber-Fsymb} relates these families
by \(F\)-moves, so they are bases of the same subspace.
The two sequences of \(F\)-moves in the pentagon
connect the same initial and final basis, and therefore
their compositions coincide.}

The square equations, together with the requirement that
Eq.~\eqref{eq:fiber-Fsymb} admit \(F\)-symbols~---~whose
existence is guaranteed when
Eq.~\eqref{eq:bootstrap-group-order} holds~---~capture
all pairs \((\mathcal C,\mathcal F)\) consisting of
a fusion category \(\mathcal C\) with the given fusion
rules and a fiber functor \(\mathcal F\).
Indeed, every such pair provides a solution, and conversely
every solution satisfying the conditions above defines
such a pair.
Different solutions may nevertheless describe equivalent
pairs, so we must compare the resulting presentations.
For this comparison, we use the equivalences described
in Section~\ref{Sec:FiberFunctor}.
Fusion-space basis transformations change the gauge representative
of the associator, whereas basis transformations in the \(V_a\)
identify monoidally isomorphic fiber functors.

Accordingly, after reconstructing the $F$-symbols associated with each
parametric form, we first determine which of them are related by
fusion-space basis transformations and relabelings  of the simple objects induced by
automorphisms of the fusion ring.
Parametric forms whose $F$-symbols are not related in this way define
different fusion categories. When they are related, we bring their
$F$-symbols to a common representative and then compare the corresponding
maps $J_{ab}$ modulo changes of basis in the spaces $V_a$. This determines
the monoidal-isomorphism classes of fiber functors on the fixed category.
Transformations that leave the chosen $F$-symbols invariant define
monoidal autoequivalences of the category and may permute these classes.

\section{Fiber functors from the TY\texorpdfstring{$(\mathbb{Z}_2\times \mathbb{Z}_2)$}{(Z2 x Z2)} fusion ring}
\label{sec:TYZ2}

In this section, we apply the bootstrap procedure to the Tambara--Yamagami fusion ring \cite{Tambara:1998vmj} associated with the abelian group $\mathbb{Z}_2\times\mathbb{Z}_2$. Recall that, for any finite abelian group $\Gamma$, the simple objects of $\mathrm{TY}(\Gamma)$ are the elements of $\Gamma$, together with a non-invertible object $m$, and obey the fusion rules
\begin{equation}\label{eq:fusTY}
    g\otimes h=gh\,,
    \qquad
    g\otimes m=m\otimes g=m\,,
    \qquad
    m\otimes m=\bigoplus_{g\in\Gamma}g\,,
    \qquad g,h\in\Gamma\,.
\end{equation}

This example provides a useful intermediate case: unlike $\operatorname{Rep}(S_3)$, it already exhibits several of the genuinely nontrivial features of the method. At the same time, the ring is sufficiently small that the relevant equations can be written explicitly and the various steps of the computation can be followed in detail. Moreover, this case has been studied from several complementary viewpoints, ranging from the classification of fusion categories with these fusion rules \cite{Tambara:1998vmj} to the classification of their fiber functors and corresponding Hopf-algebra realizations \cite{Tambara:2000qlz} (see also \cite{MasuokaDim68}). The cited classifications provide independent checks of the solutions obtained through our bootstrap procedure.

\subsection{From the fusion ring to the square equations}

We now specialize to
\begin{equation}
\Gamma=\{\mathsf{1},a,b,c=ab\}\simeq\mathbb{Z}_2\times\mathbb{Z}_2 \:.
\end{equation}
For this choice of $\Gamma$, the fusion rules become
\begin{equation}
    g\otimes h=gh,
    \qquad
    g\otimes m=m\otimes g=m,
    \qquad
    m\otimes m=\mathsf{1}\oplus a\oplus b\oplus c,
    \qquad
    g,h\in\Gamma \:.
    \label{eq:BootDQ-fusion}
\end{equation}
The fusion rules determine the dimensions of the corresponding vector spaces, so that
\begin{equation}
    \mathcal{F}(g)\equiv V_g\simeq\mathbb C,
    \qquad\mbox{and}\qquad
    \mathcal{F}(m)\equiv V_m\simeq\mathbb C^2 \:.
\end{equation}

For \(g,h\in\Gamma\), the monoidal structure maps have the form
\begin{equation}
    J_{gh}\in\mathbb C^\ast,
    \qquad
    J_{gm},J_{mg}\in\operatorname{GL}(2,\mathbb C) \:.
\end{equation}
For the fusion product \(m\otimes m\), we assemble the four fusion-channel maps into
\begin{equation}
    J_{mm}
    =
    \begin{pmatrix}
        J_{mm}^\mathsf{1}\\
        J_{mm}^a\\
        J_{mm}^b\\
        J_{mm}^c
    \end{pmatrix}
    \in\operatorname{GL}(4,\mathbb C) \:,
    \label{eq:BootDQ-Jmm}
\end{equation}
where the basis of \(V_m\otimes V_m\) is ordered as
\((11,12,21,22)\), and the fusion channels as \((\mathsf{1},a,b,c)\).

The components of a tensor automorphism $\tau\in \operatorname{Aut}^{\otimes}(\mathcal F)$ are
\begin{equation}
    \tau=(\tau_\mathsf{1},\tau_a,\tau_b,\tau_c,\tau_m),
\end{equation}
with $\tau_g\in\mathbb C^\ast$ ($g\in\Gamma$) and $\tau_m\in\operatorname{GL}(2,\mathbb C)$.

For the fusion ring under consideration, the square equations encoding the
compatibility between \(\tau\) and \(J\) separate into three families:
\begin{itemize}
\item \textbf{Pointed sector:}
\begin{equation}
    J_{gh}\tau_g\tau_h
    =
    \tau_{gh}J_{gh},
    \qquad
    g,h\in\Gamma\:.
    \label{eq:BootDQ-pointed}
\end{equation}
\item \textbf{Mixed sector:}
\begin{equation}
    \tau_mJ_{gm}
    =
    \tau_gJ_{gm}\tau_m,
    \qquad
    \tau_mJ_{mg}
    =
    \tau_gJ_{mg}\tau_m \:, \qquad
    g\in\Gamma \:.
    \label{eq:BootDQ-mixed}
\end{equation}
\item \textbf{Non-invertible sector:}
\begin{equation}
    J_{mm}(\tau_m\otimes\tau_m)
    =
    \operatorname{diag}
    (\tau_\mathsf{1},\tau_a,\tau_b,\tau_c)\,
    J_{mm} \:,
    \label{eq:BootDQ-big}
\end{equation}
or equivalently, channel by channel,
\begin{equation}
    J_{mm}^g(\tau_m\otimes\tau_m)
    =
    \tau_g J_{mm}^g \,,
    \qquad
    g\in\Gamma \:.
    \label{eq:BootDQ-bigrows}
\end{equation}
\end{itemize}
Together, Eqs.~\eqref{eq:BootDQ-pointed}, \eqref{eq:BootDQ-mixed},
and~\eqref{eq:BootDQ-bigrows} constitute the complete set of compatibility
conditions between \(J\) and \(\tau\) for this fusion ring.

\subsection{Solving the square equations: common constraints}\label{ssec:BootDQ-common}

We carry out the bootstrap analysis by treating both the monoidal
structure maps \(J\) and the tensor-automorphism components \(\tau\) as
unknowns and solving the square equations simultaneously, sector by sector.

Let us start from the pointed sector. Since all \(J_{gh}\) are nonzero scalars, they cancel from
Eq.~\eqref{eq:BootDQ-pointed}, leaving
\begin{equation}
    \tau_{gh}=\tau_g\tau_h \:.
\end{equation}
Thus, the pointed-sector components
\((\tau_\mathsf{1},\tau_a,\tau_b,\tau_c)\) of a tensor automorphism
are determined by a character
\begin{equation}
    \chi:\Gamma\longrightarrow\{\pm1\}\:, \qquad\mbox{ such that }\,\tau_g=\chi(g)\:.
\end{equation}
There are four such characters, which we denote by
\(\chi_\mathsf{1},\chi_a,\chi_b,\chi_c\), with
\begin{equation}
\begin{split}
    \chi_\mathsf{1}&=(1,\,\,\,1,\,\,\,\,\,1,\,\,\,\,1),\\
    \chi_a&=(1,\,\,\,\,1,-1,-1),\\
    \chi_b&=(1,-1,\,\,\,\,1,-1),\\
    \chi_c&=(1,-1,-1,\,\,\,\,1),
\end{split}
\qquad\qquad
\text{on}\quad(\mathsf{1},a,b,c)\:.
\label{eq:BootDQ-characters}
\end{equation}
The sixteen pointed maps \(J_{gh}\) remain completely unconstrained by
these equations.

A tensor automorphism extending the character \(\chi\) must therefore be of the form
\begin{equation}
    (\,\chi,\,\,\tau_m^\chi\,) \:,
\end{equation}
where \(\tau_m^\chi\in\operatorname{GL}(2,\mathbb C)\). For a given character \(\chi\), we now determine the possible matrices \(\tau_m^\chi\in\operatorname{GL}(2,\mathbb C)\), together with compatible \(J_{gm}\), \(J_{mg}\), and \(J_{mm}\), by solving the remaining square equations.

Let us first consider the trivial character $\chi_\mathsf{1}(g)=1$ $\forall g\in\Gamma$. In this sector
Eq.~\eqref{eq:BootDQ-big}, together with the invertibility of \(J_{mm}\),
gives
\begin{equation}
    \tau_m^\mathsf{1}\otimes \tau_m^\mathsf{1}=\mathbbm{1}_4 \:,
\end{equation}
whose invertible solutions are
\begin{equation}
    \tau_m^\mathsf{1}=\pm\mathbbm{1}_2 \:.
\end{equation}
Thus the trivial character sector always contains a central $\mathbb{Z}_2$ subgroup of $ \mathrm{Aut}^\otimes(\mathcal{F})$, with elements 
\begin{equation}
    (\chi_\mathsf{1},\mathbbm{1}_2)
    \qquad\mbox{and}\qquad
    (\chi_\mathsf{1},-\mathbbm{1}_2) \:.
    \label{eq:BootDQ-center}
\end{equation}

Let \(\chi_g\) be one of the nontrivial characters in \eqref{eq:BootDQ-characters}, and choose \(h\in\Gamma\) such that \(\chi_g(h)=-1\). The mixed square \eqref{eq:BootDQ-mixed} then gives
\begin{equation}
    \tau_m^{\chi_g}J_{hm}
    =
    -J_{hm}\tau_m^{\chi_g} \:.
\end{equation}
Since \(J_{hm}\) is invertible, \(\tau_m^{\chi_g}\) is conjugate to
\(-\tau_m^{\chi_g}\) and is therefore traceless. The Cayley--Hamilton theorem then implies
\begin{equation}
    \left(\tau_m^{\chi_g}\right)^2
    =
    \varepsilon_g\,\mathbbm 1_2 \, \qquad
\mbox{with}\qquad\varepsilon_g=-\det\left(\tau_m^{\chi_g}\right)\in\mathbb C^\ast.
\end{equation}
On the other hand, applying
Eq.~\eqref{eq:BootDQ-big} twice and using \(\chi_g^2=\chi_\mathsf{1}\) gives
\begin{equation}
    \mathbbm 1_4
    =
    \left(\tau_m^{\chi_g}\otimes\tau_m^{\chi_g}\right)^2
    =
    \varepsilon_g^2\,\mathbbm 1_4.
\end{equation}
Hence \(\varepsilon_g=\pm1\).

Since the trivial-character sector consists of $\{(\chi_\mathsf{1},\mathbbm{1}_2), (\chi_\mathsf{1},-\mathbbm{1}_2) \}$
each character \(\chi_g\) that can be lifted to a tensor automorphism
gives precisely two elements:
\begin{equation}
    (\chi_g,\tau_m^{\chi_g}),
    \qquad
    (\chi_g,-\tau_m^{\chi_g}) \:.
\end{equation}
As \(\widehat{\Gamma}\) has four elements, the tensor automorphism group
has at most eight elements, with equality if and only if every character
of \(\Gamma\) extends. Moreover, the characters that extend form a subgroup of
\(\widehat{\Gamma}\), so the only possible group orders are two, four, and
eight. The order-two case is impossible.\footnote{
Here \(\dim H_{\mathcal F}=4\cdot1^2+2^2=8\), and
\(H_{\mathcal F}^{*}\) is semisimple by the Larson--Radford theorem
\cite{LarsonRadford1988}. Its one-dimensional simple modules correspond
to the group-like elements of \(H_{\mathcal F}\). If there were exactly
two, the remaining six dimensions would have to be a sum of squares
\(d_i^2\) with \(d_i\geq2\), which is impossible.}
Hence the tensor automorphism group has order four or eight.

\subsection{The maximal case: \texorpdfstring{$\operatorname{Rep}(G)$}{Rep(G)} categories}\label{Sec:maximalCaseTY}

We first consider the maximal case, in which the tensor automorphism group has order eight. As noted above, this occurs precisely when all three nontrivial characters in Eq.~\eqref{eq:BootDQ-characters} extend. The elements of the group are then
\begin{equation}
    (\chi_\mathsf{1},\pm\mathbbm{1}),\qquad
    (\chi_a,\pm \tau_m^{\chi_a}),\qquad
    (\chi_b,\pm \tau_m^{\chi_b}),\qquad
    (\chi_c,\pm \tau_m^{\chi_c}).
\end{equation}

The product of representatives in two different nontrivial sectors belongs
to the third sector. Thus, one must have
\begin{equation}
    \tau_m^{\chi_a}\tau_m^{\chi_b}
    =
    \pm \tau_m^{\chi_c},
\end{equation}
and similarly under cyclic permutations of \(a,b,c\). In particular, the fact that \(\tau_m^{\chi_c}\) is traceless implies that 
\(\tau_m^{\chi_a}\tau_m^{\chi_b}\) must also be traceless.
It follows\footnote{For two traceless \(2\times2\)
matrices \(A,B\) one has
\(
    AB+BA=\operatorname{tr}(AB)\,\mathbbm{1}_2.
\)} that
\begin{equation}
    \left\{\tau_m^{\chi_g},\tau_m^{\chi_h}\right\}
    = 0,
    \qquad \mbox{when }
    g\neq h \:.
\end{equation}
Moreover, squaring
\(\tau_m^{\chi_a}\tau_m^{\chi_b}=\pm \tau_m^{\chi_c}\) gives
\(
    -\varepsilon_a\varepsilon_b
    =
    \varepsilon_c,
\)
and hence
\begin{equation}
    \varepsilon_a\varepsilon_b\varepsilon_c=-1.
    \label{eq:BootDQ-epsilonconstraint}
\end{equation}
The four allowed sign triples for the $(\varepsilon_a,\varepsilon_b,\varepsilon_c)$ are therefore
\begin{equation}
    (-,+,+),
    \qquad
    (+,-,+),
    \qquad
    (+,+,-),
    \qquad
    (-,-,-).
    \label{eq:BootDQ-signclasses}
\end{equation}

We next use the relations satisfied by the tensor automorphisms to constrain the monoidal structure maps, starting with the mixed square equations involving \(\tau_m^{\chi_c}\). Since this matrix is traceless and invertible, the mixed square equations imply that
\begin{equation}\label{Eq:JwithTauc}
    J_{am}\tau_m^{\chi_c}=-\tau_m^{\chi_c}J_{am}
    \qquad\mbox{and}\qquad
    J_{bm}\tau_m^{\chi_c}=-\tau_m^{\chi_c}J_{bm},
\end{equation}
while \(J_{cm}\) and \(J_{\mathsf{1}m}\) commute with \(\tau_m^{\chi_c}\). These relations determine only the commutation type of the four mixed maps:
\(J_{am}\) and \(J_{bm}\) lie in the anticommutant of
\(\tau_m^{\chi_c}\), whereas \(J_{cm}\) and \(J_{\mathsf{1}m}\) lie in its
commutant. Their components within these subspaces remain undetermined at
this stage.

In the maximal-automorphism case, the existence of the automorphisms associated with
\(\chi_a\) and \(\chi_b\) further restricts this residual freedom. Imposing
the mixed-square constraints associated with all three nontrivial characters
simultaneously, one finds\footnote{Here we also use that a matrix commuting with all three
\(\tau_m^{\chi_g}\) is proportional to the identity.}

\begin{equation}
    J_{\mathsf{1}m}\propto\mathbbm{1}_2,
    \qquad
    J_{gm}\propto \tau_m^{\chi_g},
    \qquad
    g=a,b,c.
    \label{eq:BootDQ-JproptoTau}
\end{equation}

The proportionality relations above determine the mixed maps up to nonzero
normalizations and a common change of basis on \(V_m\). Using this remaining
basis freedom, we choose the following canonical form for the matrices \(J_{gm}\):
\begin{equation}\label{eq:BootDQ-Acanonical}
\begin{aligned}
    J_{\mathsf{1}m}&=\mathrm{x}_1
    \begin{pmatrix}1&0\\0&1\end{pmatrix},
    &
    J_{am}&=\mathrm{x}_a
    \begin{pmatrix}0&1\\-1&0\end{pmatrix},
    \\
    J_{bm}&=\mathrm{x}_b
    \begin{pmatrix}0&1\\1&0\end{pmatrix},
    &
    J_{cm}&=\mathrm{x}_c
    \begin{pmatrix}1&0\\0&-1\end{pmatrix},
\end{aligned}
\end{equation}
where $\mathrm{x}_g\in\mathbb C^\ast$ for $g\in\Gamma$.
The maps $J_{mg}$
take the same parametric forms,
with independent nonzero normalizations \(\widetilde{\mathrm{x}}_g\).

In this basis, the eight tensor automorphisms can be arranged into the following four pairs:
\begin{equation}
    (\chi_\mathsf{1},\pm\mathbbm{1}_2),\qquad
(\chi_a,\pm\sqrt{\varepsilon_a}\, \boldsymbol{\sigma}_2),\qquad
    (\chi_b,\pm\sqrt{\varepsilon_b}\, \boldsymbol{\sigma}_1),\qquad
    (\chi_c,\pm\sqrt{\varepsilon_c}\, \boldsymbol{\sigma}_3)\:,
\label{eq:BootDQ-eightautomorphisms}
\end{equation}
where $\boldsymbol{\sigma}_i$ ($i=1,2,3$) are the three Pauli matrices.

It remains to solve the \(m\otimes m\) square. Since the anticommutation signs cancel upon taking tensor squares, the four linearly independent $4\times 4$ matrices
\begin{equation}
    \tau_m^{\chi_g}\otimes\tau_m^{\chi_g}\:,
    \qquad g\in\Gamma,
\end{equation}
commute and have one-dimensional joint eigenspaces. The square equation~\eqref{eq:BootDQ-bigrows} requires each row \(J_{mm}^u\), corresponding to the channel \(u\in\{\mathsf{1},a,b,c\}\), to be the joint left eigenvector with eigenvalues \(\chi_g(u)\):
\begin{equation}
    J_{mm}^u
    \bigl(\tau_m^{\chi_g}\otimes\tau_m^{\chi_g}\bigr)
    =
    \chi_g(u)J_{mm}^u,
    \qquad
    g\in\Gamma.
    \label{eq:BootDQ-joint}
\end{equation}
Thus, \(J_{mm}\) is determined up to independent nonzero rescalings of its four rows. For the four allowed sign triples \((\varepsilon_a,\varepsilon_b,\varepsilon_c)\), one obtains:
\begin{itemize}
\item For
\(
    (\varepsilon_a,\varepsilon_b,\varepsilon_c)=(-1,+1,+1)
\):
\begin{equation}
    J_{mm}^{(1)}
    =
    \begin{pmatrix}
        \mathrm{z}_\mathsf{1}^{(1)}&0&0&\mathrm{z}_\mathsf{1}^{(1)}\\
        0&\mathrm{z}_a^{(1)}&-\mathrm{z}_a^{(1)}&0\\
        0&\mathrm{z}_b^{(1)}&\mathrm{z}_b^{(1)}&0\\
        \mathrm{z}_c^{(1)}&0&0&-\mathrm{z}_c^{(1)}
    \end{pmatrix}.
    \label{eq:BootDQ-B1}
\end{equation}

\item For 
\(
    (\varepsilon_a,\varepsilon_b,\varepsilon_c)=(+1,-1,+1)
\):
\begin{equation}
    J_{mm}^{(2)}
    =
    \begin{pmatrix}
        \mathrm{z}_\mathsf{1}^{(2)}&0&0&-\mathrm{z}_\mathsf{1}^{(2)}\\
        0&\mathrm{z}_a^{(2)}&\mathrm{z}_a^{(2)}&0\\
        0&\mathrm{z}_b^{(2)}&-\mathrm{z}_b^{(2)}&0\\
        \mathrm{z}_c^{(2)}&0&0&\mathrm{z}_c^{(2)}
    \end{pmatrix}.
    \label{eq:BootDQ-B2}
\end{equation}

\item For
\(
    (\varepsilon_a,\varepsilon_b,\varepsilon_c)=(+1,+1,-1)
\):
\begin{equation}
    J_{mm}^{(3)}
    =
    \begin{pmatrix}
        0&\mathrm{z}_\mathsf{1}^{(3)}&\mathrm{z}_\mathsf{1}^{(3)}&0\\
        \mathrm{z}_a^{(3)}&0&0&-\mathrm{z}_a^{(3)}\\
        \mathrm{z}_b^{(3)}&0&0&\mathrm{z}_b^{(3)}\\
        0&\mathrm{z}_c^{(3)}&-\mathrm{z}_c^{(3)}&0
    \end{pmatrix},
    \label{eq:BootDQ-B3}
\end{equation}
\item For
\(
    (\varepsilon_a,\varepsilon_b,\varepsilon_c)=(-1,-1,-1)
\):
\begin{equation}
    J_{mm}^{(4)}
    =
    \begin{pmatrix}
        0&\mathrm{z}_\mathsf{1}^{(4)}&-\mathrm{z}_\mathsf{1}^{(4)}&0\\
        \mathrm{z}_a^{(4)}&0&0&\mathrm{z}_a^{(4)}\\
        \mathrm{z}_b^{(4)}&0&0&-\mathrm{z}_b^{(4)}\\
        0&\mathrm{z}_c^{(4)}&\mathrm{z}_c^{(4)}&0
    \end{pmatrix}.
    \label{eq:BootDQ-B4}
\end{equation}
\end{itemize}
Taking all \(\mathrm{z}_u^{(q)}\neq 0\) (with $q=1,2,3,4$ and \({u\in\{\mathsf{1},a,b,c\}}\)) ensures that \(J_{mm}\) is invertible in each case.

\

Finally, the group structure can be determined directly from the signs \(\varepsilon_g\). An element in the \(\chi_g\) sector satisfies
\begin{equation}
    (\chi_g,\tau_m^{\chi_g})^2
    =
    \begin{cases}
        (\chi_\mathsf{1},\mathbbm{1}_2),&\varepsilon_g=+1\,,\\
        (\chi_\mathsf{1},-\mathbbm{1}_2),&\varepsilon_g=-1\:.
    \end{cases}
    \label{eq:BootDQ-orders}
\end{equation}
Since the three nontrivial \(\tau_m\)-components anticommute, the corresponding tensor automorphisms do not commute, and the group is therefore non-abelian. Moreover, it has order eight, so the only possibilities are \(D_8\) and \(Q_8\).

For solutions \(1\)--\(3\), one nontrivial sector has \(\varepsilon_g=-1\), while the other two have \(\varepsilon_g=+1\). The resulting multiset of element orders is
\begin{equation}
    \{1;\,2,2,2,2,2;\,4,4\} \:,
\end{equation}
which identifies the group as
\begin{equation}
    \operatorname{Aut}^{\otimes}(\mathcal{F})\simeq D_8 \:.
\end{equation}
For solution \(4\), all three \(\varepsilon_g\) are negative, giving the multiset
\begin{equation}
    \{1;\,2;\,4,4,4,4,4,4\} \:.
\end{equation}
This identifies the group as
\begin{equation}
    \operatorname{Aut}^{\otimes}(F)\simeq Q_8 \:.
\end{equation}

In each case, the spaces
\(
    V_\mathsf{1},\ V_a,\ V_b,\ V_c,\ V_m
\)
form a complete set of irreducible representations of \(G\), since their squared dimensions add up to
\(
    4\cdot1^2+2^2=8=|G|.
\)
Moreover, the square equations state precisely that the maps \(J\) intertwine the corresponding \(G\)-actions. The resulting monoidal data therefore define a fiber functor on \(\operatorname{Rep}(G)\). Thus, up to relabeling, the first three solutions reconstruct
\(\operatorname{Rep}(D_8)\), while the fourth reconstructs
\(\operatorname{Rep}(Q_8)\).

For each of the four solutions, the square equations leave the nonzero coefficients \(J_{gh}\), \(\mathrm{x}_g\), \(\widetilde{\mathrm{x}}_g\), and \(\mathrm{z}_g\) unfixed. This freedom follows from the homogeneity of the equations. Since the fusion rules are multiplicity-free, all fusion spaces are one-dimensional, and these coefficients can be modified independently by rescaling the corresponding fusion-space bases. 

\

Summarizing, we obtain four parametric forms and their tensor automorphism groups:
\begin{equation}
\begin{array}{c|c|c}
    \text{solution}
    &(\varepsilon_a,\varepsilon_b,\varepsilon_c)
    &\operatorname{Aut}^\otimes(\mathcal F)
    \\
    \hline
    1&(-,+,+)&D_8\\
    2&(+,-,+)&D_8\\
    3&(+,+,-)&D_8\\
    4&(-,-,-)&Q_8
\end{array}
\label{eq:BootDQ-summary}
\end{equation}
These four forms fall into two orbits under the automorphism group of the pointed subring,\footnote{Throughout, \(\mathbb F_p\) denotes the field with
\(p\) elements, whose additive group is identified with
\(\mathbb Z/p\mathbb Z\). We write
\(\mathbb F_p^\ast=\mathbb F_p\setminus\{0\}\) for its
set of nonzero elements, equipped with multiplication.}
\begin{equation}
    \operatorname{Aut}(\Gamma)
    \simeq\operatorname{GL}(2,\mathbb F_2)
    \simeq S_3,
\end{equation}
which acts by permuting the three nontrivial invertible objects \(a,b,c\).
This action is transitive on the first three sign assignments, whereas
\((-,-,-)\) is fixed and therefore forms a separate orbit.

The four parametric forms are distinguished by a basis-independent invariant
\(\epsilon_q\in\Gamma\). Namely, \(\epsilon_q\) is the unique fusion channel for which
\begin{equation}
    J_{mm}^{\epsilon_q}P=-J_{mm}^{\epsilon_q} \:,
\end{equation}
where \(P\) is the flip on \(V_m\otimes V_m\). For the four solutions, one finds
\begin{equation}
    \epsilon_1=a\,,\qquad
    \epsilon_2=b\,,\qquad
    \epsilon_3=c\,,\qquad
    \epsilon_4=\mathsf{1}\:.
\end{equation}
Thus, the first three solutions form a single orbit under
\(\operatorname{Aut}(\Gamma)\), whereas the fourth forms a separate orbit.

This invariant has a direct group-theoretical interpretation. Since the
\((-1)\)-eigenspace of \(P\) is \(\bigwedge^2V_m\), one has
\(
    \epsilon_q=\det\rho_m.
\)
The nontrivial orbit corresponds to \(D_8\), while the trivial orbit corresponds to \(Q_8\). Further details and the comparison with Tambara's classification \cite{Tambara:2000qlz} are given in Appendix~\ref{app:Tambara-comparison}.

\subsubsection*{Reconstructing the \(F\)-symbols in the Tambara--Yamagami basis}

We now reconstruct the associator from the four solutions obtained above,
following the same procedure as for \(\operatorname{Rep}(S_3)\).
The scalar parameters cannot all be fixed independently by changes of
basis in the representation spaces \(V_x\), since each such transformation
acts simultaneously on several \(J\)-maps. At this stage, however, the
associator has not yet been fixed, and we may also rescale independently
the bases of the one-dimensional fusion spaces, as described above.
These rescalings change the gauge representative of the reconstructed
\(F\)-symbols.

We choose the fusion-space bases so that the resulting \(F\)-symbols are
in the standard Tambara--Yamagami gauge. This requirement fixes the
relative values and signs displayed below. For all four solutions, we take
\(\mathrm{x}_g=1\) for every \(g\in\Gamma\), set
\(\widetilde{\mathrm{x}}_\mathsf{1}=\mathrm{z}_\mathsf{1}^{(r)}=1\), and choose the pointed
maps and the remaining parameters as follows:
\begin{equation}
\begin{gathered}
 (J_{gh})_{g,h\in\Gamma}=
 \begin{pmatrix}
  1&1&1&1\\
  1&-1&1&-1\\
  1&-1&1&-1\\
  1&1&1&1
 \end{pmatrix},
 \qquad
 \begin{array}{c|c|c}
  q&(\widetilde{\mathrm{x}}_a,\widetilde{\mathrm{x}}_b,\widetilde{\mathrm{x}}_c)
   &(\mathrm{z}_a^{(q)},\mathrm{z}_b^{(q)},\mathrm{z}_c^{(q)})\\
  \hline
  1&(-1,\,\,\,\,1,\,\,\,\,1)&(\,\,\,\,1,\,\,\,\,1,\,\,\,\,1)\\
  2&(\,\,\,\,1,-1,\,\,\,\,1)&(-1,-1,\,\,\,\,1)\\
  3&(\,\,\,\,1,\,\,\,\,1,-1)&(\,\,\,\,1,\,\,\,\,1,\,\,\,\,1)\\
  4&(-1,-1,-1)&(-1,-1,\,\,\,\,1)
 \end{array}.
\end{gathered}
\label{eq:BootDQ-TY-normalizations}
\end{equation}
Here rows and columns of the pointed matrix are ordered as
\((\mathsf{1},a,b,c)\); the other parameters refer to
Eq.~\eqref{eq:BootDQ-Acanonical} and
Eqs.~\eqref{eq:BootDQ-B1}--\eqref{eq:BootDQ-B4}.
These values specify a choice of bases in the one-dimensional fusion
spaces; their relative signs are chosen to place the reconstructed
associator in the standard Tambara--Yamagami gauge.
In particular, the nontrivial signs of \(J_{gh}\) form a normalized
two-cocycle and hence give a trivial pointed associator, consistently
with the discussion of fiber functors on \(\mathrm{Vec}_G^\omega\)
following Eq.~\eqref{eq:vecG}.

We compute the \(F\)-symbols by substituting the normalized \(J\)-maps
into Eq.~\eqref{eq:fiber-Fsymb} and solving for its coefficients.
For example, for the triple \((m,m,m)\), with output \(m\) and left
intermediate channel \(\mathsf{1}\), this equation reads
\begin{equation}
 J_{\mathsf{1}m}(J_{mm}^{\mathsf{1}}\otimes\mathbbm{1}_m)
 =
 \sum_{h\in\Gamma}
 [F_{mmm}^{m}]_{\mathsf{1},h}\,
 J_{mh}(\mathbbm{1}_m\otimes J_{mm}^{h}).
 \label{eq:BootDQ-TY-firstrow}
\end{equation}
The four unknown coefficients are precisely the first row of
\(F_{mmm}^{m}\). Substituting the matrices above and comparing their
entries gives \([F_{mmm}^{m}]_{\mathsf{1},h}=+\tfrac12\) for solutions \(1,2,3\)
and \([F_{mmm}^{m}]_{\mathsf{1},h}=-\tfrac12\) for solution \(4\), for every
\(h\in\Gamma\).

The same equation determines the remaining \(F\)-symbols.
In particular, for the triple \((g,m,h)\) with output \(m\), it reduces to
\begin{equation}
J_{mh}J_{gm}=[F_{gmh}^{m}]_{m,m}\,J_{gm}J_{mh} \:.
\end{equation}
Thus \([F_{gmh}^{m}]_{m,m}\) is \(+1\) when the two mixed maps commute
and \(-1\) when they anticommute.

Repeating this calculation for all admissible channels, we obtain
the following \(F\)-symbols in the Tambara--Yamagami basis:
\begin{equation}
 {}[F_{ghk}^{ghk}]_{gh,hk}
   =[F_{ghm}^{m}]_{gh,m}
   =[F_{mgh}^{m}]_{m,gh}=
 [F_{gmm}^{h}]_{m,gh}
   =[F_{mmg}^{h}]_{hg,m}=1\:,
   \label{eq:BootDQ-TY-Fsymbols1}
\end{equation}
and
\begin{equation}
 [F_{gmh}^{m}]_{m,m}
   =[F_{mgm}^{h}]_{m,m}=\xi(g,h)\,,\qquad\qquad
 [F_{mmm}^{m}]_{g,h}=\frac{\tau_{\rm TY}}{\xi(g,h)}\:.
\label{eq:BootDQ-TY-Fsymbols2}
\end{equation}
Here \(g,h,k\in\Gamma\), and \(\tau_{\rm TY}=[F_{mmm}^{m}]_{\mathsf{1},\mathsf{1}}\) is
\(+\tfrac12\) for solutions \(1,2,3\) and \(-\tfrac12\) for solution \(4\).
The signs computed in the mixed sectors give the same bicharacter
in all four solutions:
\(\xi(a^{x_1}b^{x_2},a^{y_1}b^{y_2})
=(-1)^{x_1y_2+x_2y_1}\), with \(x_i,y_i\in\mathbb Z_2\).
This bicharacter is symmetric, nondegenerate and alternating:
\(\xi(g,g)=1\) for every \(g\in\Gamma\), whereas
\(\xi(g,h)=-1\) for distinct nontrivial \(g,h\).
Thus the only difference between the associators of the first three
solutions and that of the fourth is the overall sign of \(F_{mmm}^{m}\).

By the Tambara--Yamagami classification \cite{Tambara:1998vmj},
the pair \((\xi,\tau_{\rm TY})\) determines the fusion category up to isometry
of \(\xi\). Our results give
\begin{equation}
 (\xi,\tau_{\rm TY})=
 \begin{cases}
  (\xi_{\mathrm{alt}},+\tfrac12),
     &q=1,2,3:\quad\operatorname{Rep}(D_8),\\
  (\xi_{\mathrm{alt}},-\tfrac12),
     &q=4:\quad\operatorname{Rep}(Q_8),
 \end{cases}
 \label{eq:BootDQ-TY-invariants}
\end{equation}
consistently with the groups reconstructed from
\(\operatorname{Aut}^{\otimes}(\mathcal F)\).
The three \(D_8\) solutions give exactly the same \(F\)-symbols
in this basis. Moreover, since \(\xi_{\mathrm{alt}}\) is invariant
under every permutation of \(a,b,c\), the \(F\)-symbols in~\eqref{eq:BootDQ-TY-Fsymbols1} and \eqref{eq:BootDQ-TY-Fsymbols2} are invariant under all
automorphisms of the fusion ring.

\subsection{The non-maximal case: \texorpdfstring{$\operatorname{Rep}(H)$}{Rep(H)} categories}
\label{ssec:BootDQ-Hopf}

We now return to the square equations before imposing the requirement
that all three nontrivial characters of \(\Gamma\) extend to tensor
automorphisms. The constraints derived in Section~\ref{ssec:BootDQ-common}
continue to hold. In particular, the
trivial character sector contains precisely
\((\chi_\mathsf{1},\pm\mathbbm{1}_2)\), and each nontrivial character that extends
admits two extensions, with traceless components satisfying
\((\tau_m^{\chi_g})^2=\varepsilon_g\mathbbm{1}_2\),
where \(\varepsilon_g=\pm1\).

We require only that one nontrivial character extend and, since the
three choices are related by relabeling \(a,b,c\), take it to be
\(\chi_c\). We solve the corresponding square equations without assuming extensions of
\(\chi_a\) and \(\chi_b\). Since the equations for a single character leave
part of \(J\) unconstrained, we then determine which candidate monoidal
structures admit compatible \(F\)-symbols. Only at this stage is the full
tensor automorphism group fixed, since the other characters may also extend.

Since \(\tau_m^{\chi_c}\) is traceless and invertible, it has two
distinct eigenvalues. We can therefore choose a basis of \(V_m\)
in which
\begin{equation}
 \tau_m^{\chi_c}
 =\pm\sqrt{\varepsilon_c}
 \begin{pmatrix}1&0\\0&-1\end{pmatrix},
 \qquad \varepsilon_c=\pm1 \: .
 \label{eq:BootH-tau}
\end{equation}
The mixed square equations require \(J_{\mathsf{1}m},J_{m\mathsf{1}},J_{cm},J_{mc}\)
to commute with \(\tau_m^{\chi_c}\), whereas
\(J_{am},J_{ma},J_{bm},J_{mb}\) must anticommute with it.
In the chosen basis, these maps therefore take the form
\begin{equation}
\begin{aligned}
 J_{\mathsf{1}m}&=\begin{pmatrix}\mathrm{x}_{\mathsf{1},1}&0\\0&\mathrm{x}_{\mathsf{1},2}\end{pmatrix},
 &J_{m\mathsf{1}}&=\begin{pmatrix}\widetilde{\mathrm{x}}_{1,1}&0\\0&\widetilde{\mathrm{x}}_{1,2} \end{pmatrix},\\
 J_{cm}&=\begin{pmatrix}\mathrm{x}_{c,1}&0\\0&\mathrm{x}_{c,2}\end{pmatrix},
 &J_{mc}&=\begin{pmatrix}\widetilde{\mathrm{x}}_{c,1}&0\\0&\widetilde{\mathrm{x}}_{c,2}\end{pmatrix},\\
 J_{am}&=\begin{pmatrix}0&\mathrm{x}_{a,1}\\\mathrm{x}_{a,2}&0\end{pmatrix},
 &J_{ma}&=\begin{pmatrix}0&\widetilde{\mathrm{x}}_{a,1}\\ \widetilde{\mathrm{x}}_{a,2}&0\end{pmatrix},\\
 J_{bm}&=\begin{pmatrix}0&\mathrm{x}_{b,1}\\\mathrm{x}_{b,2}&0\end{pmatrix},
 &J_{mb}&=\begin{pmatrix}0&\widetilde{\mathrm{x}}_{b,1}\\ \widetilde{\mathrm{x}}_{b,2}&0\end{pmatrix}.
\end{aligned}
\label{eq:basisVm_1}
\end{equation}
All parameters displayed above are independent and nonzero.\footnote{At this stage, the square equations do not
require \(J_{\mathsf{1}m}\) or \(J_{m\mathsf{1}}\) to be scalar.}
Moreover, since
\(\tau_m^{\chi_c}\otimes\tau_m^{\chi_c}
=\varepsilon_c\operatorname{diag}(1,-1,-1,1)\),
Eq.~\eqref{eq:BootDQ-bigrows} gives
\begin{equation}
\begin{aligned}
 J_{mm}
 &=\begin{pmatrix}
 \mathrm{z}_{\mathsf{1},1}&0&0&\mathrm{z}_{\mathsf{1},2}\\
 0&\mathrm{z}_{a,1}&\mathrm{z}_{a,2}&0\\
 0&\mathrm{z}_{b,1}&\mathrm{z}_{b,2}&0\\
 \mathrm{z}_{c,1}&0&0&\mathrm{z}_{c,2}
 \end{pmatrix},
 &&\qquad\varepsilon_c=+1,\\
 J_{mm}
 &=\begin{pmatrix}
0&\mathrm{z}_{\mathsf{1},1}&\mathrm{z}_{\mathsf{1},2}&0\\
\mathrm{z}_{a,1}&0&0&\mathrm{z}_{a,2}\\
\mathrm{z}_{b,1}&0&0&\mathrm{z}_{b,2}\\
0&\mathrm{z}_{c,1}&\mathrm{z}_{c,2}&0
\end{pmatrix},
 &&\qquad\varepsilon_c=-1.
\end{aligned}
\label{eq:BootH-Jmm-initial}
\end{equation}
All these matrices are required to be invertible.
The pointed maps \(J_{gh}\) remain arbitrary nonzero scalars.

The square equations associated with the extension of \(\chi_c\)
leave the parameters displayed above unconstrained. We now determine which choices admit
\(F\)-symbols compatible with Eq.~\eqref{eq:fiber-Fsymb}.

We first consider the maps involving the unit object.
The sectors \((\mathsf{1},\mathsf{1},m)\) and \((m,\mathsf{1},\mathsf{1})\), both with output \(m\), give
\begin{equation}
 J_{\mathsf{1}\mathsf{1}}J_{\mathsf{1}m}
 = [F_{\mathsf{1}\mathsf{1}m}^{m}]_{\mathsf{1},m}\,(J_{\mathsf{1}m})^2\qquad\mbox{ and }\qquad
 (J_{m\mathsf{1}})^2
 = [F_{m\mathsf{1}\mathsf{1}}^{m}]_{m,\mathsf{1}}\,J_{\mathsf{1}\mathsf{1}}J_{m\mathsf{1}} \:.
\label{eq:BootH-unit-compatibility}
\end{equation}
Since the maps are invertible and the displayed \(F\)-coefficients
are nonzero scalars, these equations imply
\begin{equation}
 J_{\mathsf{1}m}
 =\frac{J_{\mathsf{1}\mathsf{1}}}{[F_{\mathsf{1}\mathsf{1}m}^{m}]_{\mathsf{1},m}}\,\mathbbm{1}_m \,,
 \qquad
 J_{m\mathsf{1}}
 =[F_{m\mathsf{1}\mathsf{1}}^{m}]_{m,\mathsf{1}}J_{\mathsf{1}\mathsf{1}}\,\mathbbm{1}_m \:.
\end{equation}
Thus, \(\mathrm{x}_{\mathsf{1},1}=\mathrm{x}_{\mathsf{1},2}\) and \(\widetilde{\mathrm{x}}_{\mathsf{1},1}=\widetilde{\mathrm{x}}_{\mathsf{1},2}\).
We absorb these nonzero factors into the corresponding
fusion-space bases and henceforth set
\(J_{\mathsf{1}m}=J_{m\mathsf{1}}=\mathbbm{1}_{V_m}\).

Next, for the triple \((h,m,m)\) with output \(g\),
Eq.~\eqref{eq:fiber-Fsymb} reads
\begin{equation}
 J_{mm}^{g}(J_{hm}\otimes\mathbbm{1}_m)
 =
 [F_{hmm}^{g}]_{m,hg}\,
 J_{h,hg}\,J_{mm}^{hg} \,,
 \qquad g,h\in\Gamma \:.
 \label{eq:BootH-row-compatibility}
\end{equation}
The right intermediate channel is \(hg\), as \(h(hg)=g\).
The unknown coefficient is \([F_{hmm}^{g}]_{m,hg}\).
Since \(J_{h,hg}\) is a nonzero scalar, this coefficient exists
if and only if
\begin{equation}
 J_{mm}^{g}(J_{hm}\otimes\mathbbm{1}_m)
 \;\propto\;
 J_{mm}^{hg} \:.
 \label{Eq:proportion-p2RepH}
\end{equation}
We now analyze these proportionality conditions separately
for the two cases \(\varepsilon_c=+1\) and \(\varepsilon_c=-1\).

\paragraph{The case \(\varepsilon_c=+1\).}
The rows of \(J_{mm}\) labelled by \(\mathsf{1},c\)
can have nonzero entries only in the first and fourth columns,
while those labelled by \(a,b\) can have nonzero entries only
in the second and third columns,
see Eq.~\eqref{eq:BootH-Jmm-initial}.
The \((c,m,m)\) proportionality conditions pair the rows
\(\mathsf{1}\leftrightarrow c\) and \(a\leftrightarrow b\).
Using this form of \(J_{mm}\), together with invertibility,
we obtain
\begin{equation}
 \mathrm{x}_{c,2}=-\mathrm{x}_{c,1}\,,\qquad
 \mathrm{z}_{\mathsf{1},1}\mathrm{z}_{c,2}+\mathrm{z}_{\mathsf{1},2}\mathrm{z}_{c,1}=0\,,\qquad
 \mathrm{z}_{a,1}\mathrm{z}_{b,2}+\mathrm{z}_{a,2}\mathrm{z}_{b,1}=0\:.
 \label{eq:BootH-c-constraints}
\end{equation}
Together with invertibility, these relations force all
\(\mathrm{z}\)-parameters to be nonzero.
We retain the parametrization in
Eq.~\eqref{eq:BootH-Jmm-initial}, subject to
Eq.~\eqref{eq:BootH-c-constraints}.

The \((a,m,m)\) and \((b,m,m)\) sectors constrain the remaining
left mixed maps. For instance, the output \(\mathsf{1}\) in the
\((a,m,m)\) sector requires
\begin{equation}
 J_{mm}^{\mathsf{1}}(J_{am}\otimes\mathbbm{1}_m)
 =(0,\mathrm{z}_{\mathsf{1},2}\mathrm{x}_{a,2},\mathrm{z}_{\mathsf{1},1}\mathrm{x}_{a,1},0)
 \;\propto\;
 J_{mm}^{a}=(0,\mathrm{z}_{a,1},\mathrm{z}_{a,2},0) \:.
\end{equation}
Using all output channels, and similarly imposing compatibility
in the \((m,m,a)\) and \((m,m,b)\) sectors for the right mixed
maps, gives
\begin{equation}
\begin{aligned}
 \mathrm{z}_{\mathsf{1},1}\mathrm{z}_{a,1}\mathrm{x}_{a,1}&=\mathrm{z}_{\mathsf{1},2}\mathrm{z}_{a,2}\mathrm{x}_{a,2}\:,
 &\qquad\mathrm{z}_{c,1}\mathrm{z}_{b,1}\mathrm{x}_{a,1}&=\mathrm{z}_{c,2}\mathrm{z}_{b,2}\mathrm{x}_{a,2}\:,\\
 \mathrm{z}_{\mathsf{1},1}\mathrm{z}_{b,1}\mathrm{x}_{b,1}&=\mathrm{z}_{\mathsf{1},2}\mathrm{z}_{b,2}\mathrm{x}_{b,2}\:,
 &\qquad\mathrm{z}_{c,1}\mathrm{z}_{a,1}\mathrm{x}_{b,1}&=\mathrm{z}_{c,2}\mathrm{z}_{a,2}\mathrm{x}_{b,2}\:,\\
 \mathrm{z}_{\mathsf{1},1}\mathrm{z}_{a,2}\widetilde{\mathrm{x}}_{a,1}&=\mathrm{z}_{\mathsf{1},2}\mathrm{z}_{a,1}\widetilde{\mathrm{x}}_{a,2}\:,
 &\qquad\mathrm{z}_{c,1}\mathrm{z}_{b,2}\widetilde{\mathrm{x}}_{a,1}&=\mathrm{z}_{c,2}\mathrm{z}_{b,1}\widetilde{\mathrm{x}}_{a,2}\:,\\
 \mathrm{z}_{\mathsf{1},1}\mathrm{z}_{b,2}\widetilde{\mathrm{x}}_{b,1}&=\mathrm{z}_{\mathsf{1},2}\mathrm{z}_{b,1}\widetilde{\mathrm{x}}_{b,2}\:,
 &\qquad\mathrm{z}_{c,1}\mathrm{z}_{a,2}\widetilde{\mathrm{x}}_{b,1}&=\mathrm{z}_{c,2}\mathrm{z}_{a,1}\widetilde{\mathrm{x}}_{b,2}\:.
\end{aligned}
\label{eq:BootH-mixed-constraints}
\end{equation}
The \((m,m,c)\) sector also gives \(\mathrm{x}_{c,2}=-\mathrm{x}_{c,1}\).
In each line of Eq.~\eqref{eq:BootH-mixed-constraints},
the two relations are equivalent once
Eq.~\eqref{eq:BootH-c-constraints} is imposed.
These relations are homogeneous linear equations in
\((\mathrm{x}_{a,1},\mathrm{x}_{a,2})\), \((\mathrm{x}_{b,1},\mathrm{x}_{b,2})\), \((\widetilde{\mathrm{x}}_{a,1},\widetilde{\mathrm{x}}_{a,2})\), and \((\widetilde{\mathrm{x}}_{b,1},\widetilde{\mathrm{x}}_{b,2})\),
with coefficients given by the entries of \(J_{mm}\).
Each pair satisfies one independent equation with nonzero
coefficients, which determines the ratio of its entries
and leaves an overall nonzero scalar free.

We next impose Eq.~\eqref{eq:fiber-Fsymb} on the triples
\((g,m,h)\), with \(g,h\in\Gamma\) and output \(m\). In these sectors,
\begin{equation}
    J_{mh}J_{gm}
    =
    [F_{gmh}^{m}]_{m,m}\,J_{gm}J_{mh} \:.
    \label{eq:BootH-mixed-commutator}
\end{equation}
Thus the two products must be proportional. Consider first \(g=h=a\).
Using the matrices in Eq.~\eqref{eq:basisVm_1}, we find
\begin{equation}
    J_{ma}J_{am}
    =
    \begin{pmatrix}
        \widetilde{\mathrm{x}}_{a,1}\mathrm{x}_{a,2} & 0\\
        0 & \widetilde{\mathrm{x}}_{a,2}\mathrm{x}_{a,1}
    \end{pmatrix} \,,
    \qquad
    J_{am}J_{ma}
    =
    \begin{pmatrix}
        \mathrm{x}_{a,1}\widetilde{\mathrm{x}}_{a,2} & 0\\
        0 & \mathrm{x}_{a,2}\widetilde{\mathrm{x}}_{a,1}
    \end{pmatrix} \:.
\end{equation}
These matrices are proportional if and only if $(\mathrm{x}_{a,2}\widetilde{\mathrm{x}}_{a,1})^2=(\mathrm{x}_{a,1}\widetilde{\mathrm{x}}_{a,2})^2$.
The corresponding relations in
Eq.~\eqref{eq:BootH-mixed-constraints} give
\begin{equation}
    \frac{\mathrm{x}_{a,2}}{\mathrm{x}_{a,1}}
    =
    \frac{\mathrm{z}_{\mathsf{1},1}\mathrm{z}_{a,1}}
         {\mathrm{z}_{\mathsf{1},2}\mathrm{z}_{a,2}}\,,
    \qquad
    \frac{\widetilde{\mathrm{x}}_{a,2}}
         {\widetilde{\mathrm{x}}_{a,1}}
    =
    \frac{\mathrm{z}_{\mathsf{1},1}\mathrm{z}_{a,2}}
         {\mathrm{z}_{\mathsf{1},2}\mathrm{z}_{a,1}}\,,
\qquad
\mbox{and hence}
\qquad
    \frac{\mathrm{x}_{a,2}\widetilde{\mathrm{x}}_{a,1}}
         {\mathrm{x}_{a,1}\widetilde{\mathrm{x}}_{a,2}}=\frac{\mathrm{z}_{a,1}^{\,2}}
              {\mathrm{z}_{a,2}^{\,2}}\:.
\end{equation}
Putting all together, we obtain 
\(\mathrm{z}_{a,1}^{\,4}=\mathrm{z}_{a,2}^{\,4}\), i.e. 
\begin{equation}
    \zeta^4=1\,, \qquad \mbox{with} \qquad \zeta:=
    \frac{\mathrm{z}_{a,2}}{\mathrm{z}_{a,1}} \:.
    \label{eq:BootH-zeta}
\end{equation}
Using Eqs.~\eqref{eq:BootH-c-constraints}
and~\eqref{eq:BootH-mixed-constraints}, one verifies that all remaining
choices of \(g,h\) impose no additional conditions.

We now consider the residual change of basis
\(\varphi_m=\operatorname{diag}(t_1,t_2)\), which preserves
the diagonal form of \(\tau_m^{\chi_c}\) chosen in
Eq.~\eqref{eq:BootH-tau}.
With the convention of Eq.~\eqref{eq:fiber_equiv_base}, $\varphi_m$ acts as
\begin{equation}
 \frac{\mathrm{z}_{\mathsf{1},2}}{\mathrm{z}_{\mathsf{1},1}}\longmapsto
 \left(\frac{t_1}{t_2}\right)^2\frac{\mathrm{z}_{\mathsf{1},2}}{\mathrm{z}_{\mathsf{1},1}},
 \qquad
 \frac{\mathrm{z}_{a,2}}{\mathrm{z}_{a,1}}\longmapsto\frac{\mathrm{z}_{a,2}}{\mathrm{z}_{a,1}}.
 \label{eq:BootH-residual-basis}
\end{equation}
We may then conveniently choose \(\mathrm{z}_{\mathsf{1},2}/\mathrm{z}_{\mathsf{1},1}=-1/\zeta\).
Absorbing the remaining nonzero factors into the normalizations
of the individual fusion maps gives
\begin{equation}
\begin{aligned}
 J_{cm}&=\mathrm{x}_c\begin{pmatrix}1&0\\0&-1\end{pmatrix},
 &J_{am}&=\mathrm{x}_a\begin{pmatrix}0&1\\-1&0\end{pmatrix},
 &J_{bm}&=\mathrm{x}_b\begin{pmatrix}0&1\\1&0\end{pmatrix},\\
 J_{mc}&=\widetilde{\mathrm{x}}_c\begin{pmatrix}1&0\\0&-1\end{pmatrix},
 &J_{ma}&=\widetilde{\mathrm{x}}_a
          \begin{pmatrix}0&1\\-\zeta^2&0\end{pmatrix},
 &J_{mb}&=\widetilde{\mathrm{x}}_b
          \begin{pmatrix}0&1\\\zeta^2&0\end{pmatrix}.
\end{aligned}
\label{eq:BootH-mixed-final}
\end{equation}
The same choice of basis, together with
Eq.~\eqref{eq:BootH-c-constraints} and the definition of \(\zeta\),
gives
\begin{equation}
 \mathrm{z}_{\mathsf{1},2}=-\frac{\mathrm{z}_{\mathsf{1},1}}{\zeta},
 \qquad
 \mathrm{z}_{c,2}=-\frac{\mathrm{z}_{\mathsf{1},2}\mathrm{z}_{c,1}}{\mathrm{z}_{\mathsf{1},1}}=\frac{\mathrm{z}_{c,1}}{\zeta},
 \qquad
 \mathrm{z}_{a,2}=\zeta \mathrm{z}_{a,1},
 \qquad
 \mathrm{z}_{b,2}=-\frac{\mathrm{z}_{a,2}\mathrm{z}_{b,1}}{\mathrm{z}_{a,1}}=-\zeta \mathrm{z}_{b,1}.
\end{equation}
Renaming the remaining nonzero parameters as
\(\mathrm{z}_\mathsf{1}:=\mathrm{z}_{\mathsf{1},1}\), \(\mathrm{z}_a:=\mathrm{z}_{a,1}\), \(\mathrm{z}_b:=\mathrm{z}_{b,1}\), and \(\mathrm{z}_c:=\mathrm{z}_{c,1}\),
we obtain
\begin{equation}
 J_{mm}=
 \begin{pmatrix}
 \mathrm{z}_\mathsf{1}&0&0&-\mathrm{z}_\mathsf{1}/\zeta\\
 0&\mathrm{z}_a&\zeta \mathrm{z}_a&0\\
 0&\mathrm{z}_b&-\zeta\mathrm{z}_b&0\\
 \mathrm{z}_c&0&0&\mathrm{z}_c/\zeta
 \end{pmatrix}.
 \label{eq:BootH-Jmm-plus}
\end{equation}
Substitution into Eq.~\eqref{eq:fiber-Fsymb} shows that,
after the preceding constraints have been imposed, the
\((m,m,m)\) sector introduces no further restriction.
Hence all four values \(\zeta=\pm1,\pm i\) are allowed.
The nonzero scalar parameters
\(\mathrm{x}_a, \widetilde{\mathrm{x}}_a,\mathrm{x}_b, \widetilde{\mathrm{x}}_b,\mathrm{x}_c, \widetilde{\mathrm{x}}_c,
\mathrm{z}_\mathsf{1},\mathrm{z}_a,\mathrm{z}_b,\mathrm{z}_c\), together with the pointed coefficients
\(J_{gh}\) for \(g,h\neq\mathsf{1}\), can be changed independently
by rescaling the corresponding fusion-space bases.

\paragraph{The case \(\varepsilon_c=-1\).}
Applying the same procedure to the \(\varepsilon_c=-1\) matrix in
Eq.~\eqref{eq:BootH-Jmm-initial} yields the mixed maps in
Eq.~\eqref{eq:BootH-mixed-final}. In this branch, we define
\(
    \zeta:=\frac{\mathrm{z}_{\mathsf{1},2}}{\mathrm{z}_{\mathsf{1},1}}
\)
and use the residual basis freedom to set
\begin{equation}
    \frac{\mathrm{z}_{a,2}}{\mathrm{z}_{a,1}}
    =-\frac{1}{\zeta} \:.
\end{equation}
Renaming \(\mathrm{z}_u:=\mathrm{z}_{u,1}\) for \(u\in\Gamma\), we obtain
\begin{equation}
 J_{mm}=
 \begin{pmatrix}
0&\mathrm{z}_\mathsf{1}&\zeta\mathrm{z}_\mathsf{1}&0\\
\mathrm{z}_a&0&0&-\mathrm{z}_a/\zeta\\
\mathrm{z}_b&0&0&\mathrm{z}_b/\zeta\\
0&\mathrm{z}_c&-\zeta\mathrm{z}_c&0
\end{pmatrix},
\qquad
\mathrm{z}_u\in\mathbb C^\ast.
\label{eq}
\end{equation}
The sectors with at most two copies of \(m\) again require
\(\zeta^4=1\). In this branch, however, the \((m,m,m)\) equation
admits a solution for its \(F\)-coefficients only when
\(\zeta^2=1\). Thus only \(\zeta=\pm1\) remain.

\paragraph{Identification of the solutions.}
For \(\varepsilon_c=+1\), the values \(\zeta=-1,+1\) recover
respectively the \(D_8\) solutions in
Eqs.~\eqref{eq:BootDQ-B1} and~\eqref{eq:BootDQ-B2}.
For \(\varepsilon_c=-1\), the values \(\zeta=+1,-1\) recover
respectively the \(D_8\) and \(Q_8\) solutions in
Eqs.~\eqref{eq:BootDQ-B3} and~\eqref{eq:BootDQ-B4}.
In these cases the other nontrivial characters also extend,
and the full tensor automorphism group has order eight.

The new solutions are therefore
\(\varepsilon_c=+1\), \(\zeta=\pm i\).
The change of basis
\(\varphi_m=\left(\begin{smallmatrix}0&i\\i&0\end{smallmatrix}\right)\)
exchanges \(\zeta\) with \(\zeta^{-1}\), with corresponding
redefinitions of the free scalar parameters.
The resulting fiber functors are monoidally isomorphic;
thus the two values of \(\zeta\) do not give rise to
inequivalent fiber functors.

\paragraph{Reconstructing the \(F\)-symbols.}
We now compute the \(F\)-symbols from
Eq.~\eqref{eq:fiber-Fsymb} and identify the resulting fusion
category through its Tambara--Yamagami invariants.
We choose \(\zeta=i\) and normalize the \(J\)-maps to obtain
the \(F\)-symbols in the same Tambara--Yamagami basis used
in Section~\ref{Sec:maximalCaseTY}.
Explicitly, we set
\begin{equation}
 \mathrm{x}_c=\widetilde{\mathrm{x}}_c=\mathrm{z}_\mathsf{1}=\mathrm{z}_c=\mathrm{x}_a=\mathrm{x}_b=1\,,\qquad
 \mathrm{z}_a=\mathrm{z}_b=i\,,\qquad
 \widetilde{\mathrm{x}}_a=-i\,,\qquad
 \widetilde{\mathrm{x}}_b=i\,,
 \label{eq:BootH-TY-normalizations}
\end{equation}
with the pointed maps \(J_{gh}\) chosen as in
Eq.~\eqref{eq:BootDQ-TY-normalizations}.
Computing the \(F\)-symbols from Eq.~\eqref{eq:fiber-Fsymb}
then gives exactly the Tambara--Yamagami form
in Eqs.~\eqref{eq:BootDQ-TY-Fsymbols1}
and~\eqref{eq:BootDQ-TY-Fsymbols2}, now with
\begin{equation}
 \xi(a^{x_1}b^{x_2},a^{y_1}b^{y_2})
 =(-1)^{x_1y_1+x_2y_2} \,,
 \qquad
 \tau_{\rm TY}=+\frac12 \:.
 \label{eq:BootH-TY-invariants}
\end{equation}
This bicharacter is symmetric and nondegenerate but not alternating:
\(\xi(a,a)=\xi(b,b)=-1\).
The resulting pair \((\xi,\tau)\) identifies the category as
\(\operatorname{Rep}(H_8)\), where \(H_8\) is the
Kac--Paljutkin Hopf algebra \cite{Tambara:1998vmj}.

For this solution, the full tensor automorphism group is
\begin{equation}
 \operatorname{Aut}^{\otimes}(\mathcal F)
 =
 \left\{
 (\chi_\mathsf{1},\pm\mathbbm{1}_2)\,,\
 \left(\chi_c,\pm
 \begin{pmatrix}1&0\\0&-1\end{pmatrix}\right)
 \right\}
 \simeq\mathbb Z_2\times\mathbb Z_2 \:.
 \label{eq:BootH-automorphisms}
\end{equation}
This group agrees with the group-like elements of \(H_8\).
Thus the additional associator constraints recover a
non-group representation category from the same fusion ring.

\paragraph{Choosing a different nontrivial character.}
Recall that \(\operatorname{Aut}(\Gamma)\simeq S_3\), already
used to compare the \(\operatorname{Rep}(G)\) solutions,
permutes \(a,b,c\) and preserves the full fusion ring.
It also permutes the nontrivial characters.
Thus the choices of extending \(\chi_a\), \(\chi_b\), or
\(\chi_c\) are related by relabeling the square equations
and their solutions.
We now examine how these relabelings act on the
\(F\)-symbols of the \(H_8\) solutions.
Let \(\xi_g\) denote the bicharacter obtained when the
extending nontrivial character is \(\chi_g\).
Starting from \(\xi_c=\xi\) in
Eq.~\eqref{eq:BootH-TY-invariants}, the transpositions
\(a\leftrightarrow c\) and \(b\leftrightarrow c\) give
\(\xi_a\) and \(\xi_b\), respectively.
Their values on the generators are
\begin{equation}
\begin{array}{c|ccc}
 g & \xi_g(a,a) & \xi_g(a,b) & \xi_g(b,b)\\
 \hline
 c & -1 & +1 & -1\\
 a & +1 & -1 & -1\\
 b & -1 & -1 & +1
\end{array},
\qquad \tau_{\rm TY}=+\frac12.
\label{eq:BootH-relabelled-bicharacters}
\end{equation}
Symmetry and multiplicativity determine all remaining values.

These three associators cannot be related by changes of basis
in the fusion spaces while keeping the object labels unchanged.
Indeed, under a fusion-space basis transformation,
\begin{equation}
 [\widetilde F_{gmh}^{m}]_{m,m}
 =
 \frac{M_{gm}^{m}M_{mh}^{m}}
      {M_{mh}^{m}M_{gm}^{m}}
 [F_{gmh}^{m}]_{m,m}
 =
 [F_{gmh}^{m}]_{m,m}\:.
\end{equation}
Thus the coefficients
\([F_{gmh}^{m}]_{m,m}=\xi(g,h)\) are gauge invariant
and distinguish the three representatives.

The three representatives nevertheless form a single orbit
under automorphisms of the fusion ring and describe the same
fusion category \(\operatorname{Rep}(H_8)\).
Applying the relabelings simultaneously to the \(F\)-symbols
and the \(J\)-maps, as in Eq.~\eqref{eq:fiber_relabeling},
also identifies the constructed fiber functors after transport
to a common associator representative.
The three choices therefore do not yield three inequivalent
fiber functors on that representative.
This distinction between labeled associators and fiber functors
on a chosen representative will reappear in
Section~\ref{ssec:Bootp3-parametric-comparison}.

\subsection{Counting fiber functors}
\label{ssec:BootTY-fiber-count}

The preceding subsections determine the admissible parametric
forms of the \(J\)-maps and reconstruct their associators.
It remains to determine how many inequivalent fiber functors
these forms describe on each resulting fusion category.
As explained in Section~\ref{ssec:bootstrap-code}, this requires
choosing a common representative of the associator and comparing
the compatible \(J\)-maps up to monoidal isomorphism.

We therefore use the TY \(F\)-symbols obtained above and solve
Eq.~\eqref{eq:fiber-Fsymb} for the scalar parameters in each
parametric form, keeping these \(F\)-symbols unchanged.

Let \(\widetilde J\) denote the normalized representative and write
\begin{equation}
 J_{xy}^{z}=\lambda_{xy}^{z}\widetilde J_{xy}^{z},
 \qquad \lambda_{xy}^{z}\in\mathbb C^\ast \:.
\end{equation}
The unit maps remain normalized to identities.
Since every admissible entry of the TY \(F\)-symbols is nonzero,
comparison of the independent right-channel maps in
Eq.~\eqref{eq:fiber-Fsymb} gives
\begin{equation}
 \lambda_{xy}^{e}\lambda_{ez}^{w}
 =
 \lambda_{yz}^{f}\lambda_{xf}^{w}
 \qquad
 \text{for all admissible channels }e,f \:.
 \label{eq:BootTY-scalar-compatibility}
\end{equation}
The sectors \((g,h,m)\), \((g,m,m)\), and \((m,m,g)\)
determine these factors in the form
\begin{equation}
 \lambda_{gh}^{gh}=\frac{t_{gh}}{t_gt_h}\,,
 \qquad
 \lambda_{gm}^{m}=\lambda_{mg}^{m}=t_g^{-1}\,,
 \qquad
 \lambda_{mm}^{g}=\frac{t_g}{t_m^2}\,,
 \qquad t_1=1\:.
 \label{eq:BootTY-scalar-solutions}
\end{equation}
with \(t_a,t_b,t_c,t_m\in\mathbb C^\ast\).
These expressions also satisfy all remaining sectors.
They are precisely the factors induced by
\(\varphi_x=t_x \mathbbm{1}_{V_x}\) in
Eq.~\eqref{eq:fiber_equiv_base}.
Thus \emph{each parametric} form contributes \emph{one fiber functor}
up to monoidal isomorphism.

The three \(D_8\) forms share the same associator but remain
inequivalent: their antisymmetric channels
\(\epsilon_r=a,b,c\) are invariant under changes of basis
in the spaces \(V_x\).
Since the \(D_8\) associator is invariant under permutations
of \(a,b,c\), these relabelings define monoidal autoequivalences
of \(\operatorname{Rep}(D_8)\).
They act on fiber functors by precomposition, permuting
transitively the three monoidal-isomorphism classes.
Thus the three fiber functors form a single orbit under
autoequivalences, while remaining pairwise non-isomorphic
as monoidal functors.
For \(H_8\), the choices \(\zeta=\pm i\) and the different
extending characters give one class after transport to
the associator in Eq.~\eqref{eq:BootH-TY-invariants}.
Hence the numbers of inequivalent fiber functors are
\begin{equation}
 \operatorname{Rep}(D_8):\ 3\,,
 \qquad
 \operatorname{Rep}(Q_8):\ 1\,,
 \qquad
 \operatorname{Rep}(H_8):\ 1\:.
\end{equation}

For the group categories, these numbers agree with the
classification by conjugacy classes of pairs \((K,[\psi])\),
where \(K\leq G\) and \([\psi]\) is nondegenerate
\cite{ostrik2006modulecategoriesdrinfelddouble}.
For \(D_8\), the trivial subgroup and the two Klein four
subgroups contribute one class each; for \(Q_8\), only the
trivial subgroup contributes.
The \(H_8\) count agrees with Tambara's classification
\cite{Tambara:2000qlz}.

Finally, the bootstrap recovers three of the four fusion categories with
the fusion ring $\mathrm{TY}(\mathbb Z_2\times\mathbb Z_2)$. The remaining
one has the same bicharacter as $\operatorname{Rep}(H_8)$ and the opposite
sign of $\tau$, and is the unique category with these fusion rules that is
not the representation category of a semisimple Hopf algebra
\cite{Tambara:1998vmj,Tambara:2000qlz}. It therefore admits no fiber
functor, and its absence from our list is due to the fact that the square equations can only produce categories for which a fiber
functor exists.

\section{Extraspecial groups: fiber functors and \texorpdfstring{$F$}{F}-symbols}\label{sec:extraspecial}

We now extend the analysis of Section~\ref{sec:TYZ2}
to fusion rings with nontrivial fusion multiplicities.
These arise from the representation categories of extraspecial
groups of order \(p^3\), with \(p\) an odd prime.
We consider the fusion ring with $p^2$
invertible objects \(g\in\Gamma_p\simeq\mathbb Z_p^2\) and $p-1$
non-invertible objects \(X_m\), $m\in \mathbb{F}_p^*$,
with fusion rules
\begin{equation}
\begin{aligned}
 g\otimes h&=gh,
 &g\otimes X_m&=X_m\otimes g=X_m,\\
 X_m\otimes X_n&=
 \begin{cases}
  \displaystyle\bigoplus_{g\in\Gamma_p}g,&m+n=0,\\[2mm]
  p\,X_{m+n},&m+n\neq0.
 \end{cases}
\end{aligned}
\label{eq:Fus_Esp}
\end{equation}
This is a Jordan--Larson fusion ring~\cite{Jordan_2009};
see also~\cite{czenky2026nimrepresentationstambarayamagamigeneralizations}
for a recent study of its module categories.

Let \(H_{\mathcal F}\) denote the finite-dimensional semisimple Hopf
algebra reconstructed from the fiber functor \(\mathcal F\). We restrict
to the maximal case,
\begin{equation}
    \bigl|\operatorname{Aut}^{\otimes}(\mathcal F)\bigr|
    =
    \dim H_{\mathcal F}
    =
    p^3.
    \label{eq:Bootp-maximal-branch}
\end{equation}
Under this condition, \(H_{\mathcal F}\) is a group algebra.
Thus $H_{\mathcal F}\simeq\mathbb C[G]$
for a group \(G\) of order \(p^3\), and the reconstructed category is
equivalent to \(\operatorname{Rep}(G)\). Since the fusion ring is not
pointed, \(G\) must be non-abelian. For odd \(p\), the two possibilities
are the groups of exponents \(p\) and \(p^2\), which we denote by
\(\mathrm{Esp}_p\) and \(\mathrm{Esp}_{p^2}\), respectively. For $p=3$, $\mathrm{Esp}_3 \simeq 3^{1+2}_{+}$ is the Heisenberg group over $\mathbb{F}_3$, while $\mathrm{Esp}_9 \simeq 3^{1+2}_{-}\simeq M_{27}$, the member of order $27$ of the family $M_{p^3}$ of modular maximal-cyclic groups. They admit
the presentations
\begin{equation}
\begin{aligned}
 \mathrm{Esp}_p
 &=
 \left\langle x,y,z\ \middle|\
 x^p=y^p=z^p=\mathsf{1},\;
 xz=zx,\;
 yz=zy,\;
 xy=yxz
 \right\rangle,\\
 \mathrm{Esp}_{p^2}
 &=
 \left\langle x,y,z\ \middle|\
 x^p=y^p=z,\;
 z^p=\mathsf{1},\;
 xz=zx,\;
 yz=zy,\;
 xy=yxz
 \right\rangle.
\end{aligned}
\label{eq:extraspecial_pres}
\end{equation}
Their representation categories both realize the fusion ring
in Eq.~\eqref{eq:Fus_Esp}.\footnote{For \(p=2\), the same presentations give
\(D_8\) and \(Q_8\), whose common fusion ring was studied in
Section~\ref{sec:TYZ2}.}

Solutions with fewer tensor automorphisms need not reconstruct group
algebras, as illustrated by the \(\operatorname{Rep}(H_8)\) solution of
Section~\ref{ssec:BootDQ-Hopf}. We leave the analysis of such semisimple
Hopf-algebra realizations for odd \(p\) to future work.

For \(p=3\), the bootstrap yields eighteen parametric forms
with the simple-object labels fixed.
We obtain uniform expressions for their complete \(F\)-symbols,
including the multiplicity-three blocks, in terms of two
gauge invariants.
At fixed associator, we find nine and three inequivalent fiber
functors, respectively.
Explicit \(J\)-matrices and further computational details are
provided in Appendix~\ref{app:p3-mixed-matrices} and the ancillary
Mathematica notebooks.

\subsection{From the fusion ring to the square equations}

For \(p=3\), write \(A:=X_1\), \(B:=X_2\), and
\(\Gamma\simeq\mathbb Z_3^2\). Choosing \(g_2,g_3\) as
generators, we use the following fixed ordering:
\begin{equation}
\begin{gathered}
 g_1=\mathsf{1}\,,\qquad
 g_2\,,\qquad
 g_3\,,\qquad
 g_4=g_2^2g_3^2\,,\qquad
 g_5=g_2g_3\,,\\
 g_6=g_3^2\,,\qquad
 g_7=g_2^2\,,\qquad
 g_8=g_2g_3^2\,,\qquad
 g_9=g_2^2g_3\:.
\end{gathered}\label{Eq:p3FixedOrderg}
\end{equation}

The fusion rules become
\begin{equation}
    g\otimes h=gh\,,
    \qquad
    g\otimes A=A\otimes g=A\,,
    \qquad
    g\otimes B=B\otimes g=B\:,
    \label{eq:Bootp3-fusion1}
\end{equation}
and
\begin{equation}
    A\otimes A=3B\,,
    \qquad
    B\otimes B=3A\,,
    \qquad
    A\otimes B=B\otimes A
    =\bigoplus_{g\in\Gamma}g\:.
    \label{eq:Bootp3-fusion2}
\end{equation}
For a fiber functor \(\mathcal F\), the fusion rules require
\begin{equation}
 V_g=\mathcal F(g)\simeq\mathbb C\,,
 \qquad
 V_A=\mathcal F(A)\simeq\mathbb C^3\,,
 \qquad
 V_B=\mathcal F(B)\simeq\mathbb C^3\:.
\end{equation}

The pointed and mixed maps are
\begin{equation}
 J_{gh}\in\mathbb C^\ast\,,
 \qquad
 J_{gA},J_{Ag},J_{gB},J_{Bg}\in GL(3,\mathbb C)\:.
\end{equation}
The remaining monoidal structure maps are organized as follows. Let
\begin{equation}
    J_{AB}^{g}\colon V_A\otimes V_B\longrightarrow V_g
\end{equation}
denote the component in the \(g\)-channel, and assemble these components
in the fixed ordering of~\(\Gamma\) given in \eqref{Eq:p3FixedOrderg}:
\begin{equation}
    J_{AB}
    =
    \begin{pmatrix}
        J_{AB}^{g_1}\\
        \vdots\\
        J_{AB}^{g_9}
    \end{pmatrix}
    \in GL(9,\mathbb C).
    \label{eq:Bootp3-JAB}
\end{equation}
We define \(J_{BA}\) analogously.\footnote{The bases of
\(V_A\otimes V_B\) and \(V_B\otimes V_A\) are ordered lexicographically,
with the first tensor-factor index varying more slowly than the second.}
For the multiplicity-three channels, we write
\begin{equation}
    J_{AA}
    =
    \begin{pmatrix}
        [J_{AA}^{B}]_1\\
        [J_{AA}^{B}]_2\\
        [J_{AA}^{B}]_3
    \end{pmatrix}
    \in GL(9,\mathbb C),
    \label{eq:Bootp3-JAAJBB}\qquad\qquad
    J_{BB}
    =
    \begin{pmatrix}
        [J_{BB}^{A}]_1\\
        [J_{BB}^{A}]_2\\
        [J_{BB}^{A}]_3
    \end{pmatrix}
    \in GL(9,\mathbb C),
\end{equation}
where
\[
    [J_{AA}^{B}]_\alpha
    \colon V_A\otimes V_A\longrightarrow V_B,
    \qquad
    [J_{BB}^{A}]_\alpha
    \colon V_B\otimes V_B\longrightarrow V_A,
    \qquad
    \alpha=1,2,3,
\]
are \(3\times9\) matrices. Thus the rows of \(J_{AA}\) and \(J_{BB}\) are grouped first by the multiplicity index \(\alpha\), then by the basis index
in \(V_B\) and \(V_A\), respectively.

A tensor automorphism has components
\begin{equation}
 \tau_g\in\mathbb C^\ast\quad(g\in\Gamma)
 \qquad\mbox{and}\qquad
 \tau_A,\tau_B\in GL(3,\mathbb C)\:.
\end{equation}

The square equations encoding the
compatibility between \(\tau\) and \(J\) are organized by fusion sector:
\begin{itemize}
\item \textbf{Pointed sector}:
\begin{equation}
    J_{gh}\tau_g\tau_h
    =
    \tau_{gh}J_{gh},
    \qquad
    g,h\in\Gamma\:.
    \label{eq:Bootp3-pointed}
\end{equation}
\item \textbf{Mixed sectors}:
\begin{equation}
\begin{aligned}
    \tau_AJ_{gA}
    &=
    \tau_gJ_{gA}\tau_A\,,
    &
    \qquad\tau_AJ_{Ag}
    &=
    \tau_gJ_{Ag}\tau_A\,,
    \\
    \tau_BJ_{gB}
    &=
    \tau_gJ_{gB}\tau_B\,,
    &
    \qquad\tau_BJ_{Bg}
    &=
    \tau_gJ_{Bg}\tau_B\,,
\end{aligned}\qquad g\in\Gamma\:.
\label{eq:Bootp3-mixedAB}
\end{equation}
\item \textbf{Multiplicity-free products} $A\otimes B$ and $B\otimes A$:
\begin{align}
    J_{AB}(\tau_A\otimes\tau_B)
    &=
    \operatorname{diag}\bigl(\tau_g\bigr)_{g\in\Gamma}J_{AB}\,,
    \label{eq:Bootp3-AB}
    \\
    J_{BA}(\tau_B\otimes\tau_A)
    &=
    \operatorname{diag}\bigl(\tau_g\bigr)_{g\in\Gamma}J_{BA}\:.
    \label{eq:Bootp3-BA}
\end{align}
Equivalently, channel by channel,
\begin{align}
    J^g_{AB}(\tau_A\otimes\tau_B)
    &=\tau_g J^g_{AB}\,,
    &
    J^g_{BA}(\tau_B\otimes\tau_A)
    &=\tau_g J^g_{BA},
    \qquad g\in\Gamma\:.
    \label{eq:Bootp3-ABrows}
\end{align}

\item \textbf{Multiplicity-three products \(A\otimes A\) and
\(B\otimes B\)}:
\begin{align}
    J_{AA}(\tau_A\otimes\tau_A)
    &=
    (\mathbbm 1_3\otimes\tau_B)\,J_{AA}\,,
    \label{eq:Bootp3-AA}
    \\
    J_{BB}(\tau_B\otimes\tau_B)
    &=
    (\mathbbm 1_3\otimes\tau_A)\,J_{BB}\:.
    \label{eq:Bootp3-BB}
\end{align}
Here \(\mathbbm{1}_3\) acts on the fusion multiplicity space, on which a
tensor automorphism acts trivially.
Equivalently, for each multiplicity label $\alpha=1,2,3$,
\begin{align}
 [J_{AA}^{B}]_\alpha(\tau_A\otimes\tau_A)
 &=\tau_B[J_{AA}^{B}]_\alpha\,,
 &
 [J_{BB}^{A}]_\alpha(\tau_B\otimes\tau_B)
 &=\tau_A[J_{BB}^{A}]_\alpha\:.
 \label{eq:Bootp3-multrows}
\end{align}
\end{itemize}

\subsection{Solving the square equations: reconstruction of \textit{J} and \texorpdfstring{$\tau$}{tau}}\label{Sec:SquareEqp3}

We solve the square equations sector by sector, using the basis freedoms
described in Section~\ref{sec:FusCat}. In particular, independent
\(\operatorname{GL}(3,\mathbb C)\) transformations on \(V_A\) and \(V_B\)
are used to put selected mixed maps in canonical form, with the remaining
freedom given by their centralizers. We keep changes of basis in the fusion
multiplicity spaces separate, since they also transform the \(F\)-symbols
reconstructed below.

\subsubsection*{Pointed sector}

Since \(J_{gh}\) is a nonzero scalar,
Eq.~\eqref{eq:Bootp3-pointed} reduces to
\begin{equation}
 \tau_{gh}=\tau_g\tau_h,
 \qquad g,h\in\Gamma.
 \label{eq:Bootp3-character}
\end{equation}
Hence the pointed components of a tensor automorphism form a
character
\begin{equation}
    \chi\colon\Gamma\longrightarrow\mathbb Z_3,
    \qquad
    \tau_g=\chi(g),
    \label{eq:Bootp3-pointed-character}
\end{equation}
where $\mathbb{Z}_3=\{1,\omega,\omega^2\}$, with $\omega=e^{2\pi i/3}$.
There are nine such characters, denoted by
\(\chi_{r,s}\), where \(r,s\in\{0,1,2\}\) are understood modulo \(3\),
and determined by
\begin{equation}
\chi_{r,s}(g_2^m\,g_3^n)=\omega^{rm+sn}\:.
\end{equation}
For a tensor automorphism whose restriction to the pointed
sector is \(\chi\), we write
\begin{equation}
 (\,\chi,\,\,\tau_A^\chi,\,\,\tau_B^\chi\,) \:,
\end{equation}
where \(\tau_A^\chi\) and \(\tau_B^\chi\) are its components
on the non-invertible objects \(A\) and \(B\), respectively.
These matrices will be determined by the remaining square
equations.
The scalars \(J_{gh}\in\mathbb C^\ast\) remain unconstrained
by these equations.

\subsubsection*{Mixed sector}

The mixed equations \eqref{eq:Bootp3-mixedAB} have the same form, and the
\(A\)- and \(B\)-sectors are independent at this stage.
We therefore describe the simultaneous determination of
\(J_{gA}\) and \(\tau_A^\chi\); the other mixed maps
are obtained by the same procedure.

We begin by using one of the mixed equations to fix a convenient basis of
\(V_A\). The character \(\chi_{0,1}\) is nontrivial on \(g_3\), with
\(\chi_{0,1}(g_3)=\omega\), so Eq.~\eqref{eq:Bootp3-mixedAB} for
\(J_{g_3A}\) becomes
\begin{equation}
 \tau_A^{\chi_{0,1}}J_{g_3A}
 \bigl(\tau_A^{\chi_{0,1}}\bigr)^{-1}
 =\omega J_{g_3A} \:.
\end{equation}
Because conjugate matrices have the same spectrum, this relation
implies that the spectrum of \(J_{g_3A}\) is invariant under
multiplication by \(\omega\). Since \(J_{g_3A}\) is invertible and three-dimensional,
it is therefore diagonalizable, with
\[
 J_{g_3A}=\operatorname{diag}(\lambda_1,\lambda_2,\lambda_3),
 \qquad
 \{\lambda_1,\lambda_2,\lambda_3\}
 =\{\ell,\ell\omega,\ell\omega^2\},
 \qquad \ell\neq0,
\]
in a suitable basis. The ordering of the eigenvectors will be fixed at a later stage.

The characters \(\chi_{1,0}\) and \(\chi_{2,0}\) are trivial on \(g_3\),
so Eq.~\eqref{eq:Bootp3-mixedAB} implies that the corresponding
\(A\)-sector components
\(\tau_A^{\chi_{1,0}}\) and \(\tau_A^{\chi_{2,0}}\)
commute with \(J_{g_3A}\). Since its eigenvalues are distinct,
these matrices are diagonal in the same basis.
We use these diagonal automorphisms to constrain
the remaining mixed maps.\footnote{See ``A[3] diagonal'',
in particular ``Exploiting A[3] diagonal'' and
``Exploiting tau[5],tau[9] diagonal'', in \texttt{1\_ParametricForms.nb}.}

For \(r=1,2\), the square equations involving these
diagonal automorphisms read
\begin{equation}
 \tau_A^{\chi_{r,0}}J_{gA}
 =\chi_{r,0}(g)\,J_{gA}\tau_A^{\chi_{r,0}},
 \qquad g\in\Gamma.
 \label{eq:Bootp3-first-branch-squares}
\end{equation}
Writing, for each \(r=1,2\),
\begin{equation}
    \tau_A^{\chi_{r,0}}
    =
    \operatorname{diag}
    \bigl(\lambda_1^{(r)},\lambda_2^{(r)},\lambda_3^{(r)}\bigr),
\end{equation}
we obtain
\begin{equation}
    \bigl(
    \lambda_i^{(r)}
    -
    \chi_{r,0}(g)\lambda_j^{(r)}
    \bigr)
    (J_{gA})_{ij}
    =
    0.
    \label{eq:Bootp3-diagonal-support}
\end{equation}
Thus \((J_{gA})_{ij}\) can be nonzero only if
\begin{equation}
    \lambda_i^{(r)}
    =
    \chi_{r,0}(g)\lambda_j^{(r)}.
\end{equation}
For \(g=g_2\), one has
\(\chi_{r,0}(g_2)=\omega^r\). Since \(J_{g_2A}\) is invertible, the same argument used above for
\(J_{g_3A}\) shows that the eigenvalues of
\(\tau_A^{\chi_{r,0}}\) form a three-element orbit under multiplication
by \(\omega^r\). For each column of \(J_{g_2A}\),
Eq.~\eqref{eq:Bootp3-diagonal-support} selects a unique row, so its
nonzero entries form a cyclic permutation pattern. The two possible orientations are
\begin{equation}\label{Eq:formep3initinere}
    \begin{pmatrix}
        0&*&0\\
        0&0&*\\
        *&0&0
    \end{pmatrix}
    \qquad\mbox{and}\qquad
    \begin{pmatrix}
        0&0&*\\
        *&0&0\\
        0&*&0
    \end{pmatrix},
\end{equation}
where every star denotes a nonzero entry.

For each pattern in \eqref{Eq:formep3initinere}, we impose
Eq.~\eqref{eq:Bootp3-first-branch-squares} for both
\(r=1,2\) and all \(g\in\Gamma\).
This determines the allowed positions of the nonzero
entries of the remaining mixed maps.
In particular, \(J_{1A}\), \(J_{g_3A}\), and \(J_{g_6A}\)
are diagonal, since both characters are trivial on
\(\{1,g_3,g_6\}\).

We next impose the equations for a character
\(\chi_{r,s}\) with \(s\neq0\). In this case
 \(\chi_{r,s}(g_3)\neq1\) and so its automorphism matrix
\(\tau_A^{\chi_{r,s}}\) is non-diagonal in the chosen basis. Consider then 
\begin{equation}
 \tau_A^{\chi_{r,s}}J_{gA}
 =\chi_{r,s}(g)\,J_{gA}\tau_A^{\chi_{r,s}},
 \qquad g\in\{\mathsf{1},g_3,g_6\}\:.
 \label{eq:Bootp3-second-branch-squares}
\end{equation}
For each of the two patterns in \eqref{Eq:formep3initinere}, these equations yield
two possible parametric forms, giving four intermediate
solutions.\footnote{See ``Second turning point'' in
\texttt{1\_ParametricForms.nb}. The automorphism matrix used
in this step is denoted there by \texttt{tau[2]}.}
Each solution specifies both the mixed maps and the
automorphism matrices; these are determined simultaneously.
Imposing the remaining mixed square equations determines
the positions of the nonzero entries and relations among
their values in each of the four forms.

We now identify which of the four mixed-sector forms differ only by a
change of basis on \(V_A\). First, diagonal changes of basis preserve
the chosen \(J_{g_3A}\) and allow us to fix the relative entries of the
other matrices, leaving overall nonzero scalar coefficients. The ordering
of the eigenvectors of \(J_{g_3A}\), however, is arbitrary: exchanging
the second and third eigenvectors maps the four forms into pairs, after
redefinitions of those coefficients. Thus only two classes of
\(A\)-sector solutions remain, which we denote by \(\mathsf F\) and
\(\mathsf S\).\footnote{
In \texttt{1\_ParametricForms.nb}, the four forms are numbered \(1,2,3,4\).
The exchange, denoted there by \texttt{P}, identifies \(1\sim4\) and
\(2\sim3\); representatives \(1\) and \(2\) correspond to
\(\mathsf F\) and \(\mathsf S\), respectively.}

Representative matrices for the two classes are shown below.
Here \(\mathrm{t}_{\chi,\mathsf X}^A,\mathrm{x}_{g,\mathsf X}\in\mathbb C^\ast\),
with \(\mathsf X\in\{\mathsf F,\mathsf S\}\).
We display the automorphism associated with
\(\chi_{1,2}\) and the mixed maps for the generators
\(g_2,g_3\):\footnote{The character \(\chi_{1,2}\)
corresponds to the second pointed solution in
\texttt{1\_ParametricForms.nb}.}
\begin{equation}
\begin{aligned}
\left.\tau_A^{\chi_{1,2}}\right|_{\mathsf F}
&=
\mathrm{t}_{\chi_{1,2},\mathsf F}^A
\begin{pmatrix}
0&1&0\\
0&0&1\\
1&0&0
\end{pmatrix},
&
\left.\tau_A^{\chi_{1,2}}\right|_{\mathsf S}
&=
\mathrm{t}_{\chi_{1,2},\mathsf S}^A
\begin{pmatrix}
0&0&1\\
1&0&0\\
0&1&0
\end{pmatrix},
\\[3mm]
\left.J_{g_2A}\right|_{\mathsf F}
&=
\mathrm{x}_{g_2,\mathsf F}
\begin{pmatrix}
0&1&0\\
0&0&\omega\\
\omega^2&0&0
\end{pmatrix},
&
\left.J_{g_2A}\right|_{\mathsf S}
&=
\mathrm{x}_{g_2,\mathsf S}
\begin{pmatrix}
0&1&0\\
0&0&\omega^2\\
\omega&0&0
\end{pmatrix},
\\[3mm]
\left.J_{g_3A}\right|_{\mathsf F}
&=
\mathrm{x}_{g_3,\mathsf F}
\begin{pmatrix}
1&0&0\\
0&\omega^2&0\\
0&0&\omega
\end{pmatrix},
&
\left.J_{g_3A}\right|_{\mathsf S}
&=
\mathrm{x}_{g_3,\mathsf S}
\begin{pmatrix}
1&0&0\\
0&\omega&0\\
0&0&\omega^2
\end{pmatrix}.
\end{aligned}
\end{equation}
Every remaining matrix is likewise a nonzero scalar
multiple of a monomial matrix whose nonzero entries are
powers of \(\omega\).
The complete matrix forms are collected in
Appendix~\ref{app:p3-mixed-matrices}.\footnote{
The mixed-sector solutions and their comparison under
change of basis are given in
\texttt{1\_ParametricForms.nb}, under ``Second turning point''
and ``Isomorphic solutions''.}
Suppressing the class labels and using analogous notation
for the \(B\)-sector, we write
\begin{equation}
 \tau_A^\chi=\mathrm{t}_\chi^A\mathfrak M_{A,\chi},
 \qquad
 \tau_B^\chi=\mathrm{t}_\chi^B\mathfrak M_{B,\chi}.
 \label{eq:Bootp3-aut-scalars}
\end{equation}
and
\begin{equation}
J_{gA}=\mathrm{x}_g\,\mathfrak M_{gA},
 \qquad
 J_{Ag}=\widetilde{\mathrm{x}}_g\,\mathfrak M_{gA}.
\end{equation}
The matrix factors \(\mathfrak M_{A,\chi}\) and
\(\mathfrak M_{gA}\) are determined by the chosen class.
The same factor occurs in \(J_{gA}\) and \(J_{Ag}\)
because they satisfy identical equations with
the same automorphism matrices.

The \(B\)-sector is obtained in the same way, with
independent coefficients
\(\mathrm{t}_\chi^B,\mathrm{y}_g,\widetilde{\mathrm{y}}_g\in\mathbb C^\ast\).
All these scalar coefficients remain unconstrained
by the mixed equations.
The parameters \(\mathrm{x}_g,\widetilde{\mathrm{x}}_g,\mathrm{y}_g,\widetilde{\mathrm{y}}_g\)
belong to the monoidal structure, whereas
\(\mathrm{t}_\chi^A,\mathrm{t}_\chi^B\) normalize the automorphism matrices.
The latter will be restricted to finitely many values
by the non-invertible square equations.

The independent choices for the \(A\)- and \(B\)-sectors
therefore give four candidate combinations,
\begin{equation}
 \mathsf{FF},\qquad
 \mathsf{FS},\qquad
 \mathsf{SF},\qquad
 \mathsf{SS},
 \label{eq:Bootp3-four-combinations}
\end{equation}
where the first and second letters refer to
the \(A\)- and \(B\)-sectors, respectively.
We next determine which combinations satisfy
the remaining square equations.

\subsubsection*{Non-invertible sector}

We now impose the square equations involving
\(
    J_{AA},\, J_{BB},\, J_{AB},\, J_{BA},
\)
in this order.\footnote{In
\texttt{1\_ParametricForms.nb}, the corresponding matrices
\(J_{AA},J_{BB},J_{AB},J_{BA}\) are denoted by
\texttt{G[1]}, \texttt{G[2]}, \texttt{G[3]}, and
\texttt{G[4]}, respectively.}
Besides determining these \(J\)-maps, the equations
constrain the scalar factors \(\mathrm{t}_\chi^A,\mathrm{t}_\chi^B\)
left undetermined by the mixed sector.

The combinations \(\mathsf{FF}\) and \(\mathsf{SS}\)
are excluded by the \(AA\)-equations
\eqref{eq:Bootp3-AA}:
for \(\mathsf{FF}\), imposing the equations for the two
nontrivial diagonal automorphisms gives three candidate
invertible forms of \(J_{AA}\).
The equation for \(\chi_{1,2}\) then forces
\(J_{AA}=0\) in every case, contradicting invertibility.
The same calculation excludes \(\mathsf{SS}\).\footnote{
See ``Case FF'' and ``Case SS'' in
\texttt{1\_ParametricForms.nb}, where the equations for
\texttt{G[1]} are imposed successively.}

We next consider \(\mathsf{FS}\).
The \(AA\)-equation for \(\chi_{1,0}\), together with
invertibility of \(J_{AA}\), gives three parametric
matrix forms for \(J_{AA}\), related by cyclic permutations
of the basis of \(V_B\).
After choosing one representative, the equation for
\(\chi_{1,2}\) gives three further parametric forms
for \(J_{AA}\), related by diagonal changes of basis.
The two identifications are generated, respectively, by
\begin{equation}
 U=
 \begin{pmatrix}
  0&1&0\\
  0&0&1\\
  1&0&0
 \end{pmatrix}
 \qquad\mbox{and}\qquad
 D=\begin{pmatrix}
     1 && \\ & \omega & \\ && \omega^2 \\
 \end{pmatrix}\:.
\end{equation}
Since \(A\otimes A=3B\), a change of basis on \(V_B\)
acts on the target of each component \([J_{AA}^{B}]_\alpha\).
The changes \(\mathbbm{1},U,U^2\) relate the first three forms,
while \(\mathbbm{1},D,D^2\) relate the three obtained at the next step.
Each change of basis is also applied consistently to
\(\tau_B^\chi\) and the mixed maps involving \(B\).
Thus only one representative is needed at each step.\footnote{
See ``Case FS'', subsection ``G[1]'', in
\texttt{1\_ParametricForms.nb}, for the successive solutions
and the changes of basis relating them.}

Its multiplicity components, assembled as in
Eq.~\eqref{eq:Bootp3-JAAJBB}, are
\begin{equation}
 [J_{AA}^{B}]_\alpha
 =
 \begin{pmatrix}
  0&0&\mathrm{r}_{\alpha1}&0&\mathrm{r}_{\alpha2}&0&\mathrm{r}_{\alpha3}&0&0\\
  \mathrm{r}_{\alpha2}&0&0&0&0&\mathrm{r}_{\alpha3}&0&\mathrm{r}_{\alpha1}&0\\
  0&\mathrm{r}_{\alpha3}&0&\mathrm{r}_{\alpha1}&0&0&0&0&\mathrm{r}_{\alpha2}
 \end{pmatrix},
 \qquad \alpha=1,2,3.
 \label{eq:Bootp3-FS-JAA-example}
\end{equation}
The square equations determine the displayed pattern,
leaving the nine entries of the $3\times 3$ matrix
\begin{equation}\label{eq_Rmatrixp3}
\mathrm{R}:=(\mathrm{r}_{\alpha j})_{\alpha,j=1,2,3}
\end{equation}
as free parameters.
After reordering its rows and columns, the assembled
matrix \(J_{AA}\) is block diagonal, with three blocks
equal to \(\mathrm{R}\). Hence
\[
 \det J_{AA}\neq0
 \quad\Longleftrightarrow\quad
 \det \mathrm{R}\neq0.
\]
Individual entries of \(\mathrm{R}\) may therefore vanish,
provided \(\mathrm{R}\) remains invertible.\footnote{
In \texttt{1\_ParametricForms.nb}, under ``Case FS'',
subsection ``G[1]'', invertibility is imposed directly
on the assembled \(9\times9\) matrix.
The matrix \(\mathrm{R}\) packages its remaining parameters
and makes the invertibility condition explicit.}

For \(J_{BB}\), the equation associated with
\(\chi_{2,0}\) gives three parametric matrix forms
analogous to Eq.~\eqref{eq:Bootp3-FS-JAA-example}.
Their free parameters can be collected into the matrix
\[
 \mathrm{S}:=(\mathrm{s}_{\alpha j})_{\alpha,j=1,2,3}.
\]
For each of these three forms, imposing the equation
associated with \(\chi_{2,1}\) gives three further
possibilities. Thus, with the representative \(J_{AA}\)
fixed above, we obtain \emph{nine candidate pairs}
\((J_{AA},J_{BB})\).\footnote{
See ``Case FS'', subsection ``G[2]'', in
\texttt{1\_ParametricForms.nb}.}
We retain all nine candidates at this stage. Their
inequivalence as completed solutions will be established
in Section~\ref{ssec:Bootp3-parametric-comparison}
using the invariants introduced there.

For each resulting pair \((J_{AA},J_{BB})\), we impose
the \(AA\)- and \(BB\)-equations for all nine characters
\(\chi\in\widehat{\Gamma}\).
These equations restrict the scalar factors
\(\mathrm{t}_\chi^A,\mathrm{t}_\chi^B\) introduced in
Eq.~\eqref{eq:Bootp3-aut-scalars}
to exactly three compatible pairs for each \(\chi\).\footnote{
See ``Case FS'', subsection ``Automorphisms'', in
\texttt{1\_ParametricForms.nb}.}
Thus each character has three extensions, giving
\(9\cdot3=27\) candidate tensor automorphisms.

The remaining \(AB\)- and \(BA\)-equations are compatible
with all \(27\) candidate automorphisms in each of the
nine forms.
They determine each row \(J_{AB}^g\) and \(J_{BA}^g\)
up to a scalar coefficient, that we name respectively $\mathrm{z}_g,\widetilde{\mathrm{z}}_g$ ($g\in\Gamma$).
The assembled matrices are invertible precisely when
all these coefficients are nonzero.\footnote{
See ``Case FS'', subsections ``G[3]'' and
``Morphism G[4]'', in \texttt{1\_ParametricForms.nb}.}
This completes the \(\mathsf{FS}\) solutions.

Interchanging \(A\) and \(B\) gives nine further
parametric forms of type \(\mathsf{SF}\).\footnote{
The corresponding calculation is presented under
``Case SF'' in \texttt{1\_ParametricForms.nb},
with the same subsection headings as for \(\mathsf{FS}\).}
The completed square system therefore yields \(18\)
parametric forms, nine of each type, all with \(27\)
tensor automorphisms and hence satisfying
Eq.~\eqref{eq:Bootp-maximal-branch}. 
To identify the reconstructed groups, it suffices to
consider their action on \(V_A\).
Indeed, every simple object occurs as a direct summand
of a tensor power of \(A\): the object \(B\) occurs in
\(A\otimes A=3B\), while every \(g\in\Gamma\) occurs in
\[
 A\otimes B=\bigoplus_{g\in\Gamma}g
\]
and hence in \(A^{\otimes3}\).
The square equations therefore determine every component
of a tensor automorphism from \(\tau_A\), so this action
is faithful.
The order of each automorphism $(\chi,\tau_A,\tau_B)$  is consequently the
smallest positive integer \(n\) such that
\((\tau_A)^n=\mathbbm{1}_3\).
The calculation gives two order distributions:
\begin{equation}
\begin{array}{c|ccc}
    & \text{order }1 & \text{order }3 & \text{order }9\\
    \hline
    \text{first order distribution}  & 1 & 26 & 0\\
    \text{second order distribution} & 1 & 8  & 18
\end{array} \:.
\label{eq:Bootp3-order-distributions}
\end{equation}
The reconstructed groups are non-abelian. Indeed, after suppressing
the nonzero scalar factors, the displayed representatives satisfy
\[
    \tau_A^{\chi_{1,0}}\tau_A^{\chi_{1,2}}
    =
    \omega\,
    \tau_A^{\chi_{1,2}}\tau_A^{\chi_{1,0}}
\]
in the \(\mathsf F\) class, and the analogous relation with
\(\omega^2\) in the \(\mathsf S\) class. 

The two non-abelian groups of order \(27\) are distinguished
by their exponents.
For two of the eighteen parametric forms, every
nonidentity automorphism has order three, giving
the first distribution in
Eq.~\eqref{eq:Bootp3-order-distributions}.
This identifies the tensor-automorphism group as
\begin{equation}
 \operatorname{Aut}^{\otimes}(\mathcal F)
 \simeq \mathrm{Esp}_3.
\end{equation}
For the remaining sixteen forms, eight nonidentity
automorphisms have order three and eighteen have order
nine, giving the second distribution.
The group therefore has exponent nine, and we obtain
\begin{equation}
 \operatorname{Aut}^{\otimes}(\mathcal F)
 \simeq \mathrm{Esp}_9.
\end{equation}
We label the eighteen parametric forms by \(q=1,\ldots,18\),
assigning \(q=1,2\) to the two forms with
tensor-automorphism group \(\mathrm{Esp}_3\), and
\(q=3,\ldots,18\) to those with group \(\mathrm{Esp}_9\).
Thus the forms \(q=1,2\) reconstruct
\(\operatorname{Rep}(\mathrm{Esp}_3)\), while
\(q=3,\ldots,18\) reconstruct
\(\operatorname{Rep}(\mathrm{Esp}_9)\).\footnote{
The complete parametric \(J\)-maps and tensor-automorphism
matrices are given in \texttt{1\_ParametricForms.nb}.
}

\subsection{Reconstruction of the \texorpdfstring{\(F\)}{F}-symbols and gauge invariants}
\label{ssec:Bootp3-Fsymbols}

As in Section~\ref{Sec:maximalCaseTY}, we derive the
\(F\)-symbols from the eighteen parametric forms of the
\(J\)-maps. We identify two gauge invariants of the resulting
associators and show that the free parameters can be chosen
so that all \(F\)-symbols admit a uniform expression in terms
of these invariants.

\paragraph{From parametric \(J\)-maps to associators.}
For any invertible choice of the \(J\)-maps within one of the
eighteen parametric forms, Eq.~\eqref{eq:fiber-Fsymb} is a
linear system for the \(F\)-symbols.
Since all these forms satisfy the maximality condition
\eqref{eq:Bootp-maximal-branch}, their \(J\)-maps provide
Clebsch--Gordan decompositions, as explained in
Section~\ref{ssec:bootstrap-code}.
Their compositions therefore give two bases of the same
intertwiner space, so the \(F\)-symbols are uniquely
determined as the corresponding change-of-basis coefficients.
They satisfy the pentagon equations by the argument
given in that section.

We first determine whether varying the free parameters
within a given form changes the gauge class of its associator.
Within each parametric form obtained in
Section~\ref{Sec:SquareEqp3}, the tensor-automorphism
matrices are independent of the free parameters of the
\(J\)-maps. The multiplicity components of the invertible
\(J\)-maps therefore form bases of fixed intertwiner spaces
\(\operatorname{Hom}_G(V_a\otimes V_b,V_c)\).
Varying the parameters amounts to changing these bases in the fusion spaces, as in
Eq.~\eqref{eq:fiber_fusion_space_map}.
By Eq.~\eqref{eq:F_equiv_base}, each form therefore
determines a single gauge class of associators.

\paragraph{Gauge-invariant data.}
To compare the associators obtained from different parametric
forms, we first examine quantities that are invariant under
changes of basis in the fusion spaces.
For the \(q\)-th parametric form, consider the scalar block
\begin{equation}
 \mathit{I}_q(g,h):=F^{(q),A}_{gAh},
 \qquad g,h\in\Gamma.
 \label{eq:Bootp3-second-invariant}
\end{equation}
The intermediate objects on both sides are \(A\). Under a change of
basis in the fusion spaces, the scalar factors cancel, showing the
gauge invariance of \(\mathit{I}_q\):
\begin{equation}
 \mathit{I}'_q(g,h)
 =\frac{M_{gA}^{A}M_{Ah}^{A}}{M_{Ah}^{A}M_{gA}^{A}}
   \mathit{I}_q(g,h)
 =\mathit{I}_q(g,h).
 \label{eq:Bootp3-second-invariant-transformation}
\end{equation}

The bicharacter property follows directly from the mixed \(J\)-maps.
For this block, Eq.~\eqref{eq:fiber-Fsymb} reads
\begin{equation}
 J_{Ah}J_{gA}
 =\mathit{I}_q(g,h)J_{gA}J_{Ah}.
\end{equation}
The solutions obtained in the previous subsection satisfy
\(J_{Ag}\propto J_{gA}\) with nonzero scalar proportionality factors.
Therefore, one also has
\[
 J_{hA}J_{gA}
 =\mathit{I}_q(g,h)J_{gA}J_{hA}.
\]
Commuting \(J_{kA}\) past \(J_{gA}J_{hA}\) and using
\(J_{gA}J_{hA}\propto J_{gh,A}\) gives
\[
 \mathit{I}_q(gh,k)=\mathit{I}_q(g,k)\mathit{I}_q(h,k).
\]
The same argument in the second variable gives
\[
 \mathit{I}_q(g,hk)=\mathit{I}_q(g,h)\mathit{I}_q(g,k).
\]
Moreover, setting \(h=g\) gives \(\mathit{I}_q(g,g)=1\).
Thus \(\mathit{I}_q\) is an alternating bicharacter.

An alternating bicharacter on \(\Gamma\) is determined by its
value on the generators. Evaluating the commutation relation
for the reconstructed mixed \(J\)-maps gives
\begin{equation}
 \mathit{I}_q(g_2,g_3)=\omega^{\iota_q},
 \qquad \iota_q\in\{1,2\}.
 \label{eq:Bootp3-second-invariant-exponent}
\end{equation}
Consequently, one can write
\begin{equation}
 \mathit{I}_q(g,h)=\omega^{\iota_q(m_g n_h- n_g m_h)},
 \label{eq:Bootp3-second-invariant-bicharacter}
\end{equation}
where \(x=g_2^{m_x}g_3^{n_x}\) for \(x\in\Gamma\). 
Since \(\iota_q\neq0\) in \(\mathbb F_3\), this bicharacter
is nondegenerate.

A second invariant is obtained from the \(3\times3\) blocks
\(F^{(q),g}_{AAA}\). Their transformation is
\begin{equation}
 \widetilde F^{(q),g}_{AAA}
 =\frac{M_{BA}^{g}}{M_{AB}^{g}}\,
   M_{AA}^{B}F^{(q),g}_{AAA}(M_{AA}^{B})^{-1}.
 \label{eq:Bootp3-FAAAg-transformation}
\end{equation}
Consequently,
\begin{equation}
 \mathcal E_q(g):=
 \frac{\bigl(\operatorname{tr}F^{(q),g}_{AAA}\bigr)^3}
      {\det F^{(q),g}_{AAA}}
 \label{eq:Bootp3-first-invariant}
\end{equation}
is invariant: conjugation preserves trace and determinant, and
the scalar factor contributes the same third power to numerator
and denominator.
For each reconstructed form there is a unique \(\epsilon_q\in\Gamma\)
such that
\begin{equation}
 \mathcal E_q(g)=27\,\delta_{g,\epsilon_q}.
 \label{eq:Bootp3-epsilon-definition}
\end{equation}

\paragraph{Distinguishing the reconstructed associators.}
To evaluate these invariants, we choose admissible values of the free parameters in each of the eighteen forms and solve Eq.~\eqref{eq:fiber-Fsymb} for the F-symbols. Since varying these parameters only changes
the fusion-space bases, the resulting invariants are
independent of this choice.
Their evaluation gives a different pair
\((\epsilon_q,\iota_q)\) for each form.\footnote{
The F-symbols are computed in \texttt{2\_Fsymbols.nb} ("Computing F-symbols"); the invariants are evaluated in \texttt{4\_FusRingAut.nb} ("Invariants of the fiber functors).}
With our numbering,
\begin{equation}
 \epsilon_q=g_{\lceil q/2\rceil},\qquad q=1,\ldots,18,
 \label{eq:Bootp3-epsilon-values}
\end{equation}
and
\begin{equation}
 \iota_q=\begin{cases}1,&q\text{ odd},\\2,&q\text{ even}.
 \end{cases}
 \label{eq:Bootp3-second-invariant-values}
\end{equation}
The two forms reconstructing \(\operatorname{Rep}(\mathrm{Esp}_3)\)
have \(\epsilon_q=\mathsf{1}\), whereas the sixteen forms reconstructing
\(\operatorname{Rep}(\mathrm{Esp}_9)\) have \(\epsilon_q\neq\mathsf{1}\).
For reference, the correspondence between the labels \(q\)
and the associator invariants is summarized as:
\begin{equation}
\begin{array}{c|cc||ccccccccccccccccc}
 q&1&2&3&4&5&6&7&8&9&10&11&12&13&14&15&16&17&18\\
 \hline  
 \epsilon_q&\mathsf{1}&\mathsf{1}&g_2&g_2&g_3&g_3&g_4&g_4&g_5&g_5&g_6&g_6&g_7&g_7&g_8&g_8&g_9&g_9\\
 \iota_q&1&2&1&2&1&2&1&2&1&2&1&2&1&2&1&2&1&2
\end{array}\,.
\end{equation}

Each pair \((\epsilon,\iota)\in\Gamma\times\{1,2\}\)
occurs exactly once. Since both quantities are gauge invariant,
different forms give gauge-inequivalent associators at fixed
object labels. Together with the gauge equivalence of parameter
choices within each form, this shows that \((\epsilon,\iota)\)
completely labels the associators reconstructed in the maximal
branch.

\paragraph{A basis choice adapted to the invariants.}
Having labeled the eighteen classes by $(\epsilon,\iota)$, we now
choose the free parameters of the $J$-maps so that the corresponding
$F$-symbols admit uniform expressions in these invariants.
In what follows we suppress the label \(q\) and write
\(\epsilon=\epsilon_q\) and \(\mathit{I}=\mathit{I}_q\) (correspondingly
\(\iota=\iota_q\)). 

We write \(g=g_2^{m_g}g_3^{n_g}\),
\(h=g_2^{m_h}g_3^{n_h}\), \(\epsilon=g_2^{m_\epsilon}g_3^{n_\epsilon}\) and set 
\( u_g:=m_g-n_g \). All coordinates and exponents are understood modulo three.
Let \((\hat{e}_0,\hat{e}_1,\hat{e}_2)\) be the standard ordered basis of 
\(\mathbb C^3\), and define
\begin{equation}\label{Eq:XZdef}
  X\hat{e}_j=\hat{e}_{j+1}\,, \qquad\mbox{and}\qquad   Z\hat{e}_j=\omega^j \hat{e}_j\:.
\end{equation}

We choose the following parameter values for each form,
using its associated pair \((\epsilon,\iota)\):\footnote{
We use the matrix normalization of
\texttt{1\_ParametricForms.nb}. Remember that \(\mathrm R=(\mathrm r_{\alpha j})\) is the coefficient matrix
of \(J_{AA}\) in Eq.~\eqref{eq_Rmatrixp3} and
\(\mathrm S=(\mathrm s_{\alpha j})\) its counterpart for \(J_{BB}\).
}
\begin{equation}
\begin{aligned}
 J_{gh}&=\omega^{\iota\,m_h u_g}\,, \\
 x_g & = \widetilde{\mathrm x}_g=1 \,, \\
 y_g&=\widetilde{\mathrm y}_g=\omega^{\iota(1-m_g)u_g} \,,
\end{aligned}\qquad\quad
\begin{aligned}
\mathrm z_g&=\tfrac{1}{\sqrt{3}}\omega^{\iota u_{\epsilon} m_g}\,, \\
\widetilde{\mathrm z}_g&=\tfrac{1}{\sqrt{3}}\omega^{\iota m_{\epsilon} u_g},
\end{aligned}\qquad\quad
\begin{aligned}
 \mathrm R&=\omega^{\iota n_\epsilon}\,\mathbbm1_3 \,, \\
 \mathrm S&=\omega^{-\iota m_\epsilon}\,Z^{\iota u_\epsilon}\:.
\end{aligned}
\label{eq:Bootp3-invariant-gauge}
\end{equation}
All scalar parameters are nonzero and \(\mathrm R,\mathrm S\) are invertible,
so these \(J\)-maps are invertible. The factor \(1/\sqrt3\) in
\(\mathrm z_g,\widetilde{\mathrm z}_g\) is chosen so that, moreover, all
\(J\)-matrices are unitary with respect to the standard inner products on the
\(V_a\) and on the fusion spaces; consequently the resulting F-symbols are
unitary as well.
Substituting \eqref{eq:Bootp3-invariant-gauge} into Eq.~\eqref{eq:fiber-Fsymb} gives
the uniform expressions
\eqref{eq:Bootp3-invariant-F-pointed}--
\eqref{eq:Bootp3-projective-F-nine} below, in which the
dependence on the parametric form enters only through
\((\epsilon,\iota)\).

The pointed coefficients \(J_{gh}\) form a normalized
bilinear two-cocycle, so the pointed associator is trivial.
The two forms with the same \(\epsilon\) give
complex-conjugate \(F\)-symbols.\footnote{The symbolic J-maps of the two forms with the same $\epsilon$ are entrywise complex conjugate, and all parameters in Eq.~\eqref{eq:Bootp3-invariant-gauge} are of the form $\omega^{\iota(\dots)}$ up to a real factor $\tfrac{1}{\sqrt{3}}$; hence the corresponding J-maps, and therefore the F-symbols, are complex conjugate.}

\paragraph{Explicit \(F\)-symbols.}

We follow the component convention of
Section~\ref{sec:FusCat}, omitting labels of
one-dimensional fusion spaces and uniquely determined
intermediate objects.

The non-scalar blocks are expressed through a three-dimensional projective
representation \(U\) of \(\Gamma\simeq\mathbb Z_3\times \mathbb Z_3\) with multiplier
\(\mathit{I}^{-1}\), i.e.
\begin{equation}
 U(g)U(h)=\mathit{I}(g,h)^{-1}U(gh),
 \qquad U(\mathsf{1})=\mathbbm1_3.
 \label{eq:Bootp3-projective-multiplier}
\end{equation}
Since \(\mathit{I}\) is nondegenerate, this
three-dimensional representation is irreducible and,
for this fixed multiplier, unique up to a change of
basis in \(V^B_{AA}\).
Thus \(U\) introduces no additional data beyond~\(\iota\).
In the basis fixed above, it is given by
\begin{equation}
 U(g)=
 \omega^{\iota u_g(m_g-1)}
 Z^{\iota(m_g-n_g)}X^{m_g},
 \qquad\mbox{i.e.}\quad
 U(g)_{jk}=
 \delta_{j,k+m_g}\,
 \omega^{\iota u_g(j+m_g-1)}\:,
 \label{eq:Bootp3-projective-basis}
\end{equation}
where $X,Z$ are defined in \eqref{Eq:XZdef} and  \(j,k\in\{0,1,2\}\), with index arithmetic understood
modulo three.\footnote{
The matrix entry \(U(g)_{jk}\) corresponds to position
\texttt{[[j+1,k+1]]} in the notebook arrays.}
For \(\iota=2\), \(U(g)\) is the entrywise complex conjugate
of the matrix for \(\iota=1\).
The bar below denotes entrywise complex conjugation.

\subparagraph{Scalar blocks.}
The pointed blocks are
\begin{equation}
 F^{ghk}_{g,h,k}=1,\qquad g,h,k\in\Gamma,
 \label{eq:Bootp3-invariant-F-pointed}
\end{equation}
and the remaining scalar blocks are
\begin{equation}
\begin{aligned}
 F^A_{ghA}&=\mathit{I}(g,h),
 &F^A_{gAh}&=\mathit{I}(g,h),
 &F^A_{Agh}&=1,\\
 F^B_{ghB}&=1,
 &F^B_{gBh}&=\mathit{I}(g,h)^{-1},
 &F^B_{Bgh}&=\mathit{I}(g,h)^{-1},\\[2pt]
 F^h_{gAB}&=\mathit{I}(g,h)^{-1},
 &F^h_{AgB}&=\mathit{I}(\epsilon,g)^{-1}\mathit{I}(g,h)^{-1},
 &F^h_{ABg}&=\mathit{I}(\epsilon,g)^{-1}\mathit{I}(g,h)^{-1},\\
 F^h_{gBA}&=1,
 &F^h_{BgA}&=\mathit{I}(\epsilon,g)\mathit{I}(g,h),
 &F^h_{BAg}&=\mathit{I}(\epsilon,g),
\end{aligned}
\label{eq:Bootp3-invariant-F-scalars}
\end{equation}
with \(g,h\in\Gamma\). Thus the scalar sector involves only
\(\mathit{I}(g,h)\) and \(\mathit{I}(\epsilon,g)\).

\subparagraph{The \(3\times3\) blocks.}
The six families with one invertible object are
\begin{equation}
 F^B_{gAA}=F^B_{AAg}=U(g),
 \qquad F^B_{AgA}=U(g^{-1}),
 \label{eq:Bootp3-projective-F-mixed-A}
\end{equation}
and
\begin{equation}
 F^A_{gBB}=\overline{U(g)},\qquad
 F^A_{BgB}=\mathit{I}(\epsilon,g)^{-1}
 \,\overline{U(g^{-1})},\qquad
 F^A_{BBg}=\mathit{I}(\epsilon,g)^{-1}
 \,\overline{U(g)}.
 \label{eq:Bootp3-projective-F-mixed-B}
\end{equation}
The two families with three non-invertible objects and invertible
outcome are
\begin{equation}
 F^g_{AAA}=U(g)\,U(\epsilon^{-1}),
 \qquad
 F^g_{BBB}=\overline{U(g)\,U(\epsilon)}.
 \label{eq:Bootp3-projective-F}
\end{equation}

\subparagraph{The \(9\times9\) blocks.}Writing \(g\) for the invertible intermediate object of the left tree
(\(a\otimes b\to g\)) and \(h\) for that of the right tree
(\(b\otimes c\to h\)), the six blocks of size \(9\times9\) are
\begin{equation}
\begin{aligned}
 \bigl(F^A_{AAB}\bigr)_{(\alpha,\beta),h}&=\tfrac{1}{\sqrt{3}}\,U(h)_{\alpha\beta},
 &\bigl(F^B_{BBA}\bigr)_{(\alpha,\beta),h}&=\tfrac{1}{\sqrt{3}}\,\mathit{I}(\epsilon,h)\,\overline{U(h)}_{\alpha\beta},\\
 \bigl(F^B_{ABB}\bigr)_{g,(\gamma,\delta)}&=\tfrac{1}{\sqrt{3}}\overline{U(g)\,U(\epsilon^{-1})}_{\delta\gamma},
 &\bigl(F^A_{BAA}\bigr)_{g,(\gamma,\delta)}&=\tfrac{1}{\sqrt{3}}\bigl(U(g)\,U(\epsilon)\bigr)_{\delta\gamma},\\
 \bigl(F^A_{ABA}\bigr)_{g,h}&=\tfrac13\,\mathit{I}(g,h)^{-1}\,\mathit{I}(\epsilon,h),
 &\bigl(F^B_{BAB}\bigr)_{g,h}&=\tfrac13\,\mathit{I}(g,h)\,\mathit{I}(\epsilon,h)^{-1}.
\end{aligned}
\label{eq:Bootp3-projective-F-nine}
\end{equation}
In these bases, all \(F\)-symbols are determined by
\((\epsilon,\iota)\), up to the choice of basis in which
\(U\) is written. 
These F-symbols are unitary, as expected from the unitarity of the \(J\)-maps
in Eq.~\eqref{eq:Bootp3-invariant-gauge}.\footnote{The normalization of the
\(9\times9\) blocks parallels the Tambara--Yamagami case. There, the only
non-scalar block \([F^{m}_{mmm}]_{g,h}=\tau_{\mathrm{TY}}/\xi(g,h)\) in
Eq.~\eqref{eq:BootDQ-TY-Fsymbols2} has \(|\tau_{\mathrm{TY}}|=|\Gamma|^{-1/2}=\tfrac12\), and
is unitary because the characters \(\xi(g,\cdot)\) are orthogonal. Here the
blocks \(F^{A}_{ABA}\) and \(F^{B}_{BAB}\), in which both fusion trees pass
through invertible intermediate objects, play the same role: all their
\(|\Gamma|^2\) entries are nonzero, and the factor
\(|\Gamma|^{-1/2}=\tfrac13\) together with the orthogonality of the
bicharacter makes them unitary. The remaining four \(9\times9\) blocks have
no counterpart in \(\mathrm{TY}(\mathbb Z_2\times\mathbb Z_2)\), since one of
their fusion trees passes through the multiplicity-three channel
\(A\otimes A=3B\) or \(B\otimes B=3A\).}

For each \((\epsilon,\iota)\), these formulas specify one matrix
\(F^d_{abc}\) for every choice of \(a,b,c,d\) with
\(\operatorname{Hom}(a\otimes b\otimes c,d)\neq0\).
Its rows and columns are indexed by the admissible intermediate
objects and multiplicity labels in the two fusion trees.
The fusion rules therefore determine both the number and the
dimensions of these matrices, giving \(1779\) blocks:
\begin{equation}
 1701\text{ scalars},\qquad
 72\text{ blocks of size }3\times3,\qquad
 6\text{ blocks of size }9\times9,
 \label{eq:Bootp3-F-blockcount}
\end{equation}
for a total of \(2835\) entries.

For all eighteen forms, we substitute the parameter
values in Eq.~\eqref{eq:Bootp3-invariant-gauge},
with \((\epsilon,\iota)=(\epsilon_q,\iota_q)\), into the
corresponding \(J\)-maps and solve the linear system
\eqref{eq:fiber-Fsymb} for the \(F\)-symbols.
The results agree entry by entry, including vanishing
entries, with
Eqs.~\eqref{eq:Bootp3-invariant-F-pointed}--
\eqref{eq:Bootp3-projective-F-nine}.
We also verify all \(64881\) scalar components of the
pentagon for each form, using exact algebraic
simplification.\footnote{
See ``Computing F-symbols'', ``Checking pentagon identities'',
and ``Parametrization'' in \texttt{2\_Fsymbols.nb}.}

\paragraph{Comparison with the triality solution of
Kaidi--Shi--Shimamori--Sun.}
For \(\epsilon=\mathsf{1}\), our \(F\)-symbols for
\(\operatorname{Rep}(\mathrm{Esp}_3)\) are equivalent to 
the corresponding triality solution in Eq.~(6.14)
of~\cite{Kaidi:2025hyr}.
As noted there, that formula gives one solution for
the chosen bicharacter, corresponding to trivial
fractionalization.
For the same bicharacter, our construction also gives
the associators of \(\operatorname{Rep}(\mathrm{Esp}_9)\),
with \(\epsilon\neq\mathsf{1}\), describing a second,
monoidally inequivalent category not included in
that explicit solution.\footnote{
For the alternating bicharacter considered here,
the condition in their Eq.~(6.13) forces their
intertwining matrix to be scalar. Their three-defect
block therefore has nonzero trace only in the identity
channel, whereas ours has nonzero trace at
\(g=\epsilon\).}

\subsection{Relabeling orbits and representative forms}
\label{ssec:Bootp3-parametric-comparison}

Different gauge classes, defined by changes of basis in
the fusion spaces, can describe equivalent monoidal
categories when fusion-ring relabelings are allowed,
as in Eq.~\eqref{eq:F_equiv_auto}.
This parallels the labeled associators of
\(\operatorname{Rep}(H_8)\), which are gauge-inequivalent
but related by relabelings. For \(\operatorname{Rep}(D_8)\),
instead, the three forms share the same associator and yield
distinct fiber functors on that fixed category.
We now study these relabelings for the eighteen forms.

\paragraph{Relabelings of the fusion ring.}
The automorphisms of the fusion ring form the group
\begin{equation}
 \operatorname{Aut}(\mathcal K(\mathcal C))
 \simeq GL(2,\mathbb F_3)\times\mathbb Z_2,
 \label{eq:Bootp3-fusion-ring-aut}
\end{equation}
where the first factor acts on \(\Gamma\) and fixes \(A,B\),
while the second exchanges \(A\) and \(B\).

Let \(\sigma\in GL(2,\mathbb F_3)\), and relabel every
\(g\in\Gamma\) as \(\sigma(g)\).
The invariants of the relabeled associator are
\begin{equation}
 \epsilon'=\sigma(\epsilon),
 \qquad
 \iota'=\det(\sigma)^{-1}\iota.
 \label{eq:Bootp3-invariant-relabeling}
\end{equation}
Indeed, the distinguished channel \(\epsilon\) is relabeled
as \(\sigma(\epsilon)\), while the bicharacter becomes
\[
 \mathit{I}'(g,h)
 =\mathit{I}(\sigma^{-1}(g),\sigma^{-1}(h)).
\]
The transformation of \(\iota\) follows from
Eq.~\eqref{eq:Bootp3-second-invariant-bicharacter} and
\[
 \det[\sigma^{-1}x\ \sigma^{-1}y]
 =\det(\sigma)^{-1}\det[x\ y].
\]
The determinant and its inverse are taken in
\(\mathbb F_3^\ast\).

Relabeling replaces \(\epsilon,\mathit{I},U\) by
\(\epsilon',\mathit{I}',U\circ\sigma^{-1}\) in the formulas
of Section~\ref{ssec:Bootp3-Fsymbols}.
Let \(U'\) be the realization in
Eq.~\eqref{eq:Bootp3-projective-basis} with parameter \(\iota'\).
There is a unitary matrix \(W\) such that\footnote{Both representations have two-cocycle
\((\mathit{I}')^{-1}\). Nondegeneracy of \(\mathit{I}'\)
ensures uniqueness of the irreducible unitary projective
representation up to unitary equivalence.}
\[
 W U(\sigma^{-1}(g))W^{-1}=U'(g),
 \qquad g\in\Gamma.
\]
Thus the fusion-space transformations
\(M_{AA}^{B}=W\), \(M_{BB}^{A}=\overline W\), with all other
\(M\)'s equal to the identity, restore the chosen
representative for \((\epsilon',\iota')\).

Combining \(\sigma\) with the exchange \(A\leftrightarrow B\)
transforms the associator invariants as
\begin{equation}
 \epsilon'=\sigma(\epsilon)^{-1},
 \qquad
 \iota'=-\det(\sigma)^{-1}\iota.
 \label{eq:Bootp3-invariant-relabeling-swap}
\end{equation}
The exchange inverts the determinant object and the
alternating bicharacter, as explained below.

The pairs \((\epsilon_q,\iota_q)\) form two relabeling
orbits: \(q=1,2\), with \(\epsilon_q=\mathsf{1}\), and
\(q=3,\ldots,18\), with \(\epsilon_q\neq\mathsf{1}\).
Indeed, \(GL(2,\mathbb F_3)\) acts transitively on the
nonidentity elements of \(\Gamma\), and their stabilizers
contain matrices of both determinant values.
Thus \(\epsilon\) can be transported to any nonidentity
element while \(\iota\) can independently take either value.
For \(\epsilon=\mathsf{1}\), a determinant-\(2\) transformation
exchanges the two values of \(\iota\).
The two orbits correspond to
\(\operatorname{Rep}(\mathrm{Esp}_3)\) and
\(\operatorname{Rep}(\mathrm{Esp}_9)\), respectively.\footnote{
See ``Automorphisms of the fusion ring'' in
\texttt{4\_FusRingAut.nb}.
The code enumerates the \(48\) elements of
\(GL(2,\mathbb F_3)\), with and without exchanging \(A,B\),
and compares \texttt{epsilon[ParamForm]} and the
bicharacter matrices \texttt{Invariant[ParamForm]}.}

For example, combining \(\sigma=-\mathrm{id}_{\Gamma}\)
with the exchange \(A\leftrightarrow B\) gives
\[
 (\epsilon,\iota)\longmapsto(\epsilon,-\iota),
\]
and therefore exchanges the two forms with the same
\(\epsilon\).
Explicit relabelings relating the remaining forms, together
with the basis changes needed to recover the chosen matrix
representatives, are described in
Appendix~\ref{app:p3-mixed-matrices}.

\paragraph{Reduction to two representative forms.}
The equivalences established above allow us to work with
a single parametric form in each orbit: transport along
a monoidal equivalence gives a bijection between
monoidal-isomorphism classes of fiber functors.
For the calculations in the next subsection, we therefore
choose \(q=1,3\), with
\((\epsilon_1,\iota_1)=(\mathsf{1},1)\) and
\((\epsilon_3,\iota_3)=(g_2,1)\), and fix one associator
representative for each.

This reduction also applies to the parametric \(J\)-maps.
Transporting \(J\) and \(\tau\) by a fusion-ring automorphism
preserves the square equations, invertibility, and maximality.
Up to changes of basis in the \(V_a\), the transported data
therefore belong to the classified form identified by the
transformed invariants. A fusion-space basis transformation
can then bring the associator to the chosen gauge.

The fiber functors obtained for \(q=1,3\) can thus be
transported to all other forms in their respective orbits.
At each fixed associator, we classify them up to monoidal
isomorphism, without further identifying them under
source-category autoequivalences.

\subsection{Fiber functors for a fixed associator}
\label{ssec:Bootp3-fixed-associator}

In this subsection, following the approach of
Section~\ref{ssec:BootTY-fiber-count}, we fix an explicit
associator for each of the two relabeling orbits and
classify the compatible \(J\)-maps up to monoidal isomorphism.

\begin{table}[t!]
\centering\small\setlength{\tabcolsep}{5pt}\renewcommand{\arraystretch}{1.15}
\begin{tabular}{c|ccccccccc}
\toprule
$g\backslash h$ & $g_1$ & $g_2$ & $g_3$ & $g_4$ & $g_5$ & $g_6$ & $g_7$ & $g_8$ & $g_9$\\
\midrule
$g_{1}$ & $0$ & $0$ & $0$ & $0$ & $0$ & $0$ & $0$ & $0$ & $0$\\
$g_{2}$ & $0$ & $1$ & $0$ & $2$ & $1$ & $0$ & $2$ & $1$ & $2$\\
$g_{3}$ & $0$ & $2$ & $0$ & $1$ & $2$ & $0$ & $1$ & $2$ & $1$\\
$g_{4}$ & $0$ & $0$ & $0$ & $0$ & $0$ & $0$ & $0$ & $0$ & $0$\\
$g_{5}$ & $0$ & $0$ & $0$ & $0$ & $0$ & $0$ & $0$ & $0$ & $0$\\
$g_{6}$ & $0$ & $1$ & $0$ & $2$ & $1$ & $0$ & $2$ & $1$ & $2$\\
$g_{7}$ & $0$ & $2$ & $0$ & $1$ & $2$ & $0$ & $1$ & $2$ & $1$\\
$g_{8}$ & $0$ & $2$ & $0$ & $1$ & $2$ & $0$ & $1$ & $2$ & $1$\\
$g_{9}$ & $0$ & $1$ & $0$ & $2$ & $1$ & $0$ & $2$ & $1$ & $2$\\
\bottomrule\end{tabular}
\caption{Exponents of \(\omega\) for the pointed coefficients \(J_{gh}=\omega^{\iota m_hu_g}\) at \(\iota=1\), common to the forms \(q=1\) and \(q=3\); rows are labelled by \(g\), columns by \(h\). }
\label{tab:Bootp3-appendix-Jgh}
\end{table}
By the orbit reduction in
Section~\ref{ssec:Bootp3-parametric-comparison}, we need only
\(q=1\) (\(\operatorname{Rep}(\mathrm{Esp}_3)\),
\((\epsilon,\iota)=(\mathsf{1},1)\)) and
\(q=3\) (\(\operatorname{Rep}(\mathrm{Esp}_9)\),
\((\epsilon,\iota)=(g_2,1)\)), printed in
Appendix~\ref{app:p3-mixed-matrices}.
To make the calculation reproducible by hand, we give the
corresponding specializations of
Eq.~\eqref{eq:Bootp3-invariant-gauge}.
In both cases \(\mathrm x_g=\widetilde{\mathrm x}_g=1\) and
\(\mathrm R=\mathbbm1_3\). The pointed coefficients \(J_{gh}=\omega^{m_hu_g}\)
are the same for the two forms and are listed in
Table~\ref{tab:Bootp3-appendix-Jgh}; the remaining parameters are given in
Table~\ref{tab:Bootp3-appendix-parameters}.
\begin{table}[t!]
\centering\small\setlength{\tabcolsep}{5pt}\renewcommand{\arraystretch}{1.2}
\begin{tabular}{c|c|ccccccccc}
\toprule
$\quad q\quad$ & Parameters & $g_1$ & $g_2$ & $g_3$ & $g_4$ & $g_5$ & $g_6$ & $g_7$ & $g_8$ & $g_9$\\
\midrule
\multirow{3}{*}{$1$}
& $\mathrm y_g=\widetilde{\mathrm y}_g$ & $0$ & $0$ & $2$ & $0$ & $0$ & $1$ & $1$ & $0$ & $2$\\
& $\mathrm z_g$ & $0$ & $0$ & $0$ & $0$ & $0$ & $0$ & $0$ & $0$ & $0$\\
& $\widetilde{\mathrm z}_g$ & $0$ & $0$ & $0$ & $0$ & $0$ & $0$ & $0$ & $0$ & $0$\\
\midrule
\multirow{3}{*}{$3$}
& $\mathrm y_g=\widetilde{\mathrm y}_g$ & $0$ & $0$ & $2$ & $0$ & $0$ & $1$ & $1$ & $0$ & $2$\\
& $\mathrm z_g$ & $0$ & $1$ & $0$ & $2$ & $1$ & $0$ & $2$ & $1$ & $2$\\
& $\widetilde{\mathrm z}_g$ & $0$ & $1$ & $2$ & $0$ & $0$ & $1$ & $2$ & $2$ & $1$\\
\bottomrule
\end{tabular}
\caption{Exponents of \(\omega\) for the parameters of
Eq.~\eqref{eq:Bootp3-invariant-gauge} whose phases are not identically
trivial, for the parametric forms \(q=1\) and \(q=3\) printed in
Appendix~C. An entry \(k\) means \(\mathrm y_g=\widetilde{\mathrm y}_g=\omega^{k}\)
and \(\mathrm z_g=\omega^{k}/\sqrt3\), \(\widetilde{\mathrm z}_g=\omega^{k}/\sqrt3\). In both cases
\(\mathrm x_g=\widetilde{\mathrm x}_g=1\), \(\mathrm R=\mathbbm 1_3\), and
\(J_{gh}\) is given in Table~\ref{tab:Bootp3-appendix-Jgh}; 
moreover \(\mathrm S=\mathbbm 1_3\)
for \(q=1\) and \(\mathrm S=\operatorname{diag}(\omega^2,1,\omega)\) for \(q=3\).}
\label{tab:Bootp3-appendix-parameters}
\end{table}
Inserting these values into the $J$-matrices of Appendices~\ref{sec:AppEsp3}
and~\ref{sec:AppEsp9} and solving Eq.~\eqref{eq:fiber-Fsymb} gives the
\(F\)-symbols
\eqref{eq:Bootp3-invariant-F-pointed}--\eqref{eq:Bootp3-projective-F-nine}
with \((\epsilon,\iota)=(\mathsf{1},1)\) and \((g_2,1)\), respectively.
As a first check, the blocks \(F^B_{gAA}\) for the two generators are
\[
 U(g_2)=ZX,\qquad U(g_3)=\omega Z^2,
\]
in agreement with Eq.~\eqref{eq:Bootp3-projective-basis}, and they
satisfy \(U(g_2)U(g_3)=\omega\,U(g_3)U(g_2)\) and \(U(g_2)U(g_3)=\mathit{I}(g_2,g_3)^{-1}U(g_5)\), with \(g_5=g_2g_3\).

For either of the two representative forms, denote the
reference maps chosen above by \(\widetilde J\) and their
associator by \(F\).
We now keep \(F\) fixed and determine the remaining
choices of \(J\), up to monoidal isomorphism.

Applying a fusion-space transformation \(M\) to the
reference maps \(\widetilde J\) transforms their
\(F\)-symbols according to Eq.~\eqref{eq:F_equiv_base}.
Consequently, transformations in the stabilizer of \(F\)
produce \(J\)-maps compatible with the same associator.
Every compatible solution within the chosen parametric
form arises in this way: it is related to the reference
maps by a fusion-space transformation, which must preserve
\(F\), since invertible \(J\)-maps determine their
\(F\)-symbols uniquely.

For \(\operatorname{Rep}(D_8)\), fusion-ring relabelings
relate the three parametric forms of \(J\) while preserving
the same \(F\)-symbols in the chosen fusion-space bases.
Here, instead, distinct forms have different invariants
\((\epsilon,\iota)\), so a relabeling can preserve \(F\),
possibly after a fusion-space basis transformation,
only if it preserves this pair.
Let \(J'\) be obtained from \(\widetilde J\) by a relabeling
and a fusion-space transformation restoring \(F\).
Since these operations preserve maximality, \(J'\) can
be brought to a classified parametric form by changes
of basis in the \(V_a\).
These changes leave \(F\), and hence \((\epsilon,\iota)\),
unchanged, so this must be the same form as
\(\widetilde J\).
Within that form, every solution compatible with \(F\)
is obtained by a fusion-space transformation in its
stabilizer, as shown above.
Thus including relabelings yields no additional
monoidal-isomorphism classes.

The solutions obtained by acting with this stabilizer
may still define monoidally isomorphic fiber functors.
We now determine the stabilizer and classify the resulting
\(J\)-maps modulo changes of basis in the spaces \(V_a\).

\paragraph{Pointed and mixed sectors.}
Write \(M_{g,h}:=M_{g,h}^{gh}\in\mathbb C^\ast\).
Preservation of the pointed \(F\)-symbols gives
\[
 M_{g,h}M_{gh,k}=M_{h,k}M_{g,hk},
 \qquad g,h,k\in\Gamma.
\]
Comparing the equations for the triples \((A,g,h)\) and \((A,h,g)\)
also gives \(M_{g,h}=M_{h,g}\).
A symmetric \(\mathbb C^\ast\)-valued \(2\)-cocycle on the
finite Abelian group \(\Gamma\) is a coboundary:
\[
 M_{g,h}=\frac{t_g t_h}{t_{gh}},
 \qquad t_g\in\mathbb C^\ast.
\]
The changes of basis \(\phi_g=t_g\) in the spaces \(V_g\)
then set
\begin{equation}
 M_{g,h}=1,
 \qquad g,h\in\Gamma.
 \label{eq:Bootp3-pointed-gauge}
\end{equation}
The remaining rescalings satisfy \(\phi_{gh}=\phi_g\phi_h\),
so they are characters of \(\Gamma\).

Requiring the transformations \(M\) to preserve the mixed
\(F\)-symbols in Eq.~\eqref{eq:Bootp3-invariant-F-scalars}
gives
\begin{equation}
 M_{Ag}^{A}=M_{gB}^{B},
 \qquad M_{gA}^{A}=M_{Bg}^{B},
 \label{eq:Bootp3-mixed-identifications}
\end{equation}
and
\begin{equation}
 M_{Ah}^{A}M_{Ag}^{A}=M_{A,hg}^{A},
 \qquad M_{hA}^{A}M_{gA}^{A}=M_{hg,A}^{A}.
 \label{eq:Bootp3-mixed-characters}
\end{equation}
Thus both \(g\mapsto M_{Ag}^{A}\) and \(g\mapsto M_{gA}^{A}\)
are characters. With the convention of
Eq.~\eqref{eq:fiber_equiv_base}, choosing \(\phi_g=M_{gA}^{A}\)
and \(\phi_A=\phi_B=\mathbbm1_3\) sets
\(M_{gA}^{A}=M_{Bg}^{B}=1\). The remaining character is
\begin{equation}
 \vartheta(g):=\frac{M_{Ag}^{A}}{M_{gA}^{A}}
 =\frac{M_{gB}^{B}}{M_{Bg}^{B}} \,,
 \qquad \vartheta\in\widehat\Gamma\:.
 \label{eq:Bootp3-residual-character}
\end{equation}
Collecting the mixed-sector transformations determined so far,
we obtain
\begin{equation}
 M_{gA}^{A}=M_{Bg}^{B}=1,
 \qquad M_{Ag}^{A}=M_{gB}^{B}=\vartheta(g).
 \label{eq:Bootp3-mixed-gauge}
\end{equation}
Equivalently, the mixed maps are
\begin{equation}
\begin{aligned}
 J_{gA}&=\widetilde J^{(q)}_{gA},
 &\qquad J_{Bg}&=\widetilde J^{(q)}_{Bg},\\
 J_{Ag}&=\vartheta(g)\widetilde J^{(q)}_{Ag},
 & J_{gB}&=\vartheta(g)\widetilde J^{(q)}_{gB}.
\end{aligned}
\label{eq:Bootp3-mixed-normalized-J}
\end{equation}
Thus the normalized mixed maps are determined by the character
\(\vartheta\).

\paragraph{The remaining channels.}
Requiring \(M\) to preserve the blocks
\(F_{gAA}^{B}=U(g)\) and
\(F_{gBB}^{A}=\overline{U(g)}\), together with
Eq.~\eqref{eq:Bootp3-mixed-gauge}, gives
\begin{equation}
 \left(M_{AA}^{B}\right) U(g)\left(M_{AA}^{B}\right)^{-1}=\vartheta(g)U(g),
 \qquad
 \left(M_{BB}^{A}\right)\overline{U(g)}\left(M_{BB}^{A}\right)^{-1}=\vartheta(g)^{-1}\overline{U(g)}.
 \label{eq:Bootp3-fixed-F-intertwiners}
\end{equation}

Since \(\mathit{I}\) is nondegenerate, there is a unique \(k\in\Gamma\)
such that 
\begin{equation}
\vartheta(g)=\mathit{I}(k,g) \qquad\mbox{ for all }g \:.
\end{equation}
The projective multiplication law implies
\(U(k)U(g)U(k)^{-1}=\mathit{I}(k,g)U(g)\).
Irreducibility of \(U\) and Schur's lemma then give
\begin{equation}
 M_{AA}^{B}=c_AU(k),\qquad M_{BB}^{A}=c_B\overline{U(k)},
 \qquad c_A,c_B\in\mathbb C^\ast.
 \label{eq:Bootp3-fixed-F-PQ}
\end{equation}

The remaining equations impose a constraint on
the character \(\vartheta\), which so far could be any
element of \(\widehat\Gamma\).
All entries of \(F_{ABA}^{A}\) are nonzero, so cancellation
in each entry of its stabilizer equation gives
\begin{equation}
 M_{AB}^{g}=\vartheta(h)M_{BA}^{h},
 \qquad g,h\in\Gamma \:.
\end{equation}
In particular, taking \(g=h=\epsilon\) yields
\begin{equation}
 M_{AB}^{\epsilon}
 =\vartheta(\epsilon)M_{BA}^{\epsilon}.
\end{equation}
On the other hand, \(F_{AAA}^{\epsilon}=\mathbbm1_3\), so its
stabilizer equation reduces to
\begin{equation}
 M_{BA}^{\epsilon}\,M_{AA}^{B}
 =M_{AB}^{\epsilon}\,M_{AA}^{B}.
\end{equation}
Since \(M_{AA}^{B}\) is invertible and \(M_{BA}^{\epsilon}\neq0\),
these two equations imply
\begin{equation}
 \vartheta(\epsilon)=1 \:.
 \label{eq:Bootp3-character-constraint}
\end{equation}
Thus the allowed characters must be trivial on
the subgroup generated by \(\epsilon\).

To complete the determination of the remaining coefficients,
notice that \(M_{AB}^{g}=\vartheta(h)M_{BA}^{h}\) holds for
all independently varying \(g,h\).
Thus \(M_{AB}^{g}\) is independent of \(g\), and
\begin{equation}
 M_{BA}^{g}=\vartheta(g)^{-1}M_{AB}^{\mathsf{1}}.
\end{equation}
To determine \(M_{AB}^{\mathsf{1}}\), use the
\(h=\mathsf{1}\) column of \(F_{AAB}^{A}\) in
Eq.~\eqref{eq:Bootp3-projective-F-nine}.
Since \(U(\mathsf{1})=\mathbbm1_3\), its entries are
\(\delta_{\alpha\beta}/3\).
Acting with \(M_{AA}^{B},M_{BB}^{A}\) on its two multiplicity
indices gives
\(\frac13 M_{AA}^{B}(M_{BB}^{A})^{\mathsf T}\).
Using \(\vartheta(\mathsf{1})=1\), the stabilizer equation
for this column therefore reads
\[
 M_{AB}^{\mathsf{1}}\mathbbm1_3 = M_{AA}^{B}(M_{BB}^{A})^{\mathsf T}
 =c_Ac_B\,U(k)U(k)^\dagger=c_Ac_B\mathbbm1_3 \:,
\]
where in the last steps we have used Eq.~\eqref{eq:Bootp3-fixed-F-PQ} and unitarity of \(U(k)\).
Hence
\begin{equation}
 M_{AB}^{g}=c_Ac_B,
 \qquad
 M_{BA}^{g}=c_Ac_B\,\vartheta(g)^{-1}.
 \label{eq:Bootp3-fixed-F-ABBA}
\end{equation}
These coefficients multiply the reference maps:
\(J_{AB}^{g}=M_{AB}^{g}\widetilde J_{AB}^{(q),g}\) and
\(J_{BA}^{g}=M_{BA}^{g}\widetilde J_{BA}^{(q),g}\).
Conversely, applying Eq.~\eqref{eq:F_equiv_base} with the
transformations obtained above leaves all the \(F\)-symbols
of Section~\ref{ssec:Bootp3-Fsymbols} unchanged whenever
\(\vartheta(\epsilon)=1\).
This follows by substitution, using the projective
multiplication law for \(U\).
Thus \(\vartheta(\epsilon)=1\) is necessary and sufficient
for these transformations to preserve \(F^{(q)}\).

\paragraph{Normalization and isomorphism classes.}
For a fixed character \(\vartheta\), the solutions obtained
above depend only on the nonzero scalars \(c_A,c_B\).
Their coefficient matrices are
\begin{equation}
 \mathrm R=c_AU(k)\, \widetilde{\mathrm R} \,,
 \qquad
 \mathrm S=c_B\overline{U(k)}\, \widetilde{\mathrm S} \:,
 \label{eq:Bootp3-fixed-F-RS}
\end{equation}
where \(k\in\Gamma\) is uniquely determined by
\(\vartheta(g)=\mathit{I}(k,g)\), and
\(\widetilde{\mathrm R},\widetilde{\mathrm S}\) are the reference matrices
in Eq.~\eqref{eq:Bootp3-invariant-gauge}.
We now remove this scalar freedom by changes of basis
in \(V_A,V_B\).

In the basis chosen in Eq.~\eqref{eq:Bootp3-projective-basis},
\(U(k)\) has exactly one nonzero entry in each row and column.
Since the reference matrices \(\widetilde{\mathrm R},\widetilde{\mathrm S}\)
are invertible diagonal matrices in this gauge,
\(\mathrm R,\mathrm S\) also have exactly
one nonzero entry in each row. Their first-row sums
\[
 r:=\sum_{j=1}^{3}\mathrm r_{1j}
 \qquad\mbox{and}\qquad
 s:=\sum_{j=1}^{3}\mathrm s_{1j}
\]
are therefore nonzero for every solution. Hence, the normalization
\begin{equation}
 r=s=1
 \label{eq:Bootp3-row-sum-normalization}
\end{equation}
can be achieved by scalar changes of basis
\(\phi_A=\lambda_A\mathbbm1_3\),
\(\phi_B=\lambda_B\mathbbm1_3\), with \(\phi_g=1\).
Indeed, these changes of basis preserve the normalized
pointed and mixed maps and act as
\[
 \mathrm R\longmapsto
 \frac{\lambda_B}{\lambda_A^2}\mathrm R,
 \qquad
 \mathrm S\longmapsto
 \frac{\lambda_A}{\lambda_B^2}\mathrm S.
\]
Choosing $\lambda_A^3=r^2s$ and $\lambda_B=\frac{\lambda_A^2}{r}$,
where \(r,s\) are the first-row sums before the change of basis,
sets both sums to \(1\), so this normalization excludes no solutions.
The choice of cube root does not affect the resulting
\(J\)-maps: replacing
\((\lambda_A,\lambda_B)\) by
\((\alpha\lambda_A,\alpha^2\lambda_B)\), with \(\alpha^3=1\),
acts trivially on every channel.

For a fixed \(\vartheta\), the first-row normalization
fixes \(c_A,c_B\) uniquely, and
Eq.~\eqref{eq:Bootp3-fixed-F-ABBA} then fixes the remaining
maps. Hence all solutions with the same \(\vartheta\)
are monoidally isomorphic and admit a unique representative
satisfying the pointed, mixed, and first-row normalizations.

To distinguish the solutions associated with different
characters, recall that
\(\widetilde J_{Ag}=\widetilde J_{gA}\).
After the canonical identifications involving
the one-dimensional space \(V_g\),
Eq.~\eqref{eq:Bootp3-mixed-normalized-J} gives
\[
 J_{Ag}J_{gA}^{-1}=\vartheta(g)\mathbbm1_3.
\]
Under a change of basis in the \(V_a\), the left-hand side
is conjugated by \(\phi_A\).
Its scalar value \(\vartheta(g)\) is therefore unchanged.
Distinct characters consequently define monoidally
non-isomorphic fiber functors.

The monoidal-isomorphism classes obtained at fixed \(F\)
are thus parametrized by
\begin{equation}
 \{\vartheta\in\widehat\Gamma:
   \vartheta(\epsilon)=1\}\:.
\end{equation}
For the representative \(q=1\), with
\((\epsilon,\iota)=(\mathsf{1},1)\), all nine characters
are allowed.
For the representative \(q=3\), with
\((\epsilon,\iota)=(g_2,1)\), the element \(\epsilon=g_2\)
has order three, so exactly three characters satisfy
\(\vartheta(\epsilon)=1\).
Transporting these classifications to the other forms
in the same relabeling orbits gives
\begin{equation}
\begin{aligned}
 \epsilon=\mathsf{1} &: &&9\ \text{isomorphism classes of fiber functors},\\
 \epsilon\neq\mathsf{1} &: &&3\ \text{isomorphism classes of fiber functors}.
\end{aligned}
\label{eq:Bootp3-fixed-F-count}
\end{equation}
These counts agree with the solutions obtained
by fixing the chosen associator of each form and solving
Eq.~\eqref{eq:fiber-Fsymb} for the free parameters of
\(J\), with the normalizations described above.
Although the orbit reduction allows us to restrict
the analytical classification to \(q=1,3\), we performed
this numerical check for all eighteen forms.\footnote{
See ``Computing J from F-symbols'' in
\texttt{3\_FiberFunctors.nb}.
The comparisons of the solutions and the fusion-ring
automorphisms are implemented in
\texttt{4\_FusRingAut.nb}; the common fusion-ring
definitions are in \texttt{0\_Functions.wl}.}
We now describe how the character labels are transported
between the chosen associator representatives.

\paragraph{Transport of the fiber-functor characters.}
Let \(\sigma\in GL(2,\mathbb F_3)\).
We use the restoring fusion-space transformations described
in Section~\ref{ssec:Bootp3-parametric-comparison} to compare
the transported maps with the chosen associator representatives.
For the mixed \(J\)-maps, the corresponding transformation
of the character is
\begin{equation}
 \vartheta'(\sigma(g))=\vartheta(g).
 \label{eq:Bootp3-character-relabeling}
\end{equation}
The changes of basis \(W,\overline W\) act only on the
multiplicity-three channels and leave the mixed maps
unchanged, so this relation also holds after restoring
the chosen associator representative.

For the \(J\)-maps, the exchange replaces the ratio
\(J_{Ag}J_{gA}^{-1}\) by \(J_{Bg}J_{gB}^{-1}\).
Using Eq.~\eqref{eq:Bootp3-mixed-normalized-J}, the latter
equals \(\vartheta(g)^{-1}\mathbbm1_3\).
Thus the relabeled maps have character
\(\vartheta\circ\sigma^{-1}\) inverted pointwise.
To compare them with the chosen associator representative,
we take a restoring fusion-space transformation that
leaves this character unchanged.\footnote{
If a restoring transformation introduces a character
factor \(\eta\), both the initial and final characters
are trivial on \(\epsilon'\), so \(\eta(\epsilon')=1\).
Composing with the transformation preserving the target
associator and corresponding to \(\eta^{-1}\), constructed above, removes
this factor.}
With this choice, the transported fiber-functor character
satisfies
\begin{equation}
 \vartheta'(\sigma(g))=\vartheta(g)^{-1}.
 \label{eq:Bootp3-character-relabeling-swap}
\end{equation}

Including the fiber-functor character gives the triples
\[
 (\epsilon,\iota,\vartheta),
 \qquad \vartheta(\epsilon)=1,
\]
which distinguish the \(2\cdot9+16\cdot3=66\)
normalized labeled solutions.

\subsection{Expected generalization to arbitrary odd primes}
\label{ssec:Bootp-general-p}

The computation for \(p=3\) suggests a uniform extension
to the fusion rings in Eq.~\eqref{eq:Fus_Esp}
for arbitrary odd primes.
We continue to impose the maximality condition
\eqref{eq:Bootp-maximal-branch}.
The square equations now involve multiplicity-\(p\)
channels; once their solutions are determined,
the \(F\)-symbols can again be reconstructed linearly
from Eq.~\eqref{eq:fiber-Fsymb}.

The representation-theoretic data below motivate
a conjectural description of these solutions.
What remains to be established for \(p>3\) is that
the proposed invariants classify all parametric forms
and their compatible fiber functors.

\paragraph{Group-theoretical interpretation at \(p=3\).}
The invariants \(\epsilon\) and \(\iota\) encode the
determinant and central character of the representation
associated with \(A\). This interpretation motivates
the extension to arbitrary odd primes.

Let \(G=\operatorname{Aut}^{\otimes}(\mathcal F)\)
and let \(\rho_A\colon G\to GL(V_A)\) be the
three-dimensional representation associated with \(A\).
The first invariant is its determinant:
\begin{equation}
 \epsilon=\det\rho_A,
 \label{eq:Bootp3-epsilon-determinant}
\end{equation}
where the one-dimensional representation
\(\det\rho_A\) is identified with an object of \(\Gamma\).

Indeed, up to a nonzero scalar, \(F_{AAA}^{g}\)
represents the cyclic permutation of the three tensor
factors on
\(\operatorname{Hom}_G(V_A^{\otimes3},V_g)\).
Its trace is \(3\) in the channel
\(\bigwedge^3V_A=\det V_A\) and zero in every other
channel. 
Since a cyclic permutation has order three, its spectrum
is \(\{1,1,1\}\) in the determinant channel and
\(\{1,\omega,\omega^2\}\) otherwise.
This gives precisely the values of \(\mathcal E\)
in Eq.~\eqref{eq:Bootp3-epsilon-definition}.

The second invariant specifies the central character.
With the generators \(x,y,z\) of
Eq.~\eqref{eq:extraspecial_pres} chosen compatibly with
the conventions for \(g_2,g_3\), the central element \(z\)
acts as
\begin{equation}
 \rho_A(z)=\omega^{\iota}\mathbbm{1}_3.
 \label{eq:Bootp3-b-central-character}
\end{equation}
The relation \(xy=yxz\) then gives
\begin{equation}
 \rho_A(x)\rho_A(y)
 =
 \omega^{\iota}\rho_A(y)\rho_A(x).
 \label{eq:Bootp3-b-commutator}
\end{equation}
Thus \(\iota=1,2\) records the two nontrivial central
characters.

These identifications explain the two and sixteen
labeled forms.
For \(\mathrm{Esp}_3\), the determinant character is
trivial, leaving only the two choices of \(\iota\).
For \(\mathrm{Esp}_9\), it is nontrivial and can receive
any of the eight nonidentity labels in \(\Gamma\);
together with \(\iota=1,2\), this gives \(8\cdot2=16\)
possibilities.

Finally, \(B\) is dual to \(A\), so
\begin{equation}
 \det\rho_B=\epsilon^{-1},
 \qquad
 \rho_B(z)=\omega^{-\iota}\mathbbm{1}_3.
 \label{eq:Bootp3-dual-invariants}
\end{equation}
This explains the inversion of \(\epsilon\) and
the sign change of \(\iota\) under \(A\leftrightarrow B\).

\paragraph{Conjectural classification of the parametric forms.}
We now propose an extension of the classification obtained
at \(p=3\): the labeling of parametric forms by associator
invariants, their relabeling orbits, and the classification
of fiber functors at fixed associator.
The group-theoretical interpretation above suggests
the following invariants for the distinguished object \(X_1\).

The first is
\begin{equation}
 \epsilon:=\det\rho_{X_1}\in\Gamma_p,
 \label{eq:Bootp-general-epsilon}
\end{equation}
which is trivial for \(\mathrm{Esp}_p\) and nontrivial
for \(\mathrm{Esp}_{p^2}\).
It is carried by \(\bigwedge^pV_{X_1}\); in terms of the
central character \(\psi\) of \(\rho_{X_1}\),
\[
 \epsilon(t)=\det\rho_{X_1}(t)=\psi(t^p),
 \qquad t\in G.
\]
This interpretation extends to every odd prime,
whereas the explicit formula
\((\operatorname{tr}F)^3/\det F\) used above is specific
to \(p=3\); an analogous extraction from the
\(F\)-symbols for \(p>3\) remains to be established.

The second is a nontrivial alternating bicharacter,
expected to take the form
\begin{equation}
 \mathit{I}(g,h)
 =\omega_p^{\,\iota(m_g n_h-n_g m_h)},
 \qquad
 \iota\in\mathbb F_p^\ast,
 \qquad
 \omega_p=e^{2\pi i/p}.
 \label{eq:Bootp-general-bicharacter}
\end{equation}
Here \(g,h\in\Gamma_p\) are written in the coordinates
determined by a fixed pair of generators.

We conjecture that each pair
\((\epsilon,\iota)\in\Gamma_p\times\mathbb F_p^\ast\)
labels exactly one parametric form and one gauge class
of the reconstructed associators.
The cases \(\epsilon=\mathsf{1}\) and
\(\epsilon\neq\mathsf{1}\) would therefore give
\begin{equation}
\begin{aligned}
 \operatorname{Rep}(\mathrm{Esp}_p)&:
 &&p-1\ \text{forms},\\
 \operatorname{Rep}(\mathrm{Esp}_{p^2})&:
 &&(p^2-1)(p-1)\ \text{forms},
\end{aligned}
\label{eq:Bootp-general-form-count}
\end{equation}
for a total of \(p^2(p-1)\).
At \(p=3\), this recovers the decomposition \(2+16=18\).

\paragraph{Relabeling orbits.}
The fusion-ring relabeling group is
\begin{equation}
 \operatorname{Aut}(\mathcal{K}(\mathcal{C}))
 \simeq GL(2,\mathbb F_p)\times\mathbb F_p^\ast,
 \label{eq:Bootp-general-ring-aut}
\end{equation}
where the first factor acts on \(\Gamma_p\), while
\(u\in\mathbb F_p^\ast\) sends \(X_m\) to \(X_{um}\).
Under \(GL(2,\mathbb F_p)\), the proposed invariant
pairs transform as
\[
 (\epsilon,\iota)\longmapsto
 \bigl(\sigma(\epsilon),\det(\sigma)^{-1}\iota\bigr).
\]
More generally, combining this relabeling with the
choice of the old \(X_k\) as the new distinguished
object \(X_1\) gives
\[
 (\epsilon,\iota)\longmapsto
 \bigl(\sigma(\epsilon)^k,\,
 k^{-1}\det(\sigma)^{-1}\iota\bigr).
\]
The pairs form two orbits, distinguished by whether
\(\epsilon\) is trivial, matching the two representation
categories in Eq.~\eqref{eq:Bootp-general-form-count}.
Their stabilizers in the full relabeling group have
orders \(|GL(2,\mathbb F_p)|\) for
\(\epsilon=\mathsf{1}\) and \(p(p-1)\) otherwise.

\paragraph{Fiber functors at fixed associator.}
As in the \(p=3\) classification, we conjecture that
the monoidal-isomorphism classes of fiber functors
at fixed associator are parametrized by
\begin{equation}
 \{\vartheta\in\widehat\Gamma_p:
   \vartheta(\epsilon)=1\}.
 \label{eq:Bootp-general-character-condition}
\end{equation}
For \(\epsilon=\mathsf{1}\), all \(p^2\) characters
are allowed. For \(\epsilon\neq\mathsf{1}\), the element
\(\epsilon\) has order \(p\), so exactly \(p\) characters
satisfy this condition.
These counts agree with the categorical classification
in Section~\ref{ssec:Bootp3-counting-fiber-functors};
the explicit parametrization remains conjectural
for \(p>3\).

\subsection{Counting fiber functors}
\label{ssec:Bootp3-counting-fiber-functors}

We now use the group-theoretical classification recalled
in Section~\ref{ssec:BootTY-fiber-count} to establish
the fiber-functor counts for arbitrary odd primes.
At \(p=3\), this confirms the completeness of the lists
obtained in Section~\ref{ssec:Bootp3-fixed-associator};
for \(p>3\), it verifies the counts predicted by the
conjectural character parametrization above.

Recall that fiber functors on \(\operatorname{Rep}(G)\),
up to monoidal natural isomorphism, are classified by
\(G\)-conjugacy classes of pairs
\begin{equation}
 (H,[\psi]),
 \qquad H\leq G,
 \qquad [\psi]\in H^2(H,\mathbb C^\ast),
 \label{eq:Bootp3-Ostrik-pairs}
\end{equation}
with \([\psi]\) non-degenerate, meaning that
\(\mathbb C_\psi[H]\) is a full matrix
algebra~\cite{ostrik2006modulecategoriesdrinfelddouble}.
Thus \(|H|\) must be a square.
For \(|G|=p^3\), the possibilities are \(H=\{1\}\),
which contributes with the trivial class, and subgroups of order \(p^2\).
Among the latter, only \(H\simeq\mathbb Z_p^2\)
admit non-degenerate classes: indeed,
\begin{equation}
 H^2(\mathbb Z_p^2,\mathbb C^\ast)
 \simeq\mathbb Z_p,
 \label{eq:Bootp3-elementary-abelian-H2}
\end{equation}
with all \(p-1\) nonzero classes non-degenerate,
whereas \(H^2(\mathbb Z_{p^2},\mathbb C^\ast)=0\).

For either group in Eq.~\eqref{eq:extraspecial_pres},
the \(p+1\) subgroups of order \(p^2\) are
\begin{equation}
 H_{[m:n]}:=\langle a^m b^n,c\rangle,
 \qquad [m:n]\in\mathbb P^1(\mathbb F_p).
 \label{eq:Bootp3-maximal-subgroups}
\end{equation}

All these subgroups are normal and conjugation acts as
\[
 a^m b^n\longmapsto a^m b^n c^s,
 \qquad c\longmapsto c,
 \qquad s\in\mathbb F_p.
\]
This automorphism preserves
every alternating bicharacter and hence every class
in \(H^2(H_{[m:n]},\mathbb C^\ast)\).
Consequently, distinct subgroups and distinct
non-degenerate classes on each subgroup remain distinct
under the conjugacy quotient in
Eq.~\eqref{eq:Bootp3-Ostrik-pairs}.

It remains to determine for which choices $[m:n]\in \mathbb{P}^1(\mathbb{F}_p)$ the subgroup \(H_{[m:n]}\) is isomorphic to $\mathbb{Z}_p^2$.
For both groups in Eq.~\eqref{eq:extraspecial_pres}, the commutation relations give
\begin{equation}
 (a^m b^n)^r
 =a^{mr}b^{nr}c^{-mn\binom{r}{2}}.
 \label{eq:extraspecial_powers}
\end{equation}
If \(p\) is odd, the central factor is trivial
when \(r=p\).

For \(G=\mathrm{Esp}_p\), we have \(a^p=b^p=1\),
so every \(H_{[m:n]}\) is isomorphic to \(\mathbb Z_p^2\).
Including the trivial subgroup with its non-degenerate trivial class, the number of
inequivalent fiber functors is therefore
\begin{equation}
 1+(p+1)(p-1)=p^2.
 \label{eq:Bootp3-Ostrik-count-Esp-p}
\end{equation}

For \(G=\mathrm{Esp}_{p^2}\), the relations \(a^p=b^p=c\)
instead give
\begin{equation}
 (a^m b^n)^p=c^{m+n}.
 \label{eq:Bootp3-Esp-p2-powers}
\end{equation}
Thus \(H_{[m:n]}\cong \mathbb{Z}_p^2\) precisely when
\(m+n=0\) in \(\mathbb F_p\).
This selects a single point of \(\mathbb P^1(\mathbb F_p)\);
the remaining \(p\) subgroups are cyclic of order \(p^2\).
The number of inequivalent fiber functors is consequently
\begin{equation}
 1+(p-1)=p.
 \label{eq:Bootp3-Ostrik-count-Esp-p2}
\end{equation}

At \(p=3\), these counts give nine inequivalent fiber
functors on \(\operatorname{Rep}(\mathrm{Esp}_3)\)
and three on \(\operatorname{Rep}(\mathrm{Esp}_9)\),
in agreement with Eq.~\eqref{eq:Bootp3-fixed-F-count}.
For one associator representative of each category,
our solutions are distinguished respectively by all
nine characters of \(\Gamma\), or by the three characters
satisfying \(\vartheta(\epsilon)=1\).
The categorical classification therefore confirms
that these lists exhaust the monoidal-isomorphism
classes of fiber functors on the two categories.
The eighteen parametric forms and their combined list
of sixty-six solutions also record the choice of
labeled associator representative; they do not describe
additional fiber functors on either category.

This completeness statement concerns the two reconstructed
representation categories.
Solutions of the original fusion-ring problem with fewer
tensor automorphisms remain to be classified.
As in the \(H_8\) analysis of
Section~\ref{ssec:BootDQ-Hopf}, they can be investigated
by relaxing the maximality condition and imposing
the square and associativity equations.

\paragraph{Characters and pairs $(H,[\psi])$.}
The two parametrizations of the fiber functors are related explicitly. Let
$\pi\colon G\to G/Z(G)\simeq\mathbb Z_p^2$ and let $\lambda_A$ be the central character of
$\rho_A$. Since $\Gamma\simeq\widehat{G/Z(G)}$, there is a canonical isomorphism
$\widehat\Gamma\simeq G/Z(G)$, $\vartheta(g)=g(t_\vartheta)$. Twisting the reference fiber
functor by $\Omega_\alpha$ supported on $H$, Eq.~\eqref{eq:app-twisted-J} gives, with
$b(\chi,\eta)=\alpha(\chi,\eta)/\alpha(\eta,\chi)=\chi(h_\eta)$,
\begin{equation*}
  J_{Ag}J_{gA}^{-1}=\sum_{\chi\in\widehat H} b\big(\chi,g|_H\big)\,\rho_A(e_\chi)
  =\rho_A\big(h_{g|_H}\big)=\lambda_A\big(h_{g|_H}\big)\,\mathbbm 1_p ,
\end{equation*}
where $h_{g|_H}\in Z(G)$ because $g|_H$ is trivial on $Z(G)$ and $b$ is alternating.
Comparing with Eq.~\eqref{eq:Bootp3-mixed-normalized-J}, $\vartheta(g)=\lambda_A(h_{g|_H})$ and
$\ker\vartheta=\mathrm{Ann}(H/Z(G))$. Hence $\vartheta=1$ labels the pair $(\{1\},[1])$, while
$\vartheta\neq1$ labels $H_\vartheta=\pi^{-1}\langle t_\vartheta\rangle$, with the position of
$t_\vartheta$ on this line fixing the non-degenerate class $[\psi]$. This is consistent with
Section~\ref{Sec:ApplicationRepEspGr}, where the center of the interface algebra is spanned by the $\mathcal O_g$ with
$g\in\ker\vartheta$. Using $\epsilon(t)=\lambda_A(t^p)$ (Section~\ref{ssec:Bootp-general-p}), the selection
rule~\eqref{eq:Bootp3-character-constraint} becomes
\begin{equation*}
  \vartheta(\epsilon)=\epsilon(t_\vartheta)=\lambda_A\big(t_\vartheta^{\,p}\big)=1
  \iff t_\vartheta^{\,p}=1 \iff H_\vartheta\simeq\mathbb Z_p^2 .
\end{equation*}
The associator constraint therefore reproduces the non-degeneracy condition of
Eq.~\eqref{eq:extraspecial_powers}: the pairs with $H\simeq\mathbb Z_{p^2}$, which carry no non-degenerate class,
are exactly those excluded by $\vartheta(\epsilon)\neq1$.

\section{Non-invertible SPT phases and anomalous interfaces}\label{sec:physics}

In this section, we interpret the mathematical results of this paper
in terms of bosonic, unitary theories in $1+1$ dimensions.
We review the established physical interpretations of $F$-symbols
and fiber functors, and then use the results of
Section~\ref{sec:extraspecial} to generalize previous work.

\subsection{Review of boundaries, phases and interfaces}

The input of the bootstrap algorithm is a finite set of simple objects $a \in \mathscr{I}$ with a fusion ring $\mathcal{K}$. The former label simple topological lines of a two-dimensional QFT, and the latter describes their parallel fusion:
\begin{center}
    \tikzset{every picture/.style={line width=0.75pt}} 

\begin{tikzpicture}[x=0.75pt,y=0.75pt,yscale=-1,xscale=1]

\draw    (99.67,62.04) -- (100,163.04) ;
\draw   (95.33,118.04) .. controls (97.84,113.47) and (99.34,108.9) .. (99.83,104.33) .. controls (100.34,108.9) and (101.84,113.47) .. (104.33,118.04) ;
\draw    (224.67,63.04) -- (225,164.04) ;
\draw   (220.33,119.04) .. controls (222.84,114.47) and (224.34,109.9) .. (224.83,105.33) .. controls (225.34,109.9) and (226.84,114.47) .. (229.33,119.04) ;
\draw    (61.67,62.04) -- (62,163.04) ;
\draw   (57.33,118.04) .. controls (59.84,113.47) and (61.34,108.9) .. (61.83,104.33) .. controls (62.34,108.9) and (63.84,113.47) .. (66.33,118.04) ;

\draw (122,103) node [anchor=north west][inner sep=0.75pt]   [align=left] {$\displaystyle =\, \bigoplus _{c} \ N_{ab}^{c}$ \ \ \ \ \ \ \ \ \ \ \  .};
\draw (46.67,147.67) node [anchor=north west][inner sep=0.75pt]   [align=left] {$\displaystyle a$};
\draw (85,148.33) node [anchor=north west][inner sep=0.75pt]   [align=left] {$\displaystyle b$};
\draw (209.67,147) node [anchor=north west][inner sep=0.75pt]   [align=left] {$\displaystyle c$};

\end{tikzpicture}
\end{center}
The output of our algorithm is both a set of $F$-symbols solving the pentagon equations of $\mathcal{K}$ and a set of $J$-maps satisfying the compatibility equations \eqref{eq:fiber-Fsymb}. The $F$-symbols encode the associativity of the fusion of topological line defects. In the physical setting considered here, the following graphical representation, introduced in Section~\ref{sec:FusCat}, shows how the $F$-symbols relate different ways of fusing three topological line defects labeled by $a,b,c$:
\begin{center}
\tikzset{every picture/.style={line width=0.75pt}} 

\begin{tikzpicture}[x=0.75pt,y=0.75pt,yscale=-1,xscale=1]

\draw  [draw opacity=0][fill={rgb, 255:red, 0; green, 0; blue, 0 }  ,fill opacity=1 ] (131.07,145.36) .. controls (131.07,143.43) and (132.54,141.86) .. (134.37,141.86) .. controls (136.19,141.86) and (137.67,143.43) .. (137.67,145.36) .. controls (137.67,147.3) and (136.19,148.86) .. (134.37,148.86) .. controls (132.54,148.86) and (131.07,147.3) .. (131.07,145.36) -- cycle ;
\draw    (134.37,145.36) -- (159.33,176) ;
\draw    (176.33,94.67) -- (109.33,175) ;
\draw  [draw opacity=0][fill={rgb, 255:red, 0; green, 0; blue, 0 }  ,fill opacity=1 ] (153.07,119.36) .. controls (153.07,117.43) and (154.54,115.86) .. (156.37,115.86) .. controls (158.19,115.86) and (159.67,117.43) .. (159.67,119.36) .. controls (159.67,121.3) and (158.19,122.86) .. (156.37,122.86) .. controls (154.54,122.86) and (153.07,121.3) .. (153.07,119.36) -- cycle ;
\draw    (156.37,119.36) -- (204.33,176) ;
\draw  [draw opacity=0][fill={rgb, 255:red, 0; green, 0; blue, 0 }  ,fill opacity=1 ] (480.07,145.36) .. controls (480.07,143.43) and (481.54,141.86) .. (483.37,141.86) .. controls (485.19,141.86) and (486.67,143.43) .. (486.67,145.36) .. controls (486.67,147.3) and (485.19,148.86) .. (483.37,148.86) .. controls (481.54,148.86) and (480.07,147.3) .. (480.07,145.36) -- cycle ;
\draw    (483.37,145.36) -- (461.33,176) ;
\draw    (481.33,94.67) -- (414.33,175) ;
\draw  [draw opacity=0][fill={rgb, 255:red, 0; green, 0; blue, 0 }  ,fill opacity=1 ] (458.07,119.36) .. controls (458.07,117.43) and (459.54,115.86) .. (461.37,115.86) .. controls (463.19,115.86) and (464.67,117.43) .. (464.67,119.36) .. controls (464.67,121.3) and (463.19,122.86) .. (461.37,122.86) .. controls (459.54,122.86) and (458.07,121.3) .. (458.07,119.36) -- cycle ;
\draw    (461.37,119.36) -- (509.33,176) ;
\draw   (114.03,163.37) .. controls (118.05,161.79) and (121.42,159.67) .. (124.1,157.02) .. controls (122.08,160.21) and (120.72,163.94) .. (120.04,168.21) ;
\draw   (137.36,135.37) .. controls (141.39,133.79) and (144.75,131.68) .. (147.44,129.02) .. controls (145.41,132.21) and (144.06,135.94) .. (143.37,140.21) ;
\draw   (160.69,107.37) .. controls (164.72,105.79) and (168.08,103.68) .. (170.77,101.02) .. controls (168.74,104.21) and (167.39,107.94) .. (166.71,112.21) ;
\draw   (180.79,154.56) .. controls (180.3,150.27) and (179.11,146.48) .. (177.24,143.2) .. controls (179.8,145.97) and (183.06,148.24) .. (187.02,149.99) ;
\draw   (148.46,169.23) .. controls (147.96,164.93) and (146.78,161.15) .. (144.9,157.87) .. controls (147.47,160.64) and (150.73,162.9) .. (154.68,164.66) ;
\draw   (429.36,151.37) .. controls (433.39,149.79) and (436.75,147.68) .. (439.44,145.02) .. controls (437.41,148.21) and (436.06,151.94) .. (435.37,156.21) ;
\draw   (464.22,166.26) .. controls (468.1,164.34) and (471.26,161.94) .. (473.71,159.06) .. controls (471.98,162.42) and (470.94,166.25) .. (470.63,170.57) ;
\draw   (467.36,105.37) .. controls (471.39,103.79) and (474.75,101.68) .. (477.44,99.02) .. controls (475.41,102.21) and (474.06,105.94) .. (473.37,110.21) ;
\draw   (474.79,142.23) .. controls (474.3,137.93) and (473.11,134.14) .. (471.24,130.87) .. controls (473.8,133.64) and (477.06,135.9) .. (481.02,137.66) ;
\draw   (496.46,167.9) .. controls (495.97,163.6) and (494.78,159.82) .. (492.9,156.53) .. controls (495.47,159.31) and (498.73,161.57) .. (502.68,163.32) ;

\draw (100,178) node [anchor=north west][inner sep=0.75pt]   [align=left] {$\displaystyle a$};
\draw (159.33,178) node [anchor=north west][inner sep=0.75pt]   [align=left] {$\displaystyle b$};
\draw (206.33,178) node [anchor=north west][inner sep=0.75pt]   [align=left] {$\displaystyle c$};
\draw (116.33,129) node [anchor=north west][inner sep=0.75pt]   [align=left] {$\displaystyle \alpha $};
\draw (150.33,131) node [anchor=north west][inner sep=0.75pt]   [align=left] {$\displaystyle e$};
\draw (183.33,76) node [anchor=north west][inner sep=0.75pt]   [align=left] {$\displaystyle d$};
\draw (140.07,102.36) node [anchor=north west][inner sep=0.75pt]   [align=left] {$\displaystyle \beta $};
\draw (212,101) node [anchor=north west][inner sep=0.75pt]   [align=left] {$\displaystyle =\sum _{f\in \mathcal{I}}\sum _{\gamma =1}^{N_{bc}^{f}}\sum _{\delta =1}^{N_{af}^{d}}\left[ F_{abc}^{d}\right]_{( e\alpha \beta )( f\gamma \delta )} $};
\draw (405,178) node [anchor=north west][inner sep=0.75pt]   [align=left] {$\displaystyle a$};
\draw (460.33,178) node [anchor=north west][inner sep=0.75pt]   [align=left] {$\displaystyle b$};
\draw (511.33,178) node [anchor=north west][inner sep=0.75pt]   [align=left] {$\displaystyle c$};
\draw (492.33,129) node [anchor=north west][inner sep=0.75pt]   [align=left] {$\displaystyle \gamma $};
\draw (455.33,131) node [anchor=north west][inner sep=0.75pt]   [align=left] {$\displaystyle f$};
\draw (488.33,76) node [anchor=north west][inner sep=0.75pt]   [align=left] {$\displaystyle d$};
\draw (445.07,102.36) node [anchor=north west][inner sep=0.75pt]   [align=left] {$\displaystyle \delta $};

\end{tikzpicture}
\end{center}
Junctions between these defects correspond to morphisms in the fusion category. Together, these data fully specify the symmetry of the QFT and its anomaly.

The fiber functor $\mathcal{F}$, introduced in Section~\ref{Sec:FiberFunctor}, has three physical interpretations:
\begin{itemize}
\item It corresponds to a strongly-symmetric boundary condition $\mathcal{B}$ \cite{Zhang:2023wlu,Choi:2023xjw}. Every topological line $a$ has topological junctions with the boundary, labeled by an index $i=1,\ldots,d_a$, where $d_a=\mathrm{dim}(V_a)$ and $V_a=\mathcal{F}(a)$, see Fig. \ref{fig:boundary-junction}. In this interpretation, the monoidal structure of the fiber functor describes the fusion of junctions of two lines $a,b$ with the boundary, see Fig. \ref{fig:boundary-J}.
\begin{figure}[ht]
    \centering
    \begin{subfigure}{\linewidth}
        \centering
        \tikzset{every picture/.style={line width=0.75pt}} 

\begin{tikzpicture}[x=0.75pt,y=0.75pt,yscale=-1,xscale=1]

\draw    (240.11,50.44) -- (240.11,183.11) ;
\draw  [draw opacity=0][fill={rgb, 255:red, 155; green, 155; blue, 155 }  ,fill opacity=0.3 ] (240.56,50.44) -- (272.11,50.44) -- (272.11,183.11) -- (240.56,183.11) -- cycle ;
\draw  [draw opacity=0][fill={rgb, 255:red, 0; green, 0; blue, 0 }  ,fill opacity=1 ] (237.07,115.39) .. controls (237.07,113.46) and (238.54,111.89) .. (240.37,111.89) .. controls (242.19,111.89) and (243.67,113.46) .. (243.67,115.39) .. controls (243.67,117.32) and (242.19,118.89) .. (240.37,118.89) .. controls (238.54,118.89) and (237.07,117.32) .. (237.07,115.39) -- cycle ;
\draw    (134.44,115.89) -- (240.37,115.39) ;
\draw   (170.27,110.89) .. controls (175,113.4) and (179.72,114.91) .. (184.44,115.41) .. controls (179.72,115.91) and (175,117.42) .. (170.27,119.93) ;
\draw   (235.84,84.5) .. controls (238.35,79.77) and (239.86,75.05) .. (240.36,70.32) .. controls (240.86,75.05) and (242.37,79.77) .. (244.88,84.5) ;
\draw   (235.84,158.5) .. controls (238.35,153.77) and (239.86,149.05) .. (240.36,144.32) .. controls (240.86,149.05) and (242.37,153.77) .. (244.88,158.5) ;

\draw (134,92.33) node [anchor=north west][inner sep=0.75pt]   [align=left] {$\displaystyle a$};
\draw (249,98.39) node [anchor=north west][inner sep=0.75pt]   [align=left] {$\displaystyle i$};
\draw (220.67,171) node [anchor=north west][inner sep=0.75pt]   [align=left] {$\displaystyle \mathcal{B}$};
\draw (221,46.33) node [anchor=north west][inner sep=0.75pt]   [align=left] {$\displaystyle \mathcal{B}$};

\end{tikzpicture}
        \caption{Topological junction of a line $a$ ending on the strongly-symmetric boundary $\mathcal{B}$, labeled by $i=1,\ldots,d_a$, with $d_a=\dim V_a$ and $V_a=\mathcal{F}(a)$.}
        \label{fig:boundary-junction}
    \end{subfigure}

    \vspace{1em}

    \begin{subfigure}{\linewidth}
        \centering
        \tikzset{every picture/.style={line width=0.75pt}} 

\begin{tikzpicture}[x=0.75pt,y=0.75pt,yscale=-1,xscale=1]

\draw    (162.11,49.44) -- (162.11,186.11) ;
\draw  [draw opacity=0][fill={rgb, 255:red, 155; green, 155; blue, 155 }  ,fill opacity=0.3 ] (162.56,49.33) -- (194.11,49.33) -- (194.11,186.11) -- (162.56,186.11) -- cycle ;
\draw  [draw opacity=0][fill={rgb, 255:red, 0; green, 0; blue, 0 }  ,fill opacity=1 ] (159.26,138.78) .. controls (159.26,136.84) and (160.73,135.28) .. (162.56,135.28) .. controls (164.38,135.28) and (165.86,136.84) .. (165.86,138.78) .. controls (165.86,140.71) and (164.38,142.28) .. (162.56,142.28) .. controls (160.73,142.28) and (159.26,140.71) .. (159.26,138.78) -- cycle ;
\draw    (56.44,138.89) -- (162.37,138.39) ;
\draw   (92.27,133.89) .. controls (97,136.4) and (101.72,137.91) .. (106.44,138.41) .. controls (101.72,138.91) and (97,140.42) .. (92.27,142.93) ;
\draw   (157.84,121.5) .. controls (160.35,116.77) and (161.86,112.05) .. (162.36,107.32) .. controls (162.86,112.05) and (164.37,116.77) .. (166.88,121.5) ;
\draw   (157.84,166.5) .. controls (160.35,161.77) and (161.86,157.05) .. (162.36,152.32) .. controls (162.86,157.05) and (164.37,161.77) .. (166.88,166.5) ;
\draw    (56.44,88.89) -- (162.37,88.39) ;
\draw   (92.27,83.89) .. controls (97,86.4) and (101.72,87.91) .. (106.44,88.41) .. controls (101.72,88.91) and (97,90.42) .. (92.27,92.93) ;
\draw  [draw opacity=0][fill={rgb, 255:red, 0; green, 0; blue, 0 }  ,fill opacity=1 ] (159.26,88.78) .. controls (159.26,86.84) and (160.73,85.28) .. (162.56,85.28) .. controls (164.38,85.28) and (165.86,86.84) .. (165.86,88.78) .. controls (165.86,90.71) and (164.38,92.28) .. (162.56,92.28) .. controls (160.73,92.28) and (159.26,90.71) .. (159.26,88.78) -- cycle ;
\draw   (157.84,74.5) .. controls (160.35,69.77) and (161.86,65.05) .. (162.36,60.32) .. controls (162.86,65.05) and (164.37,69.77) .. (166.88,74.5) ;
\draw    (544.78,52.44) -- (544.78,185.11) ;
\draw  [draw opacity=0][fill={rgb, 255:red, 155; green, 155; blue, 155 }  ,fill opacity=0.3 ] (545.22,52.44) -- (576.78,52.44) -- (576.78,185.11) -- (545.22,185.11) -- cycle ;
\draw  [draw opacity=0][fill={rgb, 255:red, 0; green, 0; blue, 0 }  ,fill opacity=1 ] (541.73,117.39) .. controls (541.73,115.46) and (543.21,113.89) .. (545.03,113.89) .. controls (546.86,113.89) and (548.33,115.46) .. (548.33,117.39) .. controls (548.33,119.32) and (546.86,120.89) .. (545.03,120.89) .. controls (543.21,120.89) and (541.73,119.32) .. (541.73,117.39) -- cycle ;
\draw    (483.11,117.89) -- (545.03,117.39) ;
\draw   (507.94,112.89) .. controls (512.66,115.4) and (517.39,116.91) .. (522.11,117.41) .. controls (517.39,117.91) and (512.66,119.42) .. (507.94,121.93) ;
\draw   (540.5,86.5) .. controls (543.01,81.78) and (544.52,77.05) .. (545.02,72.32) .. controls (545.52,77.05) and (547.03,81.78) .. (549.55,86.5) ;
\draw   (540.5,160.5) .. controls (543.01,155.78) and (544.52,151.05) .. (545.02,146.32) .. controls (545.52,151.05) and (547.03,155.78) .. (549.55,160.5) ;
\draw  [draw opacity=0][fill={rgb, 255:red, 0; green, 0; blue, 0 }  ,fill opacity=1 ] (479.73,117.39) .. controls (479.73,115.46) and (481.21,113.89) .. (483.03,113.89) .. controls (484.86,113.89) and (486.33,115.46) .. (486.33,117.39) .. controls (486.33,119.32) and (484.86,120.89) .. (483.03,120.89) .. controls (481.21,120.89) and (479.73,119.32) .. (479.73,117.39) -- cycle ;
\draw    (435.44,79.22) -- (483.03,117.39) ;
\draw    (437.18,157.63) -- (483.03,117.39) ;
\draw   (449.81,140.36) .. controls (454.67,139.49) and (458.86,137.85) .. (462.39,135.45) .. controls (459.52,138.61) and (457.32,142.54) .. (455.78,147.22) ;
\draw   (454.64,89.14) .. controls (456.47,93.73) and (458.91,97.5) .. (461.97,100.48) .. controls (458.3,98.3) and (454.02,96.94) .. (449.11,96.37) ;

\draw (50.67,66) node [anchor=north west][inner sep=0.75pt]   [align=left] {$\displaystyle a$};
\draw (171.67,78.72) node [anchor=north west][inner sep=0.75pt]   [align=left] {$\displaystyle i$};
\draw (142.67,174) node [anchor=north west][inner sep=0.75pt]   [align=left] {$\displaystyle \mathcal{B}$};
\draw (54,118) node [anchor=north west][inner sep=0.75pt]   [align=left] {$\displaystyle b$};
\draw (170.33,130.06) node [anchor=north west][inner sep=0.75pt]   [align=left] {$\displaystyle j$};
\draw (220,81) node [anchor=north west][inner sep=0.75pt]   [align=left] {$\displaystyle =\sum _{c}\sum _{\gamma =1}^{N_{ab}^{c}}\sum _{k=1}^{d_{c}}\left[( J_{ab})^{-1\ c}\right]_{( i,j)( \gamma ;k)}$};
\draw (511.33,93) node [anchor=north west][inner sep=0.75pt]   [align=left] {$\displaystyle c$};
\draw (553.67,100.39) node [anchor=north west][inner sep=0.75pt]   [align=left] {$\displaystyle k$};
\draw (527,172) node [anchor=north west][inner sep=0.75pt]   [align=left] {$\displaystyle \mathcal{B}$};
\draw (441,60.67) node [anchor=north west][inner sep=0.75pt]   [align=left] {$\displaystyle a$};
\draw (446.33,154) node [anchor=north west][inner sep=0.75pt]   [align=left] {$\displaystyle b$};
\draw (483.33,91) node [anchor=north west][inner sep=0.75pt]   [align=left] {$\displaystyle \gamma $};

\end{tikzpicture}
        \caption{Fusion of the junctions of two lines $a$ and $b$ at the boundary $\mathcal{B}$, implemented by the monoidal structure $J_{a,b}\colon V_a\otimes V_b\to V_{a\otimes b}$ of the fiber functor.}
        \label{fig:boundary-J}
    \end{subfigure}
    \caption{Fiber functor $\mathcal{F}$ as a strongly-symmetric boundary condition $\mathcal{B}$: (a) the spaces $V_a=\mathcal{F}(a)$ are spaces of boundary junctions; (b) the monoidal structure $J$ determines their fusion.}
    \label{fig:fiber-functor-boundary}
\end{figure}
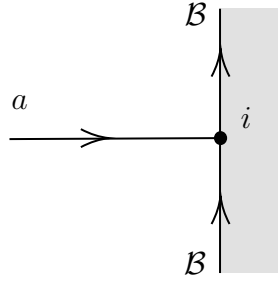
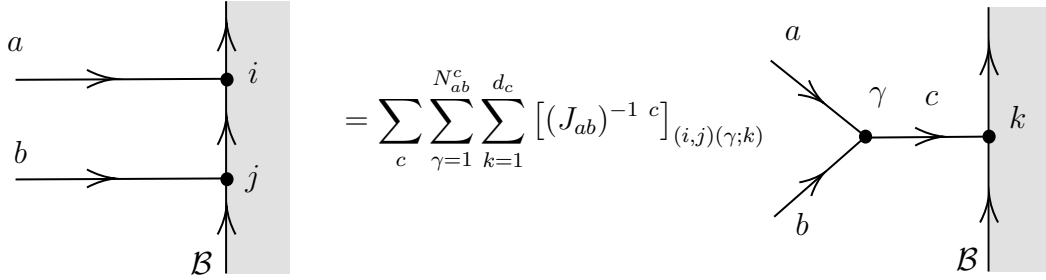
\item It corresponds to a $\mathcal{C}$-symmetric trivially gapped phase. In the deep IR, this is described by a trivial two-dimensional TQFT with topological lines labeled by objects of $\mathcal{C}$ \cite{Huang:2021zvu,Inamura:2021wuo}. The vector spaces $\mathcal{F}(a)=V_a$ correspond to the twisted sector Hilbert spaces and the monoidal structure maps represent pair-of-pants amplitudes, as shown in Fig. \ref{fig:TQFT-fiber-functor}.
\item It corresponds to the monoidal functor mapping the UV symmetry $\mathcal{C}$ to the trivial IR symmetry $\mathrm{Vec}$ \cite{Antinucci:2025fjp}. The image of each object describes the fate of the corresponding topological line under this RG flow. The monoidal maps ensure that the associativity of defect fusion in the UV is compatible with the (trivial) associativity in the IR.
\end{itemize}

\begin{figure}[ht]
    \centering
    \begin{subfigure}[b]{0.45\textwidth}
        \centering
        \resizebox{!}{3.5cm}{\tikzset{every picture/.style={line width=0.75pt}} 

\begin{tikzpicture}[x=0.75pt,y=0.75pt,yscale=-1,xscale=1]

\draw    (178.22,50.22) -- (178.22,186.22) ;
\draw    (265.22,50.22) -- (265.22,186.22) ;
\draw  [color={rgb, 255:red, 0; green, 0; blue, 0 }  ,draw opacity=0.2 ] (178.22,50.22) .. controls (178.22,39.18) and (197.7,30.22) .. (221.72,30.22) .. controls (245.75,30.22) and (265.22,39.18) .. (265.22,50.22) .. controls (265.22,61.27) and (245.75,70.22) .. (221.72,70.22) .. controls (197.7,70.22) and (178.22,61.27) .. (178.22,50.22) -- cycle ;
\draw  [draw opacity=0][dash pattern={on 4.5pt off 4.5pt}] (178.33,186.22) .. controls (178.33,186.22) and (178.33,186.22) .. (178.33,186.22) .. controls (178.33,175.18) and (197.78,166.22) .. (221.78,166.22) .. controls (245.77,166.22) and (265.22,175.18) .. (265.22,186.22) -- (221.78,186.22) -- cycle ; \draw  [dash pattern={on 4.5pt off 4.5pt}] (178.33,186.22) .. controls (178.33,186.22) and (178.33,186.22) .. (178.33,186.22) .. controls (178.33,175.18) and (197.78,166.22) .. (221.78,166.22) .. controls (245.77,166.22) and (265.22,175.18) .. (265.22,186.22) ;  
\draw  [draw opacity=0] (265.11,186.22) .. controls (265.11,186.22) and (265.11,186.22) .. (265.11,186.22) .. controls (265.11,197.27) and (245.66,206.22) .. (221.67,206.22) .. controls (197.67,206.22) and (178.22,197.27) .. (178.22,186.22) -- (221.67,186.22) -- cycle ; \draw   (265.11,186.22) .. controls (265.11,186.22) and (265.11,186.22) .. (265.11,186.22) .. controls (265.11,197.27) and (245.66,206.22) .. (221.67,206.22) .. controls (197.67,206.22) and (178.22,197.27) .. (178.22,186.22) ;  
\draw    (221.72,70.22) -- (221.22,205.89) ;
\draw   (216.84,130.5) .. controls (219.35,125.77) and (220.86,121.05) .. (221.36,116.32) .. controls (221.86,121.05) and (223.37,125.77) .. (225.88,130.5) ;

\draw (228,131) node [anchor=north west][inner sep=0.75pt]   [align=left] {$\displaystyle a$};

\end{tikzpicture}}
        \caption{Twisted sector Hilbert space $V_a=\mathcal{F}(a)$: states on a circle pierced by the line $a$.}
        \label{fig:TQFT-V}
    \end{subfigure}
    \hfill
    \begin{subfigure}[b]{0.45\textwidth}
        \centering
        \resizebox{!}{3.5cm}{\tikzset{every picture/.style={line width=0.75pt}} 

\begin{tikzpicture}[x=0.75pt,y=0.75pt,yscale=-1,xscale=1]

\draw  [draw opacity=0][dash pattern={on 4.5pt off 4.5pt}] (121.67,280.78) .. controls (121.67,280.78) and (121.67,280.78) .. (121.67,280.78) .. controls (121.67,280.78) and (121.67,280.78) .. (121.67,280.78) .. controls (121.67,269.73) and (141.12,260.78) .. (165.11,260.78) .. controls (189.1,260.78) and (208.56,269.73) .. (208.56,280.78) -- (165.11,280.78) -- cycle ; \draw  [dash pattern={on 4.5pt off 4.5pt}] (121.67,280.78) .. controls (121.67,280.78) and (121.67,280.78) .. (121.67,280.78) .. controls (121.67,280.78) and (121.67,280.78) .. (121.67,280.78) .. controls (121.67,269.73) and (141.12,260.78) .. (165.11,260.78) .. controls (189.1,260.78) and (208.56,269.73) .. (208.56,280.78) ;  
\draw  [draw opacity=0] (208.44,280.78) .. controls (208.44,280.78) and (208.44,280.78) .. (208.44,280.78) .. controls (208.44,291.82) and (188.99,300.78) .. (165,300.78) .. controls (141.01,300.78) and (121.56,291.82) .. (121.56,280.78) -- (165,280.78) -- cycle ; \draw   (208.44,280.78) .. controls (208.44,280.78) and (208.44,280.78) .. (208.44,280.78) .. controls (208.44,291.82) and (188.99,300.78) .. (165,300.78) .. controls (141.01,300.78) and (121.56,291.82) .. (121.56,280.78) ;  
\draw  [draw opacity=0][dash pattern={on 4.5pt off 4.5pt}] (263.67,280.78) .. controls (263.67,280.78) and (263.67,280.78) .. (263.67,280.78) .. controls (263.67,280.78) and (263.67,280.78) .. (263.67,280.78) .. controls (263.67,269.73) and (283.12,260.78) .. (307.11,260.78) .. controls (331.1,260.78) and (350.56,269.73) .. (350.56,280.78) -- (307.11,280.78) -- cycle ; \draw  [dash pattern={on 4.5pt off 4.5pt}] (263.67,280.78) .. controls (263.67,280.78) and (263.67,280.78) .. (263.67,280.78) .. controls (263.67,280.78) and (263.67,280.78) .. (263.67,280.78) .. controls (263.67,269.73) and (283.12,260.78) .. (307.11,260.78) .. controls (331.1,260.78) and (350.56,269.73) .. (350.56,280.78) ;  
\draw  [draw opacity=0] (350.44,280.78) .. controls (350.44,280.78) and (350.44,280.78) .. (350.44,280.78) .. controls (350.44,280.78) and (350.44,280.78) .. (350.44,280.78) .. controls (350.44,291.82) and (330.99,300.78) .. (307,300.78) .. controls (283.01,300.78) and (263.56,291.82) .. (263.56,280.78) -- (307,280.78) -- cycle ; \draw   (350.44,280.78) .. controls (350.44,280.78) and (350.44,280.78) .. (350.44,280.78) .. controls (350.44,280.78) and (350.44,280.78) .. (350.44,280.78) .. controls (350.44,291.82) and (330.99,300.78) .. (307,300.78) .. controls (283.01,300.78) and (263.56,291.82) .. (263.56,280.78) ;  
\draw  [draw opacity=0] (208.56,280.81) .. controls (208.56,280.81) and (208.56,280.81) .. (208.56,280.81) .. controls (208.56,280.81) and (208.56,280.81) .. (208.56,280.81) .. controls (208.56,254.07) and (220.72,232.39) .. (235.72,232.39) .. controls (250.73,232.39) and (262.89,254.07) .. (262.89,280.81) -- (235.72,280.81) -- cycle ; \draw   (208.56,280.81) .. controls (208.56,280.81) and (208.56,280.81) .. (208.56,280.81) .. controls (208.56,280.81) and (208.56,280.81) .. (208.56,280.81) .. controls (208.56,254.07) and (220.72,232.39) .. (235.72,232.39) .. controls (250.73,232.39) and (262.89,254.07) .. (262.89,280.81) ;  
\draw    (121.67,280.78) .. controls (121.56,237.22) and (192.22,177.89) .. (192.22,134.81) ;
\draw   (192.22,134.81) .. controls (192.22,123.76) and (211.7,114.81) .. (235.72,114.81) .. controls (259.75,114.81) and (279.22,123.76) .. (279.22,134.81) .. controls (279.22,145.85) and (259.75,154.81) .. (235.72,154.81) .. controls (211.7,154.81) and (192.22,145.85) .. (192.22,134.81) -- cycle ;
\draw    (279.22,134.81) .. controls (280.89,179.89) and (350.22,236.56) .. (350.44,280.78) ;
\draw  [draw opacity=0] (166.22,299.89) .. controls (166.22,247.42) and (197.71,204.89) .. (236.56,204.89) .. controls (275.4,204.89) and (306.89,247.42) .. (306.89,299.89) -- (236.56,299.89) -- cycle ; \draw   (166.22,299.89) .. controls (166.22,247.42) and (197.71,204.89) .. (236.56,204.89) .. controls (275.4,204.89) and (306.89,247.42) .. (306.89,299.89) ;  
\draw    (235.72,154.81) -- (235.56,204.56) ;
\draw   (230.84,183.5) .. controls (233.35,178.77) and (234.86,174.05) .. (235.36,169.32) .. controls (235.86,174.05) and (237.37,178.77) .. (239.88,183.5) ;
\draw   (183.76,227.61) .. controls (188.93,226.22) and (193.41,224.1) .. (197.21,221.24) .. controls (194.1,224.83) and (191.67,229.16) .. (189.93,234.22) ;
\draw   (283.04,236.46) .. controls (282.18,231.17) and (280.53,226.5) .. (278.07,222.43) .. controls (281.33,225.89) and (285.39,228.74) .. (290.24,230.99) ;
\draw  [draw opacity=0][fill={rgb, 255:red, 0; green, 0; blue, 0 }  ,fill opacity=1 ] (232.73,204.39) .. controls (232.73,202.46) and (234.21,200.89) .. (236.03,200.89) .. controls (237.86,200.89) and (239.33,202.46) .. (239.33,204.39) .. controls (239.33,206.32) and (237.86,207.89) .. (236.03,207.89) .. controls (234.21,207.89) and (232.73,206.32) .. (232.73,204.39) -- cycle ;

\draw (222,207.44) node [anchor=north west][inner sep=0.75pt]   [align=left] {$\displaystyle \gamma $};
\draw (161.33,238.78) node [anchor=north west][inner sep=0.75pt]   [align=left] {$\displaystyle a$};
\draw (302.67,239) node [anchor=north west][inner sep=0.75pt]   [align=left] {$\displaystyle b$};
\draw (215.33,168.67) node [anchor=north west][inner sep=0.75pt]   [align=left] {$\displaystyle c$};

\end{tikzpicture}}
        \caption{Pair-of-pants amplitude $[J_{a,b}^c]_\gamma\colon V_a\otimes V_b\to V_{c}$, fusing two twisted sectors through the junction of lines $\gamma \in \mathrm{Hom}(a\otimes b,c)$.}
        \label{fig:TQFT-J}
    \end{subfigure}
    \caption{Fiber functor $\mathcal{F}$ as a trivial $\mathcal{C}$-symmetric TQFT: (a) the vector spaces $\mathcal{F}(a)=V_a$ are the twisted sector Hilbert spaces; (b) the monoidal structure $J$ is given by the pair-of-pants amplitudes.}
    \label{fig:TQFT-fiber-functor}
\end{figure}
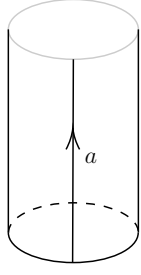
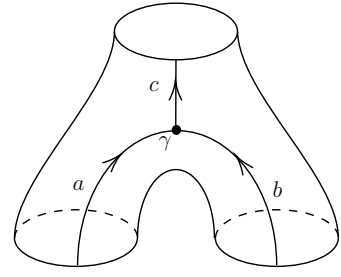

In particular, a symmetry-preserving interface between two distinct SPT phases must support nontrivial degrees of freedom. For SPT phases protected by invertible symmetries, this is understood through anomaly inflow: the interface carries the relative anomaly determined by the two bulk phases. This principle applies in any spacetime dimension.

For a finite internal symmetry group $G$ acting unitarily, bosonic SPT phases in two spacetime dimensions form a torsor over $H^2(G,U(1))$. Choosing a reference phase identifies them with classes $\psi\in H^2(G,U(1))$. The interface between phases labeled by $\psi_1$ and $\psi_2$ then carries the anomaly class $\psi_1\psi_2^{-1}$, with the inverse class corresponding to the opposite orientation. This class characterizes the projective action of $G$ on the interface quantum mechanics and, when nontrivial, enforces a symmetry-protected degeneracy.

For non-invertible symmetries, this description in terms of a relative group-cohomology class does not directly generalize. Instead, the protected interface degrees of freedom can be studied through the algebra of symmetry operators acting on them.

\begin{figure}[t]
\begin{center}
    \tikzset{every picture/.style={line width=0.75pt}} 

\begin{tikzpicture}[x=0.75pt,y=0.75pt,yscale=-1,xscale=1]

\draw [color={rgb, 255:red, 208; green, 2; blue, 27 }  ,draw opacity=1 ][line width=1.5]    (159.6,22.73) -- (159.6,172.33) ;
\draw    (79.6,96.2) -- (241.2,96.8) ;
\draw  [draw opacity=0][fill={rgb, 255:red, 0; green, 0; blue, 0 }  ,fill opacity=1 ] (156.73,96.63) .. controls (156.73,94.7) and (158.21,93.13) .. (160.03,93.13) .. controls (161.86,93.13) and (163.33,94.7) .. (163.33,96.63) .. controls (163.33,98.57) and (161.86,100.13) .. (160.03,100.13) .. controls (158.21,100.13) and (156.73,98.57) .. (156.73,96.63) -- cycle ;
\draw   (109.87,92.02) .. controls (114.6,94.53) and (119.32,96.04) .. (124.04,96.54) .. controls (119.32,97.05) and (114.6,98.55) .. (109.87,101.07) ;
\draw   (197.07,92.82) .. controls (201.8,95.33) and (206.52,96.84) .. (211.24,97.34) .. controls (206.52,97.85) and (201.8,99.35) .. (197.07,101.87) ;

\draw (81.07,76.13) node [anchor=north west][inner sep=0.75pt]   [align=left] {$\displaystyle a$};
\draw (229.67,76.33) node [anchor=north west][inner sep=0.75pt]   [align=left] {$\displaystyle a$};
\draw (162,67.13) node [anchor=north west][inner sep=0.75pt]   [align=left] {$\displaystyle [\mathcal{O}_{a}]_{( i,i')}$};
\draw (73.33,21.67) node [anchor=north west][inner sep=0.75pt]   [align=left] {SPT$\displaystyle _{1}$};
\draw (209.33,21) node [anchor=north west][inner sep=0.75pt]   [align=left] {SPT$\displaystyle _{2}$};
\draw (169,158) node [anchor=north west][inner sep=0.75pt]   [align=left] {\textcolor[rgb]{0.82,0.01,0.11}{$\displaystyle \mathcal{I}$}};

\end{tikzpicture}
\end{center}
    \caption{Topological junction $[\mathcal{O}_a]_{(i,i')}$ at the intersection between the topological defect line $a$ (black horizontal line) and the interface $\mathcal{I}$ (red vertical line) between two distinct SPT phases. The index $i$ labels a basis of the vector space of the SPT on the left, and the primed index $i'$ labels a basis of the vector space of the SPT on the right.}
    \label{fig:interf-O}
\end{figure}
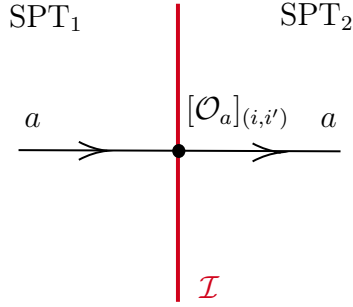

To describe this algebra, consider a topological defect line $a$ crossing a symmetry-preserving interface $\mathcal{I}$ between two distinct SPT phases, as shown in Fig.~\ref{fig:interf-O}. Let $\mathcal{F}_1$ and $\mathcal{F}_2$ be the fiber functors associated with the phases on the left and right, respectively. We denote the topological junction at the intersection by $[\mathcal{O}_a]_{(i,i')}$, where $i$ and $i'$ label bases of $\mathcal{F}_1(a)$ and $\mathcal{F}_2(a)$, respectively.
The fusion of these topological junction operators determines the symmetry algebra acting on the quantum mechanics living on the interface. Their multiplication is given by\footnote{When the fiber functors are interpreted as strongly symmetric boundary conditions $\mathcal{B}_1, \mathcal{B}_2$, the same algebra computes the fusion of topological lines stretching between the two boundaries. This interpretation is closer to the strip algebra $\mathrm{Str}_\mathcal{C}(\mathcal{M})$, but with different module categories with one simple object.} \cite{Kitaev:2011dxc,Komargodski:2020mxz,Huang:2021zvu,Cordova:2024iti,Choi:2024tri,Inamura:2024jke,Benini:2025lav}:
\begin{equation}\label{eq:interface-algebra}
    [\mathcal{O}_a]_{(i,i')}[\mathcal{O}_b]_{(j,j')}=\sum_{c,\gamma,
    k,k'} [(J_{ab}^{(1)})^{-1\,c}]_{(i,j)(\gamma;k)}[\mathcal{O}_c]_{(k,k')}[(J_{ab}^{(2)})^c]_{(\gamma;k')(i',j')} \: .
\end{equation}
The algebra is associative if both $J$-maps are in a basis with the same $F$-symbols. The statement that the interface between two different SPT phases carries an anomaly translates into the absence of one-dimensional representations for the algebra \eqref{eq:interface-algebra}.

Notice that by taking the same SPT phase on both sides, Eq. \eqref{eq:interface-algebra} is very reminiscent of Eq. \eqref{eq:tannaka_square}. In fact, for $\mathcal{C}=\operatorname{Rep}(G)$ the operators $\mathcal{O}_a$ form the algebra of functions on $G$, $\mathrm{Fun}(G)\cong \mathbb C^{|G|}$, and the automorphisms $\tau_a:V_a\rightarrow V_a$ furnish a one-dimensional representation.\footnote{The fact that the entries of the matrices of irreducible representations of $G$ form a basis for the algebra $\mathrm{Fun}(G)$ is the discrete version of the Peter-Weyl theorem. The product of the algebra is given by Clebsch-Gordan coefficients as in Eq. \eqref{eq:interface-algebra}.}
By contrast, for two distinct SPT phases, a one-dimensional
representation of \eqref{eq:interface-algebra} would define
a monoidal natural isomorphism between their fiber functors
\cite{Inamura:2024jke}.
Since the fiber functors are monoidally inequivalent,
no such representation exists.

\begin{figure}[t]
    \centering
    \tikzset{every picture/.style={line width=0.75pt}} 

\begin{tikzpicture}[x=0.75pt,y=0.75pt,yscale=-1,xscale=1]

\draw [color={rgb, 255:red, 208; green, 2; blue, 27 }  ,draw opacity=1 ][line width=1.5]    (139.6,50.4) -- (139.6,200) ;
\draw    (59.6,89.87) -- (221.2,90.47) ;
\draw    (59.6,159.87) -- (221.2,160.47) ;
\draw  [draw opacity=0][fill={rgb, 255:red, 0; green, 0; blue, 0 }  ,fill opacity=1 ] (136.73,160.3) .. controls (136.73,158.37) and (138.21,156.8) .. (140.03,156.8) .. controls (141.86,156.8) and (143.33,158.37) .. (143.33,160.3) .. controls (143.33,162.23) and (141.86,163.8) .. (140.03,163.8) .. controls (138.21,163.8) and (136.73,162.23) .. (136.73,160.3) -- cycle ;
\draw  [draw opacity=0][fill={rgb, 255:red, 0; green, 0; blue, 0 }  ,fill opacity=1 ] (136.73,90.3) .. controls (136.73,88.37) and (138.21,86.8) .. (140.03,86.8) .. controls (141.86,86.8) and (143.33,88.37) .. (143.33,90.3) .. controls (143.33,92.23) and (141.86,93.8) .. (140.03,93.8) .. controls (138.21,93.8) and (136.73,92.23) .. (136.73,90.3) -- cycle ;
\draw   (177.07,156.29) .. controls (181.8,158.8) and (186.52,160.31) .. (191.24,160.81) .. controls (186.52,161.31) and (181.8,162.82) .. (177.07,165.33) ;
\draw   (88.47,156.09) .. controls (93.2,158.6) and (97.92,160.11) .. (102.64,160.61) .. controls (97.92,161.11) and (93.2,162.62) .. (88.47,165.13) ;
\draw   (89.87,85.69) .. controls (94.6,88.2) and (99.32,89.71) .. (104.04,90.21) .. controls (99.32,90.71) and (94.6,92.22) .. (89.87,94.73) ;
\draw   (177.07,86.49) .. controls (181.8,89) and (186.52,90.51) .. (191.24,91.01) .. controls (186.52,91.51) and (181.8,93.02) .. (177.07,95.53) ;
\draw [color={rgb, 255:red, 208; green, 2; blue, 27 }  ,draw opacity=1 ][line width=1.5]    (483.6,50.4) -- (483.6,200) ;
\draw    (448.8,124.9) -- (518.4,125.5) ;
\draw   (458.87,120.69) .. controls (463.6,123.2) and (468.32,124.71) .. (473.04,125.21) .. controls (468.32,125.71) and (463.6,127.22) .. (458.87,129.73) ;
\draw  [draw opacity=0][fill={rgb, 255:red, 0; green, 0; blue, 0 }  ,fill opacity=1 ] (444.53,124.7) .. controls (444.53,122.77) and (446.01,121.2) .. (447.83,121.2) .. controls (449.66,121.2) and (451.13,122.77) .. (451.13,124.7) .. controls (451.13,126.63) and (449.66,128.2) .. (447.83,128.2) .. controls (446.01,128.2) and (444.53,126.63) .. (444.53,124.7) -- cycle ;
\draw  [draw opacity=0][fill={rgb, 255:red, 0; green, 0; blue, 0 }  ,fill opacity=1 ] (480.73,125.3) .. controls (480.73,123.37) and (482.21,121.8) .. (484.03,121.8) .. controls (485.86,121.8) and (487.33,123.37) .. (487.33,125.3) .. controls (487.33,127.23) and (485.86,128.8) .. (484.03,128.8) .. controls (482.21,128.8) and (480.73,127.23) .. (480.73,125.3) -- cycle ;
\draw   (496.07,120.49) .. controls (500.8,123) and (505.52,124.51) .. (510.24,125.01) .. controls (505.52,125.51) and (500.8,127.02) .. (496.07,129.53) ;
\draw  [draw opacity=0][fill={rgb, 255:red, 0; green, 0; blue, 0 }  ,fill opacity=1 ] (516.53,125.7) .. controls (516.53,123.77) and (518.01,122.2) .. (519.83,122.2) .. controls (521.66,122.2) and (523.13,123.77) .. (523.13,125.7) .. controls (523.13,127.63) and (521.66,129.2) .. (519.83,129.2) .. controls (518.01,129.2) and (516.53,127.63) .. (516.53,125.7) -- cycle ;
\draw    (399.84,87.42) -- (447.43,125.59) ;
\draw   (419.04,97.34) .. controls (420.87,101.93) and (423.31,105.7) .. (426.37,108.68) .. controls (422.7,106.5) and (418.42,105.14) .. (413.51,104.57) ;
\draw    (519.84,125.62) -- (567.43,163.79) ;
\draw   (539.04,135.54) .. controls (540.87,140.13) and (543.31,143.9) .. (546.37,146.88) .. controls (542.7,144.7) and (538.42,143.34) .. (533.51,142.77) ;
\draw    (400.58,166.23) -- (446.43,125.99) ;
\draw   (413.21,148.96) .. controls (418.07,148.09) and (422.26,146.45) .. (425.79,144.05) .. controls (422.92,147.21) and (420.72,151.14) .. (419.18,155.82) ;
\draw    (519.18,126.63) -- (565.03,86.39) ;
\draw   (531.81,109.36) .. controls (536.67,108.49) and (540.86,106.85) .. (544.39,104.45) .. controls (541.52,107.61) and (539.32,111.54) .. (537.78,116.22) ;
\draw    (276,125.69) .. controls (307.59,105.53) and (323.78,144.33) .. (354.6,127.12) ;
\draw [shift={(357,125.69)}, rotate = 148.78] [fill={rgb, 255:red, 0; green, 0; blue, 0 }  ][line width=0.08]  [draw opacity=0] (10.72,-5.15) -- (0,0) -- (10.72,5.15) -- (7.12,0) -- cycle    ;

\draw (61.07,69.8) node [anchor=north west][inner sep=0.75pt]   [align=left] {$\displaystyle a$};
\draw (209.67,70) node [anchor=north west][inner sep=0.75pt]   [align=left] {$\displaystyle a$};
\draw (58.07,139.6) node [anchor=north west][inner sep=0.75pt]   [align=left] {$\displaystyle b$};
\draw (209.27,140.2) node [anchor=north west][inner sep=0.75pt]   [align=left] {$\displaystyle b$};
\draw (547.67,72) node [anchor=north west][inner sep=0.75pt]   [align=left] {$\displaystyle a$};
\draw (395.07,142.2) node [anchor=north west][inner sep=0.75pt]   [align=left] {$\displaystyle b$};
\draw (562.27,140.2) node [anchor=north west][inner sep=0.75pt]   [align=left] {$\displaystyle b$};
\draw (405.07,69.8) node [anchor=north west][inner sep=0.75pt]   [align=left] {$\displaystyle a$};
\draw (142,60.8) node [anchor=north west][inner sep=0.75pt]   [align=left] {$\displaystyle [\mathcal{O}_{a}]_{( i,i')}$};
\draw (141.4,132.6) node [anchor=north west][inner sep=0.75pt]   [align=left] {$\displaystyle [\mathcal{O}_{b}]_{( j,j')}$};
\draw (485,138.4) node [anchor=north west][inner sep=0.75pt]   [align=left] {$\displaystyle [\mathcal{O}_{c}]_{( k,k')}$};
\draw (444.2,98.2) node [anchor=north west][inner sep=0.75pt]   [align=left] {$\displaystyle \gamma $};
\draw (513,98.2) node [anchor=north west][inner sep=0.75pt]   [align=left] {$\displaystyle \gamma $};
\draw (463.2,103) node [anchor=north west][inner sep=0.75pt]   [align=left] {$\displaystyle c$};
\draw (148,186) node [anchor=north west][inner sep=0.75pt]   [align=left] {\textcolor[rgb]{0.82,0.01,0.11}{$\displaystyle \mathcal{I}$}};
\draw (493,184) node [anchor=north west][inner sep=0.75pt]   [align=left] {\textcolor[rgb]{0.82,0.01,0.11}{$\displaystyle \mathcal{I}$}};

\end{tikzpicture}
    \caption{Fusion of two junctions $[\mathcal{O}_a]_{(i,i')}$ and $[\mathcal{O}_b]_{(j,j')}$ on the interface $\mathcal{I}$ (red vertical line). The defect lines $a$ and $b$ are fused into lines $c\in a\otimes b$ independently on the two sides of the interface. The resulting configuration has single junctions $[\mathcal{O}_c]_{(k,k')}$. The summation over indices and lines with the respective monoidal structure maps coefficients in Eq. \eqref{eq:interface-algebra} is omitted in this cartoon.}
    \label{fig:boundary-fusion}
\end{figure}
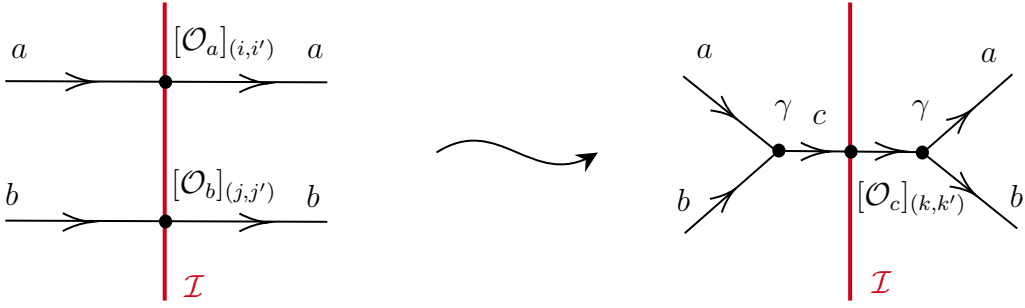

Schematically, this algebra follows from fusing the junctions independently from the left and the right using the $J^{(1)}$ and $J^{(2)}$-maps respectively, as drawn in Fig. \ref{fig:boundary-fusion}.

\subsection{Application to representations of extraspecial groups}\label{Sec:ApplicationRepEspGr}

We study a physical application of fiber functors for the fusion ring of representations of extraspecial groups, with focus on the $p=3$ case explicitly worked out in Section \ref{sec:extraspecial}. For context, we recall that this fusion ring can arise in several scenarios:
\begin{itemize}
    \item from a theory with non-anomalous extraspecial group symmetry, after gauging the entire group;
    \item from a theory with an anomalous $\mathbb Z_p \times \mathbb Z_p \times \mathbb Z_p$ symmetry, after gauging a non-anomalous $\mathbb Z_p \times \mathbb Z_p$ subgroup \cite{Lu:2024lzf};\footnote{Consider the background gauge fields $x,y,z\in H^1(Y_3,\mathbb Z_p)$ on the three-dimensional inflow manifold $Y_3$ for the symmetry $\mathbb{Z}_p^x\times \mathbb{Z}_p^y \times \mathbb{Z}_p^z$. Gauging $\mathbb Z_p^x \times \mathbb Z_p^y$ with anomaly inflow action $S_{\mathrm{inflow}}=\frac{2\pi i}{p}\int_{Y_3} x\cup y\cup z$ gives $\operatorname{Rep}(\mathrm{Esp}_p)$ symmetry, where $\cup$ is the cup product in cohomology. Gauging $\mathbb Z_p^x \times \mathbb Z_p^y$ with anomaly inflow action $S_{\mathrm{inflow}}=\frac{2\pi i}{p}\int_{Y_3} x\cup y\cup z + \frac{2\pi i }{p}\int_{Y_3}  \mathrm{Bock}(x)\cup z + \frac{2\pi i }{p}\int_{Y_3} \mathrm{Bock}(y) \cup z$ gives $\operatorname{Rep}(\mathrm{Esp}_{p^2})$ symmetry, where Bock is the Bockstein homomorphism associated to the short exact sequence $1\to\mathbb Z_p\to\mathbb Z_{p^2}\to \mathbb Z_p\to 1$.}
    \item from a theory with non-anomalous $\mathbb Z_p \times \mathbb Z_p$ symmetry, and with a non-invertible self--$p$-ality under twisted gauging of this abelian group \cite{Lu:2024ytl,Lu:2024lzf}.
\end{itemize}

The fiber functor analysis physically translates into the SPT phases summed up in Table \ref{eq:table-SPT}.
\begin{table}[ht]
    \centering
    $
\begin{array}{c|ccccc}
 \mathcal C
 & \operatorname{Rep}(D_8) & \operatorname{Rep}(Q_8)
 & \operatorname{Rep}(H_8) & \operatorname{Rep}(\mathrm{Esp}_3)
 & \operatorname{Rep}(\mathrm{Esp}_9) 
 \\[2pt]
 \hline
 \#\operatorname{SPT} & 3 & 1 & 1 & 9 & 3  
\end{array}
 $
    \caption{Number of SPT phases per symmetry category.}
    \label{eq:table-SPT}
\end{table}
The simplest example with multiple SPT phases is that of $\mathrm{Rep}(D_8)$. On the interface between any pair of different SPT phases, the algebra \eqref{eq:interface-algebra} was shown \cite{Inamura:2024jke,Choi:2024tri} to be isomorphic to $M_2(\mathbb C)^{\oplus 2}$,\footnote{The algebra $M_n(\mathbb C)$ is the algebra of $n\times n$ matrices generated by $E_{ij}E_{kl}=\delta_{jk}E_{il}$.} which does not admit one-dimensional representations. Its smallest irreducible representation is two-dimensional, i.e. the quantum mechanics on the interface has a qubit with degeneracy protected by the non-invertible symmetry. By contrast, on the interface between two copies of the same SPT phase, the algebra is isomorphic to $M_1(\mathbb C)^{\oplus 8}$, which is abelian and admits one-dimensional representations, compatible with the interface being trivial.

Using the $J$-maps computed in Section \ref{sec:extraspecial}, we compute the interface algebras between inequivalent fiber functors of $\operatorname{Rep}(\mathrm{Esp}_3)$ and $\operatorname{Rep}(\mathrm{Esp}_9)$. In both cases the generators of the algebra are $\mathcal{O}_g$ with $g\in \Gamma=\mathbb Z_3 \times \mathbb Z_3$, $[\mathcal{O}_A]_{(i,i')}$ and $[\mathcal{O}_B]_{(j,j')}$ with $i,i',j,j'=0,1,2.$ 
Recall from Section~\ref{sec:extraspecial} that $\iota$ labels the associator representative, whereas $\vartheta$ labels fiber functors compatible with a fixed associator. The two representatives with $\epsilon=1$ are related by relabeling the simple objects, as discussed in Section~\ref{ssec:Bootp3-parametric-comparison}. We choose the representative with $\iota=1$. For this fixed associator, the nine characters $\vartheta\in\widehat{\Gamma}$ give nine monoidal-isomorphism classes of fiber functors, corresponding to the nine SPT phases in Table~\ref{eq:table-SPT}. We compute the interface algebra with generic $\vartheta$ on the left and $\vartheta=1$ on the right. 
The algebra reads
\begin{equation}\label{IntAlg3_theta|1}
    \begin{split}
        &\mathcal{O}_{(m,n)}\mathcal{O}_{(m',n')}=\mathcal{O}_{(m+m',n+n')}\\
        &\mathcal{O}_{(m,n)}[\mathcal{O}_A]_{(i,i')}=\omega^{-(n-m)(i'-i)}[\mathcal{O}_A]_{(i-m,i'-m)}\\
        &[\mathcal{O}_A]_{(i,i')}\mathcal{O}_{(m,n)}=\vartheta(m,n)^{-1}\omega^{-(n-m)(i'-i)}[\mathcal{O}_A]_{(i-m,i'-m)}\\
        &\mathcal{O}_{(m,n)}[\mathcal{O}_B]_{(i,i')}=\vartheta(m,n)^{-1}\omega^{(n-m)(i'-i)}[\mathcal{O}_B]_{(i-m,i'-m)}\\
        &[\mathcal{O}_B]_{(i,i')}\mathcal{O}_{(m,n)}=\omega^{(n-m)(i'-i)}[\mathcal{O}_B]_{(i-m,i'-m)}\\
        &[\mathcal{O}_A]_{(i,i')}[\mathcal{O}_A]_{(j,j')}=\delta_{j,\,j'+i-i'-e_3}\,\omega^{\eta(j'-i'+1)}[\mathcal{O}_B]_{(j'-i-i'+1-e_3,\,i'+j'+1)}\\
        &[\mathcal{O}_B]_{(i,i')}[\mathcal{O}_B]_{(j,j')}=\delta_{j,\,j'+i-i'-e_3}\,\omega^{-\eta(j'-i'+1)}[\mathcal{O}_A]_{(j'-i-i'+1-e_3,\,i'+j'+1)}\\
        &[\mathcal{O}_A]_{(i,i')}[\mathcal{O}_B]_{(j,j')}=\delta_{j,\,j'+i'-i}\,\frac13\sum_{t=0}^{2}\omega^{(i'-i)(t-1+i'+j')}\,\mathcal{O}_{(1-i'-j',\,t)}\\
        &[\mathcal{O}_B]_{(i,i')}[\mathcal{O}_A]_{(j,j')}=\delta_{j,\,j'+i'-i}\,\frac13\sum_{t=0}^{2}\omega^{(i'-i)(1-i'-j'-t)+e_2(1-i'-j')+e_3t}\,\mathcal{O}_{(1-i'-j',\,t)} \: .
    \end{split}
\end{equation}
where we write $g=g_2^mg_3^n$ as $(m,n)$, set
$\vartheta(g_2)=\omega^{e_2}$ and $\vartheta(g_3)=\omega^{e_3}$, so that
$\vartheta(m,n)=\omega^{e_2m+e_3n}$, and define $\eta=-(e_2+e_3)$; all
indices $m,n,i,i',j,j'$ are taken mod~$3$, and $\delta_{x,y}$ denotes the
Kronecker delta mod~$3$.
When $\vartheta\neq 1$, the algebra \eqref{IntAlg3_theta|1} is isomorphic to the non-abelian algebra $M_3(\mathbb C)^{\oplus3}$ and therefore does not admit one-dimensional representations. Its center is spanned by the elements $\mathcal O_g$ with
$g\in\ker\vartheta\cong\mathbb Z_3$.
The three characters of this subgroup distinguish
the three copies of $M_3(\mathbb C)$. The explicit isomorphism to $M_3(\mathbb C)^{\oplus3}$ is given by the map $\Phi$ acting on the generators as
\begin{equation}\label{Phi3_theta|1}
    \begin{split}
        &\Phi\bigl(\mathcal{O}_{(m,n)}\bigr)=\sum_{r,s=0}^{2}\omega^{rm+sn}\,E_{(r,s),(r,s)}\\
        &\Phi\bigl([\mathcal{O}_A]_{(i,i')}\bigr)=\kappa^{\eta(1+e_3)}\,\omega^{(i'-i)i}\sum_{r=0}^{2}\omega^{-ri}\,E_{(r,\,i-i'),\,(r-e_2,\,i-i'-e_3)}\\
        &\Phi\bigl([\mathcal{O}_B]_{(i,i')}\bigr)=\kappa^{2\eta(1+e_3)}\,\omega^{(i'-i)(1-i)+e_3(1+e_3)-e_2(i+1-e_3)}\sum_{r=0}^{2}\omega^{r(1-i)}\,E_{(r,\,i'-i-e_3),\,(r+e_2,\,i'-i)} \: ,
    \end{split}
\end{equation}
where $\kappa=e^{2\pi i/9}$ and $E_{(r,s),(r',s')}$, with labels $(r,s)\in\mathbb Z_3\times\mathbb Z_3$, satisfy $E_{(r,s),(r',s')}E_{(u,v),(u',v')}=\delta_{r'u}\delta_{s'v}E_{(r,s),(u',v')}$.\footnote{Notice how the image of invertible objects contains matrix elements with the same row and column indices, the image of the object $A$ contains matrix elements where the column index is shifted by $-(e_2,e_3)$ with respect to the row index, while the image of the $B$ object contains matrix elements where the column index is shifted by $+(e_2,e_3)$ with respect to the row index.}
Each three-dimensional irreducible
representation describes a symmetry-protected qutrit
of interface states.

For $\operatorname{Rep}(\mathrm{Esp}_9)$, the pair $(\epsilon,\iota)$,
with $\epsilon\neq1$, labels the associator representative.
For a fixed representative, the compatible fiber-functor classes
are labeled by the three characters $\vartheta\in\widehat{\Gamma}$
that are trivial on the subgroup generated by $\epsilon$.
The compatibility condition $\vartheta(\epsilon)=1$, derived in
Section~\ref{ssec:Bootp3-fixed-associator}, accounts for the reduction
from nine phases to three. We fix $\iota=1$ and $\epsilon=g_2$, and compute interface algebras between SPT phases with a generic $\vartheta$ satisfying $\vartheta(g_2)=1$  on the left and $\vartheta=1$ on the right. Using the same convention as before we have $e_2=0$ and $\eta=-e_3$ and we obtain the algebra
\begin{equation}\label{IntAlg9_theta|1}
    \begin{split}
        &\mathcal{O}_{(m,n)}\mathcal{O}_{(m',n')}=\mathcal{O}_{(m+m',n+n')}\\
        &\mathcal{O}_{(m,n)}[\mathcal{O}_A]_{(i,i')}=\omega^{-(n-m)(i'-i)}[\mathcal{O}_A]_{(i-m,i'-m)}\\
        &[\mathcal{O}_A]_{(i,i')}\mathcal{O}_{(m,n)}=\vartheta(m,n)^{-1}\omega^{-(n-m)(i'-i)}[\mathcal{O}_A]_{(i-m,i'-m)}\\
        &\mathcal{O}_{(m,n)}[\mathcal{O}_B]_{(i,i')}=\vartheta(m,n)^{-1}\omega^{(n-m)(i'-i)}[\mathcal{O}_B]_{(i-m,i'-m)}\\
        &[\mathcal{O}_B]_{(i,i')}\mathcal{O}_{(m,n)}=\omega^{(n-m)(i'-i)}[\mathcal{O}_B]_{(i-m,i'-m)}\\
        &[\mathcal{O}_A]_{(i,i')}[\mathcal{O}_A]_{(j,j')}=\delta_{j,\,j'+i-i'-e_3}\,\omega^{-e_3(j'-i')}[\mathcal{O}_B]_{(j'-i-i'+1-e_3,\,i'+j'+1)}\\
        &[\mathcal{O}_B]_{(i,i')}[\mathcal{O}_B]_{(j,j')}=\delta_{j,\,j'+i-i'-e_3}\,\omega^{(i-i')+e_3(j'-i')}[\mathcal{O}_A]_{(j'-i-i'-e_3,\,i'+j')}\\
        &[\mathcal{O}_A]_{(i,i')}[\mathcal{O}_B]_{(j,j')}=\delta_{j,\,j'+i'-i}\,\frac13\sum_{t=0}^{2}\omega^{(i'-i)(t-1+i'+j')-e_3}\,\mathcal{O}_{(-1-i'-j',\,t)}\\
        &[\mathcal{O}_B]_{(i,i')}[\mathcal{O}_A]_{(j,j')}=\delta_{j,\,j'+i'-i}\,\frac13\sum_{t=0}^{2}\omega^{(i'-i)(1-i'-j'-t)+e_3(t-1)}\,\mathcal{O}_{(-1-i'-j',\,t)} \: .
    \end{split}
\end{equation}
For $\vartheta\neq1$, equivalently $e_3=1,2$, the algebra
\eqref{IntAlg9_theta|1} is again isomorphic to
$M_3(\mathbb C)^{\oplus3}$ via the map
\begin{equation}\label{Phi9_theta|1}
    \begin{split}
        &\Phi\bigl(\mathcal{O}_{(m,n)}\bigr)=\sum_{r,s=0}^{2}\omega^{rm+sn}\,E_{(r,s),(r,s)}\\
        &\Phi\bigl([\mathcal{O}_A]_{(i,i')}\bigr)=\kappa^{-(e_3+1)}\omega^{(i'-i)\,i}\sum_{r=0}^{2}\kappa^{r}\,\omega^{-ri}\,E_{(r,\,i-i'),\,(r,\,i-i'-e_3)}\\
        &\Phi\bigl([\mathcal{O}_B]_{(i,i')}\bigr)=\kappa^{-2(e_3+1)}\omega^{(i'-i)(1-i)+1}\sum_{r=0}^{2}\kappa^{2r}\,\omega^{r(1-i)}\,E_{(r,\,i'-i-e_3),\,(r,\,i'-i)} \: .
    \end{split}
\end{equation}
Thus, interfaces between distinct SPT phases of either
extraspecial category have three irreducible sectors,
each carrying a symmetry-protected triplet.
Symmetry-preserving perturbations may assign different energies
to these sectors but cannot split a triplet.
The reconstructed $J$-maps therefore determine both the protected
multiplet structure and the explicit symmetry action on the
interface states, including their invertible charges.

\section{Conclusions and outlook}
\label{sec:conclusions}

The examples studied in this work show how the passage from fusion
rules to non-invertible SPT phases can be carried out explicitly.
The bootstrap determines both the associators and their compatible
fiber functors, so that the dependence of the SPT classification
on data beyond the fusion ring can be examined directly.

For the Tambara--Yamagami ring associated with $\mathbb Z_2^2$,
the reconstruction agrees with the known classification of
categorifications admitting a fiber functor. The example of $\operatorname{Rep}(H_8)$ shows that relaxing
the maximality condition can allow the bootstrap to recover
fiber functors for categories that are not representation
categories of finite groups.

For the extraspecial fusion ring, the eighteen parametric forms
obtained under the maximality condition describe two representation
categories up to fusion-ring relabelings. This classification must
be distinguished from that of fiber functors within each category,
which label distinct SPT phases. The fiber functors compatible with the reconstructed associator
are labeled by characters satisfying $\vartheta(\epsilon)=1$,
yielding nine classes for $\operatorname{Rep}(\mathrm{Esp}_3)$
and three for $\operatorname{Rep}(\mathrm{Esp}_9)$, in agreement
with the group-theoretical classification.

The invariants $(\epsilon,\iota)$ give a uniform description of
all reconstructed $F$-symbols, including the multiplicity-three
blocks. For $\epsilon=1$, these reproduce the triality associator
of Ref.~\cite{Kaidi:2025hyr} up to gauge.
For $\epsilon\neq1$, we obtain explicit $F$-symbols for
$\operatorname{Rep}(\mathrm{Esp}_9)$, including all multiplicity-three
blocks, a case not covered in that work.
We verify the pentagon equations exactly for all eighteen
labeled forms.

We use the monoidal structure maps to explicitly compute the algebras $M_3(\mathbb C)^{\oplus3}$ acting
on interfaces between distinct SPT phases.  Each algebra has three irreducible sectors, each supporting
a symmetry-protected
qutrit of states. The explicit representations determine the
invertible charges of these states and the action of the
non-invertible symmetry operators.

These results leave several questions for future work.
A first direction is to extend the explicit bootstrap analysis
to extraspecial groups of order $p^3$, with $p$ an odd prime
greater than three. The $p=3$ results suggest that the forms
satisfying the maximality condition are parametrized by
$(\epsilon,\iota)\in\Gamma_p\times\mathbb F_p^*$,
where $\Gamma_p\simeq\mathbb Z_p^2$ is the group of invertible objects.
Although the group-theoretical fiber-functor counts are known,
deriving this parametrization from the bootstrap equations would
explain how the reconstruction extends to arbitrary odd primes.
A further goal is to extract $\epsilon$ from the associator
and establish the parametrization of compatible fiber functors
by characters satisfying $\vartheta(\epsilon)=1$.

A second direction is to study systematically the solutions
obtained by relaxing the maximality condition. The $H_8$ example motivates a search for further Hopf-algebra
categorifications of the extraspecial fusion ring, guided by the
classification of semisimple Hopf algebras of dimension $p^3$.
The case $p=3$ will be treated in a forthcoming paper.
Applying the same procedure to other integral fusion rings would
also clarify the scope and computational efficiency of the bootstrap,
both in separating categorifications and in detecting obstructions
to a fiber functor.

A further physical objective is to realize the reconstructed data
in lattice models and matrix-product-operator constructions.
For \(\operatorname{Rep}(\mathrm{Esp}_3)\) and
\(\operatorname{Rep}(\mathrm{Esp}_9)\), a common microscopic realization
of their nine and three SPT phases, respectively, would allow their
interfaces to be studied within a single framework. The triplet degeneracies
and explicit symmetry action derived here provide concrete
predictions for such models, connecting the categorical
reconstruction to observable properties of non-invertible SPT phases.

\section*{Acknowledgements}

We are deeply indebted to Andrés Collinucci for introducing us to Tannaka duality and its potential applications, as well as for his involvement in the early stages of this project. 
We thank Andrea Antinucci, Philip Boyle Smith, Matteo Dell'Acqua, Johann Quenta-Raygada, Rajath Radhakrishnan, Zhengdi Sun, Michele Tortora, Marco Venuti, and Alison Warman for stimulating discussions.
We also acknowledge the use of ChatGPT (OpenAI) and Claude (Anthropic) for assistance in checking intermediate calculations, improving the presentation, and preparing the manuscript. All AI-assisted output was independently reviewed and verified by the authors, who take full responsibility for the content of this work.
A.S., R.V., and M.Z. acknowledge support from INFN through the Iniziativa Specifica ST\&FI.

\appendix

\section{The transported flip and Tambara's quadratic forms}
\label{app:Tambara-comparison}

In Section~\ref{sec:TYZ2}, the maximal bootstrap analysis produced the four parametric forms summarized in Eq.~\eqref{eq:BootDQ-summary}. In this appendix, we give a basis-independent characterization of these solutions and relate it to Tambara's classification of fiber functors~\cite{Tambara:2000qlz}.

\subsubsection*{The transported flip}

Let
\[
    P:V_m\otimes V_m\longrightarrow V_m\otimes V_m,
    \qquad
    P(v\otimes w)=w\otimes v,
\]
be the flip operator. For \(g=a,b,c\), define
\[
    \mathbb T_g
    :=
    \tau_m^{\chi_g}\otimes\tau_m^{\chi_g},
    \qquad
    \mathbb T_1:=\mathbbm{1}_4.
\]
These matrices are independent of the choice of sign for
\(\tau_m^{\chi_g}\). The relations derived in
Section~\ref{sec:TYZ2} imply
\[
    \mathbb T_g^2=\mathbbm{1}_4,
    \qquad
    \mathbb T_g\mathbb T_h
    =
    \mathbb T_h\mathbb T_g.
\]
They are also linearly independent. More precisely,
\[
    \operatorname{tr}(\mathbb T_g\mathbb T_h)
    =
    4\delta_{g,h},
    \qquad
    g,h\in\Gamma.
\]
Their simultaneous diagonalization therefore gives four one-dimensional joint eigenspaces.

The flip commutes with every \(\mathbb T_g\). Since each row
\(J_{mm}^u\) spans a joint left eigenspace, it has a definite parity under \(P\):
\begin{equation}
    J_{mm}^uP=s_uJ_{mm}^u,
    \qquad
    s_u\in\{\pm1\}.
    \label{eq:app-TY-parity}
\end{equation}
Equivalently, the transported flip
\[
    \widehat P
    :=
    J_{mm}PJ_{mm}^{-1}
\]
acts diagonally on the fusion-channel decomposition:
\begin{equation}
    \widehat P
    =
    \bigoplus_{u\in\Gamma}
    s_u\,\operatorname{id}_{V_u}.
    \label{eq:app-TY-transported-flip}
\end{equation}

The signs \(s_u\) are invariant under changes of basis in the vector spaces \(V_u\). Indeed, under the transformation in
Eq.~\eqref{eq:fiber_equiv_base},
\[
    J_{mm}'
    =
    \Phi\,J_{mm}
    (\phi_m\otimes\phi_m)^{-1},
    \qquad
    \Phi:=\bigoplus_{u\in\Gamma}\phi_u,
\]
one has
\[
    \widehat P'
    =
    \Phi\,\widehat P\,\Phi^{-1}
    =
    \widehat P.
\]
Here we used the fact that \(P\) commutes with
\(\phi_m\otimes\phi_m\), while both \(\Phi\) and \(\widehat P\) act by scalars on each one-dimensional space \(V_u\). Rescaling the individual fusion-channel maps \(J_{mm}^u\) also leaves the signs unchanged. Thus, the collection \((s_u)_{u\in\Gamma}\) is intrinsic up to a relabeling of the elements of \(\Gamma\).

The \((-1)\)-eigenspace of \(P\) is the one-dimensional space
\(\bigwedge^2V_m\), while its \(+1\)-eigenspace is
\(\operatorname{Sym}^2V_m\). Consequently, there is a unique channel
\(u\in\Gamma\) for which \(s_u=-1\). Since
\(\bigwedge^2V_m\) carries the determinant character of the two-dimensional representation \(V_m\), this channel is naturally identified with \(\det(m)\):
\begin{equation}
    s_u=-1
    \quad\Longleftrightarrow\quad
    u=\det(m).
    \label{eq:app-TY-determinant}
\end{equation}

\subsubsection*{Quadratic refinements}

The signs \(s_u\) can be expressed directly in terms of the parameters
\(\varepsilon_g\). The matrices
\[
    \mathbbm{1}_2,\qquad
    \tau_m^{\chi_a},\qquad
    \tau_m^{\chi_b},\qquad
    \tau_m^{\chi_c}
\]
form a trace-orthogonal basis of \(\operatorname{End}(V_m)\), with
\[
    \operatorname{tr}
    \left[
        \left(\tau_m^{\chi_g}\right)^2
    \right]
    =
    2\varepsilon_g.
\]
The corresponding trace-dual completeness relation gives
\begin{equation}
    P
    =
    \frac12
    \left[
        \mathbbm{1}_4
        +
        \sum_{g=a,b,c}
        \varepsilon_g\,
        \tau_m^{\chi_g}\otimes\tau_m^{\chi_g}
    \right].
    \label{eq:app-TY-flip-completeness}
\end{equation}
Combining this identity with
Eq.~\eqref{eq:app-TY-parity} and the joint-eigenvalue equations
\eqref{eq:BootDQ-joint}, we obtain
\begin{equation}
    s_u
    =
    \frac12
    \left[
        1
        +\varepsilon_a\chi_a(u)
        +\varepsilon_b\chi_b(u)
        +\varepsilon_c\chi_c(u)
    \right].
    \label{eq:app-TY-s-epsilon}
\end{equation}
This is consistent with Eq.~\eqref{eq:app-TY-determinant}: since
\(\varepsilon_g=-\det\tau_m^{\chi_g}\) (see Section~\ref{ssec:BootDQ-common}),
one has \(\chi_g(\det m)=-\varepsilon_g\), and
Eq.~\eqref{eq:app-TY-s-epsilon} reduces to
\(s_u=1-2\,\delta_{u,\det(m)}\).

Identifying
\[
    \Gamma\simeq\mathbb F_2^2,
    \qquad
    1=(0,0),\quad
    a=(1,0),\quad
    b=(0,1),\quad
    c=(1,1),
\]
define \(\mathsf{q}:\Gamma\to\mathbb F_2\) by
\begin{equation}
    \frac{s_u}{s_1}=(-1)^{\mathsf{q}(u)}.
    \label{eq:app-TY-q-definition}
\end{equation}
For the four parametric forms, Eqs.~\eqref{eq:app-TY-s-epsilon} and
\eqref{eq:app-TY-q-definition} give
\begin{equation}
\begin{array}{c|c|c|c}
    \text{solution}
    &(\varepsilon_a,\varepsilon_b,\varepsilon_c)
    &(s_1,s_a,s_b,s_c)
    &\mathsf{q}(x,y)
    \\
    \hline
    1&(-,+,+)&(+,-,+,+)&xy+x\\
    2&(+,-,+)&(+,+,-,+)&xy+y\\
    3&(+,+,-)&(+,+,+,-)&xy\\
    4&(-,-,-)&(-,+,+,+)&xy+x+y
\end{array}
\label{eq:app-TY-quadratic-forms}
\end{equation}
In particular,
\[
    \det(m)=a,\ b,\ c
\]
for solutions \(1\), \(2\), and \(3\), respectively, whereas
\(\det(m)=1\) for solution \(4\).

All four functions in Eq.~\eqref{eq:app-TY-quadratic-forms} are quadratic refinements of the same nondegenerate alternating bilinear form
\begin{equation}
\begin{split}
    B\bigl((x,y),(x',y')\bigr)
    &:=
    \mathsf{q}(x+x',y+y')
    +\mathsf{q}(x,y)
    +\mathsf{q}(x',y')
    \\
    &=xy'+x'y.
\end{split}
\label{eq:app-TY-bilinear-form}
\end{equation}
Equivalently, \((-1)^{B(u,v)}=\xi(u,v)\), with \(\xi\) the alternating bicharacter
of Eq.~\eqref{eq:BootDQ-TY-invariants}.
The first three refinements have Arf invariant zero and are permuted transitively by
\(\operatorname{GL}(2,\mathbb F_2)\). The fourth has Arf invariant one and forms a separate orbit.

\subsubsection*{Comparison with Tambara's classification}

In~\cite{Tambara:2000qlz}, a fiber functor is encoded in the strongly
\(\Gamma\)-graded algebra \(R=\bigoplus_{u\in\Gamma}V_u\), with multiplication
given by the pointed maps \(J_{gh}\), and in an involutive anti-automorphism
\(f\) of \(R\) defined by the right action on \(V_m\), \(w.x=f(x).w\).
Writing \(f(x_u)=\nu(u)\,x_{\sigma(u)}\) for \(0\neq x_u\in V_u\) defines an
involution \(\sigma\in\operatorname{Aut}(\Gamma)\), which is fixed by the
pointed data through
\begin{equation}
    \frac{J_{gh}}{J_{hg}}
    =
    \xi\bigl(g,\sigma(h)\bigr).
    \label{eq:app-TY-sigma}
\end{equation}
For the pointed maps in Eq.~\eqref{eq:BootDQ-TY-normalizations} the
left-hand side is the alternating bicharacter \(\xi\) of
Eq.~\eqref{eq:BootDQ-TY-invariants}, so that \(\sigma=\operatorname{id}\) and
\[
    A^\sigma/A_\sigma\simeq\Gamma.
\]
The component \(J^{\mathsf 1}_{mm}\) is Tambara's pairing \(\gamma\), and
\(J^u_{mm}(w\otimes w')\propto\gamma(x_u^{-1}.w,w')\). The two properties
\(\gamma(x.w,w')=\gamma(w,f(x).w')\) and \(\gamma(w',w)=e\,\gamma(w,w')\), with
\(e=\operatorname{sgn}\tau_{\rm TY}\), give
\begin{equation}
    J^u_{mm}P
    =
    e\,\nu(u)^{-1}\,J^{\sigma(u)}_{mm},
    \label{eq:app-TY-flip-sigma}
\end{equation}
in Tambara's normalization of the channel maps. For
\(\sigma=\operatorname{id}\), comparison with
Eq.~\eqref{eq:app-TY-parity} shows that \(s_{\mathsf 1}=e\). Tambara's
multiplicative quadratic refinement \(\mu:\Gamma\to\{\pm1\}\), given by
\(\mu=\nu\) here, is therefore related to the transported flip by
\begin{equation}
    \mu(u)
    =
    \frac{s_u}{s_1}
    =
    (-1)^{\mathsf{q}(u)},
    \qquad
    s_1=\operatorname{sgn}\tau_{\rm TY}.
    \label{eq:app-TY-Tambara-mu}
\end{equation}
Tambara's condition \(\partial\mu=\bar\xi\) is
Eq.~\eqref{eq:app-TY-bilinear-form}. Since
\(\sum_{u}(-1)^{\mathsf q(u)}=2\operatorname{sgn}(\mu)\), his condition
\(\operatorname{sgn}(\mu)=e\) is equivalent to
\(\operatorname{tr}\widehat P=\sum_u s_u=\operatorname{tr}P=2\). This holds
automatically, because \(\widehat P\) is conjugate to \(P\).
Under the identification of \(\Gamma\) with \(\mathbb F_2^2\) used above, the first three solutions lie in the orbit of
\[
    \mathsf{q}_+(x,y)=xy,
\]
whereas the fourth is represented by
\[
    \mathsf{q}_-(x,y)
    =
    x^2+y^2+xy
    =
    x+y+xy.
\]
Thus, the \(\mathsf{q}_+\) and \(\mathsf{q}_-\) classes reproduce, respectively, the
\(D_8\) and \(Q_8\) realizations obtained from the bootstrap analysis.
By Theorem~3.5 of~\cite{Tambara:2000qlz}, isomorphism classes of fiber
functors correspond to the pairs \((\sigma,\mu)\). There are three Arf-zero
refinements and one Arf-one refinement of \(B\). This gives three classes for
\(\operatorname{Rep}(D_8)\) and one for \(\operatorname{Rep}(Q_8)\), in agreement
with Section~\ref{ssec:BootTY-fiber-count}.

\subsubsection*{The transported flip for \(\operatorname{Rep}(H_8)\)}

For the solution of Section~\ref{ssec:BootDQ-Hopf}, only \(\chi_{\mathsf 1}\)
and \(\chi_c\) extend, see Eq.~\eqref{eq:BootH-automorphisms}. The only
available operator is
\(\mathbb T_c=\operatorname{diag}(1,-1,-1,1)\), whose eigenspaces are
two-dimensional, so Eq.~\eqref{eq:app-TY-parity} no longer follows. The flip
still commutes with \(\mathbb T_c\), so \(\widehat P\) preserves the channel
blocks \(\{\mathsf 1,c\}\) and \(\{a,b\}\). From
Eq.~\eqref{eq:BootH-Jmm-plus}, the rows \(J^{\mathsf 1}_{mm}\) and \(J^c_{mm}\)
are supported on \(11,22\) and are therefore \(P\)-even. On the second block,
one finds
\begin{equation}
    \widehat P
    =
    \mathrm{id}_{V_{\mathsf 1}}
    \oplus
    \begin{pmatrix}
        \alpha & \beta\,\mathrm z_a/\mathrm z_b\\
        -\beta\,\mathrm z_b/\mathrm z_a & -\alpha
    \end{pmatrix}_{V_a\oplus V_b}
    \oplus
    \mathrm{id}_{V_c},
    \qquad
    \alpha=\frac{\zeta+\zeta^{-1}}{2},\quad
    \beta=\frac{\zeta-\zeta^{-1}}{2}.
    \label{eq:app-H8-flip}
\end{equation}
The diagonal entries are basis independent. For \(\zeta=\mp1\) they give
\(s_a=-1\) or \(s_b=-1\), reproducing solutions~1 and~2 of
Eq.~\eqref{eq:app-TY-quadratic-forms}. For the \(H_8\) values \(\zeta=\pm i\),
instead, \(\alpha=0\) and \(\beta=\pm i\). In this case \(\widehat P\) is a
monomial matrix exchanging the channels \(a\leftrightarrow b\), and only its
diagonal entries \(s_{\mathsf 1}=s_c=+1\) are intrinsic. The flip is not a
morphism of \(\operatorname{Rep}(H_8)\), since \(H_8\) is not cocommutative, so
Schur's lemma does not force \(\widehat P\) to be diagonal. Accordingly, no
channel carries \(\bigwedge^2V_m\) and \(\det(m)\) is not defined.

Comparing Eq.~\eqref{eq:app-H8-flip} with Eq.~\eqref{eq:app-TY-flip-sigma},
the permutation of channels induced by \(\widehat P\) is Tambara's involution,
\(\sigma=(a\,b)\), with \(\mathsf 1\) and \(c\) fixed. The same \(\sigma\) follows
independently from Eq.~\eqref{eq:app-TY-sigma}. The pointed maps are those
of Eq.~\eqref{eq:BootDQ-TY-normalizations}, so \(J_{gh}/J_{hg}\) is still the
alternating bicharacter of the \(D_8\) case, while \(\xi\) is now given by
Eq.~\eqref{eq:BootH-TY-invariants}, e.g.\ \(\xi(a,\sigma(b))=\xi(a,a)=-1\).
Moreover, \((a\,b)\) is the only involution of \(\Gamma\) satisfying Tambara's
condition \(\xi\bigl(u,\sigma(u)\bigr)=1\) for all \(u\in\Gamma\), i.e.\ for which
the bicharacter \((u,v)\mapsto\xi\bigl(u,\sigma(v)\bigr)\) is alternating.

The corresponding subgroups are \(A^\sigma=\{\mathsf 1,c\}\) and
\(A_\sigma=\{u\,\sigma(u)\}=\{\mathsf 1,c\}\), so \(A^\sigma/A_\sigma\) is
trivial and \(\mu\) is the zero form, with \(\operatorname{sgn}(\mu)=+1\). On
the bootstrap side, Eq.~\eqref{eq:app-TY-flip-sigma} gives \(s_{\mathsf 1}=e\)
and \(s_c=e\,\nu(c)^{-1}=e\), because \(c\in A_\sigma\). The off-diagonal
block does not contribute to the trace, so
\(\operatorname{tr}\widehat P=s_{\mathsf 1}+s_c=2e\). Equating this with
\(\operatorname{tr}P=2\) forces \(e=+1\), i.e.\ \(\tau_{\rm TY}=+\tfrac12\), in
agreement with Eq.~\eqref{eq:BootH-TY-invariants}.
This reproduces Tambara's datum \((\mathbb F_2^2,\xi,+\tfrac12,(a\,b),0)\) for
\(H_8\). Since \(\sigma\) and \(\mu\) are unique, Theorem~3.5 of~\cite{Tambara:2000qlz} gives a single fiber functor, as found in
Section~\ref{ssec:BootTY-fiber-count}.

\section{Other fiber functors from twists}
\label{app:twists_and_fiber_functors}

From the fusion ring alone, the bootstrap of the main text produces every
associator admitting a fiber functor in the maximal branch, each with a
fiber functor \((\mathcal F,J^0)\), and hence, by Tannaka reconstruction,
its own finite group
\[
    G=\operatorname{Aut}^\otimes(\mathcal F,J^0)
\]
together with representations
\(\rho_a:G\to GL(V_a)\).  Once one choice of the parameters in \(J^0\)
has been fixed, the remaining fiber functors on the same category need not
be found by solving the bootstrap equations again.  They can instead be
obtained from the standard classification of fiber functors on
\(\operatorname{Rep}(G)\); see
\cite{ostrik2006modulecategoriesdrinfelddouble} for the module-category
formulation and \cite{Movshev,AljadeffEtingofGelakiNikshych,etingof2015tensor}
for the equivalent description in terms of twists.

Up to monoidal natural isomorphism, these fiber functors are classified by
\(G\)-conjugacy classes of pairs
\begin{equation}
    (A,[\psi]),
    \qquad
    A\leq G,
    \qquad
    [\psi]\in H^2(A,\mathbb C^\ast),
    \label{eq:app-twist-classification}
\end{equation}
for which the twisted group algebra \(\mathbb C_\psi[A]\) is simple.  This is
the non-degeneracy condition on \([\psi]\); in particular, \(|A|\) must be a
square.  Thus the classification first gives a finite list of candidates,
as used explicitly in Sections~\ref{ssec:BootTY-fiber-count} and
\ref{ssec:Bootp3-counting-fiber-functors}.
The trivial pair \((\{1\},[1])\) gives back the seed \(J^0\); the other
pairs label the remaining fiber functors.

For completeness, we recall how a class in
Eq.~\eqref{eq:app-twist-classification} is converted into the matrices used
in this paper.  We use the equivalence with \(\operatorname{Rep}(G)\)
determined by the seed \(J^0\).  Let
\(\Omega\in\mathbb C[G]\otimes\mathbb C[G]\) be a
normalized Drinfeld twist representing the corresponding gauge class.  In
the convention of Eq.~\eqref{eq:monoidal_structure}, the new monoidal
structure is
\begin{equation}
    \boxed{
    J^{\Omega}_{a,b}
    =
    J^0_{a,b}\,
    (\rho_a\otimes\rho_b)(\Omega).
    }
    \label{eq:app-twisted-J}
\end{equation}
The multiplication morphisms \([J^{\Omega,c}_{ab}]_\lambda\) are simply the
blocks of this matrix in the chosen fusion-channel decomposition.  With the
opposite convention for the direction of the tensorator, \(\Omega^{-1}\)
appears instead.  Cohomologous cocycles and conjugate pairs give
gauge-equivalent twists and hence monoidally isomorphic fiber functors;
these are precisely the identifications already built into
Eq.~\eqref{eq:app-twist-classification}.  The twist--Galois-object
dictionary and its gauge equivalences are reviewed systematically in
Secs.~4--5 of~\cite{AljadeffEtingofGelakiNikshych}.

For the Abelian support subgroups occurring in our examples, the twist can
be evaluated directly in the character basis.  If
\(\widehat A=\operatorname{Hom}(A,\mathbb C^\ast)\) and
\(\alpha\in Z^2(\widehat A,\mathbb C^\ast)\) represents the twist class
associated with \((A,[\psi])\), then \cite{Movshev}
\begin{equation}
    e_\chi=\frac{1}{|A|}\sum_{g\in A}\chi(g)^{-1}g,
    \qquad
    \Omega_\alpha
    =\sum_{\chi,\eta\in\widehat A}
      \alpha(\chi,\eta)e_\chi\otimes e_\eta,
    \label{eq:app-abelian-twist}
\end{equation}
and Eq.~\eqref{eq:app-twisted-J} becomes the finite matrix computation
\begin{equation}
    J^\alpha_{a,b}
    =J^0_{a,b}
      \sum_{\chi,\eta\in\widehat A}
      \alpha(\chi,\eta)\,
      \rho_a(e_\chi)\otimes\rho_b(e_\eta).
    \label{eq:app-abelian-twisted-J}
\end{equation}
The class \([\alpha]\) is the character-basis representative of the twist
class associated with \([\psi]\) by the Movshev bijection; it should not in
general be identified with \([\psi]\) without fixing the duality and twist
conventions.

\paragraph{The case \(A\simeq\mathbb Z_3\times\mathbb Z_3\).}
Write \(A=\langle x,y\rangle\), set \(\omega=e^{2\pi i/3}\), and label the
characters by
\[
    \chi_{r,s}(x^m y^n)=\omega^{rm+sn},
    \qquad r,s,m,n\in\mathbb F_3.
\]
Then
\begin{equation}
    e_{r,s}
    =\frac{1}{9}\sum_{m,n\in\mathbb F_3}
      \omega^{-rm-sn}x^m y^n,
    \qquad
    \alpha_k\bigl((r,s),(r',s')\bigr)
    =\omega^{ksr'},
    \label{eq:app-Z3-cocycles}
\end{equation}
with \(k=0,1,2\).  Since
\begin{equation}
    \frac{\alpha_k((r,s),(r',s'))}
         {\alpha_k((r',s'),(r,s))}
    =\omega^{k(sr'-s'r)},
    \label{eq:app-Z3-commutator}
\end{equation}
the commutator bicharacter is non-degenerate precisely for \(k=1,2\).
Equivalently,
\(H^2(\mathbb Z_3^2,\mathbb C^\ast)\simeq\mathbb Z_3\), and its two
nonzero classes are the non-degenerate ones.  The two corresponding twists
and monoidal structures are therefore
\begin{equation}
    \Omega_k
    =\sum_{r,s,r',s'\in\mathbb F_3}
      \omega^{ksr'}e_{r,s}\otimes e_{r',s'},
    \qquad
    J^{(k)}_{a,b}
    =J^0_{a,b}(\rho_a\otimes\rho_b)(\Omega_k),
    \qquad k=1,2.
    \label{eq:app-Z3-twisted-J}
\end{equation}
Thus each subgroup \(A\simeq\mathbb Z_3^2\) contributes two nontrivial
fiber functors before the conjugation quotient. In the extraspecial
examples the conjugation action preserves both classes, as shown in
Section~\ref{ssec:Bootp3-counting-fiber-functors}.  Hence the four such
subgroups of \(\mathrm{Esp}_3\) give
\(1+4\cdot2=9\) fiber functors, while the unique such subgroup of
\(\mathrm{Esp}_9\) gives \(1+1\cdot2=3\), including in each case the seed
associated with the trivial subgroup.

This provides an alternative to rerunning the bootstrap: bootstrap one
seed, enumerate the pairs \((A,[\psi])\), and twist the seed once for each
gauge class.

\newpage

\section{Multiplicity-three cases: \texorpdfstring{$\tau$}{tau} and \texorpdfstring{$J$}{J} matrices}\label{app:p3-mixed-matrices}

This appendix gives explicit matrix representatives of two
parametric forms obtained from the maximal bootstrap in
Section~\ref{sec:extraspecial}.
We present form \(q=1\), associated with
\(\operatorname{Rep}(\mathrm{Esp}_3)\), and form \(q=3\),
associated with \(\operatorname{Rep}(\mathrm{Esp}_9)\).
Their associator invariants are, respectively,
\((\epsilon,\iota)=(\mathsf{1},1)\) and \((g_2,1)\).
These examples make the matrix structure of the solutions explicit
and provide representatives from which other labeled forms
can be obtained by fusion-ring automorphisms.

For each form, we first list the \(27\) tensor automorphisms.
The notation
\[
    \chi[n]=
    \bigl(\tau_{g_1}[n],\ldots,\tau_{g_9}[n]\bigr)
\]
collects their scalar components on the invertible objects,
while \(\tau_A[n]\) and \(\tau_B[n]\) are their components
on the two three-dimensional simple objects.
The index \(n\) only enumerates the group elements.

We then give the mixed and non-invertible \(J\)-maps.
The coefficients of the \(J\)-maps are left as parameters,
subject to invertibility.
Thus the lists describe parametric forms rather than particular
fiber functors for a prescribed associator.
In both forms, \(J_{Ag}\) is proportional to \(J_{gA}\),
and \(J_{Bg}\) is proportional to \(J_{gB}\);
only the latter maps in each pair are displayed.

\subsection{Rep(Esp\texorpdfstring{$_3$}{3}): explicit solutions}\label{sec:AppEsp3}
We begin with parametric form \(q=1\).
Throughout this appendix, we set
\(\kappa=e^{2\pi i/9}\) and \(\omega=\kappa^3\).
For this form, all tensor-automorphism components involve only
powers of \(\omega\).
The matrices below realize the two three-dimensional
irreducible representations of \(\mathrm{Esp}_3\);
their determinant characters are trivial, in agreement
with \(\epsilon=\mathsf{1}\).

\paragraph{Automorphisms ($\chi,\tau_A,\tau_B$)}

{\footnotesize
\begin{equation*}
\begin{array}{rccc}
\chi[1]=(1,1,1,1,1,1,1,1,1), &  
\tau_A[1]=\begin{pmatrix}  \,\,1\,\, & 0 & 0 \\ 0 & \,\,1\,\, & 0 \\ 0 & 0 & \,\,1\,\, \end{pmatrix}, & 
\tau_B[1]=\begin{pmatrix}  \,\,1\,\, & 0 & 0 \\ 0 & \,\,1\,\, & 0 \\ 0 & 0 & \,\,1\,\, \end{pmatrix}. \\
\chi[2]=(1,1,1,1,1,1,1,1,1), & 
\tau_A[2]=\begin{pmatrix}  \omega^2 & 0 & 0 \\ 0 & \omega^2 & 0 \\ 0 & 0 & \omega^2 \end{pmatrix}, & 
\tau_B[2]=\begin{pmatrix}  \,\,\omega\,\, & 0 & 0 \\ 0 & \,\,\omega\,\, & 0 \\ 0 & 0 & \,\,\omega\,\, \end{pmatrix}. \\
\chi[3]=(1,1,1,1,1,1,1,1,1), & 
\tau_A[3]=\begin{pmatrix}  \,\,\omega\,\, & 0 & 0 \\ 0 & \,\,\omega\,\, & 0 \\ 0 & 0 & \,\,\omega\,\, \end{pmatrix},  & 
\tau_B[3]=\begin{pmatrix}  \omega^2 & 0 & 0 \\ 0 & \omega^2 & 0 \\ 0 & 0 & \omega^2 \end{pmatrix}. \\
\chi[4]=(1,\omega,\omega^2,1,1,\omega,\omega^2,\omega^2,\omega), &
\tau_A[4]=\begin{pmatrix} 0 & \,\,1\,\, & 0 \\ 0 & 0 & \,\,1\,\, \\ \,\,1\,\, & 0 & 0 \end{pmatrix}, &
\tau_B[4]=\begin{pmatrix} 0 & 0 & \,\,1\,\, \\ \,\,1\,\, & 0 & 0 \\ 0 & \,\,1\,\, & 0 \end{pmatrix}.
\end{array}
\end{equation*}

\begin{equation*}
\begin{array}{rccc}
\chi[5]=(1,\omega,\omega^2,1,1,\omega,\omega^2,\omega^2,\omega), &
\tau_A[5]=\begin{pmatrix} 0 & \omega^2 & 0 \\ 0 & 0 & \omega^2 \\ \omega^2 & 0 & 0 \end{pmatrix}, &
\tau_B[5]=\begin{pmatrix} 0 & 0 & \omega \\ \omega & 0 & 0 \\ 0 & \omega & 0 \end{pmatrix}. \\
\chi[6]=(1,\omega,\omega^2,1,1,\omega,\omega^2,\omega^2,\omega), &
\tau_A[6]=\begin{pmatrix} 0 & \omega & 0 \\ 0 & 0 & \omega \\ \omega & 0 & 0 \end{pmatrix}, & 
\tau_B[6]=\begin{pmatrix} 0 & 0 & \omega^2 \\ \omega^2 & 0 & 0 \\ 0 & \omega^2 & 0 \end{pmatrix}. \\
\chi[7]=(1,\omega^2,\omega,1,1,\omega^2,\omega,\omega,\omega^2), &
\tau_A[7]=\begin{pmatrix} 0 & 0 & \,\,1\,\, \\ \,\,1\,\, & 0 & 0 \\ 0 & \,\,1\,\, & 0 \end{pmatrix}, &
\tau_B[7]=\begin{pmatrix} 0 & \,\,1\,\, & 0 \\ 0 & 0 & \,\,1\,\, \\ \,\,1\,\, & 0 & 0 \end{pmatrix}. \\
\chi[8]=(1,\omega^2,\omega,1,1,\omega^2,\omega,\omega,\omega^2), &
\tau_A[8]=\begin{pmatrix} 0 & 0 & \omega^2 \\ \omega^2 & 0 & 0 \\ 0 & \omega^2 & 0 \end{pmatrix}, & 
\tau_B[8]=\begin{pmatrix} 0 & \omega & 0 \\ 0 & 0 & \omega \\ \omega & 0 & 0 \end{pmatrix}. \\
\chi[9]=(1,\omega^2,\omega,1,1,\omega^2,\omega,\omega,\omega^2), &
\tau_A[9]=\begin{pmatrix} 0 & 0 & \omega \\ \omega & 0 & 0 \\ 0 & \omega & 0 \end{pmatrix}, &
\tau_B[9]=\begin{pmatrix} 0 & \omega^2 & 0 \\ 0 & 0 & \omega^2 \\ \omega^2 & 0 & 0 \end{pmatrix}. \\
\chi[10]=(1,1,\omega,\omega^2,\omega,\omega^2,1,\omega^2,\omega), &
\tau_A[10]=\begin{pmatrix} 0 & 0 & \,\,1\,\, \\ \omega^2 & 0 & 0 \\ 0 & \omega & 0 \end{pmatrix}, &
\tau_B[10]=\begin{pmatrix} 0 & \omega & 0 \\ 0 & 0 & \,\,1\,\, \\ \omega^2 & 0 & 0 \end{pmatrix}. \\
\chi[11]=(1,1,\omega,\omega^2,\omega,\omega^2,1,\omega^2,\omega), &
\tau_A[11]=\begin{pmatrix} 0 & 0 & \omega^2 \\ \omega & 0 & 0 \\ 0 & \,\,1\,\, & 0 \end{pmatrix}, & 
\tau_B[11]=\begin{pmatrix} 0 & \omega^2 & 0 \\ 0 & 0 & \omega \\ \,\,1\,\, & 0 & 0 \end{pmatrix}. \\
\chi[12]=(1,1,\omega,\omega^2,\omega,\omega^2,1,\omega^2,\omega), &
\tau_A[12]=\begin{pmatrix} 0 & 0 & \omega \\ \,\,1\,\, & 0 & 0 \\ 0 & \omega^2 & 0 \end{pmatrix}, & 
\tau_B[12]=\begin{pmatrix} 0 & \,\,1\,\, & 0 \\ 0 & 0 & \omega^2 \\ \omega & 0 & 0 \end{pmatrix}. \\
\chi[13]=(1,\omega,1,\omega^2,\omega,1,\omega^2,\omega,\omega^2), &
\tau_A[13]=\begin{pmatrix} \,\,1\,\, & 0 & 0 \\ 0 & \omega^2 & 0 \\ 0 & 0 & \omega \end{pmatrix}, &
\tau_B[13]=\begin{pmatrix} \omega & 0 & 0 \\ 0 & \,\,1\,\, & 0 \\ 0 & 0 & \omega^2 \end{pmatrix}.\\
\chi[14]=(1,\omega,1,\omega^2,\omega,1,\omega^2,\omega,\omega^2), &
\tau_A[14]=\begin{pmatrix}  \omega^2 & 0 & 0 \\ 0 & \omega & 0 \\ 0 & 0 & \,\,1\,\, \end{pmatrix}, \quad &
\tau_B[14]=\begin{pmatrix}  \omega^2 & 0 & 0 \\ 0 & \omega & 0 \\ 0 & 0 & \,\,1\,\, \end{pmatrix}. \\
\chi[15]=(1,\omega,1,\omega^2,\omega,1,\omega^2,\omega,\omega^2), &
\tau_A[15]=\begin{pmatrix}  \omega & 0 & 0 \\ 0 & \,\,1\,\, & 0 \\ 0 & 0 & \omega^2 \end{pmatrix}, \quad &
\tau_B[15]=\begin{pmatrix}  \,\,1\,\, & 0 & 0 \\ 0 & \omega^2 & 0 \\ 0 & 0 & \omega \end{pmatrix}. \\
\chi[16]=(1,\omega^2,\omega^2,\omega^2,\omega,\omega,\omega,1,1), &
\tau_A[16]=\begin{pmatrix}  0 & \,\,1\,\, & 0 \\ 0 & 0 & \omega^2 \\ \omega & 0 & 0 \end{pmatrix}, \quad &
\tau_B[16]=\begin{pmatrix}  0 & 0 & \omega \\ \,\,1\,\, & 0 & 0 \\ 0 & \omega^2 & 0 \end{pmatrix}. \\
\chi[17]=(1,\omega^2,\omega^2,\omega^2,\omega,\omega,\omega,1,1), &
\tau_A[17]=\begin{pmatrix}  0 & \omega^2 & 0 \\ 0 & 0 & \omega \\ \,\,1\,\, & 0 & 0 \end{pmatrix},\quad &
\tau_B[17]=\begin{pmatrix}  0 & 0 & \omega^2 \\ \omega & 0 & 0 \\ 0 & \,\,1\,\, & 0 \end{pmatrix}.\\
\chi[18]=(1,\omega^2,\omega^2,\omega^2,\omega,\omega,\omega,1,1), &
\tau_A[18]=\begin{pmatrix} 0 & \omega & 0 \\ 0 & 0 & \,\,1\,\, \\ \omega^2 & 0 & 0 \end{pmatrix},\quad &
\tau_B[18]=\begin{pmatrix} 0 & 0 & \,\,1\,\, \\ \omega^2 & 0 & 0 \\ 0 & \omega & 0 \end{pmatrix}.\\
\chi[19]=(1,1,\omega^2,\omega,\omega^2,\omega,1,\omega,\omega^2), &
\tau_A[19]=\begin{pmatrix}  0 & \,\,1\,\, & 0 \\ 0 & 0 & \omega \\ \omega^2 & 0 & 0 \end{pmatrix},\quad &
\tau_B[19]=\begin{pmatrix}  0 & 0 & \omega^2 \\ \,\,1\,\, & 0 & 0 \\ 0 & \omega & 0 \end{pmatrix}.
\end{array}
\end{equation*}

\vspace{-10pt}
{\footnotesize
\begin{equation*}
\begin{array}{lll}
\chi[20]=(1,1,\omega^2,\omega,\omega^2,\omega,1,\omega,\omega^2), &
\tau_A[20]=\begin{pmatrix}  0 & \omega^2 & 0 \\ 0 & 0 & \,\,1\,\, \\ \omega & 0 & 0 \end{pmatrix},\quad &
\tau_B[20]=\begin{pmatrix}  0 & 0 & \,\,1\,\, \\ \omega & 0 & 0 \\ 0 & \omega^2 & 0 \end{pmatrix}. \\
\chi[21]=(1,1,\omega^2,\omega,\omega^2,\omega,1,\omega,\omega^2), &
\tau_A[21]=\begin{pmatrix}  0 & \omega & 0 \\ 0 & 0 & \omega^2 \\ \,\,1\,\, & 0 & 0 \end{pmatrix},\quad &
\tau_B[21]=\begin{pmatrix}  0 & 0 & \omega \\ \omega^2 & 0 & 0 \\ 0 & \,\,1\,\, & 0 \end{pmatrix}. \\
\chi[22]=(1,\omega,\omega,\omega,\omega^2,\omega^2,\omega^2,1,1), &
\tau_A[22]=\begin{pmatrix}  0 & 0 & \,\,1\,\, \\ \omega & 0 & 0 \\ 0 & \omega^2 & 0 \end{pmatrix},\quad &
\tau_B[22]=\begin{pmatrix}  0 & \omega^2 & 0 \\ 0 & 0 & \,\,1\,\, \\ \omega & 0 & 0 \end{pmatrix}. \\
\chi[23]=(1,\omega,\omega,\omega,\omega^2,\omega^2,\omega^2,1,1), &
\tau_A[23]=\begin{pmatrix}  0 & 0 & \omega^2 \\ \,\,1\,\, & 0 & 0 \\ 0 & \omega & 0 \end{pmatrix},\quad &
\tau_B[23]=\begin{pmatrix}  0 & \,\,1\,\, & 0 \\ 0 & 0 & \omega \\ \omega^2 & 0 & 0 \end{pmatrix}. \\
\chi[24]=(1,\omega,\omega,\omega,\omega^2,\omega^2,\omega^2,1,1), &
\tau_A[24]=\begin{pmatrix}  0 & 0 & \omega \\ \omega^2 & 0 & 0 \\ 0 & \,\,1\,\, & 0 \end{pmatrix},\quad &
\tau_B[24]=\begin{pmatrix}  0 & \omega & 0 \\ 0 & 0 & \omega^2 \\ \,\,1\,\, & 0 & 0 \end{pmatrix}. \\
\chi[25]=(1,\omega^2,1,\omega,\omega^2,1,\omega,\omega^2,\omega), &
\tau_A[25]=\begin{pmatrix} \,\,1\,\, & 0 & 0 \\ 0 & \omega & 0 \\ 0 & 0 & \omega^2 \end{pmatrix},\quad &
\tau_B[25]=\begin{pmatrix} \omega^2 & 0 & 0 \\ 0 & \,\,1\,\, & 0 \\ 0 & 0 & \omega \end{pmatrix}. \\
\chi[26]=(1,\omega^2,1,\omega,\omega^2,1,\omega,\omega^2,\omega), &
\tau_A[26]=\begin{pmatrix} \omega^2 & 0 & 0 \\ 0 & \,\,1\,\, & 0 \\ 0 & 0 & \omega \end{pmatrix},\quad &
\tau_B[26]=\begin{pmatrix} \,\,1\,\, & 0 & 0 \\ 0 & \omega & 0 \\ 0 & 0 & \omega^2 \end{pmatrix}. \\
\chi[27]=(1,\omega^2,1,\omega,\omega^2,1,\omega,\omega^2,\omega), &
\tau_A[27]=\begin{pmatrix} \omega & 0 & 0 \\ 0 & \omega^2 & 0 \\ 0 & 0 & \,\,1\,\, \end{pmatrix},\quad &
\tau_B[27]=\begin{pmatrix} \omega & 0 & 0 \\ 0 & \omega^2 & 0 \\ 0 & 0 & \,\,1\,\, \end{pmatrix}.
\end{array}
\end{equation*}
}

\paragraph{Morphisms}
The mixed maps \(J_{g_iA}\) and \(J_{g_iB}\) are
\(3\times3\) matrices, with nonzero scalar coefficients
\(\mathrm{x}_i\) and \(\mathrm{y}_i\).
The non-invertible maps are \(9\times9\) matrices.
For $J_{AA}$ and $J_{BB}$
each consecutive block of three rows corresponds to one
of the three fusion channels.
The coefficients $\mathrm{R}=(\mathrm{r}_{\alpha j})_{\alpha,j=1,2,3}$ and 
$\mathrm{S}=(\mathrm{s}_{\alpha j})_{\alpha,j=1,2,3}$
parametrize the freedom in these multiplicity spaces,
with \(\mathrm{R}\) and \(\mathrm{S}\) invertible.
For \(J_{AB}\) and \(J_{BA}\), the output is
\(\bigoplus_{i=1}^{9}V_{g_i}\).
Their rows therefore correspond, in order, to
\(g_1,\ldots,g_9\), and carry nonzero coefficients
\(\mathrm{z}_i\) and \(\tilde{\mathrm{z}}_i\), respectively.
This row ordering will be important when applying
fusion-ring automorphisms below.
{\footnotesize
\begin{equation*}
\begin{array}{lll}
J_{g_1A}=\mathrm{x}_1\begin{pmatrix} \,\,1\,\, & 0 & 0 \\ 0 & \,\,1\,\, & 0 \\ 0 & 0 & \,\,1\,\, \end{pmatrix}, &
J_{g_2A}=\mathrm{x}_2\begin{pmatrix} 0 & \,\,1\,\, & 0 \\ 0 & 0 & \omega \\ \omega^2 & 0 & 0 \end{pmatrix}, &
J_{g_3A}=\mathrm{x}_3\begin{pmatrix} \,\,1\,\, & 0 & 0 \\ 0 & \omega^2 & 0 \\ 0 & 0 & \omega \end{pmatrix}, \\
J_{g_4A}=\mathrm{x}_4\begin{pmatrix} 0 & 0 & \,\,1\,\, \\ \,\,1\,\, & 0 & 0 \\ 0 & \,\,1\,\, & 0 \end{pmatrix}, &
J_{g_5A}=\mathrm{x}_5\begin{pmatrix}  0 & \,\,1\,\, & 0 \\ 0 & 0 & \,\,1\,\, \\ \,\,1\,\, & 0 & 0 \end{pmatrix}, &
J_{g_6A}=\mathrm{x}_6\begin{pmatrix}  \,\,1\,\, & 0 & 0 \\ 0 & \omega & 0 \\ 0 & 0 & \omega^2 \end{pmatrix}, \\
J_{g_7A}=\mathrm{x}_7\begin{pmatrix} 0 & 0 & \,\,1\,\, \\ \omega^2 & 0 & 0 \\ 0 & \omega & 0 \end{pmatrix}, &
J_{g_8A}=\mathrm{x}_8\begin{pmatrix} 0 & \,\,1\,\, & 0 \\ 0 & 0 & \omega^2 \\ \omega & 0 & 0 \end{pmatrix}, &
J_{g_9A}=\mathrm{x}_9\begin{pmatrix} 0 & 0 & \,\,1\,\, \\ \omega & 0 & 0 \\ 0 & \omega^2 & 0 \end{pmatrix}.
\end{array}
\end{equation*}

\begin{equation*}
\begin{array}{lll}
J_{g_1B}=\mathrm{y}_1\begin{pmatrix} \,\,1\,\, & 0 & 0 \\ 0 & \,\,1\,\, & 0 \\ 0 & 0 & \,\,1\,\, \end{pmatrix}, &
J_{g_2B}=\mathrm{y}_2\begin{pmatrix} 0 & \,\,1\,\, & 0 \\ 0 & 0 & \omega^2 \\ \omega & 0 & 0 \end{pmatrix}, &
J_{g_3B}=\mathrm{y}_3\begin{pmatrix} \,\,1\,\, & 0 & 0 \\ 0 & \omega & 0 \\ 0 & 0 & \omega^2 \end{pmatrix}, \\
J_{g_4B}=\mathrm{y}_4\begin{pmatrix} 0 & 0 & \,\,1\,\, \\ \,\,1\,\, & 0 & 0 \\ 0 & \,\,1\,\, & 0 \end{pmatrix}, &
J_{g_5B}=\mathrm{y}_5\begin{pmatrix} 0 & \,\,1\,\, & 0 \\ 0 & 0 & \,\,1\,\, \\ \,\,1\,\, & 0 & 0 \end{pmatrix}, &
J_{g_6B}=\mathrm{y}_6\begin{pmatrix} \,\,1\,\, & 0 & 0 \\ 0 & \omega^2 & 0 \\ 0 & 0 & \omega \end{pmatrix}, \\
J_{g_7B}=\mathrm{y}_7\begin{pmatrix} 0 & 0 & \,\,1\,\, \\ \omega & 0 & 0 \\ 0 & \omega^2 & 0 \end{pmatrix}, &
J_{g_8B}=\mathrm{y}_8\begin{pmatrix} 0 & \,\,1\,\, & 0 \\ 0 & 0 & \omega \\ \omega^2 & 0 & 0 \end{pmatrix}, &
J_{g_9B}=\mathrm{y}_9\begin{pmatrix} 0 & 0 & \,\,1\,\, \\ \omega^2 & 0 & 0 \\ 0 & \omega & 0 \end{pmatrix}. 
\end{array}
\end{equation*}
}

{\scriptsize
\[
J_{AA}=\begin{pmatrix}
    0 & 0 & \mathrm{r}_{11} & 0 & \mathrm{r}_{12} & 0 & \mathrm{r}_{13} & 0 & 0 \\
    \mathrm{r}_{12} & 0 & 0 & 0 & 0 & \mathrm{r}_{13} & 0 & \mathrm{r}_{11} & 0 \\
    0 & \mathrm{r}_{13} & 0 & \mathrm{r}_{11} & 0 & 0 & 0 & 0 & \mathrm{r}_{12} \\
    0 & 0 & \mathrm{r}_{21} & 0 & \mathrm{r}_{22} & 0 & \mathrm{r}_{23} & 0 & 0 \\
    \mathrm{r}_{22} & 0 & 0 & 0 & 0 & \mathrm{r}_{23} & 0 & \mathrm{r}_{21} & 0 \\
    0 & \mathrm{r}_{23} & 0 & \mathrm{r}_{21} & 0 & 0 & 0 & 0 & \mathrm{r}_{22} \\
    0 & 0 & \mathrm{r}_{31} & 0 & \mathrm{r}_{32} & 0 & \mathrm{r}_{33} & 0 & 0 \\
    \mathrm{r}_{32} & 0 & 0 & 0 & 0 & \mathrm{r}_{33} & 0 & \mathrm{r}_{31} & 0 \\
    0 & \mathrm{r}_{33} & 0 & \mathrm{r}_{31} & 0 & 0 & 0 & 0 & \mathrm{r}_{32}  
\end{pmatrix},\,
J_{BB}=\begin{pmatrix}
    0 & 0 & \mathrm{s}_{11} & 0 & \mathrm{s}_{12} & 0 & \mathrm{s}_{13} & 0 & 0 \\
    \mathrm{s}_{12} & 0 & 0 & 0 & 0 & \mathrm{s}_{13} & 0 & \mathrm{s}_{11} & 0 \\
    0 & \mathrm{s}_{13} & 0 & \mathrm{s}_{11} & 0 & 0 & 0 & 0 & \mathrm{s}_{12} \\
    0 & 0 & \mathrm{s}_{21} & 0 & \mathrm{s}_{22} & 0 & \mathrm{s}_{23} & 0 & 0 \\
    \mathrm{s}_{22} & 0 & 0 & 0 & 0 & \mathrm{s}_{23} & 0 & \mathrm{s}_{21} & 0 \\
    0 & \mathrm{s}_{23} & 0 & \mathrm{s}_{21} & 0 & 0 & 0 & 0 & \mathrm{s}_{22} \\
    0 & 0 & \mathrm{s}_{31} & 0 & \mathrm{s}_{32} & 0 & \mathrm{s}_{33} & 0 & 0 \\
    \mathrm{s}_{32} & 0 & 0 & 0 & 0 & \mathrm{s}_{33} & 0 & \mathrm{s}_{31} & 0 \\
    0 & \mathrm{s}_{33} & 0 & \mathrm{s}_{31} & 0 & 0 & 0 & 0 & \mathrm{s}_{32} 
\end{pmatrix},
\]

\[
J_{AB}=\begin{pmatrix}
    0 & \mathrm{z}_1 & 0 & \mathrm{z}_1 & 0 & 0 & 0 & 0 & \mathrm{z}_1 \\
    \mathrm{z}_2 & 0 & 0 & 0 & 0 & \omega^2 \mathrm{z}_2 & 0 & \omega \mathrm{z}_2 & 0 \\
    0 & \mathrm{z}_3 & 0 & \omega \mathrm{z}_3 & 0 & 0 & 0 & 0 & \omega^2 \mathrm{z}_3 \\
    0 & 0 & \mathrm{z}_4 & 0 & \mathrm{z}_4 & 0 & \mathrm{z}_4 & 0 & 0 \\
    \mathrm{z}_5 & 0 & 0 & 0 & 0 & \mathrm{z}_5 & 0 & \mathrm{z}_5 & 0 \\
    0 & \mathrm{z}_6 & 0 & \omega^2 \mathrm{z}_6 & 0 & 0 & 0 & 0 & \omega \mathrm{z}_6 \\
    0 & 0 & \mathrm{z}_7 & 0 & \omega \mathrm{z}_7 & 0 & \omega^2 \mathrm{z}_7 & 0 & 0 \\
    \mathrm{z}_8 & 0 & 0 & 0 & 0 & \omega \mathrm{z}_8 & 0 & \omega^2 \mathrm{z}_8 & 0 \\
    0 & 0 & \mathrm{z}_9 & 0 & \omega^2 \mathrm{z}_9 & 0 & \omega \mathrm{z}_9 & 0 & 0 
\end{pmatrix},
\]
\[
J_{BA}=\begin{pmatrix}
    0 & \tilde{\mathrm{z}}_1 & 0 & \tilde{\mathrm{z}}_1 & 0 & 0 & 0 & 0 & \tilde{\mathrm{z}}_1 \\
    \tilde{\mathrm{z}}_2 & 0 & 0 & 0 & 0 & \omega \tilde{\mathrm{z}}_2 & 0 & \omega^2 \tilde{\mathrm{z}}_2 & 0 \\
    0 & \omega \tilde{\mathrm{z}}_3 & 0 & \tilde{\mathrm{z}}_3 & 0 & 0 & 0 & 0 & \omega^2 \tilde{\mathrm{z}}_3 \\
    0 & 0 & \tilde{\mathrm{z}}_4 & 0 & \tilde{\mathrm{z}}_4 & 0 & \tilde{\mathrm{z}}_4 & 0 & 0 \\
    \tilde{\mathrm{z}}_5 & 0 & 0 & 0 & 0 & \tilde{\mathrm{z}}_5 & 0 & \tilde{\mathrm{z}}_5 & 0 \\
    0 & \omega^2 \tilde{\mathrm{z}}_6 & 0 & \tilde{\mathrm{z}}_6 & 0 & 0 & 0 & 0 & \omega \tilde{\mathrm{z}}_6 \\
    0 & 0 & \omega^2 \tilde{\mathrm{z}}_7 & 0 & \omega \tilde{\mathrm{z}}_7 & 0 & \tilde{\mathrm{z}}_7 & 0 & 0 \\
    \tilde{\mathrm{z}}_8 & 0 & 0 & 0 & 0 & \omega^2 \tilde{\mathrm{z}}_8 & 0 & \omega \tilde{\mathrm{z}}_8 & 0 \\
    0 & 0 & \omega \tilde{\mathrm{z}}_9 & 0 & \omega^2 \tilde{\mathrm{z}}_9 & 0 & \tilde{\mathrm{z}}_9 & 0 & 0 
\end{pmatrix}.
\]
}

\subsection{Rep(Esp\texorpdfstring{$_9$}{9}): explicit solution}\label{sec:AppEsp9}
We now present parametric form \(q=3\), with invariants
\((\epsilon,\iota)=(g_2,1)\).
The notation and matrix conventions are the same as in
Subsection~\ref{sec:AppEsp3}.
Unlike the preceding form, some tensor-automorphism components
contain ninth roots of unity.
For example,
\[
    \tau_A[4]^3=\omega\,\mathbbm1_3,
    \qquad
    \tau_B[4]^3=\omega^2\,\mathbbm1_3.
\]
The corresponding group element therefore has order nine,
making the distinction from \(\mathrm{Esp}_3\) explicit
in the matrix realization.

\paragraph{Automorphisms ($\chi,\tau_A,\tau_B$)}

\vspace{-5pt}
{\footnotesize
\begin{equation*}
\begin{array}{ccc}
\chi[1]=(1,1,1,1,1,1,1,1,1), & 
\tau_A[1]=\begin{pmatrix} \,\,1\,\, & 0 & 0 \\ 0 & \,\,1\,\, & 0 \\ 0 & 0 & \,\,1\,\, \end{pmatrix}, &
\tau_B[1]=\begin{pmatrix} \,\,1\,\, & 0 & 0 \\ 0 & \,\,1\,\, & 0 \\ 0 & 0 & \,\,1\,\, \end{pmatrix}. \\
\chi[2]=(1,1,1,1,1,1,1,1,1), & 
\tau_A[2]=\begin{pmatrix} \omega^2 & 0 & 0 \\ 0 & \omega^2 & 0 \\ 0 & 0 & \omega^2 \end{pmatrix}, &
\tau_B[2]=\begin{pmatrix} \omega & 0 & 0 \\ 0 & \omega & 0 \\ 0 & 0 & \omega \end{pmatrix}. \\
\chi[3]=(1,1,1,1,1,1,1,1,1), & 
\tau_A[3]=\begin{pmatrix} \omega & 0 & 0 \\ 0 & \omega & 0 \\ 0 & 0 & \omega \end{pmatrix}, &
\tau_B[3]=\begin{pmatrix} \omega^2 & 0 & 0 \\ 0 & \omega^2 & 0 \\ 0 & 0 & \omega^2 \end{pmatrix}. \\
\chi[4]=(1,\omega,\omega^2,1,1,\omega,\omega^2,\omega^2,\omega), &
\tau_A[4]=\begin{pmatrix} 0 & \kappa & 0 \\ 0 & 0 & \kappa \\ \kappa & 0 & 0 \end{pmatrix}, &
\tau_B[4]=\begin{pmatrix} 0 & 0 & \kappa^2 \\ \kappa^2 & 0 & 0 \\ 0 & \kappa^2 & 0 \end{pmatrix}. \\
\chi[5]=(1,\omega,\omega^2,1,1,\omega,\omega^2,\omega^2,\omega), &
\tau_A[5]=\begin{pmatrix} 0 & \omega^2\kappa & 0 \\ 0 & 0 & \omega^2\kappa \\ \omega^2\kappa & 0 & 0  \end{pmatrix}, &
\tau_B[5]=\begin{pmatrix} 0 & 0 & \omega \kappa^2 \\ \omega \kappa^2 & 0 & 0 \\ 0 & \omega\kappa^2 & 0 \end{pmatrix}. \\
\chi[6]=(1,\omega,\omega^2,1,1,\omega,\omega^2,\omega^2,\omega), &
\tau_A[6]=\begin{pmatrix} 0 & \omega\kappa & 0 \\ 0 & 0 & \omega\kappa \\ \omega\kappa & 0 & 0 \end{pmatrix}, &
\tau_B[6]=\begin{pmatrix} 0 & 0 & \omega^2\kappa^2 \\ \omega^2\kappa^2 & 0 & 0 \\ 0 & \omega^2\kappa^2 & 0 \end{pmatrix}. \\
\chi[7]=(1,\omega^2,\omega,1,1,\omega^2,\omega,\omega,\omega^2), &
\tau_A[7]=\begin{pmatrix} 0 & 0 & \omega\kappa^2 \\ \omega\kappa^2 & 0 & 0 \\ 0 & \omega\kappa^2 & 0 \end{pmatrix}, &
\tau_B[7]=\begin{pmatrix} 0 & \kappa & 0 \\ 0 & 0 & \kappa \\ \kappa & 0 & 0 \end{pmatrix}.\\
\chi[8]=(1,\omega^2,\omega,1,1,\omega^2,\omega,\omega,\omega^2), &
\tau_A[8]=\begin{pmatrix} 0 & 0 & \kappa^2 \\ \kappa^2 & 0 & 0 \\ 0 & \kappa^2 & 0 \end{pmatrix}, &
\tau_B[8]=\begin{pmatrix} 0 & \omega\kappa & 0 \\ 0 & 0 & \omega\kappa \\ \omega\kappa & 0 & 0 \end{pmatrix}. \\
\chi[9]=(1,\omega^2,\omega,1,1,\omega^2,\omega,\omega,\omega^2), &
\tau_A[9]=\begin{pmatrix} 0 & 0 & \omega^2\kappa^2 \\ \omega^2\kappa^2 & 0 & 0 \\ 0 & \omega^2\kappa^2 & 0 \end{pmatrix}, &
\tau_B[9]=\begin{pmatrix} 0 & \omega^2\kappa & 0 \\ 0 & 0 & \omega^2\kappa \\ \omega^2\kappa & 0 & 0 \end{pmatrix}. \\
\chi[10]=(1,1,\omega,\omega^2,\omega,\omega^2,1,\omega^2,\omega), &
\tau_A[10]=\begin{pmatrix} 0 & 0 & \,\,1\,\, \\ \omega^2 & 0 & 0 \\ 0 & \omega & 0 \end{pmatrix}, &
\tau_B[10]=\begin{pmatrix} 0 & \omega & 0 \\ 0 & 0 & \,\,1\,\, \\ \omega^2 & 0 & 0 \end{pmatrix}. \\
\chi[11]=(1,1,\omega,\omega^2,\omega,\omega^2,1,\omega^2,\omega), &
\tau_A[11]=\begin{pmatrix} 0 & 0 & \omega^2 \\ \omega & 0 & 0 \\ 0 & \,\,1\,\, & 0 \end{pmatrix}, &
\tau_B[11]=\begin{pmatrix} 0 & \omega^2 & 0 \\ 0 & 0 & \omega \\ \,\,1\,\, & 0 & 0 \end{pmatrix}.\\
\chi[12]=(1,1,\omega,\omega^2,\omega,\omega^2,1,\omega^2,\omega), &
\tau_A[12]=\begin{pmatrix} 0 & 0 & \omega \\ \,\,1\,\, & 0 & 0 \\ 0 & \omega^2 & 0 \end{pmatrix}, &
\tau_B[12]=\begin{pmatrix} 0 & \,\,1\,\, & 0 \\ 0 & 0 & \omega^2 \\ \omega & 0 & 0 \end{pmatrix}. \\
\chi[13]=(1,\omega,1,\omega^2,\omega,1,\omega^2,\omega,\omega^2), &
\tau_A[13]=\begin{pmatrix} \kappa & 0 & 0 \\ 0 & \omega^2\kappa & 0 \\ 0 & 0 & \omega\kappa \end{pmatrix}, &
\tau_B[13]=\begin{pmatrix} \omega\kappa^2 & 0 & 0 \\ 0 & \kappa^2 & 0 \\ 0 & 0 & \omega^2\kappa^2 \end{pmatrix}.\\
\chi[14]=(1,\omega,1,\omega^2,\omega,1,\omega^2,\omega,\omega^2), &
\tau_A[14]=\begin{pmatrix} \omega^2\kappa & 0 & 0 \\ 0 & \omega\kappa & 0 \\ 0 & 0 & \kappa \end{pmatrix}, &
\tau_B[14]=\begin{pmatrix} \omega^2\kappa^2 & 0 & 0 \\ 0 & \omega\kappa^2 & 0 \\ 0 & 0 & \kappa^2 \end{pmatrix}. 
\end{array}
\end{equation*}

\begin{equation*}
\begin{array}{ccc}
\chi[15]=(1,\omega,1,\omega^2,\omega,1,\omega^2,\omega,\omega^2), &
\tau_A[15]=\begin{pmatrix} \omega\kappa & 0 & 0 \\ 0 & \kappa & 0 \\ 0 & 0 & \omega^2\kappa \end{pmatrix}, &
\tau_B[15]=\begin{pmatrix} \kappa^2 & 0 & 0 \\ 0 & \omega^2\kappa^2 & 0 \\ 0 & 0 & \omega\kappa^2 \end{pmatrix}. \\
\chi[16]=(1,\omega^2,\omega^2,\omega^2,\omega,\omega,\omega,1,1), &
\tau_A[16]=\begin{pmatrix} 0 & \omega\kappa^2 & 0 \\ 0 & 0 & \kappa^2 \\ \omega^2\kappa^2 & 0 & 0 \end{pmatrix}, &
\tau_B[16]=\begin{pmatrix} 0 & 0 & \omega\kappa \\ \kappa & 0 & 0 \\ 0 & \omega^2\kappa & 0 \end{pmatrix}. \\
\chi[17]=(1,\omega^2,\omega^2,\omega^2,\omega,\omega,\omega,1,1), &
\tau_A[17]=\begin{pmatrix} 0 & \kappa^2 & 0 \\ 0 & 0 & \omega^2\kappa^2 \\ \omega\kappa^2 & 0 & 0 \end{pmatrix}, &
\tau_B[17]=\begin{pmatrix} 0 & 0 & \omega^2\kappa \\ \omega\kappa & 0 & 0 \\ 0 & \kappa & 0 \end{pmatrix}. \\
\chi[18]=(1,\omega^2,\omega^2,\omega^2,\omega,\omega,\omega,1,1), &
\tau_A[18]=\begin{pmatrix} 0 & \omega^2\kappa^2 & 0 \\ 0 & 0 & \omega\kappa^2 \\ \kappa^2 & 0 & 0 \end{pmatrix}, &
\tau_B[18]=\begin{pmatrix} 0 & 0 & \kappa \\ \omega^2\kappa & 0 & 0 \\ 0 & \omega\kappa & 0 \end{pmatrix}.\\
\chi[19]=(1,1,\omega^2,\omega,\omega^2,\omega,1,\omega,\omega^2), &
\tau_A[19]=\begin{pmatrix} 0 & \,\,1\,\, & 0 \\ 0 & 0 & \omega \\ \omega^2 & 0 & 0 \end{pmatrix}, &
\tau_B[19]=\begin{pmatrix} 0 & 0 & \omega^2 \\ \,\,1\,\, & 0 & 0 \\ 0 & \omega & 0 \end{pmatrix}. \\
\chi[20]=(1,1,\omega^2,\omega,\omega^2,\omega,1,\omega,\omega^2), &
\tau_A[20]=\begin{pmatrix} 0 & \omega^2 & 0 \\ 0 & 0 & \,\,1\,\, \\ \omega & 0 & 0 \end{pmatrix}, &
\tau_B[20]=\begin{pmatrix} 0 & 0 & \,\,1\,\, \\ \omega & 0 & 0 \\ 0 & \omega^2 & 0 \end{pmatrix}. \\
\chi[21]=(1,1,\omega^2,\omega,\omega^2,\omega,1,\omega,\omega^2), &
\tau_A[21]=\begin{pmatrix} 0 & \omega & 0 \\ 0 & 0 & \omega^2 \\ \,\,1\,\, & 0 & 0 \end{pmatrix}, &
\tau_B[21]=\begin{pmatrix} 0 & 0 & \omega \\ \omega^2 & 0 & 0 \\ 0 & \,\,1\,\, & 0 \end{pmatrix}.\\
\chi[22]=(1,\omega,\omega,\omega,\omega^2,\omega^2,\omega^2,1,1), &
\tau_A[22]=\begin{pmatrix} 0 & 0 & \kappa \\ \omega\kappa & 0 & 0 \\ 0 & \omega^2\kappa & 0 \end{pmatrix}, &
\tau_B[22]=\begin{pmatrix} 0 & \omega^2\kappa^2 & 0 \\ 0 & 0 & \kappa^2 \\ \omega\kappa^2 & 0 & 0 \end{pmatrix}. \\
\chi[23]=(1,\omega,\omega,\omega,\omega^2,\omega^2,\omega^2,1,1), &
\tau_A[23]=\begin{pmatrix} 0 & 0 & \omega^2\kappa \\ \kappa & 0 & 0 \\ 0 & \omega\kappa & 0 \end{pmatrix}, &
\tau_B[23]=\begin{pmatrix} 0 & \kappa^2 & 0 \\ 0 & 0 & \omega\kappa^2 \\ \omega^2\kappa^2 & 0 & 0 \end{pmatrix}. \\
\chi[24]=(1,\omega,\omega,\omega,\omega^2,\omega^2,\omega^2,1,1), &
\tau_A[24]=\begin{pmatrix} 0 & 0 & \omega\kappa \\ \omega^2\kappa & 0 & 0 \\ 0 & \kappa & 0 \end{pmatrix}, &
\tau_B[24]=\begin{pmatrix} 0 & \omega\kappa^2 & 0 \\ 0 & 0 & \omega^2\kappa^2 \\ \kappa^2 & 0 & 0 \end{pmatrix}. \\
\chi[25]=(1,\omega^2,1,\omega,\omega^2,1,\omega,\omega^2,\omega), &
\tau_A[25]=\begin{pmatrix} \omega\kappa^2 & 0 & 0 \\ 0 & \omega^2\kappa^2 & 0 \\ 0 & 0 & \kappa^2 \end{pmatrix}, & 
\tau_B[25]=\begin{pmatrix} \omega^2\kappa & 0 & 0 \\ 0 & \kappa & 0 \\ 0 & 0 & \omega\kappa \end{pmatrix}. \\
\chi[26]=(1,\omega^2,1,\omega,\omega^2,1,\omega,\omega^2,\omega), &
\tau_A[26]=\begin{pmatrix} \kappa^2 & 0 & 0 \\ 0 & \omega\kappa^2 & 0 \\ 0 & 0 & \omega^2\kappa^2 \end{pmatrix}, &
\tau_B[26]=\begin{pmatrix} \kappa & 0 & 0 \\ 0 & \omega\kappa & 0 \\ 0 & 0 & \omega^2\kappa \end{pmatrix}. \\
\chi[27]=(1,\omega^2,1,\omega,\omega^2,1,\omega,\omega^2,\omega), &
\tau_A[27]=\begin{pmatrix} \omega^2\kappa^2 & 0 & 0 \\ 0 & \kappa^2 & 0 \\ 0 & 0 & \omega\kappa^2 \end{pmatrix}, &
\tau_B[27]=\begin{pmatrix} \omega\kappa & 0 & 0 \\ 0 & \omega^2\kappa & 0 \\ 0 & 0 & \kappa \end{pmatrix}.
\end{array}
\end{equation*}
}

\paragraph{Morphisms}
The following maps use the same parameter notation and output
ordering as in Subsection~\ref{sec:AppEsp3}.
{\scriptsize
\begin{equation*}
\begin{array}{lll}
J_{g_1A}=\mathrm{x}_1\begin{pmatrix} \,\,1\,\, & 0 & 0 \\ 0 & \,\,1\,\, & 0 \\ 0 & 0 & \,\,1\,\, \end{pmatrix}, &
J_{g_2A}=\mathrm{x}_2\begin{pmatrix} 0 & \,\,1\,\, & 0 \\ 0 & 0 & \omega \\ \omega^2 & 0 & 0 \end{pmatrix}, &
J_{g_3A}=\mathrm{x}_3\begin{pmatrix} \,\,1\,\, & 0 & 0 \\ 0 & \omega^2 & 0 \\ 0 & 0 & \omega \end{pmatrix}, \\
J_{g_4A}=\mathrm{x}_4\begin{pmatrix} 0 & 0 & \,\,1\,\, \\ \,\,1\,\, & 0 & 0 \\ 0 & \,\,1\,\, & 0 \end{pmatrix}, &
J_{g_5A}=\mathrm{x}_5\begin{pmatrix} 0 & \,\,1\,\, & 0 \\ 0 & 0 & \,\,1\,\, \\ \,\,1\,\, & 0 & 0 \end{pmatrix}, &
J_{g_6A}=\mathrm{x}_6\begin{pmatrix} \,\,1\,\, & 0 & 0 \\ 0 & \omega & 0 \\ 0 & 0 & \omega^2 \end{pmatrix}, \\
J_{g_7A}=\mathrm{x}_7\begin{pmatrix} 0 & 0 & \,\,1\,\, \\ \omega^2 & 0 & 0 \\ 0 & \omega & 0 \end{pmatrix}, &
J_{g_8A}=\mathrm{x}_8\begin{pmatrix} 0 & \,\,1\,\, & 0 \\ 0 & 0 & \omega^2 \\ \omega & 0 & 0 \end{pmatrix}, &
J_{g_9A}=\mathrm{x}_9\begin{pmatrix} 0 & 0 & \,\,1\,\, \\ \omega & 0 & 0 \\ 0 & \omega^2 & 0 \end{pmatrix}. 
\end{array}
\end{equation*}

\begin{equation*}
\begin{array}{lll}
J_{g_1B}=\mathrm{y}_1\begin{pmatrix} \,\,1\,\, & 0 & 0 \\ 0 & \,\,1\,\, & 0 \\ 0 & 0 & \,\,1\,\, \end{pmatrix},  &
J_{g_2B}=\mathrm{y}_2\begin{pmatrix} 0 & \,\,1\,\, & 0 \\ 0 & 0 & \omega^2 \\ \omega & 0 & 0 \end{pmatrix}, &
J_{g_3B}=\mathrm{y}_3\begin{pmatrix} \,\,1\,\, & 0 & 0 \\ 0 & \omega & 0 \\ 0 & 0 & \omega^2 \end{pmatrix}, \\
J_{g_4B}=\mathrm{y}_4\begin{pmatrix} 0 & 0 & \,\,1\,\, \\ \,\,1\,\, & 0 & 0 \\ 0 & \,\,1\,\, & 0 \end{pmatrix}, &
J_{g_5B}=\mathrm{y}_5\begin{pmatrix} 0 & \,\,1\,\, & 0 \\ 0 & 0 & \,\,1\,\, \\ \,\,1\,\, & 0 & 0 \end{pmatrix}, &
J_{g_6B}=\mathrm{y}_6\begin{pmatrix} \,\,1\,\, & 0 & 0 \\ 0 & \omega^2 & 0 \\ 0 & 0 & \omega \end{pmatrix}, \\
J_{g_7B}=\mathrm{y}_7\begin{pmatrix} 0 & 0 & \,\,1\,\, \\ \omega & 0 & 0 \\ 0 & \omega^2 & 0 \end{pmatrix}, &
J_{g_8B}=\mathrm{y}_8\begin{pmatrix} 0 & \,\,1\,\, & 0 \\ 0 & 0 & \omega \\ \omega^2 & 0 & 0 \end{pmatrix}, &
J_{g_9B}=\mathrm{y}_9\begin{pmatrix} 0 & 0 & \,\,1\,\, \\ \omega^2 & 0 & 0 \\ 0 & \omega & 0 \end{pmatrix}. \\
\end{array}
\end{equation*}

\vspace{0.5cm}
\[
J_{AA}=\begin{pmatrix}
    0 & 0 & \mathrm{r}_{11} & 0 & \mathrm{r}_{12} & 0 & \mathrm{r}_{13} & 0 & 0 \\
    \mathrm{r}_{12} & 0 & 0 & 0 & 0 & \mathrm{r}_{13} & 0 & \mathrm{r}_{11} & 0 \\
    0 & \mathrm{r}_{13} & 0 & \mathrm{r}_{11} & 0 & 0 & 0 & 0 & \mathrm{r}_{12} \\
    0 & 0 & \mathrm{r}_{21} & 0 & \mathrm{r}_{22} & 0 & \mathrm{r}_{23} & 0 & 0 \\
    \mathrm{r}_{22} & 0 & 0 & 0 & 0 & \mathrm{r}_{23} & 0 & \mathrm{r}_{21} & 0 \\
    0 & \mathrm{r}_{23} & 0 & \mathrm{r}_{21} & 0 & 0 & 0 & 0 & \mathrm{r}_{22} \\
    0 & 0 & \mathrm{r}_{31} & 0 & \mathrm{r}_{32} & 0 & \mathrm{r}_{33} & 0 & 0 \\
    \mathrm{r}_{32} & 0 & 0 & 0 & 0 & \mathrm{r}_{33} & 0 & \mathrm{r}_{31} & 0 \\
    0 & \mathrm{r}_{33} & 0 & \mathrm{r}_{31} & 0 & 0 & 0 & 0 & \mathrm{r}_{32} 
\end{pmatrix},\,
J_{BB}=\begin{pmatrix}
    \mathrm{s}_{11} & 0 & 0 & 0 & 0 & \mathrm{s}_{12} & 0 & \mathrm{s}_{13} & 0 \\
    0 & \omega \mathrm{s}_{12} & 0 & \omega \mathrm{s}_{13} & 0 & 0 & 0 & 0 & \omega \mathrm{s}_{11} \\
    0 & 0 & \omega^2 \mathrm{s}_{13} & 0 & \omega^2 \mathrm{s}_{11} & 0 & \omega^2 \mathrm{s}_{12} & 0 & 0 \\
    \mathrm{s}_{21} & 0 & 0 & 0 & 0 & \mathrm{s}_{22} & 0 & \mathrm{s}_{23} & 0 \\
    0 & \omega \mathrm{s}_{22} & 0 & \omega \mathrm{s}_{23} & 0 & 0 & 0 & 0 & \omega \mathrm{s}_{21} \\
    0 & 0 & \omega^2 \mathrm{s}_{23} & 0 & \omega^2 \mathrm{s}_{21} & 0 & \omega^2 \mathrm{s}_{22} & 0 & 0 \\
    \mathrm{s}_{31} & 0 & 0 & 0 & 0 & \mathrm{s}_{32} & 0 & \mathrm{s}_{33} & 0 \\
    0 & \omega \mathrm{s}_{32} & 0 & \omega \mathrm{s}_{33} & 0 & 0 & 0 & 0 & \omega \mathrm{s}_{31} \\
    0 &  0 & \omega^2 \mathrm{s}_{33} & 0 & \omega^2 \mathrm{s}_{31} & 0 & \omega^2 \mathrm{s}_{32} & 0 & 0  
\end{pmatrix},
\]
\vspace{0.5cm}
\[
J_{AB}=\begin{pmatrix}
    0 & 0 & \mathrm{z}_1 & 0 & \omega \mathrm{z}_1 & 0 & \omega^2 \mathrm{z}_1 & 0 & 0 \\
    0 & \mathrm{z}_2 & 0 &  \mathrm{z}_2 & 0 & 0 & 0 & 0 &  \mathrm{z}_2 \\
    0 & 0 & \mathrm{z}_3 & 0 & \omega^2 \mathrm{z}_3 & 0 & \omega \mathrm{z}_3 & 0 & 0 \\
    \mathrm{z}_4 & 0 & 0 & 0 & 0 & \omega \mathrm{z}_4 & 0 & \omega^2 \mathrm{z}_4 & 0 \\
    0 & \mathrm{z}_5 & 0 & \omega \mathrm{z}_5 & 0 & 0 & 0 & 0 & \omega^2 \mathrm{z}_5 \\
    0 & 0 & \mathrm{z}_6 & 0 & \mathrm{z}_6 & 0 & \mathrm{z}_6 & 0 & 0 \\
    \mathrm{z}_7 & 0 & 0 & 0 & 0 & \omega^2 \mathrm{z}_7 & 0 & \omega \mathrm{z}_7 & 0 \\
    0 & \mathrm{z}_8 & 0 & \omega^2 \mathrm{z}_8 & 0 & 0 & 0 & 0 & \omega \mathrm{z}_8 \\
    \mathrm{z}_9 & 0 & 0 & 0 & 0 & \mathrm{z}_9 & 0 & \mathrm{z}_9 & 0 
\end{pmatrix},
\]

\[
J_{BA}=\begin{pmatrix}
    0 & 0 & \omega^2 \tilde{\mathrm{z}}_1 & 0 & \omega \tilde{\mathrm{z}}_1 & 0 & \tilde{\mathrm{z}}_1 & 0 & 0 \\
    0 & \tilde{\mathrm{z}}_2 & 0 & \tilde{\mathrm{z}}_2 & 0 & 0 & 0 & 0 & \tilde{\mathrm{z}}_2 \\
    0 & 0 & \omega \tilde{\mathrm{z}}_3 & 0 & \omega^2 \tilde{\mathrm{z}}_3 & 0 & \tilde{\mathrm{z}}_3 & 0 & 0 \\
    \tilde{\mathrm{z}}_4 & 0 & 0 & 0 & 0 & \omega^2 \tilde{\mathrm{z}}_4 & 0 & \omega \tilde{\mathrm{z}}_4 & 0 \\
    0 & \omega \tilde{\mathrm{z}}_5 & 0 & \tilde{\mathrm{z}}_5 & 0 & 0 & 0 & 0 & \omega^2 \tilde{\mathrm{z}}_5 \\
    0 & 0 & \tilde{\mathrm{z}}_6 & 0 & \tilde{\mathrm{z}}_6 & 0 & \tilde{\mathrm{z}}_6 & 0 & 0 \\
    \tilde{\mathrm{z}}_7 & 0 & 0 & 0 & 0 & \omega \tilde{\mathrm{z}}_7 & 0 & \omega^2 \tilde{\mathrm{z}}_7 & 0 \\
    0 & \omega^2 \tilde{\mathrm{z}}_8 & 0 & \tilde{\mathrm{z}}_8 & 0 & 0 & 0 & 0 & \omega \tilde{\mathrm{z}}_8 \\
    \tilde{\mathrm{z}}_9 & 0 & 0 & 0 & 0 & \tilde{\mathrm{z}}_9 & 0 & \tilde{\mathrm{z}}_9 & 0 
\end{pmatrix}.
\]
}

\subsection{Obtaining other labeled forms by relabeling}

The two parametric forms displayed above provide representatives
of the relabeling orbits associated with
\(\operatorname{Rep}(\mathrm{Esp}_3)\) and
\(\operatorname{Rep}(\mathrm{Esp}_9)\).
We now explain how fusion-ring automorphisms generate
the remaining labeled forms.
These transformations change the presentation of the reconstructed
category; the transformed invariants identify which parametric
form has been obtained.

An automorphism in Eq.~\eqref{eq:Bootp3-fusion-ring-aut}
consists of an element
\(\sigma\in GL(2,\mathbb F_3)\), acting on the invertible
objects, and an element \(u\in\mathbb Z_2\), acting on \(A,B\).

For the \(\operatorname{Rep}(\mathrm{Esp}_3)\) orbit,
form \(q=2\) is obtained from form \(q=1\) by taking
\[
    \sigma=
    \begin{pmatrix}
        2&0\\
        0&1
    \end{pmatrix}
\]
and leaving \(A,B\) unchanged.
By Eq.~\eqref{eq:Bootp3-invariant-relabeling}, this preserves
\(\epsilon=\mathsf{1}\) and sends \(\iota=1\) to \(\iota=2\).
Likewise, the orbit of form \(q=3\) contains all sixteen
labeled forms associated with
\(\operatorname{Rep}(\mathrm{Esp}_9)\).
Thus the two representatives printed in this appendix generate
all eighteen labeled forms.

To implement these relabelings explicitly, we transport both
the tensor-automorphism components and the \(J\)-maps.
Keeping the enumeration index \(n\) unchanged, the transformed
tensor-automorphism components are
\begin{gather*} 
\chi^{\prime}=(\chi^{\prime}_{g_1},\chi^{\prime}_{g_2},\dots,\chi^{\prime}_{g_9}) := (\chi_{\sigma^{-1}(g_1)},\chi_{\sigma^{-1}(g_2)},\dots,\chi_{\sigma^{-1}(g_9)}),\\ 
\tau^{\prime}_A=\tau_{u^{-1}(A)}[n],\quad \tau^{\prime}_B=\tau_{u^{-1}(B)}[n]. 
\end{gather*} 
For the morphisms, we have 
\begin{gather*} 
J^{\prime}_{g_1 g_2}:=J_{\sigma^{-1}(g_1)\sigma^{-1}(g_2)},\\
J^{\prime}_{Ag}:=J_{u^{-1}(A)\sigma^{-1}(g)}, \quad J^{\prime}_{gA}:=J_{\sigma^{-1}(g)u^{-1}(A)},\\ 
J^{\prime}_{Bg}:=J_{u^{-1}(B)\sigma^{-1}(g)}, \quad J^{\prime}_{gB}:=J_{\sigma^{-1}(g)u^{-1}(B)},\\ 
J^{\prime}_{AA}:=J_{u^{-1}(A)u^{-1}(A)}, \quad J^{\prime}_{BB}:=J_{u^{-1}(B)u^{-1}(B)}. 
\end{gather*} 
For \(J_{AB}\) and \(J_{BA}\), one must also reorder the
output channels.
Since their rows are labeled by \(g_1,\ldots,g_9\),
the row assigned to \(g_i\) after relabeling is the row
previously assigned to \(\sigma^{-1}(g_i)\).
Thus, 
\[ 
J^{\prime}_{AB}:= \begin{pmatrix} J_{u^{-1}(A)u^{-1}(B)}^{\sigma^{-1}(g_1)} \\ 
J_{u^{-1}(A)u^{-1}(B)}^{\sigma^{-1}(g_2)} \\ 
\dots \\ J_{u^{-1}(A)u^{-1}(B)}^{\sigma^{-1}(g_9)} 
\end{pmatrix},
\quad J^{\prime}_{BA}:= \begin{pmatrix} J_{u^{-1}(B)u^{-1}(A)}^{\sigma^{-1}(g_1)} \\ 
J_{u^{-1}(B)u^{-1}(A)}^{\sigma^{-1}(g_2)} \\ 
\dots \\ 
J_{u^{-1}(B)u^{-1}(A)}^{\sigma^{-1}(g_9)} 
\end{pmatrix}.
\] 
The ring automorphism that we choose to consider is 
\[ 
\sigma= \begin{pmatrix} 1 & 1 \\ 1 & 0 \end{pmatrix} \in GL(2,\mathbb{Z}_3), \qquad u\in\mathbb{Z}_2 \text{ is the trivial element.}
\] 
This automorphism acts as 
\begin{gather*} 
(g_1,g_2,g_3,g_4,g_5,g_6,g_7,g_8,g_9) \longmapsto (g_1,g_5,g_2,g_8,g_9,g_7,g_4,g_3,g_6), \\ 
A\longmapsto A,\qquad B\longmapsto B, 
\end{gather*} 
which means that, for instance, 
\[ 
\chi^{\prime}[12] = (1,\omega,\omega^2,1,1,\omega,\omega^2,\omega^2,\omega), \qquad \tau^{\prime}_A=\tau_A, \qquad \tau^{\prime}_B=\tau_B. 
\] 
For the morphisms, we have 
\begin{gather*} 
J^{\prime}_{Ag_2}=J_{Ag_3},\quad J^{\prime}_{Ag_3}=J_{Ag_8},\quad J^{\prime}_{Ag_4}=J_{Ag_7},\quad J^{\prime}_{Ag_5}=J_{Ag_2},\\ J^{\prime}_{Bg_6}=J_{Bg_9},\quad J^{\prime}_{Bg_7}=J_{Bg_6},\quad J^{\prime}_{Bg_8}=J_{Bg_4},\quad J^{\prime}_{Bg_9}=J_{Bg_5},\\ J^{\prime}_{AA}=J_{AA},\quad J^{\prime}_{BB}=J_{BB},\\ 
J^{\prime}_{AB}= \begin{pmatrix} J_{AB}^{g_1}\\ 
J_{AB}^{g_3}\\ 
\dots\\ J_{AB}^{g_5} 
\end{pmatrix},
\quad J^{\prime}_{BA}= \begin{pmatrix} J_{BA}^{g_1}\\
J_{BA}^{g_3}\\
\dots\\ J_{BA}^{g_5} 
\end{pmatrix}. 
\end{gather*} 
As we explained in \eqref{eq:Bootp3-invariant-relabeling}, the new invariants are \[ \epsilon=g_2\longmapsto \epsilon^{\prime}=g_5, \qquad \iota=1\longmapsto \iota^{\prime}=2, \] so we see that this ring automorphism allows us to obtain a new parametric form, namely the one that is called ``parametric form 10'' in our code. As discussed in the main text, we can apply ring automorphisms to the third parametric form in order to obtain all the others associated with Rep(Esp$_9$). In the following table, we associate to each parametric form the list of ring automorphisms that must be applied to the third one. For each parametric form, there are six such automorphisms: three act only on the invertible objects, while the remaining three also involve the swap $A\leftrightarrow B$.

\begin{table}[ht]
$\begin{array}{c|c|c||c|c|c}
 q & A=A,\,B=B & A\leftrightarrow B & q & A=A,\,B=B & A\leftrightarrow B\\
 \hline  
3 & \big(\begin{smallmatrix}1 & 0 \\ 0 & 1\end{smallmatrix}\big),
    \,\big(\begin{smallmatrix}1 & 1 \\ 0 & 1\end{smallmatrix}\big),
    \,\big(\begin{smallmatrix}1 & 2 \\ 0 & 1\end{smallmatrix}\big)
& \big(\begin{smallmatrix}2 & 0 \\ 0 & 1\end{smallmatrix}\big),
    \,\big(\begin{smallmatrix}2 & 1 \\ 0 & 1\end{smallmatrix}\big),
    \,\big(\begin{smallmatrix}2 & 2 \\ 0 & 1\end{smallmatrix}\big)
& 11 & \big(\begin{smallmatrix}0 & 1 \\ 2 & 0\end{smallmatrix}\big),
    \,\big(\begin{smallmatrix}0 & 1 \\ 2 & 1\end{smallmatrix}\big),
    \,\big(\begin{smallmatrix}0 & 1 \\ 2 & 2\end{smallmatrix}\big)
& \big(\begin{smallmatrix}0 & 1 \\ 1 & 0\end{smallmatrix}\big),
    \,\big(\begin{smallmatrix}0 & 1 \\ 1 & 1\end{smallmatrix}\big),
    \,\big(\begin{smallmatrix}0 & 1 \\ 1 & 2\end{smallmatrix}\big)\\
4 & \big(\begin{smallmatrix}1 & 0 \\ 0 & 2\end{smallmatrix}\big),
    \,\big(\begin{smallmatrix}1 & 1 \\ 0 & 2\end{smallmatrix}\big),
    \,\big(\begin{smallmatrix}1 & 2 \\ 0 & 2\end{smallmatrix}\big)
& \big(\begin{smallmatrix}2 & 0 \\ 0 & 2\end{smallmatrix}\big),
    \,\big(\begin{smallmatrix}2 & 1 \\ 0 & 2\end{smallmatrix}\big),
    \,\big(\begin{smallmatrix}2 & 2 \\ 0 & 2\end{smallmatrix}\big)
& 12 & \big(\begin{smallmatrix}0 & 2 \\ 2 & 0\end{smallmatrix}\big),
    \,\big(\begin{smallmatrix}0 & 2 \\ 2 & 1\end{smallmatrix}\big),
    \,\big(\begin{smallmatrix}0 & 2 \\ 2 & 2\end{smallmatrix}\big)
& \big(\begin{smallmatrix}0 & 2 \\ 1 & 0\end{smallmatrix}\big),
    \,\big(\begin{smallmatrix}0 & 2 \\ 1 & 1\end{smallmatrix}\big),
    \,\big(\begin{smallmatrix}0 & 2 \\ 1 & 2\end{smallmatrix}\big)\\
5 & \big(\begin{smallmatrix}0 & 2 \\ 1 & 0\end{smallmatrix}\big),
    \,\big(\begin{smallmatrix}0 & 2 \\ 1 & 1\end{smallmatrix}\big),
    \,\big(\begin{smallmatrix}0 & 2 \\ 1 & 2\end{smallmatrix}\big)
& \big(\begin{smallmatrix}0 & 2 \\ 2 & 0\end{smallmatrix}\big),
    \,\big(\begin{smallmatrix}0 & 2 \\ 2 & 1\end{smallmatrix}\big),
    \,\big(\begin{smallmatrix}0 & 2 \\ 2 & 2\end{smallmatrix}\big)
& 13 & \big(\begin{smallmatrix}2 & 0 \\ 0 & 2\end{smallmatrix}\big),
    \,\big(\begin{smallmatrix}2 & 1 \\ 0 & 2\end{smallmatrix}\big),
    \,\big(\begin{smallmatrix}2 & 2 \\ 0 & 2\end{smallmatrix}\big)
& \big(\begin{smallmatrix}1 & 0 \\ 0 & 2\end{smallmatrix}\big),
    \,\big(\begin{smallmatrix}1 & 1 \\ 0 & 2\end{smallmatrix}\big),
    \,\big(\begin{smallmatrix}1 & 2 \\ 0 & 2\end{smallmatrix}\big)\\
6 & \big(\begin{smallmatrix}0 & 1 \\ 1 & 0\end{smallmatrix}\big),
    \,\big(\begin{smallmatrix}0 & 1 \\ 1 & 1\end{smallmatrix}\big),
    \,\big(\begin{smallmatrix}0 & 1 \\ 1 & 2\end{smallmatrix}\big)
& \big(\begin{smallmatrix}0 & 1 \\ 2 & 0\end{smallmatrix}\big),
    \,\big(\begin{smallmatrix}0 & 1 \\ 2 & 1\end{smallmatrix}\big),
    \,\big(\begin{smallmatrix}0 & 1 \\ 2 & 2\end{smallmatrix}\big)
& 14 & \big(\begin{smallmatrix}2 & 0 \\ 0 & 1\end{smallmatrix}\big),
    \,\big(\begin{smallmatrix}2 & 1 \\ 0 & 1\end{smallmatrix}\big),
    \,\big(\begin{smallmatrix}2 & 2 \\ 0 & 1\end{smallmatrix}\big)
& \big(\begin{smallmatrix}1 & 0 \\ 0 & 1\end{smallmatrix}\big),
    \,\big(\begin{smallmatrix}1 & 1 \\ 0 & 1\end{smallmatrix}\big),
    \,\big(\begin{smallmatrix}1 & 2 \\ 0 & 1\end{smallmatrix}\big)\\
7 & \big(\begin{smallmatrix}2 & 0 \\ 2 & 2\end{smallmatrix}\big),
    \,\big(\begin{smallmatrix}2 & 1 \\ 2 & 0\end{smallmatrix}\big),
    \,\big(\begin{smallmatrix}2 & 2 \\ 2 & 1\end{smallmatrix}\big)
& \big(\begin{smallmatrix}1 & 0 \\ 1 & 2\end{smallmatrix}\big),
    \,\big(\begin{smallmatrix}1 & 1 \\ 1 & 0\end{smallmatrix}\big),
    \,\big(\begin{smallmatrix}1 & 2 \\ 1 & 1\end{smallmatrix}\big)
& 15 & \big(\begin{smallmatrix}1 & 0 \\ 2 & 1\end{smallmatrix}\big),
    \,\big(\begin{smallmatrix}1 & 1 \\ 2 & 0\end{smallmatrix}\big),
    \,\big(\begin{smallmatrix}1 & 2 \\ 2 & 2\end{smallmatrix}\big)
& \big(\begin{smallmatrix}2 & 0 \\ 1 & 1\end{smallmatrix}\big),
    \,\big(\begin{smallmatrix}2 & 1 \\ 1 & 0\end{smallmatrix}\big),
    \,\big(\begin{smallmatrix}2 & 2 \\ 1 & 2\end{smallmatrix}\big)\\
8 & \big(\begin{smallmatrix}2 & 0 \\ 2 & 1\end{smallmatrix}\big),
    \,\big(\begin{smallmatrix}2 & 1 \\ 2 & 2\end{smallmatrix}\big),
    \,\big(\begin{smallmatrix}2 & 2 \\ 2 & 0\end{smallmatrix}\big)
& \big(\begin{smallmatrix}1 & 0 \\ 1 & 1\end{smallmatrix}\big),
    \,\big(\begin{smallmatrix}1 & 1 \\ 1 & 2\end{smallmatrix}\big),
    \,\big(\begin{smallmatrix}1 & 2 \\ 1 & 0\end{smallmatrix}\big)
& 16 & \big(\begin{smallmatrix}1 & 0 \\ 2 & 2\end{smallmatrix}\big),
    \,\big(\begin{smallmatrix}1 & 1 \\ 2 & 1\end{smallmatrix}\big),
    \,\big(\begin{smallmatrix}1 & 2 \\ 2 & 0\end{smallmatrix}\big)
& \big(\begin{smallmatrix}2 & 0 \\ 1 & 2\end{smallmatrix}\big),
    \,\big(\begin{smallmatrix}2 & 1 \\ 1 & 1\end{smallmatrix}\big),
    \,\big(\begin{smallmatrix}2 & 2 \\ 1 & 0\end{smallmatrix}\big)\\
9 & \big(\begin{smallmatrix}1 & 0 \\ 1 & 1\end{smallmatrix}\big),
    \,\big(\begin{smallmatrix}1 & 1 \\ 1 & 2\end{smallmatrix}\big),
    \,\big(\begin{smallmatrix}1 & 2 \\ 1 & 0\end{smallmatrix}\big)
& \big(\begin{smallmatrix}2 & 0 \\ 2 & 1\end{smallmatrix}\big),
    \,\big(\begin{smallmatrix}2 & 1 \\ 2 & 2\end{smallmatrix}\big),
    \,\big(\begin{smallmatrix}2 & 2 \\ 2 & 0\end{smallmatrix}\big)
& 17 & \big(\begin{smallmatrix}2 & 0 \\ 1 & 2\end{smallmatrix}\big),
    \,\big(\begin{smallmatrix}2 & 1 \\ 1 & 1\end{smallmatrix}\big),
    \,\big(\begin{smallmatrix}2 & 2 \\ 1 & 0\end{smallmatrix}\big)
& \big(\begin{smallmatrix}1 & 0 \\ 2 & 2\end{smallmatrix}\big),
    \,\big(\begin{smallmatrix}1 & 1 \\ 2 & 1\end{smallmatrix}\big),
    \,\big(\begin{smallmatrix}1 & 2 \\ 2 & 0\end{smallmatrix}\big)\\
10 & \big(\begin{smallmatrix}1 & 0 \\ 1 & 2\end{smallmatrix}\big),
    \,\big(\begin{smallmatrix}1 & 1 \\ 1 & 0\end{smallmatrix}\big),
    \,\big(\begin{smallmatrix}1 & 2 \\ 1 & 1\end{smallmatrix}\big)
& \big(\begin{smallmatrix}2 & 0 \\ 2 & 2\end{smallmatrix}\big),
    \,\big(\begin{smallmatrix}2 & 1 \\ 2 & 0\end{smallmatrix}\big),
    \,\big(\begin{smallmatrix}2 & 2 \\ 2 & 1\end{smallmatrix}\big)
& 18 & \big(\begin{smallmatrix}2 & 0 \\ 1 & 1\end{smallmatrix}\big),
    \,\big(\begin{smallmatrix}2 & 1 \\ 1 & 0\end{smallmatrix}\big),
    \,\big(\begin{smallmatrix}2 & 2 \\ 1 & 2\end{smallmatrix}\big)
& \big(\begin{smallmatrix}1 & 0 \\ 2 & 1\end{smallmatrix}\big),
    \,\big(\begin{smallmatrix}1 & 1 \\ 2 & 0\end{smallmatrix}\big),
    \,\big(\begin{smallmatrix}1 & 2 \\ 2 & 2\end{smallmatrix}\big)
\end{array}\,.$
\caption{List of ring automorphisms that we need to apply to obtain a specific parametric form starting from the third one.}
\end{table}

\newpage
\bibliography{Bibliography/ref}

@misc{ostrik2006modulecategoriesdrinfelddouble,
      title={Module categories over the Drinfeld double of a finite group}, 
      author={Victor Ostrik},
      year={2006},
      eprint={math/0202130},
      archivePrefix={arXiv},
      primaryClass={math.QA},
      url={https://arxiv.org/abs/math/0202130}, 
}

@article{AljadeffEtingofGelakiNikshych,
  author  = {Aljadeff, Eli and Etingof, Pavel and Gelaki, Shlomo and Nikshych, Dmitri},
  title   = {On twisting of finite-dimensional {H}opf algebras},
  journal = {Journal of Algebra},
  volume  = {256},
  number  = {2},
  pages   = {484--501},
  year    = {2002},
  doi     = {10.1016/S0021-8693(02)00092-3},
  eprint  = {math/0107167},
  archivePrefix = {arXiv}
}

@article{Thorngren:2019iar,
    author = "Thorngren, Ryan and Wang, Yifan",
    title = "{Fusion category symmetry. Part I. Anomaly in-flow and gapped phases}",
    eprint = "1912.02817",
    archivePrefix = "arXiv",
    primaryClass = "hep-th",
    reportNumber = "PUPT-2603",
    doi = "10.1007/JHEP04(2024)132",
    journal = "JHEP",
    volume = "04",
    pages = "132",
    year = "2024"
}

@article{Tannaka1938,
    author = "Tadao Tannaka",
    title = "{Über den Dualitätssatz der nichtkommutativen topologischen Gruppen}",
    journal = "Tohoku Math. J. 45 1 1-12",
    year = "1938"
}

@article{Tambara:1998vmj,
    author = "Tambara, Daisuke and Yamagami, Shigeru",
    title = "{Tensor Categories with Fusion Rules of Self-Duality for Finite Abelian Groups}",
    doi = "10.1006/jabr.1998.7558",
    journal = "J. Algebra",
    volume = "209",
    number = "2",
    pages = "692--707",
    year = "1998"
}

@article{Tambara:2000qlz,
    author = "Tambara, D.",
    title = "{Representations of tensor categories with fusion rules of self-duality for abelian groups}",
    doi = "10.1007/BF02803515",
    journal = "Isr. J. Math.",
    volume = "118",
    number = "1",
    pages = "29--60",
    year = "2000"
}

@article{Jordan_2009,
   title={On the classification of certain fusion categories},
   volume={3},
   ISSN={1661-6960},
   url={http://dx.doi.org/10.4171/JNCG/44},
   DOI={10.4171/jncg/44},
   number={3},
   journal={Journal of Noncommutative Geometry},
   publisher={European Mathematical Society - EMS - Publishing House GmbH},
   author={Jordan, David and Larson, Eric},
   year={2009},
   month={Sept}, pages={481–499} 
}

@misc{czenky2026nimrepresentationstambarayamagamigeneralizations,
      title={NIM-representations of Tambara-Yamagami generalizations}, 
      author={Agustina Czenky and Emily McGovern and Melody Molander and Monique Müller and Ana Ros Camacho},
      year={2026},
      eprint={2602.23502},
      archivePrefix={arXiv},
      primaryClass={math.QA},
      url={https://arxiv.org/abs/2602.23502}, 
}

@book{etingof2015tensor,
  title     = {Tensor Categories},
  author    = {Etingof, Pavel and Gelaki, Shlomo and Nikshych, Dmitri and Ostrik, Victor},
  series    = {Mathematical Surveys and Monographs},
  volume    = {205},
  year      = {2015},
  publisher = {American Mathematical Society},
  address   = {Providence, RI},
  isbn      = {978-1-4704-2024-6}
}

@article{Movshev,
  author  = {Movshev, M. V.},
  title   = {Twisting in Group Algebras of Finite Groups},
  journal = {Functional Analysis and Its Applications},
  volume  = {27},
  year    = {1993},
  pages   = {240--244}
}

@article{ENO2005,
  author  = {Etingof, Pavel and Nikshych, Dmitri and Ostrik, Viktor},
  title   = {On fusion categories},
  journal = {Annals of Mathematics},
  volume  = {162},
  number  = {2},
  pages   = {581--642},
  year    = {2005},
  doi     = {10.4007/annals.2005.162.581}
}

@incollection{Ocneanu1989,
  author    = {Ocneanu, Adrian},
  title     = {Quantized groups, string algebras, and Galois theory for algebras},
  booktitle = {Operator Algebras and Applications, Vol. 2},
  series    = {London Mathematical Society Lecture Note Series},
  volume    = {136},
  pages     = {119--172},
  publisher = {Cambridge University Press},
  year      = {1989},
  doi       = {10.1017/CBO9780511662287.008}
}

@article{MasuokaDim68,
  author  = {Masuoka, Akira},
  title   = {Semisimple Hopf Algebras of Dimension 6, 8},
  journal = {Israel Journal of Mathematics},
  volume  = {92},
  number  = {1--3},
  pages   = {361--373},
  year    = {1995},
  doi     = {10.1007/BF02762089}
}

@article{LarsonRadford1988,
  author  = {Larson, Richard G. and Radford, David E.},
  title   = {Finite dimensional cosemisimple Hopf algebras in
             characteristic 0 are semisimple},
  journal = {Journal of Algebra},
  volume  = {117},
  number  = {2},
  pages   = {267--289},
  year    = {1988},
  doi     = {10.1016/0021-8693(88)90107-X}
}

@article{Bhardwaj:2017xup,
    author = "Bhardwaj, Lakshya and Tachikawa, Yuji",
    title = "{On finite symmetries and their gauging in two dimensions}",
    eprint = "1704.02330",
    archivePrefix = "arXiv",
    primaryClass = "hep-th",
    reportNumber = "IPMU-17-0049",
    doi = "10.1007/JHEP03(2018)189",
    journal = "JHEP",
    volume = "03",
    pages = "189",
    year = "2018"
}

@article{Gaiotto:2014kfa,
    author = "Gaiotto, Davide and Kapustin, Anton and Seiberg, Nathan and Willett, Brian",
    title = "{Generalized Global Symmetries}",
    eprint = "1412.5148",
    archivePrefix = "arXiv",
    primaryClass = "hep-th",
    doi = "10.1007/JHEP02(2015)172",
    journal = "JHEP",
    volume = "02",
    pages = "172",
    year = "2015"
}

@article{Seifnashri:2024dsd,
    author = "Seifnashri, Sahand and Shao, Shu-Heng",
    title = "{Cluster State as a Noninvertible Symmetry-Protected Topological Phase}",
    eprint = "2404.01369",
    archivePrefix = "arXiv",
    primaryClass = "cond-mat.str-el",
    reportNumber = "YITP-SB-2024-05",
    doi = "10.1103/PhysRevLett.133.116601",
    journal = "Phys. Rev. Lett.",
    volume = "133",
    number = "11",
    pages = "116601",
    year = "2024"
}

@article{Bhardwaj:2024kvy,
    author = "Bhardwaj, Lakshya and Bottini, Lea E. and Schafer-Nameki, Sakura and Tiwari, Apoorv",
    title = "{Lattice models for phases and transitions with non-invertible symmetries}",
    eprint = "2405.05964",
    archivePrefix = "arXiv",
    primaryClass = "cond-mat.str-el",
    doi = "10.21468/SciPostPhys.20.5.134",
    journal = "SciPost Phys.",
    volume = "20",
    number = "5",
    pages = "134",
    year = "2026"
}

@article{Jia:2024bng,
    author = "Jia, Zhian",
    title = "{Generalized cluster states from Hopf algebras: non-invertible symmetry and Hopf tensor network representation}",
    eprint = "2405.09277",
    archivePrefix = "arXiv",
    primaryClass = "quant-ph",
    doi = "10.1007/JHEP09(2024)147",
    journal = "JHEP",
    volume = "09",
    pages = "147",
    year = "2024"
}

@article{Meng:2024nxx,
    author = "Meng, Chenqi and Yang, Xinping and Lan, Tian and Gu, Zhengcheng",
    title = "{Non-invertible SPTs: an on-site realization of (1+1)d anomaly-free fusion category symmetry}",
    eprint = "2412.20546",
    archivePrefix = "arXiv",
    primaryClass = "cond-mat.str-el",
    month = "12",
    year = "2024"
}

@article{Inamura:2024jke,
    author = "Inamura, Kansei and Ohyama, Shuhei",
    title = "{1+1d SPT phases with fusion category symmetry: interface modes and non-abelian Thouless pump}",
    eprint = "2408.15960",
    archivePrefix = "arXiv",
    primaryClass = "cond-mat.str-el",
    month = "8",
    year = "2024"
}

@article{Chang:2018iay,
    author = "Chang, Chi-Ming and Lin, Ying-Hsuan and Shao, Shu-Heng and Wang, Yifan and Yin, Xi",
    title = "{Topological Defect Lines and Renormalization Group Flows in Two Dimensions}",
    eprint = "1802.04445",
    archivePrefix = "arXiv",
    primaryClass = "hep-th",
    reportNumber = "CALT-TH-2017-067, CALT-TH 2017-067, PUPT-2546",
    doi = "10.1007/JHEP01(2019)026",
    journal = "JHEP",
    volume = "01",
    pages = "026",
    year = "2019"
}

@article{Kim:2025zdy,
    author = "Kim, Seolhwa and Sela, Orr and Sun, Zhengdi",
    title = "{Higher structure of non-invertible symmetries from Lagrangian descriptions}",
    eprint = "2509.20540",
    archivePrefix = "arXiv",
    primaryClass = "hep-th",
    doi = "10.1007/JHEP05(2026)120",
    journal = "JHEP",
    volume = "05",
    pages = "120",
    year = "2026"
}

@article{Thorngren:2021yso,
    author = "Thorngren, Ryan and Wang, Yifan",
    title = "{Fusion category symmetry. Part II. Categoriosities at c = 1 and beyond}",
    eprint = "2106.12577",
    archivePrefix = "arXiv",
    primaryClass = "hep-th",
    doi = "10.1007/JHEP07(2024)051",
    journal = "JHEP",
    volume = "07",
    pages = "051",
    year = "2024"
}

@article{Inamura:2021wuo,
    author = "Inamura, Kansei",
    title = "{Topological field theories and symmetry protected topological phases with fusion category symmetries}",
    eprint = "2103.15588",
    archivePrefix = "arXiv",
    primaryClass = "cond-mat.str-el",
    doi = "10.1007/JHEP05(2021)204",
    journal = "JHEP",
    volume = "05",
    pages = "204",
    year = "2021"
}

@article{Damia:2024xju,
    author = "Damia, Jeremias Aguilera and Galati, Giovanni and Hulik, Ondrej and Mancani, Salvo",
    title = "{Exploring duality symmetries, multicriticality and RG flows at c = 2}",
    eprint = "2401.04166",
    archivePrefix = "arXiv",
    primaryClass = "hep-th",
    doi = "10.1007/JHEP04(2024)028",
    journal = "JHEP",
    volume = "04",
    pages = "028",
    year = "2024"
}

@article{Lu:2022ver,
    author = "Lu, Da-Chuan and Sun, Zhengdi",
    title = "{On triality defects in 2d CFT}",
    eprint = "2208.06077",
    archivePrefix = "arXiv",
    primaryClass = "hep-th",
    doi = "10.1007/JHEP02(2023)173",
    journal = "JHEP",
    volume = "02",
    pages = "173",
    year = "2023"
}

@article{Lu:2024lzf,
    author = "Lu, Da-Chuan and Sun, Zhengdi and Zhang, Zipei",
    title = "{Exploring G-ality defects in 2-dim QFTs}",
    eprint = "2406.12151",
    archivePrefix = "arXiv",
    primaryClass = "hep-th",
    doi = "10.1007/JHEP11(2025)081",
    journal = "JHEP",
    volume = "11",
    pages = "081",
    year = "2025"
}

@article{Frohlich:2006ch,
    author = "Fr{\"o}hlich, Jurg and Fuchs, Jurgen and Runkel, Ingo and Schweigert, Christoph",
    title = "{Duality and defects in rational conformal field theory}",
    eprint = "hep-th/0607247",
    archivePrefix = "arXiv",
    reportNumber = "KCL-MTH-06-08, ZMP-HH-06-11",
    doi = "10.1016/j.nuclphysb.2006.11.017",
    journal = "Nucl. Phys. B",
    volume = "763",
    pages = "354--430",
    year = "2007"
}

@article{Choi:2023xjw,
    author = "Choi, Yichul and Rayhaun, Brandon C. and Sanghavi, Yaman and Shao, Shu-Heng",
    title = "{Remarks on boundaries, anomalies, and noninvertible symmetries}",
    eprint = "2305.09713",
    archivePrefix = "arXiv",
    primaryClass = "hep-th",
    doi = "10.1103/PhysRevD.108.125005",
    journal = "Phys. Rev. D",
    volume = "108",
    number = "12",
    pages = "125005",
    year = "2023"
}

@article{Chen:2011pg,
    author = "Chen, Xie and Gu, Zheng-Cheng and Liu, Zheng-Xin and Wen, Xiao-Gang",
    title = "{Symmetry protected topological orders and the group cohomology of their symmetry group}",
    eprint = "1106.4772",
    archivePrefix = "arXiv",
    primaryClass = "cond-mat.str-el",
    doi = "10.1103/PhysRevB.87.155114",
    journal = "Phys. Rev. B",
    volume = "87",
    number = "15",
    pages = "155114",
    year = "2013"
}

@article{PhysRevLett.50.1153,
  title = {Nonlinear Field Theory of Large-Spin Heisenberg Antiferromagnets: Semiclassically Quantized Solitons of the One-Dimensional Easy-Axis N\'eel State},
  author = {Haldane, F. D. M.},
  journal = {Phys. Rev. Lett.},
  volume = {50},
  issue = {15},
  pages = {1153--1156},
  numpages = {0},
  year = {1983},
  month = {Apr},
  publisher = {American Physical Society},
  doi = {10.1103/PhysRevLett.50.1153},
  url = {https://link.aps.org/doi/10.1103/PhysRevLett.50.1153}
}

@article{PhysRevLett.59.799,
  title = {Rigorous results on valence-bond ground states in antiferromagnets},
  author = {Affleck, Ian and Kennedy, Tom and Lieb, Elliott H. and Tasaki, Hal},
  journal = {Phys. Rev. Lett.},
  volume = {59},
  issue = {7},
  pages = {799--802},
  numpages = {0},
  year = {1987},
  month = {Aug},
  publisher = {American Physical Society},
  doi = {10.1103/PhysRevLett.59.799},
  url = {https://link.aps.org/doi/10.1103/PhysRevLett.59.799}
}

@article{Pollmann_2012,
   title={Symmetry protection of topological phases in one-dimensional quantum spin systems},
   volume={85},
   ISSN={1550-235X},
   url={http://dx.doi.org/10.1103/PhysRevB.85.075125},
   DOI={10.1103/physrevb.85.075125},
   number={7},
   journal={Physical Review B},
   publisher={American Physical Society (APS)},
   author={Pollmann, Frank and Berg, Erez and Turner, Ari M. and Oshikawa, Masaki},
   year={2012},
   month={Feb}
   }

@article{Chen:2010gda,
    author = "Chen, Xie and Gu, Zheng Cheng and Wen, Xiao Gang",
    title = "{Local unitary transformation, long-range quantum entanglement, wave function renormalization, and topological order}",
    eprint = "1004.3835",
    archivePrefix = "arXiv",
    primaryClass = "cond-mat.str-el",
    doi = "10.1103/PhysRevB.82.155138",
    journal = "Phys. Rev. B",
    volume = "82",
    pages = "155138",
    year = "2010"
}

@article{Schuch:2011niz,
    author = "Schuch, Norbert and P{\'e}rez-Garc{\'\i}a, David and Cirac, Ignacio",
    title = "{Classifying quantum phases using matrix product states and projected entangled pair states}",
    eprint = "1010.3732",
    archivePrefix = "arXiv",
    primaryClass = "cond-mat.str-el",
    doi = "10.1103/PhysRevB.84.165139",
    journal = "Phys. Rev. B",
    volume = "84",
    number = "16",
    pages = "165139",
    year = "2011"
}

@article{Dimofte:2024bwe,
    author = "Dimofte, Tudor and Niu, Wenjun",
    title = "{Tannakian QFT: from spark algebras to quantum groups}",
    eprint = "2411.04194",
    archivePrefix = "arXiv",
    primaryClass = "hep-th",
    month = "11",
    year = "2024"
}

@article{AliAhmad:2025bnd,
    author = "Ali Ahmad, Shadi and Klinger, Marc S. and Wang, Yifan",
    title = "{The many faces of non-invertible symmetries}",
    eprint = "2509.18072",
    archivePrefix = "arXiv",
    primaryClass = "hep-th",
    doi = "10.1007/JHEP05(2026)110",
    journal = "JHEP",
    volume = "05",
    pages = "110",
    year = "2026"
}

@article{Antinucci:2023ezl,
    author = "Antinucci, Andrea and Benini, Francesco and Copetti, Christian and Galati, Giovanni and Rizi, Giovanni",
    title = "{Anomalies of non-invertible self-duality symmetries: fractionalization and gauging}",
    eprint = "2308.11707",
    archivePrefix = "arXiv",
    primaryClass = "hep-th",
    month = "8",
    year = "2023"
}

@article{Antinucci:2022vyk,
    author = "Antinucci, Andrea and Benini, Francesco and Copetti, Christian and Galati, Giovanni and Rizi, Giovanni",
    title = "{The holography of non-invertible self-duality symmetries}",
    eprint = "2210.09146",
    archivePrefix = "arXiv",
    primaryClass = "hep-th",
    reportNumber = "SISSA 16/2022/FISI",
    doi = "10.1007/JHEP03(2025)052",
    journal = "JHEP",
    volume = "03",
    pages = "052",
    year = "2025"
}

@article{Diatlyk:2023fwf,
    author = "Diatlyk, Oleksandr and Luo, Conghuan and Wang, Yifan and Weller, Quinten",
    title = "{Gauging non-invertible symmetries: topological interfaces and generalized orbifold groupoid in 2d QFT}",
    eprint = "2311.17044",
    archivePrefix = "arXiv",
    primaryClass = "hep-th",
    doi = "10.1007/JHEP03(2024)127",
    journal = "JHEP",
    volume = "03",
    pages = "127",
    year = "2024"
}

@article{Niro:2022ctq,
    author = "Niro, Pierluigi and Roumpedakis, Konstantinos and Sela, Orr",
    title = "{Exploring non-invertible symmetries in free theories}",
    eprint = "2209.11166",
    archivePrefix = "arXiv",
    primaryClass = "hep-th",
    doi = "10.1007/JHEP03(2023)005",
    journal = "JHEP",
    volume = "03",
    pages = "005",
    year = "2023"
}

@article{Kaidi:2025hyr,
    author = "Kaidi, Justin and Shi, Xiaoyi and Shimamori, Soichiro and Sun, Zhengdi",
    title = "{The SymTFT for $N$-ality defects: Part I}",
    eprint = "2509.19429",
    archivePrefix = "arXiv",
    primaryClass = "hep-th",
    month = "9",
    year = "2025"
}

@article{Maeda:2025rxc,
    author = "Maeda, Jun and Oishi, Tsubasa",
    title = "{N-ality symmetry and SPT phases in (1+1)d}",
    eprint = "2504.20151",
    archivePrefix = "arXiv",
    primaryClass = "hep-th",
    reportNumber = "YITP-25-35, KUNS-3038",
    doi = "10.1007/JHEP12(2025)063",
    journal = "JHEP",
    volume = "12",
    pages = "063",
    year = "2025"
}

@article{Lu:2024ytl,
    author = "Lu, Da-Chuan and Sun, Zhengdi and You, Yi-Zhuang",
    title = "{Realizing triality and $p$-ality by lattice twisted gauging in (1+1)d quantum spin systems}",
    eprint = "2405.14939",
    archivePrefix = "arXiv",
    primaryClass = "cond-mat.str-el",
    doi = "10.21468/SciPostPhys.17.5.136",
    journal = "SciPost Phys.",
    volume = "17",
    number = "5",
    pages = "136",
    year = "2024"
}

@article{Barter_2022,
   title={Computing associators of endomorphism fusion categories},
   volume={13},
   ISSN={2542-4653},
   url={http://dx.doi.org/10.21468/SciPostPhys.13.2.029},
   DOI={10.21468/scipostphys.13.2.029},
   number={2},
   journal={SciPost Physics},
   publisher={Stichting SciPost},
   author={Barter, Daniel and Bridgeman, Jacob and Wolf, Ramona},
   year={2022},
   month=Aug }

@article{Perez-Lona:2023djo,
    author = "Perez-Lona, A. and Robbins, D. and Sharpe, E. and Vandermeulen, T. and Yu, X.",
    title = "{Notes on gauging noninvertible symmetries. Part I. Multiplicity-free cases}",
    eprint = "2311.16230",
    archivePrefix = "arXiv",
    primaryClass = "hep-th",
    doi = "10.1007/JHEP02(2024)154",
    journal = "JHEP",
    volume = "02",
    pages = "154",
    year = "2024"
}

@article{Perez-Lona:2024sds,
    author = "Perez-Lona, Alonso and Robbins, Daniel and Sharpe, Eric and Vandermeulen, Thomas and Yu, Xingyang",
    title = "{Notes on gauging noninvertible symmetries. Part II. Higher multiplicity cases}",
    eprint = "2408.16811",
    archivePrefix = "arXiv",
    primaryClass = "hep-th",
    doi = "10.1007/JHEP05(2025)066",
    journal = "JHEP",
    volume = "05",
    pages = "066",
    year = "2025"
}

@misc{vercleyen2024lowrankmultiplicityfreefusioncategories,
      title={On Low-Rank Multiplicity-Free Fusion Categories}, 
      author={Gert Vercleyen},
      year={2024},
      eprint={2405.20075},
      archivePrefix={arXiv},
      primaryClass={math-ph},
      url={https://arxiv.org/abs/2405.20075}, 
}

@article{McNamara:2026isz,
    author = "McNamara, Jacob and Wang, Zhencheng",
    title = "{Wormholes as red herrings: reflection positivity and the reconstruction of unitary quantum field theories}",
    eprint = "2607.01322",
    archivePrefix = "arXiv",
    primaryClass = "hep-th",
    month = "7",
    year = "2026"
}

@article{Choi:2024tri,
    author = "Choi, Yichul and Rayhaun, Brandon C. and Zheng, Yunqin",
    title = "{Generalized Tube Algebras, Symmetry-Resolved Partition Functions, and Twisted Boundary States}",
    eprint = "2409.02159",
    archivePrefix = "arXiv",
    primaryClass = "hep-th",
    doi = "10.1007/s00220-025-05543-8",
    journal = "Commun. Math. Phys.",
    volume = "407",
    number = "4",
    pages = "62",
    year = "2026"
}

@article{Cordova:2024iti,
    author = "Cordova, Clay and Holfester, Nicholas and Ohmori, Kantaro",
    title = "{Representation theory of solitons}",
    eprint = "2408.11045",
    archivePrefix = "arXiv",
    primaryClass = "hep-th",
    doi = "10.1007/JHEP06(2025)001",
    journal = "JHEP",
    volume = "06",
    pages = "001",
    year = "2025"
}

@article{Huang:2021zvu,
    author = "Huang, Tzu-Chen and Lin, Ying-Hsuan and Seifnashri, Sahand",
    title = "{Construction of two-dimensional topological field theories with non-invertible symmetries}",
    eprint = "2110.02958",
    archivePrefix = "arXiv",
    primaryClass = "hep-th",
    reportNumber = "CALT-TH-2021-035, YITP-SB-2021-19",
    doi = "10.1007/JHEP12(2021)028",
    journal = "JHEP",
    volume = "12",
    pages = "028",
    year = "2021"
}

@article{Kitaev:2011dxc,
    author = "Kitaev, Alexei and Kong, Liang",
    title = "{Models for Gapped Boundaries and Domain Walls}",
    eprint = "1104.5047",
    archivePrefix = "arXiv",
    primaryClass = "cond-mat.str-el",
    doi = "10.1007/s00220-012-1500-5",
    journal = "Commun. Math. Phys.",
    volume = "313",
    number = "2",
    pages = "351--373",
    year = "2012"
}

@article{Antinucci:2025fjp,
    author = "Antinucci, Andrea and Copetti, Christian and Gai, Yuhan and Schafer-Nameki, Sakura",
    title = "{Categorical Anomaly Matching}",
    eprint = "2508.00982",
    archivePrefix = "arXiv",
    primaryClass = "hep-th",
    month = "8",
    year = "2025"
}

@article{Zhang:2023wlu,
    author = "Zhang, Carolyn and C{\'o}rdova, Clay",
    title = "{Anomalies of (1+1)-dimensional categorical symmetries}",
    eprint = "2304.01262",
    archivePrefix = "arXiv",
    primaryClass = "cond-mat.str-el",
    doi = "10.1103/PhysRevB.110.035155",
    journal = "Phys. Rev. B",
    volume = "110",
    number = "3",
    pages = "035155",
    year = "2024"
}

@article{Li:2025wes,
    author = "Li, Yabo and Dell'acqua, Matteo and Mitra, Aditi",
    title = "{Classification of Thouless pumps with non-invertible symmetries and implications for Floquet phases}",
    eprint = "2510.01626",
    archivePrefix = "arXiv",
    primaryClass = "cond-mat.str-el",
    month = "10",
    year = "2025"
}

@article{Komargodski:2020mxz,
    author = "Komargodski, Zohar and Ohmori, Kantaro and Roumpedakis, Konstantinos and Seifnashri, Sahand",
    title = "{Symmetries and strings of adjoint QCD$_{2}$}",
    eprint = "2008.07567",
    archivePrefix = "arXiv",
    primaryClass = "hep-th",
    reportNumber = "YITP-SB-20-28",
    doi = "10.1007/JHEP03(2021)103",
    journal = "JHEP",
    volume = "03",
    pages = "103",
    year = "2021"
}

@article{Warman:2024lir,
    author = "Warman, Alison and Yang, Fan and Tiwari, Apoorv and Pichler, Hannes and Schafer-Nameki, Sakura",
    title = "{Categorical Symmetries in Spin Models with Atom Arrays}",
    eprint = "2412.15024",
    archivePrefix = "arXiv",
    primaryClass = "cond-mat.str-el",
    doi = "10.1103/249m-m8wq",
    journal = "Phys. Rev. Lett.",
    volume = "135",
    number = "20",
    pages = "206503",
    year = "2025"
}

@article{Benini:2025lav,
    author = "Benini, Francesco and Calabrese, Pasquale and Fossati, Michele and Singh, Amartya Harsh and Venuti, Marco",
    title = "{Entanglement asymmetry for higher and noninvertible symmetries}",
    eprint = "2509.16311",
    archivePrefix = "arXiv",
    primaryClass = "hep-th",
    reportNumber = "SISSA 10/2025/FISI",
    month = "9",
    year = "2025"
}
\bibliographystyle{Bibliography/utphys}
\end{document}